\documentclass[apj,twocolumn,twocolappendix,numberedappendix,appendixfloats,iop]{openjournal}
\usepackage{amsmath}
\usepackage{booktabs}

\usepackage[dvipsnames]{xcolor} 
\usepackage[breaklinks,colorlinks,citecolor=blue,urlcolor=blue]{hyperref}

\usepackage{orcidlink}
\usepackage{subfigure}

\renewcommand{\arraystretch}{1.2}  
\usepackage{listings}
\usepackage{color}
\definecolor{dkgreen}{rgb}{0,0.6,0}
\definecolor{gray}{rgb}{0.5,0.5,0.5}
\definecolor{mauve}{rgb}{0.58,0,0.82}
\definecolor{golden}{rgb}{0.86,0.65,0.01}
\usepackage{soul}
\usepackage{amsmath}
\usepackage{xspace}
\usepackage{graphicx}
\usepackage{enumitem}
\usepackage{amssymb}
\usepackage{xifthen}
\usepackage{hyperref}
\usepackage[normalem]{ulem}
\usepackage{multirow}

\usepackage[flushleft]{threeparttable}

\newcommand{\unit}[1]{\ensuremath{\mathrm{\,#1}}\xspace}
\newcommand{\feh}{\unit{[Fe/H]}}

\newcommand{\SSSSS}{\ensuremath{\mathit{S}^5}\xspace}

\newcommand{\kms}{\unit{km\,s^{-1}}}
\newcommand{\masyr}{\unit{mas\,yr^{-1}}}
\newcommand{\msun}{\unit{M_\odot}}

\newcommand{\kpc}{\unit{kpc}}
\newcommand{\gyr}{\unit{Gyr}}
\newcommand{\myr}{\unit{Myr}}
\newcommand{\pc}{\unit{pc}}
\newcommand{\dex}{\unit{dex}}
\newcommand{\rads}{\unit{km\,s^{-1}kpc^{-1}}}

\newcommand{\stream}{\ensuremath{300}S\xspace}
\newcommand{\sgr}{Sgr\xspace}
\newcommand{\streamPoss}{\ensuremath{300}S's\xspace}
\newcommand{\sgrPoss}{Sgr's\xspace}
\newcommand{\rrl}{RRL-1\xspace}
\newcommand{\rrlPoss}{RRL-1's\xspace}
\newcommand{\badrrl}{RRL-2\xspace}
\newcommand{\badrrlPoss}{RRL-2's\xspace}

\usepackage{lineno}
\usepackage{float}

\begin{document}


\author{Benjamin~Cohen\,\orcidlink{0009-0001-8988-4556}$^{1,*}$}
\author{Alexander~P.~Ji\,\orcidlink{0000-0002-4863-8842}$^{2,3}$}
\author{Peter~S.~Ferguson\,\orcidlink{0000-0001-6957-1627}$^{4}$}
\author{Sergey~E.~Koposov\orcidlink{0000-0003-2644-135X}$^{5,6}$}
\author{Alex~Drlica-Wagner\orcidlink{0000-0001-8251-933X}$^{2,3,7}$}
\author{Andrew~P.~Li\orcidlink{0009-0005-5355-5899}$^{8}$}
\author{Ting~S.~Li\orcidlink{0000-0002-9110-6163}$^{8,9,10}$}

\author{Lara~R.~Cullinane\orcidlink{0000-0001-8536-0547}$^{11}$}
\author{Gary~S.~Da~Costa\orcidlink{0000-0001-7019-649X}$^{12,13}$}
\author{Denis~Erkal\orcidlink{0000-0002-8448-5505}$^{14}$}
\author{Kyler~Kuehn\,\orcidlink{0000-0003-0120-0808}$^{15}$}
\author{Geraint~F.~Lewis\orcidlink{0000-0003-3081-9319}$^{16}$}
\author{Sarah~L.~Martell\orcidlink{0000-0002-3430-4163}$^{13,17}$}
\author{Andrew~B.~Pace\orcidlink{0000-0002-6021-8760}$^{18}$}
\author{Daniel~B.~Zucker\,\orcidlink{0000-0003-1124-8477}$^{19,20}$}

\author{Petra~Awad\orcidlink{0000-0002-0428-849X}$^{21}$}
\author{Joss~Bland-Hawthorn\orcidlink{0000-0001-7516-4016}$^{13,16}$}
\author{Guilherme~Limberg\orcidlink{0000-0002-9269-8287}$^{3}$}
\author{Clara~E.~Mart{\'i}nez-V{\'a}zquez\orcidlink{0000-0002-9144-7726}$^{22}$}
\author{Joshua~D.~Simon\orcidlink{0000-0002-4733-4994}$^{23}$}
\author{Yong~Yang\orcidlink{0000-0001-7609-1947}$^{16}$}

\affiliation{$^1$Department of Physics, University of Chicago, 5640 S Ellis Avenue, Chicago, IL 60637, USA}
\affiliation{$^2$Department of Astronomy \& Astrophysics, University of Chicago, 5640 S Ellis Avenue, Chicago, IL 60637, USA}
\affiliation{$^3$Kavli Institute for Cosmological Physics, University of Chicago, Chicago, IL 60637, USA}
\affiliation{$^4$ DIRAC Institute, Department of Astronomy, University of Washington, 3910 15th Ave NE, Seattle, WA, 98195, USA}

\affiliation{$^5$Institute for Astronomy, University of Edinburgh, Royal Observatory, Blackford Hill, Edinburgh EH9 3HJ, UK}
\affiliation{$^6$
Institute of Astronomy, University of Cambridge, Madingley Road,
Cambridge CB3 0HA, UK}
\affiliation{$^7$Fermi National Accelerator Laboratory, P.O.\ Box 500, Batavia, IL 60510, USA}
\affiliation{$^{8}$Department of Astronomy and Astrophysics, University of Toronto, 50 St. George Street, Toronto ON, M5S 3H4, Canada}
\affiliation{$^{9}$Dunlap Institute for Astronomy \& Astrophysics, University of Toronto, 50 St George Street, Toronto, ON M5S 3H4, Canada}
\affiliation{$^{10}$Data Sciences Institute, University of Toronto, 17th Floor, Ontario Power Building, 700 University Ave, Toronto, ON M5G 1Z5, Canada}

\affiliation{$^{11}$Leibniz-Institut f{\"u}r Astrophysik Potsdam (AIP), An der Sternwarte 16, D-14482 Potsdam, Germany}
\affiliation{$^{12}$Research School of Astronomy and Astrophysics, Australian National University, Canberra, ACT 2611, Australia}
\affiliation{$^{13}$Centre of Excellence for All-Sky Astrophysics in Three Dimensions (ASTRO 3D), Australia}
\affiliation{$^{14}$Department of Physics, University of Surrey, Guildford GU2 7XH, UK}
\affiliation{$^{15}$Lowell Observatory, 1400 W Mars Hill Rd, Flagstaff,  AZ 86001, USA}
\affiliation{$^{16}$Sydney Institute for Astronomy, School of Physics, A28, The University of Sydney, NSW 2006, Australia}
\affiliation{$^{17}$School of Physics, University of New South Wales, Sydney, NSW 2052, Australia}
\affiliation{$^{18}$Department of Astronomy, University of Virginia, 530 McCormick Road, Charlottesville, VA 22904, USA}
\affiliation{$^{19}$School of Mathematical and Physical Sciences, Macquarie University, Sydney, NSW 2109, Australia}
\affiliation{$^{20}$Macquarie University Research Centre for Astrophysics and Space Technologies, Sydney, NSW 2109, Australia}

\affiliation{$^{21}$Leiden Observatory, Leiden University, P.O.Box 9513, 2300RA Leiden, The Netherlands}
\affiliation{$^{22}$NSF NOIRLab, 670 N. A’ohoku Place, Hilo, HI 96720, USA}
\affiliation{$^{23}$Observatories of the Carnegie Institution for Science, 813 Santa Barbara St., Pasadena, CA 91101, USA}

\collaboration{\SSSSS Collaboration}

\email[$^*$Corresponding Author: ]{benmckcohen@uchicago.edu}
\title{Sifting for a Stream: The Morphology of the \stream Stellar Stream}

\begin{abstract}
Stellar streams are sensitive laboratories for understanding the small-scale structure in our Galaxy's gravitational field. 
Here, we analyze the morphology of the \stream stellar stream, which has an eccentric, retrograde orbit and thus could be an especially powerful probe of both baryonic and dark substructures within the Milky Way. 
Due to extensive background contamination from the Sagittarius stream (\sgr), we perform an analysis combining Dark Energy Camera Legacy Survey photometry, \textit{Gaia} DR3 proper motions, and spectroscopy from the Southern Stellar Stream Spectroscopic Survey (\SSSSS).
We redetermine the stream coordinate system and distance gradient, then 
apply two approaches to describe \streamPoss morphology.
In the first, we analyze stars from \textit{Gaia} using proper motions to remove \sgr. 
In the second, we generate a simultaneous model of \stream and \sgr based purely on photometric information. 
Both approaches agree within their respective domains and describe the stream over a region spanning $33^\circ$.
Overall, \stream has three well-defined density peaks and smooth variations in stream width. 
Furthermore, \stream has a possible gap of $\sim 4.7^\circ$ and a kink.
Dynamical modeling of the kink implies that \stream was dramatically influenced by the Large Magellanic Cloud. 
This is the first model of \streamPoss morphology across its entire known footprint, opening the door for deeper analysis to constrain the structures of the Milky Way.

\keywords{Stellar streams (2166), Milky Way Galaxy (1054), Local Group (929), Magellanic Clouds (990), Globular star clusters (656)}
\end{abstract}

\maketitle

\newpage

\section{Introduction}
\label{intro}

Stellar streams form through the tidal disruption of progenitor systems, such as globular clusters or dwarf galaxys. 
As the progenitor approaches a more massive host such as the Milky Way, tidal forces strip stars off of the progenitor near its Lagrange points. 
Stars released from the inner Lagrange point lose energy and move ahead of the progenitor to form the leading arm of the stream, while 
stars released from the outer Lagrange point gain energy and fall behind to form the trailing arm \citep[e.g.,][]{Johnston_98,Kupper_2008,Kupper_2010}. 
In a smooth potential without any external perturbations, stellar streams become long, relatively smooth structures that approximately follow the orbital track of their progenitor \citep{CarlinNewberg_2016}. 
Because of this regularity, stellar streams are especially sensitive to both the Milky Way mass profile and perturbations within that profile, making them of great interest in the study of Galactic dynamics.
They have been used to analyze the geometry of the Galactic potential \citep[e.g.,][]{Bonaca_2014,Kupper_2015,Bovy_2016,Vasiliev_2021,Ibata_2023}, constrain the mass of the Large Magellanic Cloud \citep[LMC, e.g.,][]{Erkal_2019,Shipp_2021,Koposov2023}, investigate the presence of dark matter subhalos without baryons \citep[e.g.,][]{Erkal_2016, Bonaca_2019,Banik_2021}, and as a present day example of hierarchical structure formation \citep[e.g.,][]{Bonaca_2021,Malhan_2022}. 
For the investigation of dark subhalos and other small perturbations, streams on retrograde orbits are especially useful due to the smaller influence of the Galaxy's spiral arms and bar \citep[e.g.,][]{Hattori_2016,Pearson_2017,Erkal_2017,Banik_2021,Yang_2025}.

In the past 20 years, nearly $120$ streams have been found, increasing the known number by two orders of magnitude \citep{Bonaca_2024}. 
Stream morphologies have been carefully analyzed for a few of these streams, including Pal 5 \citep{Erkal_2017}, GD-1 \citep{Price_Whelan_2018}, Orphan/Chenab \citep{Koposov_2019_orphan}, Jet \citep{Ferguson_2021}, ATLAS/Aliqa Uma \citep{Li_2021}, and Phoenix \citep{Tavangar_2022}, among others. 
Furthermore, \cite{Patrick_2022} characterized the morphology of a population of streams using data from the Dark Energy Survey (DES, \citealp{Abbott_2018}), the Dark Energy Camera Legacy Survey (DECaLS, \citealp{Dey_2019}), and Pan-STARRS \citep{Chambers_2016} which allowed them to investigate trends in stream morphology across the population.
However, many streams remain poorly characterized, as their low surface brightnesses and large distances make it difficult to identify stream members.

One such stream is \stream. \stream is a globular cluster stream on an extremely eccentric, retrograde orbit \citep{Li_2022}.
Previous matched filter analyses identified \streamPoss footprint as spanning at least $25^\circ$ across the sky with endpoints at distances of $\sim 14\kpc$ and $\sim 19\kpc$ \citep{Grillmair_2013,Bernard_2016}. 
\cite{Usman_2024} found the stellar population of \stream to match a relatively metal rich ($\feh = -1.35$) and old ($12.5\gyr$) isochrone. 
We present a more extensive review of literature measurements of \stream in Section \ref{ssec:300S History}.

Because of its eccentric, retrograde orbit, \stream could be used to set a strong constraint on the structure of the Milky Way's gravitational potential and any associated perturbations.
Indeed, because of \streamPoss structure and dynamics, \cite{Lu_2025} found it to be one of the most promising probes of small dark matter subhalos among a sample of $50$ streams. Moreover, \stream was the only member of the top three such streams with a retrograde orbit.
However, other than the qualitative descriptions from the matched filter maps of \citet{Grillmair_2013} and \citet{Bernard_2016}, little is known about the structural morphology of \stream. 
The challenge of describing \stream is exacerbated by its distance and the extensive background structure in the region from the Sagittarius stream (\sgr), which overlaps \stream in both on-sky position and distance \citep{Simon_2011}. 
This challenge has prevented further analysis of \streamPoss relationship to the Galactic potential. 

In this paper, we extend and numerically characterize \streamPoss structural morphology across $33^\circ$, extending the stream by $\sim 7^\circ$ and presenting a new look at its structure despite the background contamination from \sgr. 
To do this, we utilize two distinct methods with two distinct datasets -- \textit{Gaia} Data Release (DR) 3 \citep{Gaia_Mission,Gaia_DR3} and the Dark Energy Camera Legacy Survey (DECaLS) DR9 \citep{Dey_2019} -- to remove the background contamination and extract the stream signal. 
We find that \stream has a complex structure, including variations in its stellar density, a $\sim 4.7^\circ$ gap, and a kink/bend in the stream track.
To verify the position of the kink and investigate the influence of the LMC on \stream, we additionally perform preliminary dynamical modeling. 
This modeling reproduces the bend and indicates that the LMC had a dramatic influence on the formation of \stream.

We organize this paper as follows. 
In Section \ref{sec:data sources and preparation}, we describe the DECaLS, \textit{Gaia}, and Southern Stellar Stream Spectroscopic Survey (\SSSSS) datasets. 
In Section \ref{sec:empirical characterization}, we discuss both (1) our initial characterization of the stream, including an initial matched filter search and identification of a stream-centric coordinate system (Section \ref{ssec:naive matched filter}), and (2) our search for RR Lyrae (RRL) and Blue Horizontal Branch (BHB) stars, including our subsequent recalculation of \streamPoss distance gradient (Section \ref{ssec:distance gradient}).
We then describe our dual methodologies for accounting for the \sgr background contamination (Sections \ref{ssec:method 1 proper motion filtering} and \ref{ssec:method 2 double stream model}) and discuss the resulting models (Section \ref{ssec:discussion}). 
Motivated by our models, we perform preliminary dynamical modeling of \stream (Section \ref{sec:dynamical modeling}). 
We conclude in Section \ref{sec:conclusions}.

\subsection{History of \stream Characterization}\label{ssec:300S History}
Information on \stream has been slowly accumulated since its discovery in $\sim 2007$. In this section, we provide a brief history of the analysis of \stream as a compact reference.

\stream was first detected in multiple analyses of the Segue 1 satellite.
In \cite{Geha_2009} and \cite{Norris_2010}, \stream appeared as a small overdensity of stars with radial velocity of $\sim 300 \kms$, hence the name. 
In \cite{Belokurov_2007} and \cite{Niederste_Ostholt_2009}, the stream manifested as an extended spatial overdensity around Segue 1.
One of the first detailed analyses of \stream was presented in \cite{Simon_2011}, where the authors definitively determined its existence through a spectroscopic analysis of over 20 stars and subsequent confirmation of \streamPoss distinct radial velocity. 
By comparing their radial velocity members' Sloan Digital Sky Survey (SDSS) photometry to Galactic globular clusters, they also obtained initial estimates of $d \sim 22\kpc$ and $\feh \sim -1.3$, though with considerable uncertainties. 
Because of \streamPoss extreme radial velocity which differs from that of Segue 1 by $\sim 100\kms$, they disassociated the two objects.

Further spectroscopic analysis of a bright member of \stream was conducted by \cite{Frebel_2013}. 
Their spectroscopic analysis measured $\feh = -1.46\pm 0.05 \pm 0.23$ (random and systematic uncertainties respectively) and an isochrone fit gave $d = 18\pm 7 \kpc$, in agreement with \cite{Simon_2011}. 
They concluded from low aluminum and magnesium abundances that \stream is unlikely to have formed from a globular cluster. 
Finally, they disassociated \stream from \sgr and the Orphan stream, which are both present in the region, through their inconsistencies with \streamPoss radial velocity and metallicity. 

To further clarify the mounting characterization of this new structure, \cite{Grillmair_2013} performed a matched filter search over the region on SDSS data and identified \stream as extending over $25^\circ$ with distances ranging from $14\pm 3 \kpc$ to $18\pm 2 \kpc$. 
\cite{Bernard_2016} also identified the stream in a matched filter search of the Pan-STARRS1 $3\pi$ Survey, identifying $24^\circ$ of the stream between $\sim 14\kpc$ to $\sim 19 \kpc$. 
They further note that \streamPoss distance gradient is opposite to that of \sgr, providing further evidence for their disassociation. 

\cite{Carlin_2012} noted that \stream, termed Segue 1b in their work, had kinematics consistent with having the same origin as the Virgo Stellar Overdensity (VOD) in a recent, massive dwarf galaxy merger. This event has since been matched with the \textit{Gaia}-Enceladus-Sausage merger (GES, \citealp{Belokurov_2018,Helmi_2018,Haywood_2018}) by \cite{Simion_2019} and \cite{Perottoni_2022}. 

Building on these studies, \cite{Fu_2018} searched for \stream members in APOGEE (Apache Point Observatory Galactic Evolution Experiment, \citealp{Majewski_2017}) and SEGUE (Sloan Extension for Galactic Understanding and Exploration, \citealp{Yanny_2009}) data using kinematic, distance, and CMD filters.
They used a distance gradient based on the endpoints of $14\kpc$ and $19\kpc$ from \cite{Bernard_2016}. 
The on-sky distribution of their members confirmed \streamPoss association to the elongated overdensity found by \cite{Belokurov_2007} and \cite{Niederste_Ostholt_2009}. Their members had $\feh = -1.48$ with a dispersion of $0.21^{+0.12}_{-0.09}$. 
\cite{Fu_2018} also fitted an orbit to the stream and found it to be highly eccentric ($e=0.87$) with an peri/apocenter of $4.1/\sim60\kpc$. 
This orbit is distinct from that of \sgr and similar to VOD's, confirming \cite{Carlin_2012}'s result.
Finally, \cite{Fu_2018} argue that \streamPoss progenitor was a dwarf galaxy due to the apparent metallicity spread and lack of correlated light element abundances. 
They also find this conclusion consistent with their measured full width at half maximum of $0.94^\circ$; with \streamPoss velocity dispersion of $\sim 4-5 \kms$; and with its orbit eccentricity, all of which are higher than in standard globular cluster streams, such as Pal 5. 

Using the \textsc{streamfinder} algorithm \citep{Malhan_2018}, \cite{Ibata_2021} recovered \stream\ -- labeled by them as Gaia-10 -- in the combined \textit{Gaia} DR2 and EDR3 data set. 
Using narrow-band imaging from the \textit{Pristine} survey, \cite{Martin_2022} identified a mean metallicity of $\feh = -1.4\pm 0.06$. 
Using these \textit{Gaia} results, \cite{Malhan_2022} attempted to group objects based on possible merger associations. 
They found a potential merger that includes \stream with NGC 5466 and its associated stream, NGC 7492, and Tucana III, though the small number of objects in the association was not that significant ($<2\sigma$ significance).
They further noted that \cite{Massari_2019} and \cite{Forbes_2020} associated NGC 7492 with GES and NGC 5466 with Sequoia.

As part of \SSSSS, \cite{Li_2022} identified spectroscopic members of \stream to place the metallicity at $\feh = -1.26\pm 0.03$ with a dispersion of $0.04^{+0.04}_{-0.02}$ with a $95\%$ confidence upper limit of $0.11$. 
Due to the small metallicity dispersion, they concluded that \stream has a globular cluster progenitor. 
Additionally, they determined the orbit to be relatively unique, with a high eccentricity ($e=0.77$) and a peri/apocenter of $5.8/45.8\kpc$. 
This eccentricity was the highest of the $12$ stream sample these authors analyzed. 
Additionally, \stream was one of only two streams in their sample (the other being Jet) with a retrograde orbit and one of two (the other being AAU) at a non-extremity in its orbit. 
Because of \streamPoss orbit, they associate the stream with the GES merger and with the objects NGC 5466 and Tucana III, in agreement with the associations of \cite{Malhan_2022}.

Finally, \cite{Usman_2024} used \stream as a laboratory to study multiple stellar populations (MSPs). 
Using high resolution spectroscopy, they find no evidence of a metallicity dispersion and note that the most metal poor star found by \cite{Fu_2018} has an inconsistent radial velocity and is probably a non-member of \stream. The lower resulting velocity dispersion and the identification of one star in 300S with an abundance pattern matching the ``second population'' in globular clusters \citep[e.g.,][]{Bastian_2018} confirm the conclusion of \cite{Li_2022} that \streamPoss progenitor was a globular cluster. 
\cite{Usman_2024} argue that the low aluminum and magnesium abundances reported in \cite{Frebel_2013} can be explained from the \textit{ex situ} formation of \streamPoss progenitor. 
In fact, they find the magnesium abundances to be similar to those of GES as described by \cite{Limberg_2022}, further supporting these objects' association.
They determine the metallicity of the stream to be $\feh = -1.35$.
Using the observed luminosity of the stream and the time necessary for it to totally disrupt, they determine an initial mass range of $10^{4.5-4.8} \msun$ for the progenitor.
Interestingly, they note that \streamPoss initial mass may be close to a threshold to produce MSPs, although the value of that cutoff is still not known. 

To summarize, \stream is a relatively metal rich, globular cluster stream on an extremely eccentric, retrograde orbit. Spatially, it traces at least $25^\circ$ across the sky at distances between $14\kpc$ and $19\kpc$. 
It is likely independent of Segue 1, \sgr, and Orphan, and evidence suggests that it is associated with the GES merger. 
Its chemistry indicates that it formed \textit{ex situ}.

\section{Data}\label{sec:data sources and preparation}

\subsection{DECaLS DR9 Data}\label{ssec:decals}

We use photometric information from the Dark Energy Camera \citep[DECam;][]{Flaugher_2015} Legacy Survey (DECaLS) DR9 \citep{Dey_2019}. 
We select a rectangular region $(-20^\circ,20^\circ)\times(-15^\circ,15^\circ)$ centered roughly on \stream in an earlier rendition of the stream coordinate system.
We remove galaxies by requiring that the \textsc{type} flag is \textsc{psf}.
We additionally apply certain quality cuts. 
First, we require that \textsc{fracflux} for the $g$ and $r$ bands are both $<0.05$. 
This ensures that the measured flux is dominated by the source.
We also require that the \textit{g} and \textit{r} band \textsc{anymask} flags are $0$ and \textsc{flux} values are $>0$.  

We deredden the data using the corresponding transmission coefficients in \textsc{mw\_transmission} which represent the transmission of each band in linear units (see the discussion by \citealp{Ruiz-Macias_2021} for more details).
To avoid any contamination from far-field objects that could obfuscate the detected stream features, we perform a magnitude cut of $g<22.6$. 
As we are primarily focused on the main sequence and red giant branch, we follow \cite{Ferguson_2021} and perform a color cut of $0\leq g-r \leq 1$ except when searching for BHB candidate members (Section \ref{sssec:standard candles}). 

Finally, we apply masks to the data to prevent contamination by known field objects. 
We mask Segue 1 at $(\text{R.A.},\text{Decl.})$ of $(151.763^\circ,16.073^\circ)$ as well as Leo I and Leo II at $(152^\circ,12.5^\circ)$ and $(168.5^\circ,22.2^\circ)$ respectively. 
We mask these objects with circles of angular radii $0.3^\circ,$ $0.5^\circ,$ $1.0^\circ$ respectively. 
Of these overdensities, only Segue 1 lies anywhere near the stream track.
Its half-light radius is $29 \pc$, which corresponds to an approximate half-light angular radius of $\sim 0.07^\circ$ given its distance of $23\kpc$ \citep{Martin_2008}.
Therefore, with a $0.3^\circ$ mask radius, Segue 1 should not leak into our results. Further, as the stream full width at half maximum is $\sim 0.94^\circ$ \citep[][also see Section \ref{ssec:discussion}]{Fu_2018}, the mask itself also should not influence our results. These cuts leave 2,046,607 sources over an $1,186\,\deg^2$ region

We use this catalog to derive the initial matched filtered stellar density and stream coordinate system (Section \ref{ssec:naive matched filter}), the refined matched filtered stellar density (Section \ref{sssec:improved matched filter}), and one of our two empirical models of stream morphology (Section \ref{ssec:method 2 double stream model}). We also use it to analyze the distance gradient of \stream (Section \ref{ssec:nonlinear distance gradient}) and to cross-match against the other catalogs.

\subsection{\textit{Gaia} DR3 Data}

We use astrometric information from \textit{Gaia} DR3 \citep{Gaia_Mission,Gaia_DR3}.
We follow \cite{Ferguson_2021} and \cite{Pace_2022} by removing nearby sources with a parallax cut of $\overline{w} - 3\sigma_{\overline{w}} < 0.05$.
We additionally apply certain quality cuts. First, we ensure that the renormalized unit weight error (\textsc{ruwe}) is $<1.4$ \citep[as suggested by][]{Gaia_docs_ch7} and the goodness of fit statistic, \textsc{astrometric\_gof\_al}, is $<3$ \citep[as suggested in][]{Gaia_docs_ch20}.
We also require \textsc{astrometric\_excess\_noise\_sig} to be $<2$ (as in \citealp{Pace_2022}).
Further, we keep only sources with a corrected flux excess \citep[$C^*$, see Equation 6 and discussion in][]{Riello_2021} satisfying $<3\sigma_{C^*}(G)$ to exclude abnormal photometry. 
Finally, to prevent cross-contamination with Active Galactic Nuclei (AGN), we check that there is no overlap between our data and the \textit{Gaia} \textsc{Gaiaedr3.agn\_cross\_id} table \citep[as in][]{Ferguson_2021}. 
We then cross-match the \textit{Gaia} catalog with the DECaLS DR9 catalog using a matching radius of $0.5''$. We also apply the same object masks as we applied to the DECaLS DR9 data.
These cuts leave 635,606 sources over a $1,401\,\deg^2$ region.

We use this catalog as the basis for one of our two empirical models of stream morphology (Section \ref{ssec:method 1 proper motion filtering}).

\subsection{\SSSSS Catalog}\label{ssec:SSSSS Catalog}

We use stellar parameters and radial velocities from the \SSSSS internal data release (iDR3.7). 

\SSSSS spectra are acquired at the $3.9$m Anglo-Australian Telescope at Siding Spring Observatory in Australia. Using the AAOmega double spectrograph \citep{Smith_2004} and the 2dF fibre positioner \citep{Lewis_2002}, low-resolution blue spectra ($R\sim1300$, $3800\text{\AA}<\lambda<5800\text{\AA}$) and high-resolution red spectra ($R\sim10,000$, $8420\text{\AA}<\lambda<8840\text{\AA}$) are obtained simultaneously for up to $367$ science targets, along with $25$ sky fibers. These data are reduced with the 2dFdr software package \citep{AAO_2015}.

These data were processed using an improved pipeline from previous \SSSSS releases \citep[such as iDR1.5,][]{Li_2019}. 
Briefly, this pipeline fits a simultaneous model on both the red and blue arms of the AAOmega spectra as well as additional observations from different nights using \textsc{rvspecfit} \citep{Koposov_2019_rvspecfit}.
Details on the spectral fitting for these data are discussed in \cite{Ji_2021} and \cite{Li_2022}. 

\stream members were extracted from these data using a mixture model. For more details on the membership selection, see A.~P.~Li et al. (in prep.).
As we use the \SSSSS members as a pure ground-truth, we select stars with membership probability $>99\%$. 
Like with \textit{Gaia} DR3, we cross-match the \SSSSS catalog with the DECaLS DR9 catalog with a matching radius of $0.5''$ to create consistency between our datasets.
This cut removes $7$ stars, which all fail the cuts on the DECaLS DR9 dataset described in Section \ref{ssec:decals}, and leaves a pure, matched selection of $66$ members. 

We use this catalog to fit polynomials for the \stream proper motion filters (Section \ref{sssec:standard candles}), to rederive the \stream distance gradient (Section \ref{sssec:distance gradient calculation}), and to prepare and evaluate our orbital models (Section \ref{sec:dynamical modeling}).

\section{Preliminary Empirical Characterization}\label{sec:empirical characterization}

Here, we describe our preliminary analyses of \stream which are prerequisites to fitting models of stream morphology in Sections \ref{ssec:method 1 proper motion filtering} and \ref{ssec:method 2 double stream model}. To summarize, we begin with a na\"ive matched filter using the distance gradient of \cite{Fu_2018}. With this map, we are able to identify a stream coordinate system. We use this coordinate system to search for standard candles in the stream. We 
 find only 1 RRL star and no BHB stars. We then derive a new distance gradient using the selected \SSSSS members and the RRL star. Finally, we redo the matched filter using our improved distance gradient and use it to motivate our methods of stream fitting.

\subsection{Na\"ive Matched Filter and Coordinate System}\label{ssec:naive matched filter}

The matched filter method is an important tool for extracting stream signal from a stellar density map \citep[e.g.,][]{Odenkirchen_2001,Rockosi_2003}. 
The method consists of matching stars in color-magnitude space to an isochrone chosen to reflect the object of interest. 
This way, overdensities whose color-magnitude distribution match that of the object pass through the filter, while overdensities whose distribution do not match are suppressed, effectively increasing the signal to noise ratio of the object in the map. This method has been used extensively in the mapping of stellar streams \citep[e.g.,][]{Bonaca_2012,Grillmair_2017, Shipp_2018,Ferguson_2021}.

In this work, we adopt a $\feh = -1.35$, $12.5 \gyr$ isochrone, as was used for \stream by \cite{Usman_2024}.
We use MIST's DECam synthetic photometry \citep{Dotter_2016} to generate the isochrone.
Following \cite{Ferguson_2021}, we then consider stars to match the isochrone when they fall within a color range around it. 
As in \cite{Shipp_2018} and \cite{Ferguson_2021}, we define an upper and lower color padding as

\begin{equation}\label{eq:lower bound}
    L(g_{\text{iso}}) = E \cdot \left (0.001 + e^{(g_{\text{iso}} + \mu_0 + \Delta \mu/2-27.09)/1.09}\right ) + C_1
\end{equation}

\begin{equation} \label{eq:upper bound}
    U(g_{\text{iso}}) = E \cdot \left (0.001 + e^{(g_{\text{iso}} + \mu_0 -\Delta \mu/2 -27.09)/1.09)} \right ) + C_2
\end{equation}

In order to get a purer, though less complete, sample in the presence of the strong contamination in the region, we add a scaling factor $s$ and select stars with color

\begin{equation}\label{eq:matched filter}
    (g-r)_{\text{iso}} - s\times L(g_{\text{iso}}) < (g-r) < (g-r)_{\text{iso}} + s\times U(g_{\text{iso}})
\end{equation}

\begin{figure*}[!ht]
   \centering
   \begin{minipage}{1\linewidth}
       \centering
       \subfigure[\label{sfig:no matched filter}]{\includegraphics[width=0.46\textwidth
    ]{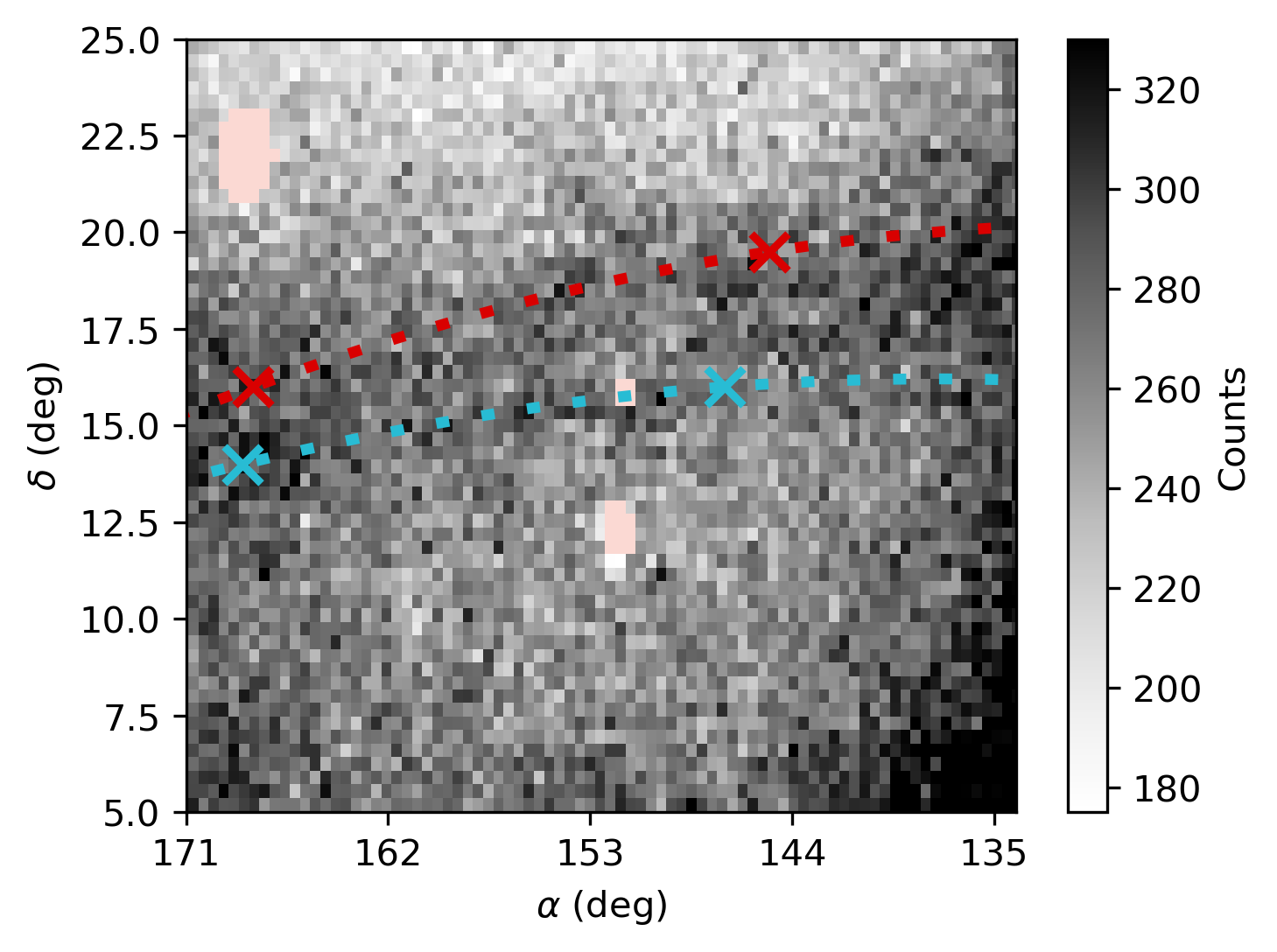}}
       \subfigure[\label{sfig:initial matched filter}]{\includegraphics[width=0.45\textwidth]{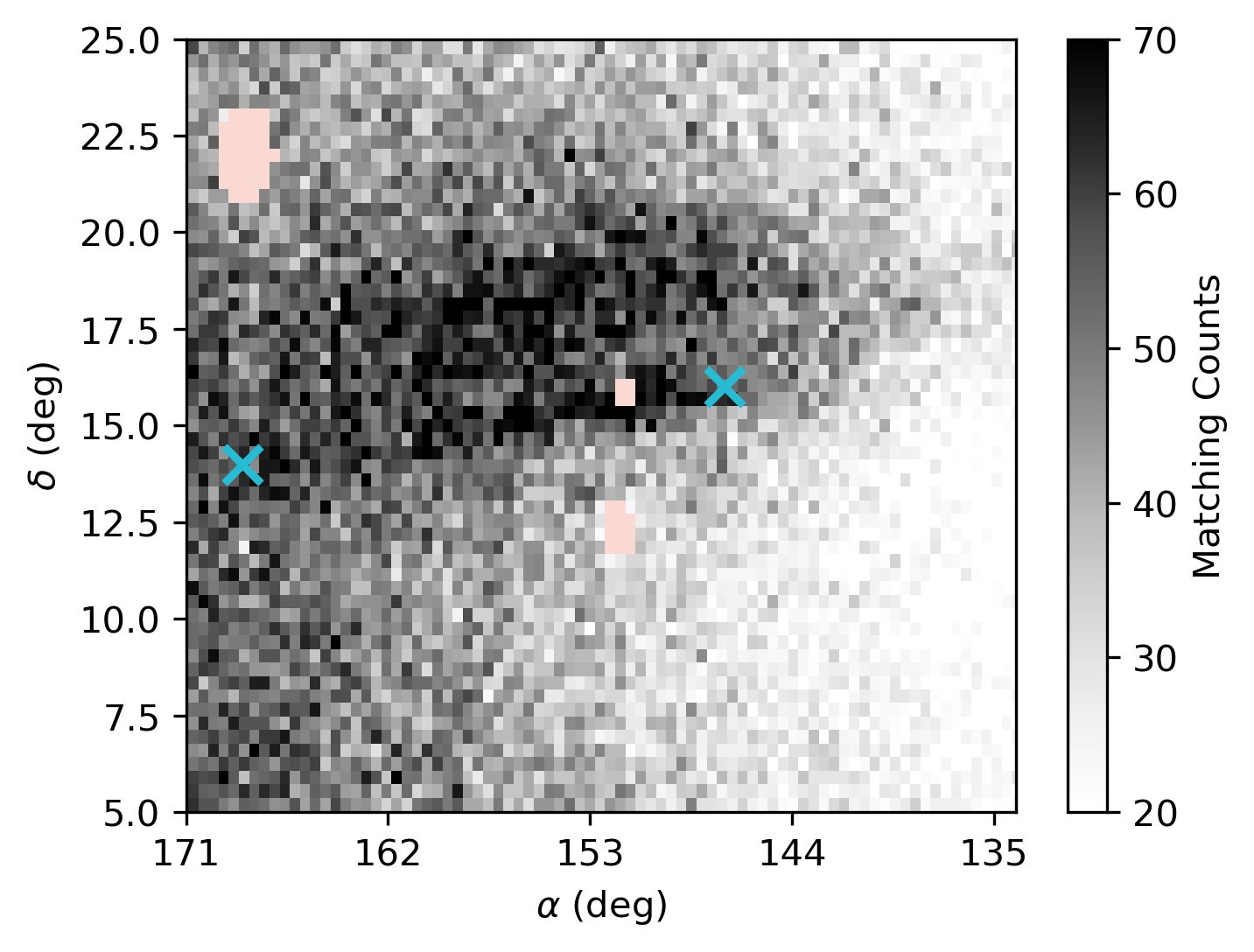}}
    \end{minipage}
    \begin{minipage}{1\linewidth}
    \centering
    \includegraphics[width=0.93\linewidth]{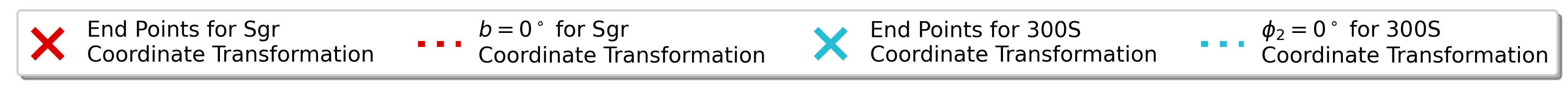}
    \end{minipage}
   \caption{Initial stellar density maps. (a) Stellar density map without matched filter application. Stars were selected using only cuts described in Section \ref{ssec:decals} including the basic color $(0<g-r < 1)$ and magnitude $(g < 22.6)$ cuts. 
   The shaded red regions are the object masks. 
   Note the presence of \sgr as the wide stripe across the unfiltered map.
   (b) Stellar density map under the na\"ive matched filter. 
   There is significant contamination from \sgr into the filtered map, as seen in the large dark region above \stream. This implies that an isochrone filter is insufficient to distinguish the two signals.}
   \label{fig:Naive Matched Filter}
\end{figure*}

We set $\Delta \mu, E,C_1,C_2$ following \cite{Shipp_2018} as $0.5,2,0.05,0.1$ respectively.
We set $\mu_0 = 16.2$ ($d = 17.4\kpc$) as the distance modulus of the stream.
This approximately corresponds to the \cite{Fu_2018} distance gradient evaluated at the position of Segue 1.
We set $s = 2/3$, although our general results are not dependent on the specific value of $s$.\footnote{We tested $s\in \left\{1/2,4/7,2/3,4/5,1\right\}$ and found that the resulting fits were not sensitive to $s$ given the uncertainties.}
We show the resulting density map of matching stars in Figure \ref{sfig:initial matched filter}. 
For comparison, we also show the unfiltered map in Figure \ref{sfig:no matched filter} where the only cuts are the quality and basic color-magnitude cuts described in Section \ref{ssec:decals}.

The presence of \sgr is clear in the unfiltered map as a wide band that extends across the field.
Without the matched filter, this signal visually dominates over \stream.
Although \stream becomes clearer in the matched filter map, there remains considerable contamination from \sgr, visible as a large, dark region above \stream.
This is unsurprising given the chemical structure of \sgr and its age \citep{Limberg_2023}.
The strong presence of \sgr within the filtered map indicates that an isochrone-based filter in color-magnitude space is insufficient to isolate \streamPoss signal. 

Due to the lack of a clear progenitor, we follow \cite{Ferguson_2021} and define a stream coordinate system based on the visual extent of the stream in the matched filter map. 
We define endpoints at $(\text{R.A.},\text{Decl.})$ of $(147.0^\circ,16.0^\circ)$ and $(168.5^\circ,14.0^\circ)$. 
We then rotate the sphere such that these endpoints fall along the $\phi_1$-axis' great circle equidistant from $0$. 
This rotation in Cartesian coordinates is the matrix

\begin{equation*}
    R = \begin{bmatrix}
     -0.89325016 &  0.36451269 &  0.26312478 \\
     -0.39958263 & -0.91195263 & -0.09314568 \\
      0.20600456 & -0.18834248 &  0.96025477 \\
    \end{bmatrix}
\end{equation*}

\noindent and the $\phi_2 = 0^\circ$ line is visible in Figure \ref{fig:Naive Matched Filter} as a dotted blue line. Under this transformation, $\phi_1$ increases with R.A. 

\subsection{\stream Distance Gradient}\label{ssec:distance gradient}

In this section, we rederive \streamPoss distance gradient with a method robust against \sgr contamination. 

\subsubsection{Proper Motion Filter}\label{sssec:proper motion filter}

\begin{figure*}[!ht]
   \centering
   \begin{minipage}{1\linewidth}
       \centering
       \subfigure[\label{sfig:pmra filter}]{\includegraphics[width=0.44\linewidth]{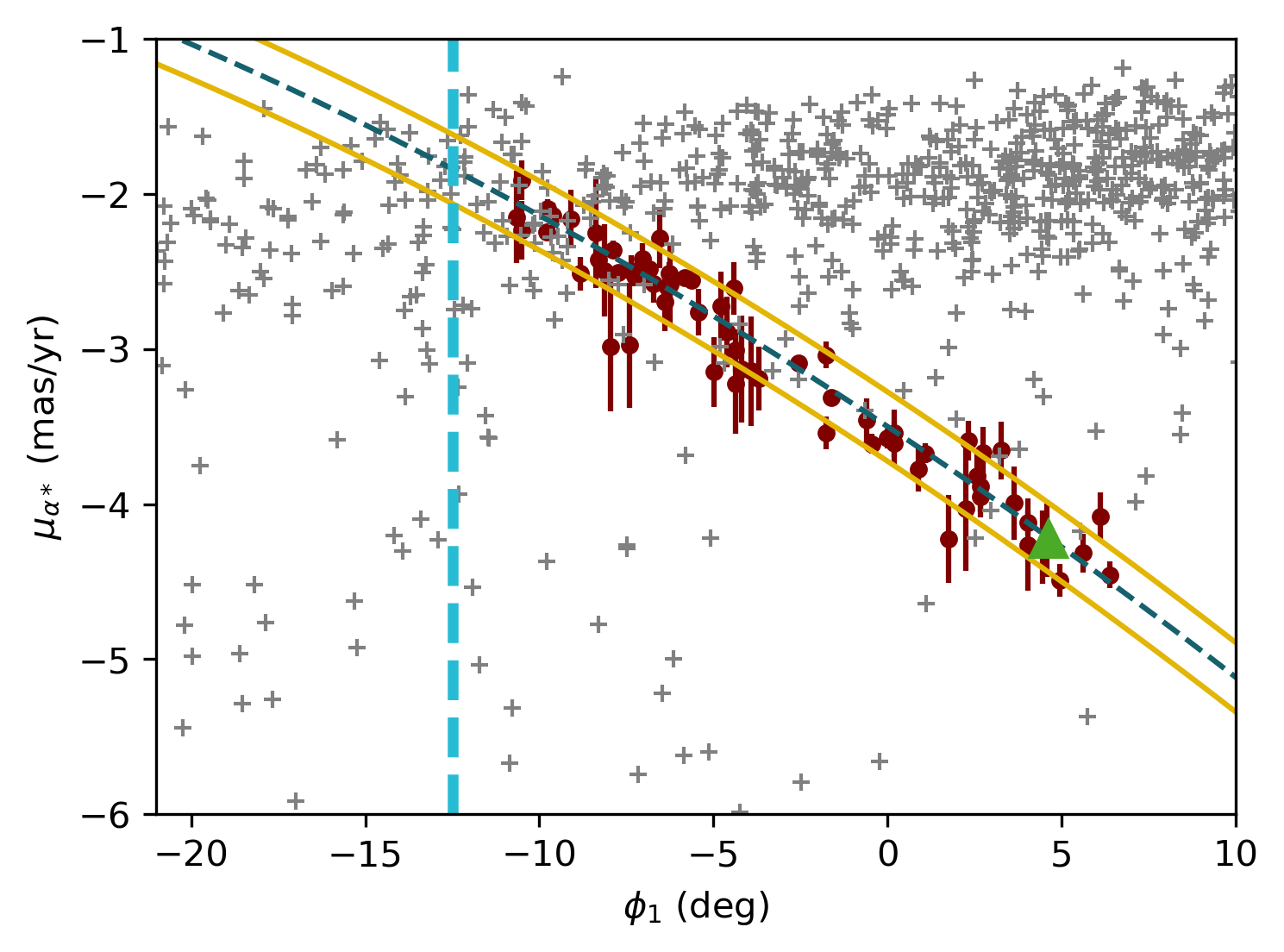}}
       \subfigure[\label{sfig:pmdec filter}]{\includegraphics[width=0.44\linewidth]{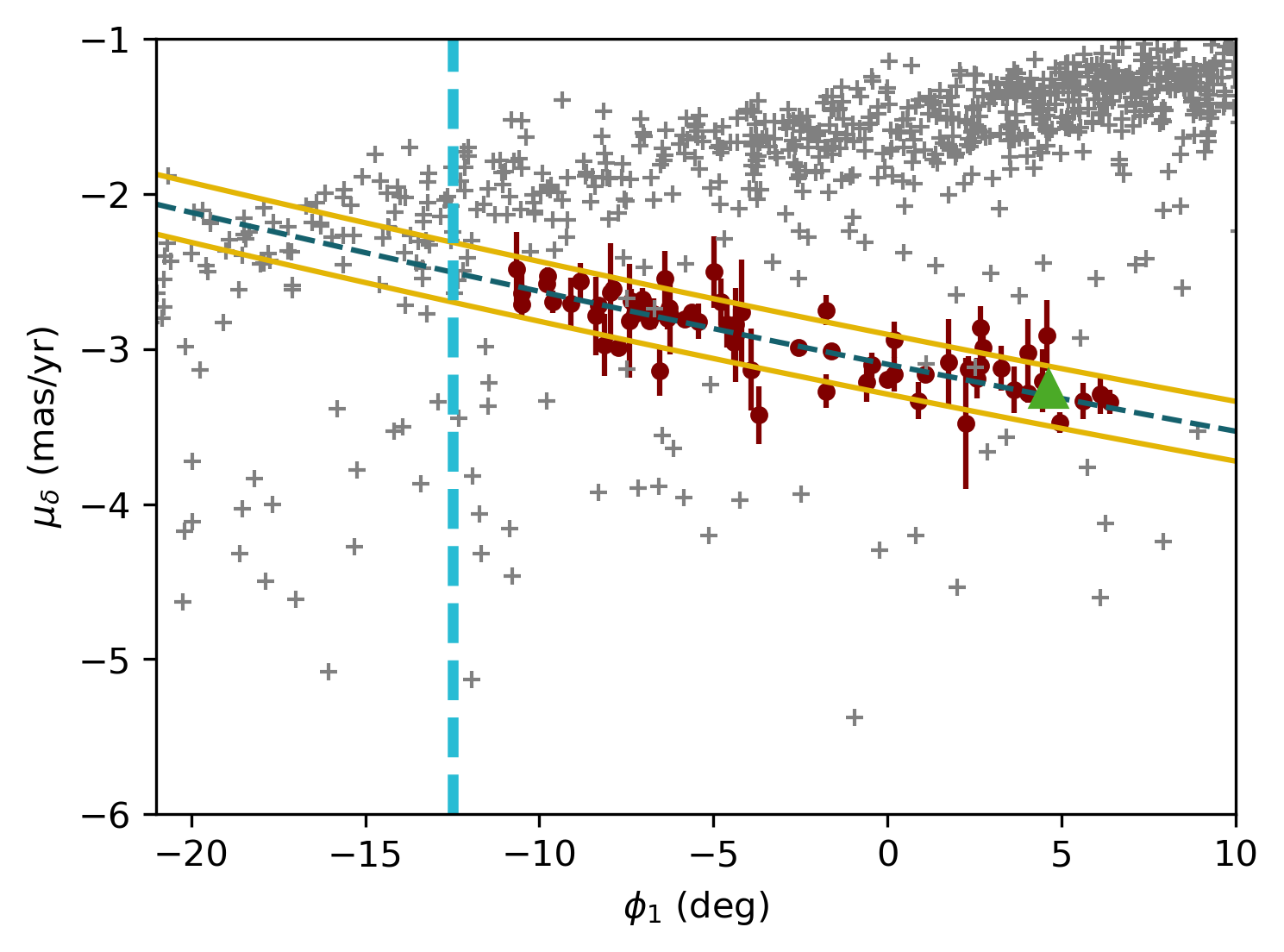}}
   \end{minipage}
    \begin{minipage}{1\linewidth}
    \centering
    \includegraphics[width=0.88\linewidth]{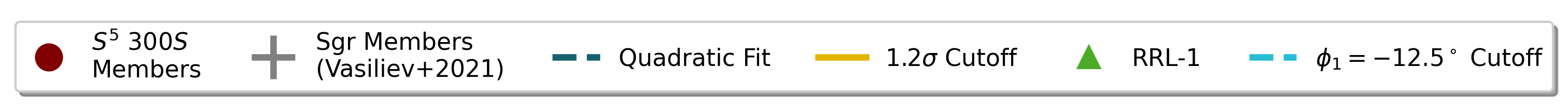}
    \end{minipage}
   \caption{Proper motion filters. \sgr simulated members are taken from the nearest wrap in the simulations of \cite{Vasiliev_2021}. (a) Filter in $\mu_{\alpha*}$. (b) Filter in $\mu_{\delta}$.}
   \label{fig:Proper Motion Filters}
\end{figure*}

We use a proper motion filter both to filter candidates in our search for standard candle members in Section \ref{sssec:standard candles} and to extract \streamPoss signal in one of our two empirical models in Section \ref{ssec:method 1 proper motion filtering}. 

To derive our proper motion filter, we use the \SSSSS spectroscopic member catalog as a pure ground truth. 
We first separately fit quadratic polynomials, $p_{\alpha*}$ and $p_\delta$, to these members in $\mu_{\alpha*}$ and $\mu_{\delta}$ respectively.\footnote{We notate $\mu_{\alpha*}\equiv \mu_\alpha\cos\delta$.}
The coefficients for these polynomials are given in Table \ref{tab:Proper Motion Quadratic Coeffs}. 
Next, we compute the standard deviations of the residuals for these fits as $\sigma_{\alpha*}$ and $\sigma_\delta$ respectively.\footnote{We find $\sigma_{\alpha*} = 0.19\masyr$ and $\sigma_\delta = 0.16\masyr$.} We then define our filters as

\begin{gather}
    |\mu_{\alpha*} - p_{\alpha*}(\phi_1)| < N\sigma_{\alpha*} \\
    |\mu_{\delta} - p_{\delta}(\phi_1)| < N\sigma_{\delta} 
\end{gather}

\noindent where we set $N=1.2$ to minimize contamination from \sgr.
However, our general results are not dependent on the specific value of $N$.\footnote{Another RRL star does not enter our filters for $N<2.32$. The star that enters at $N=2.32$, \badrrl, is likely a non-member. For further discussion, see Section \ref{sssec:standard candles}. We also examined the impact of setting $N$ to be one of $\left\{1.00,1.10,1.20,1.35,1.50\right\}$ on the result of Method 1 (see Section \ref{ssec:method 1 proper motion filtering}) and found that the resulting fits were not sensitive to $N$ given the uncertainties.}
For more discussion, see Section \ref{ssec:method 1 proper motion filtering}.

We show these filters in Figure \ref{fig:Proper Motion Filters}. 
In addition to the filters, we show simulated proper motions of \sgr from the simulation of \cite{Vasiliev_2021}. 
If $N\gtrsim1.2$, both proper motion filters would be contaminated by \sgr at $\phi_1 > -12.5^\circ$ where the peak \stream signal occurs, justifying the choice of $N=1.2$. 
It includes the majority ($61\%$) of the \SSSSS member stars while excluding the majority of \sgr in all of the region of interest. 

\begin{table}[]
    \centering
    \caption{Coefficients for Quadratic Fits to \SSSSS Member Proper Motions and Radial Velocities}
    \begin{tabular}{c|c|c|c}
        &$a_0$&$a_1$&$a_2$\\
        \hline
        $p_{\mu\alpha*}$&$-3.5007$ & $-0.1489$& $-0.0013$\\
        $p_{\mu\delta}$& $-3.0991$&$-0.0451$ &$-0.0002$\\
        $p_{vr}$ & $297.80$ & $-1.56$ & $-0.09$ \\
    \end{tabular}
    \label{tab:Proper Motion Quadratic Coeffs}
\end{table}

\subsubsection{Standard Candles}\label{sssec:standard candles}

Standard candles such as BHB and RRL stars are an effective way to determine a stream's distance gradient because of their robust color-magnitude relations.
Because of this, authors have used both BHB \citep[e.g.,][]{Deason_2011,Belokurov_2015,Ferguson_2021,Li_2021} and RRL \citep[e.g.,][]{Musella_2012, Garofalo_2013,Vivas_2020,Li_2021} stars as reliable distance tracers for streams and other Galactic structures.
We search for both BHB and RRL stars in \stream.

To find BHB stars, we use our cross-matched DECaLS DR9 and \textit{Gaia} DR3 catalogs.
We first identify BHB candidates with color $(g-r) < 0$. 
Next, we filter on-stream stars by selecting only candidates with $\left |\phi_2\right | < 1^\circ$ which corresponds to $\sim 2$ times the stream full width at half maximum of \cite{Fu_2018}. 
We also require that $\phi_1 > -12.5^\circ$ to prevent the intersection of our filters and \sgrPoss simulated proper motions.
We set an upper bound of $\phi_1 < 16^\circ$.
We then calculate $M_g$ using the relation of \cite{Belokurov_2015}

\begin{align}
    \notag M_g(g-r) = 0.398 - 0.392(g-r) + 
     2.729(g-r)^2\\ + 29.1128(g-r)^3 + 113.569(g-r)^4
\end{align}    

\noindent and select stars with distances in the range $13.5-19.5\kpc$, which allows slight variation around \streamPoss distance extent of $14-19\kpc$ found by \cite{Bernard_2016}. 
This leaves $24$ potential members. 
Finally, we select stars within our proper motion filter. 
No BHB candidates pass through this filter. 
Given the relatively high metallicity of \stream, the lack of BHB candidates follows the trend of reddening of the horizontal branch with higher metallicity \citep{Soker_2001}. 

To find RRL stars, we use the \textit{Gaia} DR3 \textsc{Gaiadr3.vari\_rrlyrae} table, which consists of cleaned and validated data on RRL stars \citep{Clementini_2023}. 
As with the BHB stars, we select members with $|\phi_2|<1^\circ$, $-12.5^\circ<\phi_1<16^\circ$, and a position outside of the object masks. We calculate $M_G$ using the PWZ relation of \cite{Garfalo_2022}:

\begin{align}
    \notag W(G,G_\text{BP},G_\text{RP}) = G - \lambda \times (G_\text{BP} - G_\text{RP})\\
    \notag =\left ( -2.49^{+0.21}_{-0.20}\right )\log(P) + \left ( 0.14^{+0.03}_{-0.03}\right )\feh\\
    + \left ( -0.88^{+0.08}_{-0.09}\right )
\end{align}

\noindent where $\lambda = 1.922$ \citep{Garfalo_2022}. We assume an RRL metallicity matching \cite{Usman_2024}'s \stream metallicity of $-1.35$ with an uncertainty of $0.11\dex$. This uncertainty is the $<95\%$ metallicity dispersion for \stream found by \cite{Li_2022}.
We again select stars with distance in the range $13.5-19.5\kpc$. This results in 8 candidates.
After application of our proper motion filter, we are left with one RRL member, henceforth referred to as \rrl.
This star was also identified as a candidate member in \cite{Fu_2018} due to its velocity of $302.9\kms$ and position.\footnote{There identified as PSO J105016.344+144644.466.} 
\cite{Fu_2018} do not classify it as a member because of its offset from a $12 \gyr$, $\feh = -1.5$ isochrone at the distance modulus given by their gradient.
However, we suspect that this can be explained by RRL photometric variability.

\rrlPoss position and velocity indicate that it is a stream member. Because its heliocentric distance is $14.9\kpc$ and galactocentric distance is $19\kpc$, it is unlikely to be associated with either the Galactic disk or bulge.
Similarly, \rrl is unlikely to be a member of \sgr. Our empirical model of \sgr (see Section \ref{sssec:sgr distance gradient through matched filter}) places its closest wrap in the region at a heliocentric distance of $\sim 20\kpc$ and simulations place it further at $\sim 25\kpc$ \citep{Vasiliev_2021} while \rrl only has a heliocentric distance of $14.9\pm 1.1\kpc$. Further, in addition to its distance and radial velocity, \rrlPoss proper motions are only $0.04\sigma_{\alpha*}$ and $0.28\sigma_{\delta}$ off Section \ref{sssec:proper motion filter}'s respective best fit curves.
Because \rrlPoss kinematics in terms of its proper motions from \textit{Gaia} and its radial velocity from \cite{Fu_2018}'s study both fit \stream well, and because its distance further rules out other possible associations, we identify \rrl as a member of \stream.

We note that one other RRL star,\footnote{
\textit{Gaia} DR3 \textsc{source\_id} 621603371140586880.}
henceforth referred to as \badrrl, passes our proper motion filter if we expand the cutoff from $1.2\sigma$ to $2.5\sigma$. 
Specifically, \badrrl has proper motions at $2.32\sigma_{\text{ra}}$ and $2.29\sigma_{\text{dec}}$ off Section \ref{sssec:proper motion filter}'s respective best fit curves. However, \badrrlPoss radial velocity was measured by \cite{Liu_2020} to be $-60.772\kms$ which is substantially different from the characteristic radial velocity of \stream of $\sim 300\kms$. Therefore, this star is a nonmember.

\begin{table}[]
    \centering
    \begin{threeparttable}
    \caption{Parameters of the Likely RRL Member: \rrl}
        \begin{tabular}{c|c}
         Parameter & Value for \rrl \\
         \hline
         \textit{Gaia} DR3 \textsc{source\_id} & $3885177800499823616$\\ 
         $\mu_{\alpha*}$ ($\masyr$)&$-4.22\pm 0.07$\\
         $\mu_\delta$ ($\masyr$)&$-3.26\pm0.06$\\
         $\#\sigma_{\alpha*}$ &$0.04$\\
         $\#\sigma_{\delta}$ &$0.28$\\
         Distance ($\kpc$)&$14.9\pm 1.1$\\
         $v_r$ $(\kms)$&$302.9\pm2.8$ \\
         $\phi_1$ (deg) &$4.63$\\
         $\phi_2$ (deg) &$0.02$ \\
    \end{tabular}
    \begin{tablenotes}
      \footnotesize
      \item The radial velocity is from the work of \cite{Fu_2018}. $\# \sigma$ is the number of standard deviations away from from the corresponding proper motion fit.
    \end{tablenotes}
    \label{tab:RRL Summary}
    \end{threeparttable}
\end{table}

\subsubsection{Distance Gradient Calculation}\label{sssec:distance gradient calculation}

We are unable to derive a distance gradient directly from standard candles because we only identify one such likely member, \rrl. To determine the slope of the gradient, we use the pure \SSSSS member catalog and derive a gradient that minimizes the red giant branch members' spread around an isochrone. Specifically, we begin by selecting \SSSSS members that are on the red giant branch by considering stars with whose matched DECaLS photometry satisfies $(g-r) > 0.35$ and $g>17.25$. Next, we calculate the distance modulus of the \SSSSS members from the $\feh=-1.35$, $12.5\gyr$ isochrone described in Section \ref{ssec:naive matched filter}, again using the matched DECaLS photometry. We then perform linear regression on these distance moduli. This regression gives a slope $k_{S^5} =-0.035 \pm 0.006$ and intercept $b_{S^5} = 16.2\pm 0.04$. 

We then set the intercept such that the distance gradient goes through \rrl. Although we could set this intercept using the $S^5$ distance moduli, this number is highly dependent on the choice of isochrone. For instance, when changing from $\feh = -1.35$ \citep[as reported in][]{Usman_2024} to $\feh = -1.26$ \citep[as reported in][]{Li_2022}, $k_{S^5}$ changes by $0.001$, or only $0.12\sigma$, while $b_{S^5}$ changes by $0.19$, or $>3\sigma$. 

This calculation leads to a distance gradient of

\begin{equation}
    \label{eq:dist_grad}
    \mu_{\stream}(\phi_1) = (-0.035\pm 0.006)\phi_1 + (16.03\pm 0.16)
\end{equation}

Note that the intercept is nevertheless in $\sim1\sigma$ agreement with $b_{S^5}$. \cite{Fu_2018} found a distance gradient as a function of R.A. of 

\begin{equation}\label{eq:fu grad}
    d(\alpha) = 48.9952-0.2083\alpha
\end{equation}

\noindent where $\alpha$ is the R.A. We cannot exactly compare Equation \ref{eq:dist_grad} to Equation \ref{eq:fu grad} because the gradient of \cite{Fu_2018} is linear in R.A. and distance while the gradient in this work is linear in $\phi_1$ and distance modulus. To make a comparison, we convert the \cite{Fu_2018} gradient into $\phi_1$ and distance modulus using our transformation matrix by converting $(\phi_1,0^\circ)$ pairs to $(\text{R.A.},\text{Decl.})$ pairs and evaluating $\mu_\text{fu}(\phi_1) = \mu\Bigl(d_\text{fu}\bigl(\alpha(\phi_1,0^\circ)\bigr)\Bigr)$. We then fit a linear function to the result in the region of interest ($\phi_1 \in [-20^\circ,13^\circ]$). After this conversion, the \cite{Fu_2018} gradient takes the form
\begin{equation}\label{eq:Fu Transformed}
    \mu_{\text{fu}}(\phi_1) \approx -0.028\phi_1 + 16.02
\end{equation}

\begin{figure}
    \centering
    \includegraphics[width=1\linewidth]{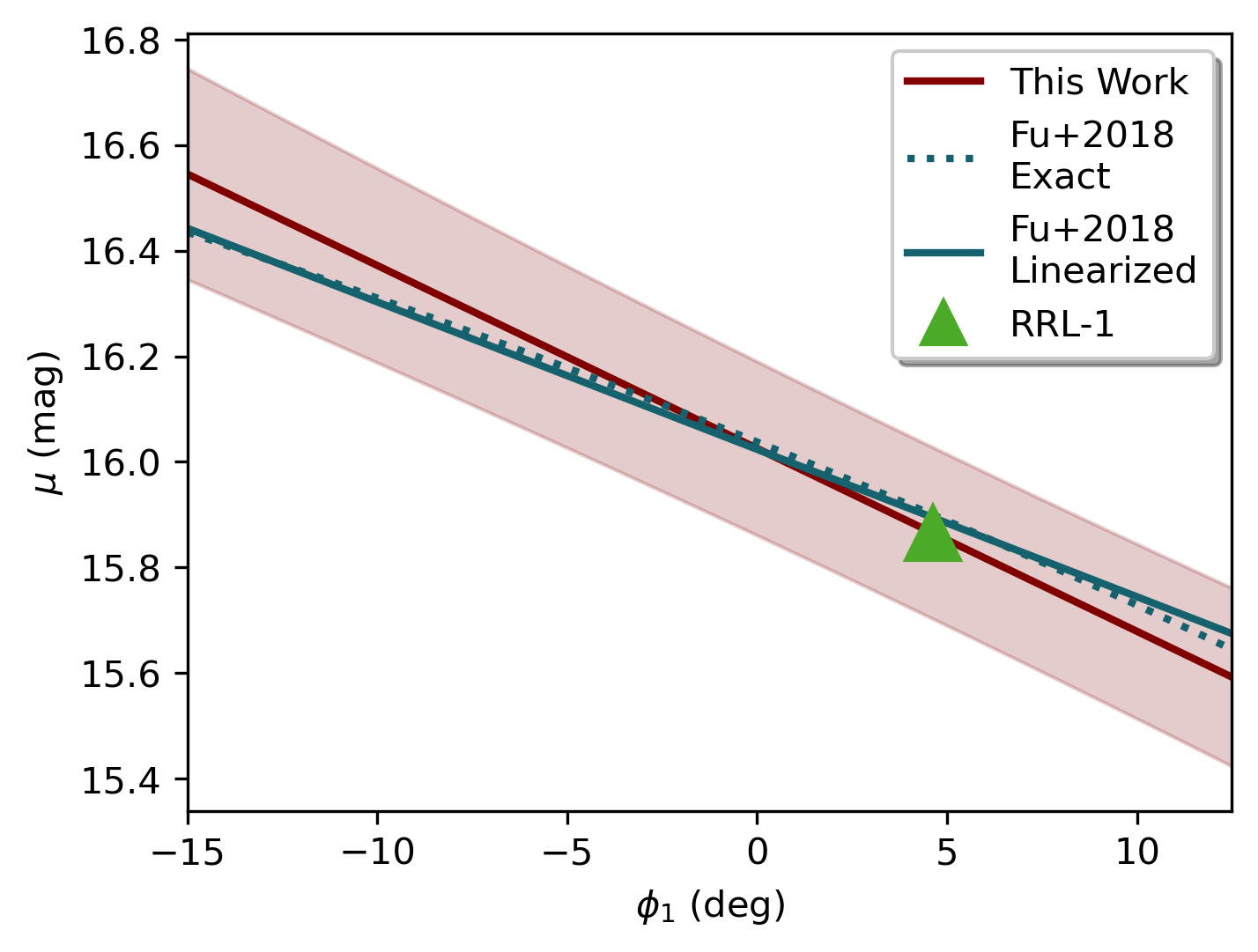}
    \caption{Comparison of our calculated distance gradient and the transformed gradient of \cite{Fu_2018} both in its linearized and exact forms. The shaded region represents $1\sigma$ uncertainty.}
    \label{fig:distance gradients}
\end{figure}

Both the slope and intercept are in reasonable agreement between these two distance gradients ($\Delta k = 1.2\sigma$ and $\Delta b = 0.06\sigma$ excluding uncertainties in the \citealp{Fu_2018} gradient).
We show the two gradients and the nonlinearized \cite{Fu_2018} result in Figure \ref{fig:distance gradients}. 
Over the region of interest where we model \stream, Equation \ref{eq:dist_grad} implies an average distance modulus of $16.15$ corresponding to a physical distance of $17\kpc$.

\begin{figure*}
    \centering
    \subfigure[\label{sfig:Hess}]{\includegraphics[width = 0.48\linewidth]{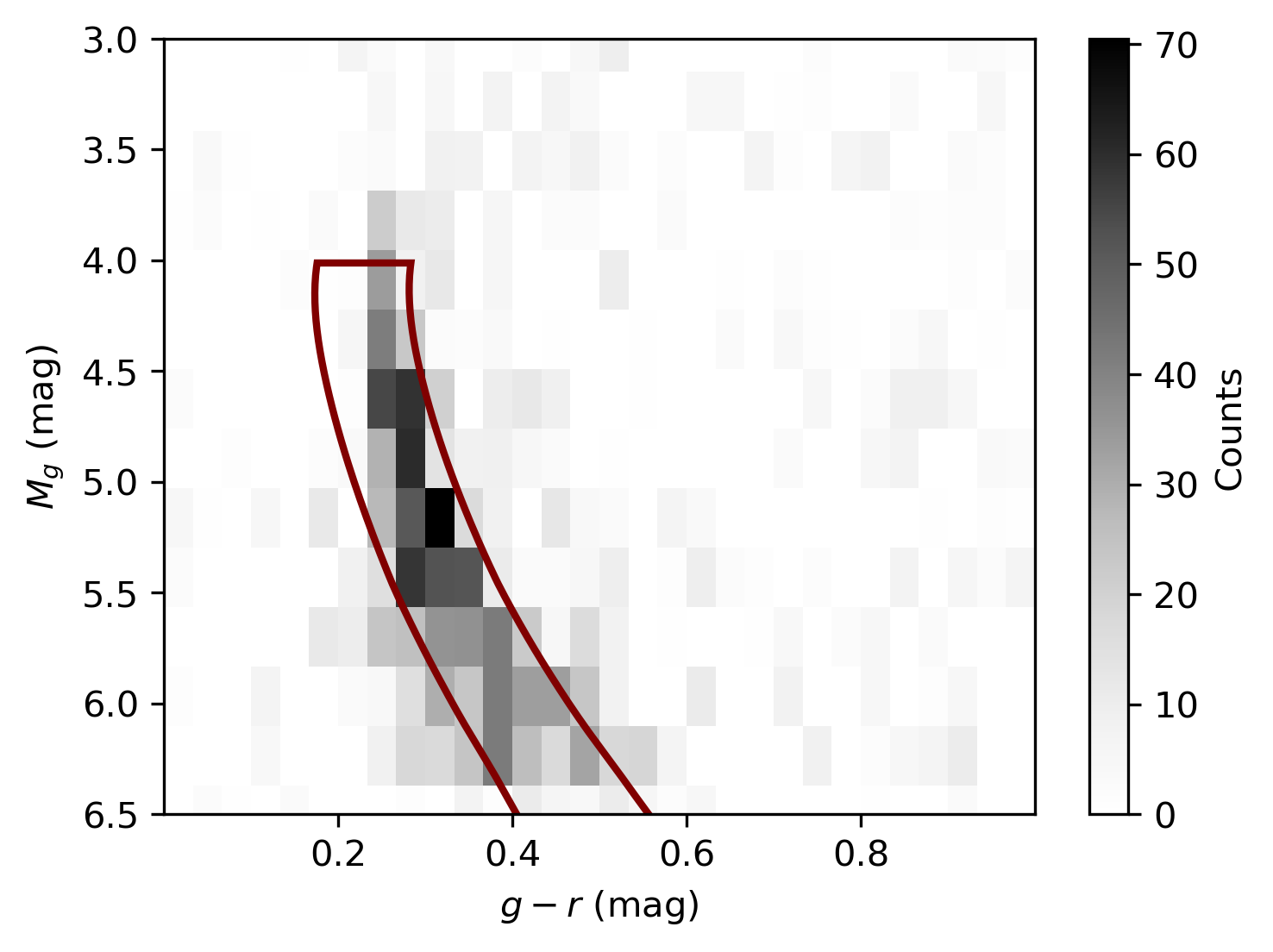}}
    \subfigure[\label{sfig:Refined Matched Filter}]{\includegraphics[width=0.48\linewidth]{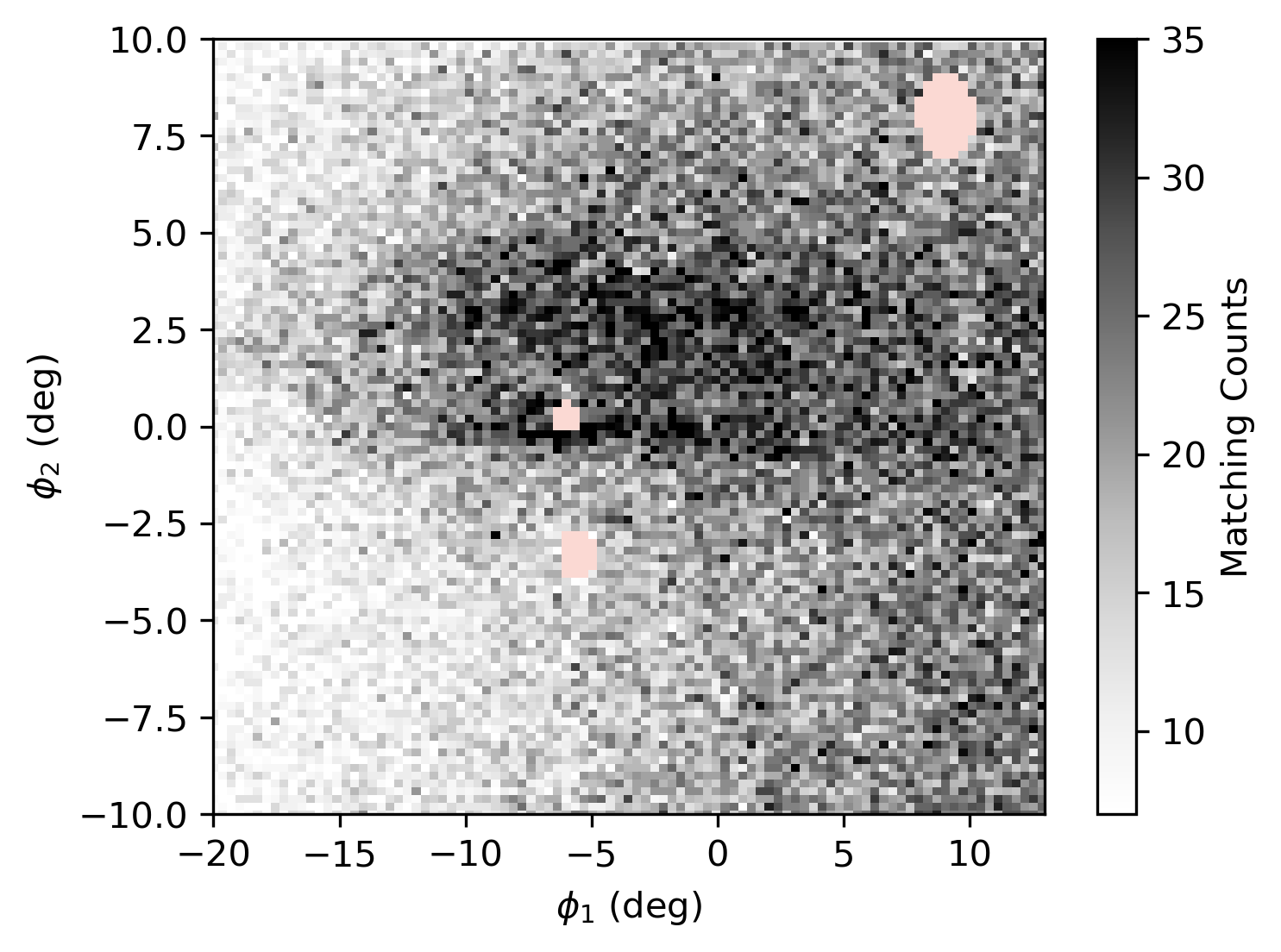}}
    \caption{(a) Hess diagram for on-stream region. $M_g$ was computed assuming a distance modulus calculated using Equation \ref{eq:dist_grad}. The red outline shows the \stream matched filter ($s = 2/3$) used in this work. (b) Refined matched filter density map. \stream is clearly visible as the thin band centered on $\phi_2=0^\circ$. The \sgr contamination seen in Figure \ref{fig:Naive Matched Filter} is still visible as a wide dark region above \stream centered at $\phi_2 \sim 2.5^\circ$.
    The red shaded regions are the object masks.
    }
    \label{fig:Hess and Refined Matched Filter}
\end{figure*}

\subsubsection{Refined Matched Filter as Motivation for Filtering Methodologies}\label{sssec:improved matched filter}

We improve our matched filter map using the distance gradient we derived in Section \ref{ssec:distance gradient}.
We show the \stream matched filter on a Hess diagram of an on-stream region with distance moduli computed using Equation \ref{eq:dist_grad} in Figure \ref{sfig:Hess}. We show the refined matched filter map in Figure \ref{sfig:Refined Matched Filter}.

\stream is visible in this map as the thin overdensity around $\phi_2 = 0^\circ$. 
The \sgr contamination is also still visible as a wide overdensity above \stream. 
This map makes the challenge of modeling \streamPoss morphology clear. 
An approach directly modeling the stream and background like that of \cite{Ferguson_2021} would be complicated by the \sgr signal.
Therefore, this map motivates the need for additional methods of signal extraction. 
The next two sections describe the two approaches we use to account for \sgrPoss influence.

\begin{figure*}[ht!]
    \centering
    \subfigure[\label{fig:CMD Gaia PM}]{\includegraphics[width=0.34\linewidth]{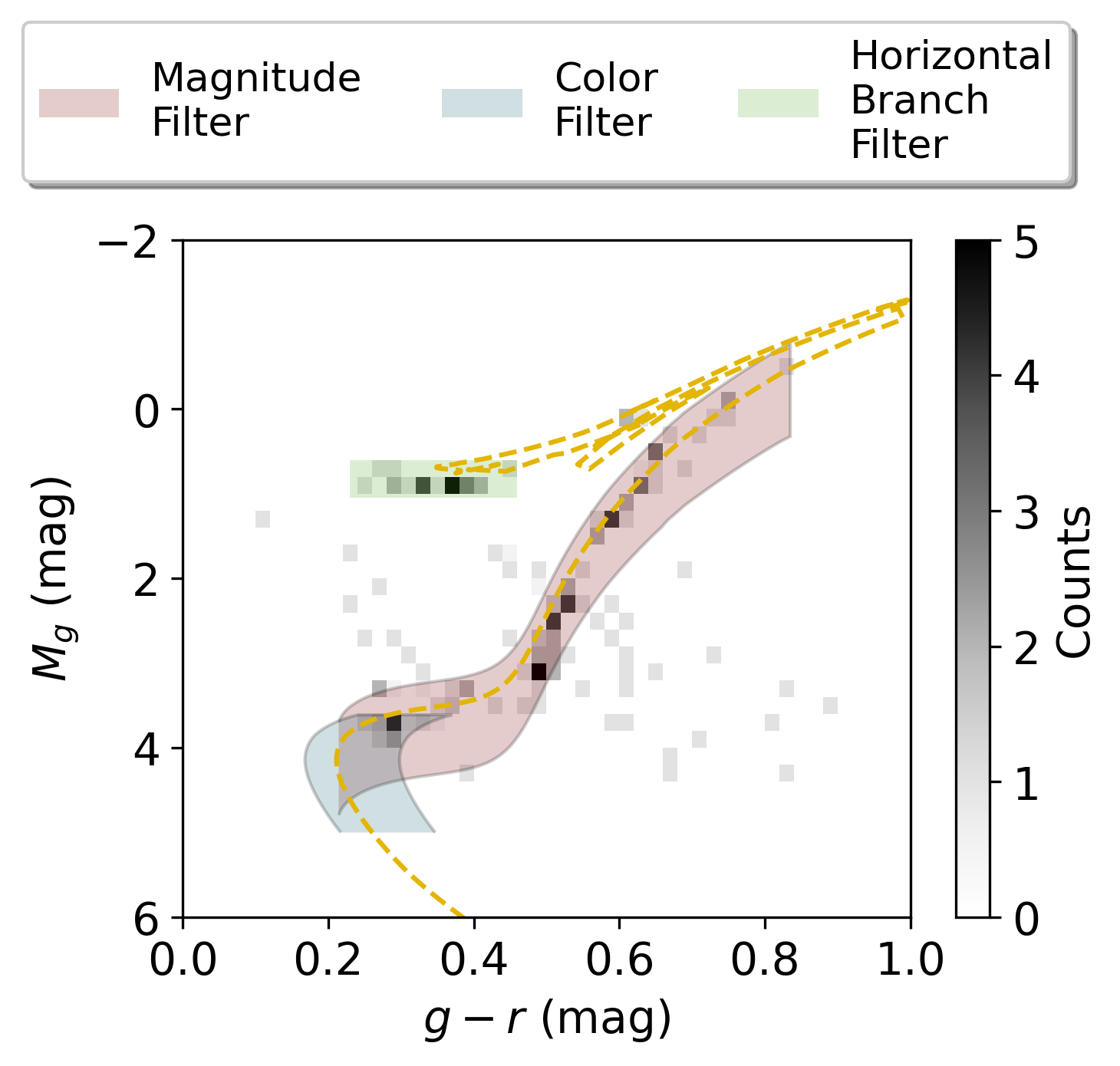}}
    \subfigure[\label{fig:p12 Gaia}]{\includegraphics[width = 0.65\linewidth]{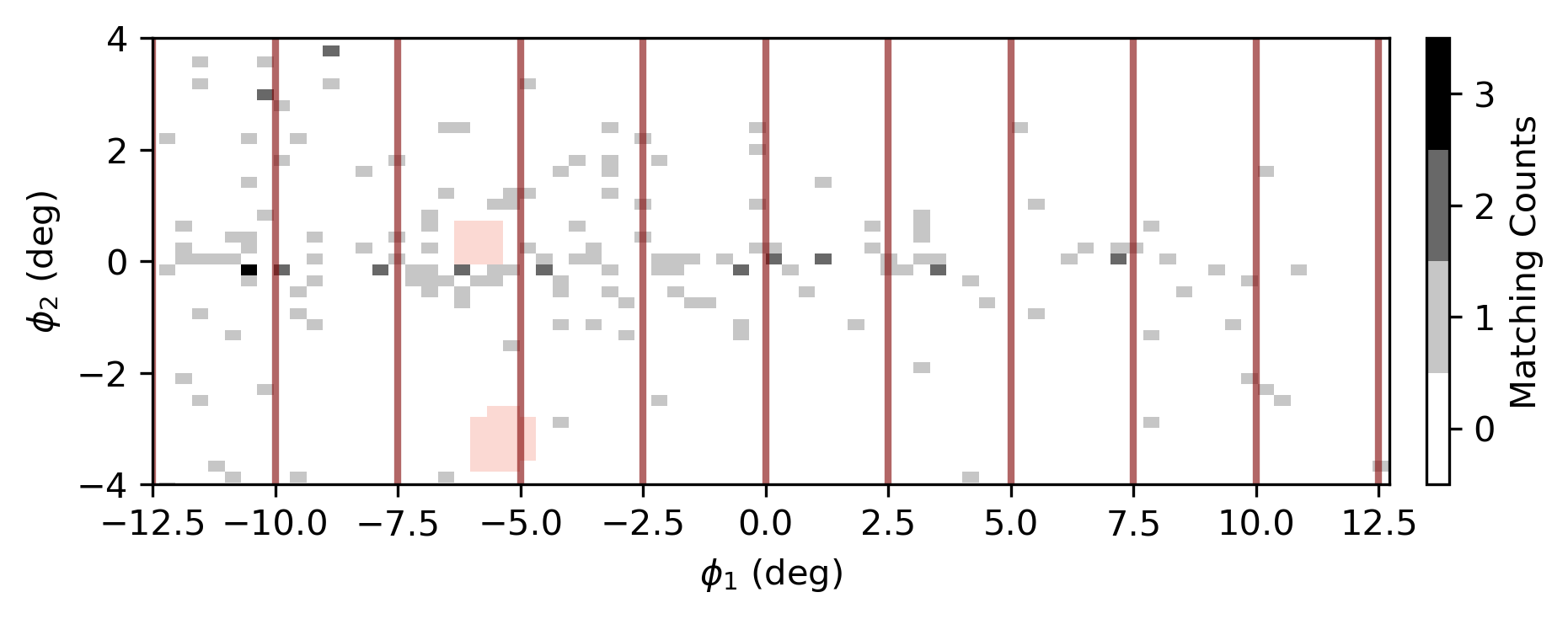}}
    \caption{Results from filtering using \textit{Gaia} proper motions in Section \ref{ssec:method 1 proper motion filtering}. (a) The matched filter used for the Method 1 of signal extraction. The yellow dashed line is the $\feh = -1.35$, $\text{Age }=12.5\gyr $ isochrone used in this work. A Hess diagram of the sample that passes the quality cuts and proper motion filters is also shown. The \stream isochrone is clearly visible.
    (b) Stellar density map in the vicinity of \stream after filtering. The red shaded regions are the object masks.}
    \label{fig:Gaia Empirical Results}
\end{figure*}

\section{Fitting Models of Stream Morphology Method 1:\\Filtering Using \textit{Gaia} DR3 Data}\label{ssec:method 1 proper motion filtering}

Our first method to filter \sgr contamination is to use kinematic information from \textit{Gaia} DR3.
Because \streamPoss main sequence turnoff is at $g \simeq 20$, many candidate members have large proper motion uncertainties and are omitted. 
Therefore, this method excludes many candidate members; leading to a relatively pure but incomplete selection.
Because of the resulting low number counts, this method alone is insufficient to extract much of the stream structure with certainty. 
Instead, we use it as a check against our second filtering method (Section \ref{ssec:method 2 double stream model}), which uses purely photometric information from DECaLS DR9 and extracts a less pure selection with much higher completeness. 

\subsection{Filtering \streamPoss Signal}\label{ssec:gaia filtering}

We use the proper motion filters derived in Section \ref{sssec:proper motion filter} as the primary filter for \streamPoss signal.
Because \stream has a main sequence turnoff at $g\simeq 20$, the proper motion uncertainties for many candidate stars are very large.
Therefore, to ensure purity, we begin by constraining the \textit{Gaia} catalog to stars with $\sqrt{\sigma_{\mu_{\alpha*}}^2 + \sigma_{\mu_\delta}^2} < 0.8 \masyr$ and with $g < 20.5$. 

Next, because \stream and \sgrPoss proper motions are very similar at some negative values of $\phi_1$ as can be seen in Figure \ref{fig:Proper Motion Filters}, we remove stars with $\phi_1 < -12.5^\circ$ to prevent contamination. 
This cut is shown in Figure \ref{fig:Proper Motion Filters} as a light blue dashed line. 
We consider $\phi_2 \in [-5^\circ,5^\circ]$ to match the region used in Method 2 (Section \ref{ssec:method 2 double stream model}).
We then employ the proper motion filters described in Section \ref{sssec:proper motion filter}.

Next, we apply a matched filter on the turnoff, sub-giant branch, red-giant branch, and horizontal branch using the cross-matched DECaLS photometry. 
We define this matched filter in three components. 
First, we define a filter over magnitudes that matches stars with $M_g^\text{iso} \in [-0.5,4]$ using upper and lower magnitude cuts of $\delta_\text{upper}$ and $\delta_\text{lower}$ respectively around the isochrone magnitude $M_g^\text{iso}(g-r)$. i.e.

\begin{equation}
    M_g^\text{iso}(g-r) - \delta_{\text{lower}} < M_g < M_g^\text{iso}(g-r) + \delta_{\text{upper}}
\end{equation}

\noindent where $M_g$ is calculated using the distance gradient in Equation \ref{eq:dist_grad}. 

Second, we define a filter over colors that matches stars with $M_g^\text{iso}\in [3.6,5]$ using right and left color cuts  of $\delta_\text{right}$ and $\delta_\text{left}$ respectively around the isochrone color $(g-r)_\text{iso}$. i.e.

\begin{equation}
    (g-r)_\text{iso} - \delta_{\text{left}} < (g-r) < (g-r)_\text{iso} + \delta_{\text{right}}
\end{equation}

Third, we define a filter around the visible horizontal branch (HB) overdensity. For this filter, we select stars within a rectangle with center $(c_{g-r},c_{M_g})$, width $w$, and height $h$. i.e.

\begin{align}\label{eq:Horizontal Branch Filter}
     \notag\bigl|(g-r) - c_{g-r}\bigr|< w/2\\
     \&\quad\bigl|M_g-c_{M_g}\bigr| < h/2
\end{align}

\noindent and we accept stars that fall into any of the three filters. We define these filters by eye in order to extract the visible isochrone on the CMD, as seen in Figure \ref{fig:CMD Gaia PM}. We use parameters $\delta_\text{lower}=0.3,\, \delta_\text{upper}=0.8,\,\delta_\text{left}=0.043,\, \delta_\text{right}=0.087,\, c_{g-r} = 0.345,\, c_{M_g} = 0.825,\, w = 0.23,$ and $h = 0.45$.
After we apply these filters, we bin the stars into $\phi_1,\phi_2$ bins of width $\Delta \phi_1 = 0.3^\circ$ and height $\Delta \phi_2 = 0.2^\circ$.
We show this filter, the underlying background subtracted color-magnitude distribution of the stars selected with proper motions, and the resulting stellar density map in Figure \ref{fig:Gaia Empirical Results}.
The \sgr overdensity that was visible in Figure \ref{sfig:Refined Matched Filter} is no longer present. 
Nevertheless, the peak of \stream around $\phi_1 = -6^\circ$ (peak A) that is visible near Segue 1 in Figure \ref{sfig:Refined Matched Filter} is still apparent in Figure \ref{fig:p12 Gaia}'s density map. 

\subsection{Modeling \streamPoss Morphology}

We model the morphology of \stream using a modified version of the approach developed in \cite{Koposov_2019_orphan}, \cite{Li_2021}, and \cite{Ferguson_2021}. In this method, the stream is modeled using three components: $\mathcal{I}(\phi_1)$, $w(\phi_1)$ and $\Phi_2(\phi_1)$ as the $\log$ central stellar density, $\log$ Gaussian width, and $\phi_2$ position, respectively. As a 2D function of these components, the stream density $\Lambda_\text{\stream}$ can be written as

\begin{equation}\label{eq:300S Density}
    \log\Lambda_\text{\stream}(\phi_1,\phi_2) = \mathcal{I}(\phi_1) -\frac{\left (\phi_2 - \Phi_2(\phi_1)\right)^2}{2e^{2w(\phi_1)}}
\end{equation}

Further, a background component $\log \Lambda_\text{background}$ is included in the model as a polynomial with coefficients $\Vec{B}$. In previous studies, quadratics with $\Vec{B}\in \mathbb{R}^3$ have been used \citep[e.g.,][]{Ferguson_2021}, i.e. $\log \Lambda_\text{background} = \begin{bmatrix}1&\phi_1&\phi_1^2\\\end{bmatrix}\cdot \Vec{B}$. 
Due to the high purity and low overall counts for the \textit{Gaia} stars here, we use a linear background model with $\Vec{B}\in \mathbb{R}^2$.

We assume that the stars are Poisson distributed in the histogram bins with a spatially varying rate defined by these models as $\Lambda(\phi_1,\phi_2) = \Lambda_\text{\stream}(\phi_1,\phi_2) + \Lambda_\text{background}(\phi_1,\phi_2)$. We then define our composite Poisson likelihood as 

\begin{equation}\label{eq:loglik}
\log \mathcal{L} = \sum_{\text{bin }i} \log\mathcal{P}\left (N^{(i)}\middle|\Lambda(\phi_1^{(i)},\phi_2^{(i)})\right)    
\end{equation}

\noindent where $N^{(i)}$ is the number count of matched stars in bin $i$ and $\phi_1^{(i)},\phi_2^{(i)}$ are its coordinates. 

We parameterize $\Vec{B}(\phi_1)$, $\mathcal{I}(\phi_1)$, $w(\phi_1)$, and $\Phi_2(\phi_1)$ as cubic splines \citep[as in e.g.][]{Koposov_2019_orphan,Li_2021,Ferguson_2021}. 
Including the endpoints, we use 13 nodes for the stream parameters $\mathcal{I}(\phi_1)$, $w(\phi_1)$, and $\Phi_2(\phi_1)$. 
We use 6 nodes for $\Vec{B}$.
Small modifications to both the number and position of these nodes do not dramatically alter the results.
Given a set of node positions in $\phi_1$, we desire to sample the posterior $y$-values of those nodes using the composite Poisson likelihood defined above. 
We follow \cite{Koposov_2019_orphan} and only set non-trivial priors on the $\Phi_2(\phi_1)$ and $w(\phi_1)$. 
We use $\mathcal{N}(0,1)$ for $\Phi_2(\phi_1)$. 
We set $\mathcal{N}(\log 0.4,0.5)$ for $w(\phi_1)$ 
This is the same width $\sigma$ as used by \cite{Koposov2023} but with a mean stream full width at half maximum of $0.94^\circ$ as found for \stream by \cite{Fu_2018}.
We sample the node coefficients using \textsc{stan} \citep{Carpenter_2017,STAN_2018}, an optimized implementation of Hamiltonian Markov Chain Monte Carlo (MCMC) \citep{Neal_2011} that applies the No U-Turn method to reduce fine-tuning and increase efficiency \citep{Hoffman_2011}.\footnote{We made use of \textsc{stan-splines}, \url{https://zenodo.org/records/14163685}, an implementation of natural cubic splines in \textsc{stan} \citep{Koposov_2019_orphan}.}
We run the sampler for $700$ warmup iterations and $800$ sampling iterations using $4$ chains. 
All of the parameters achieve a satisfactory $\hat{R}$ score of $<1.1$ indicating convergence \citep{Gelman_and_Rubin_92}. 

\begin{figure*}[!ht]
   \begin{minipage}{1\linewidth}
       \centering
       \subfigure[\label{sfig:gaia splines}Method 1 (\textit{Gaia} Based)]{\includegraphics[width=0.48\linewidth]{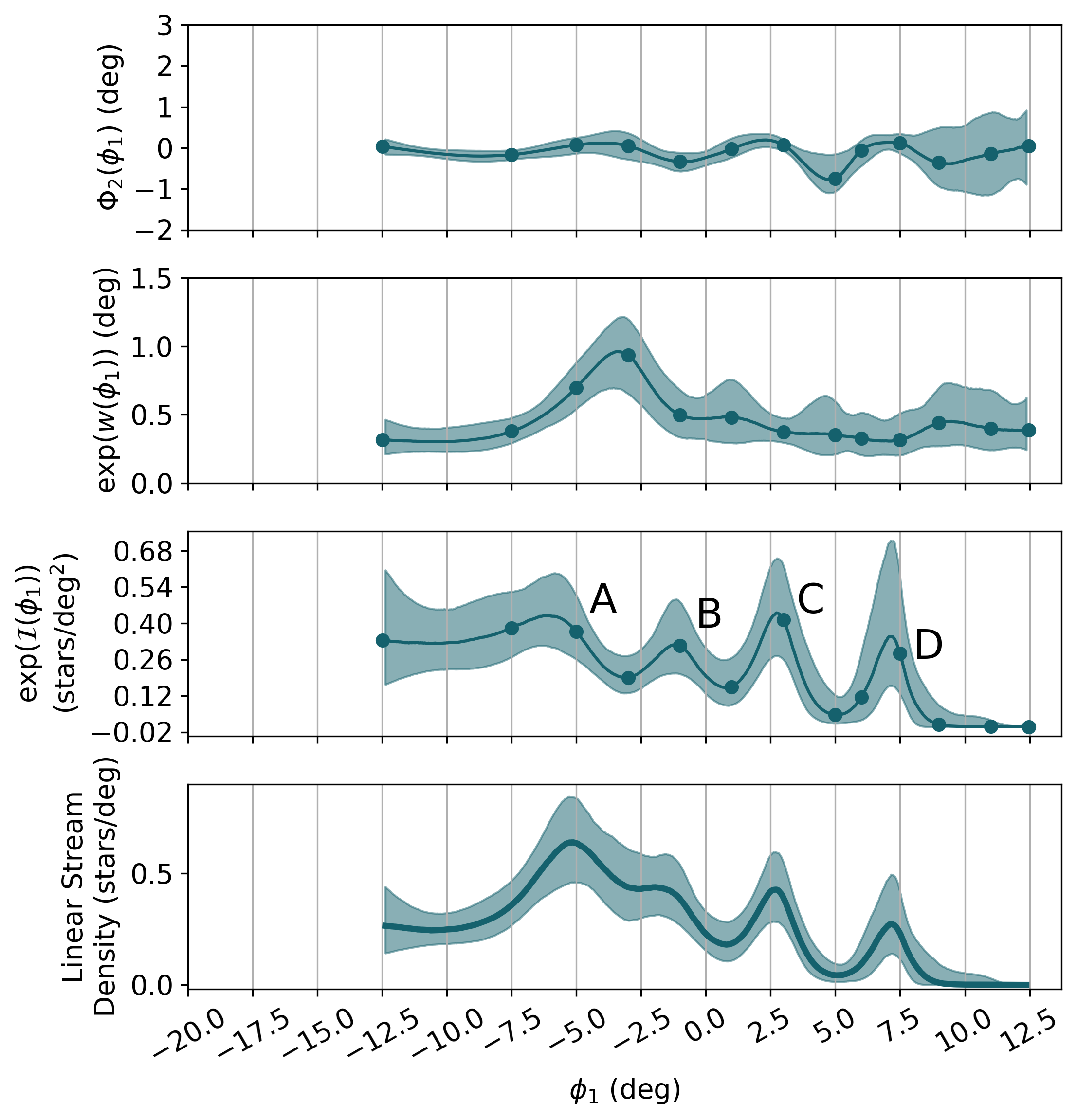}}
       \subfigure[\label{sfig:sub splines}Method 2 (DECaLS Based)]{\includegraphics[width=0.48\linewidth]{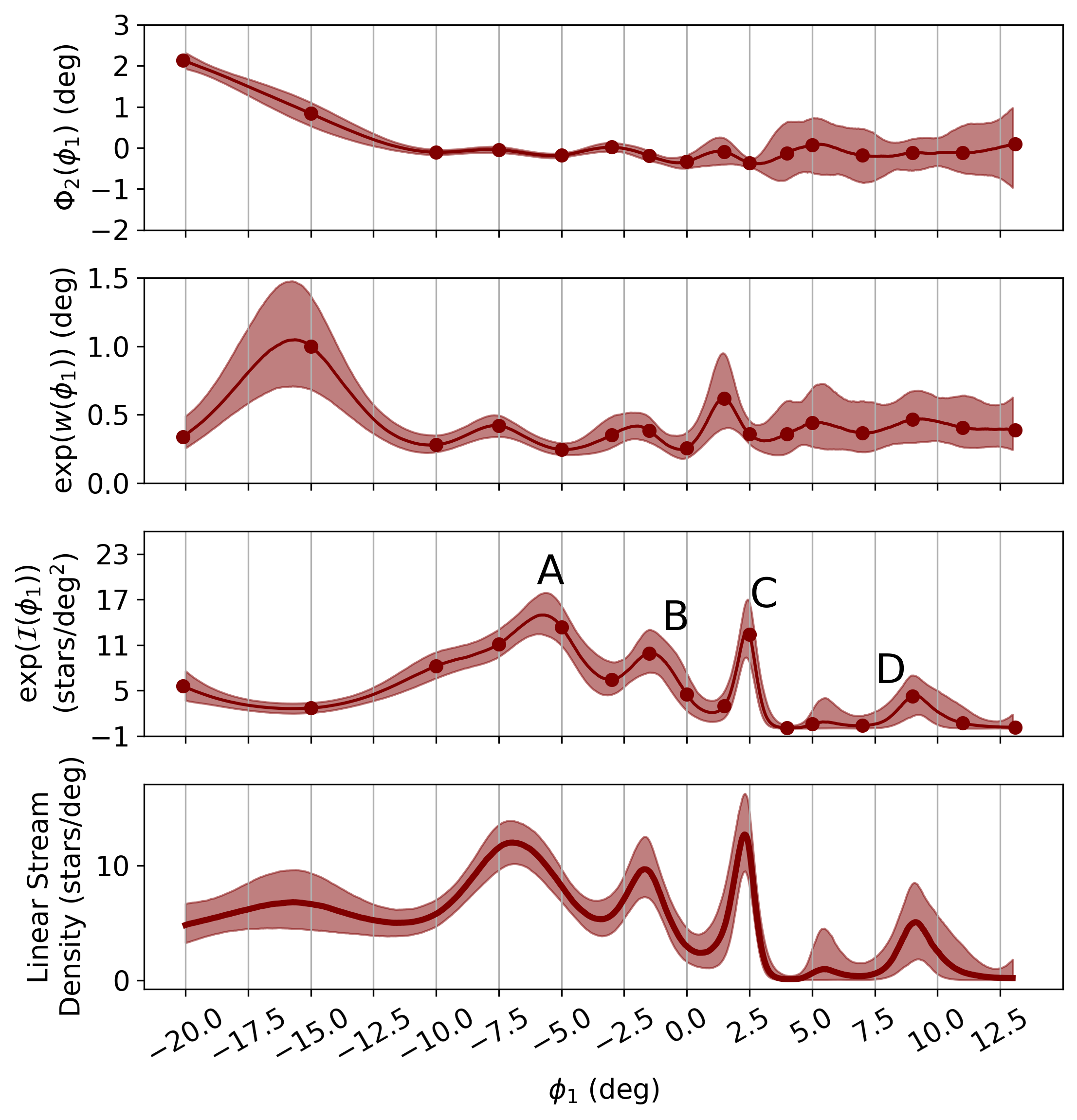}}
   \end{minipage}
    \begin{minipage}{1\linewidth}
    \centering
    \includegraphics[width=0.6\linewidth]{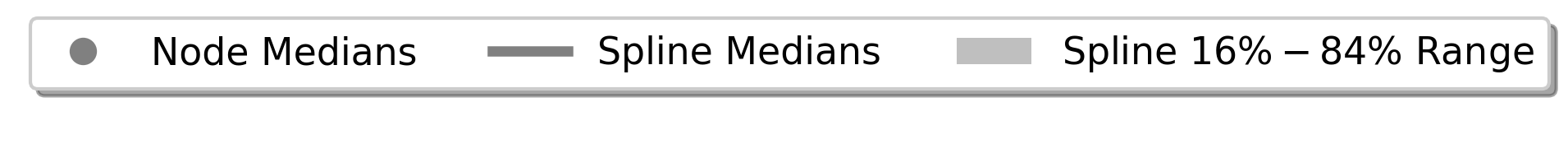}
    \end{minipage}
   \caption{Resulting spline models for both methods of signal extraction. 
   The top plots correspond to $\Phi_2(\phi_1)$ as the $\phi_2$ position of \streamPoss track. 
   The second plots show $\exp\left (w(\phi_1)\right)$ as the Gaussian width of the stream.
   The third plots are $\exp\left(\mathcal{I}(\phi_1)\right)$ as the stream's central stellar density. The locations of peaks A, B, C, and D are labeled. 
   Both $\mathcal{I}$ and $w$ are fit as splines in log space and their splines and quantile ranges are subsequently transformed to linear space. 
    Finally, the fourth plots show the linear stream density.
   (a) Method 1 of signal extraction that uses kinematic information from \textit{Gaia} DR3 in addition to DECaLS DR9 photometry (Section \ref{ssec:method 1 proper motion filtering}). 
   (b) Method 2 of signal extraction that purely uses photometric information from DECaLS DR9 (Section \ref{ssec:method 2 double stream model}).
   We compare the tracks more directly in Figure \ref{fig:stream results figure}.}
   \label{fig:Spline Values for the Two Streams}
\end{figure*}

\begin{figure*}[!ht]
   \centering
   \begin{minipage}{1\linewidth}
       \centering
       \subfigure[\label{sfig:Gaia stream component}Method 1 (\textit{Gaia} Based)]{\includegraphics[width=0.49\linewidth]{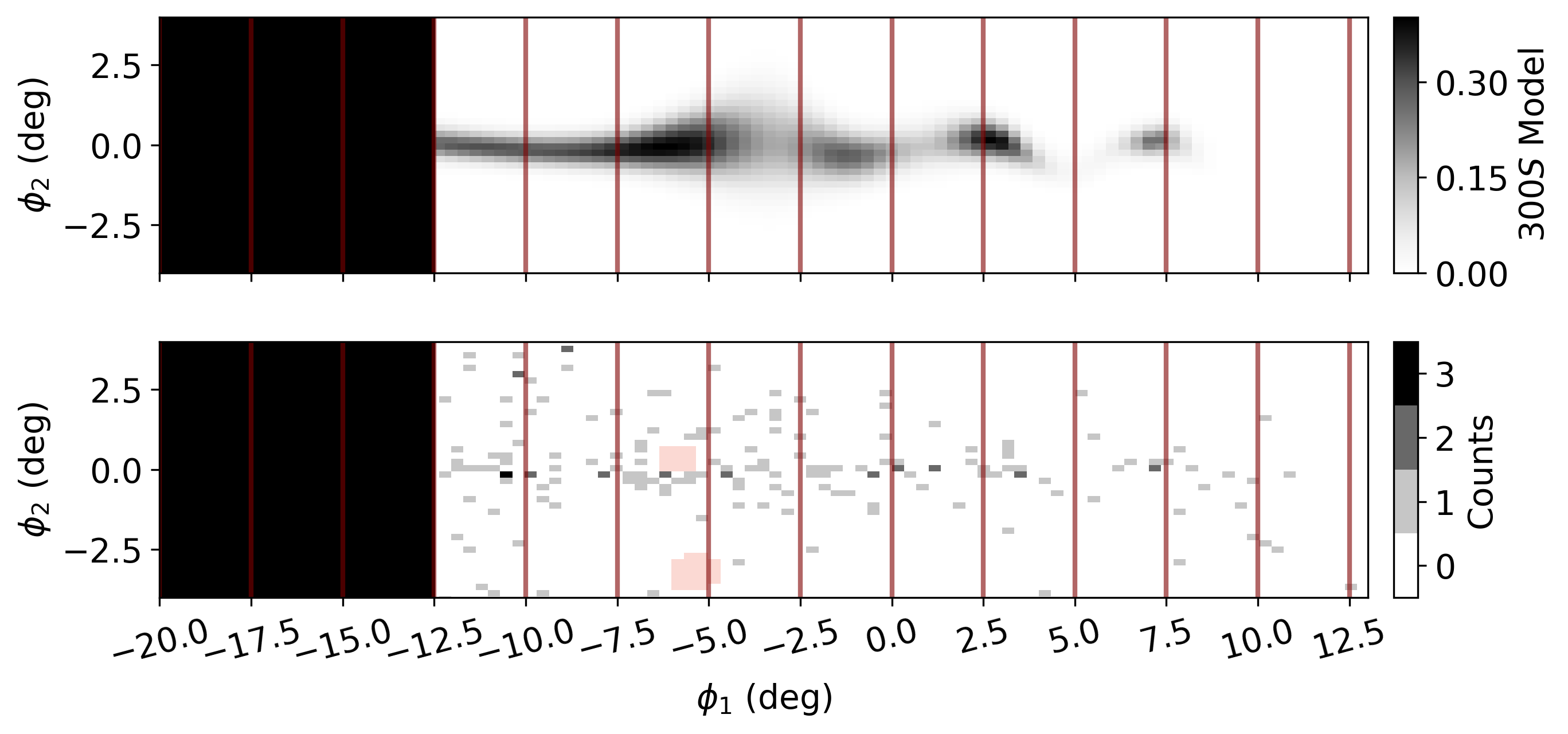}}
       \subfigure[\label{sfig:sub stream component}Method 2 (DECaLS Based)]{\includegraphics[width=0.49\linewidth]{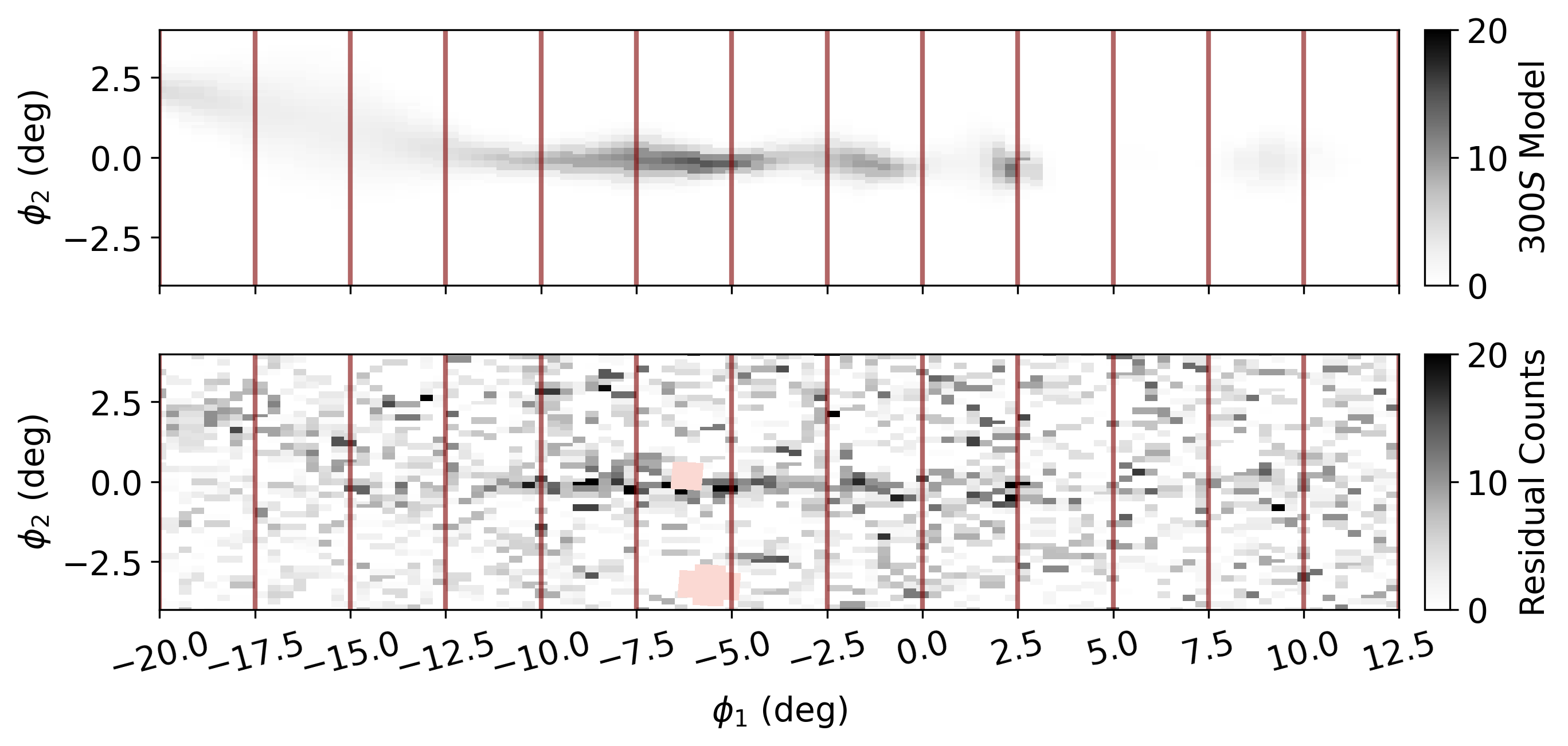}}
   \end{minipage}
   \caption{Comparison between the filtered stellar density maps and stream models produced through our two methods of filtering \sgr contamination. (a) Method 1 of signal extraction. The top panel is the \stream component of the model produced by Method 1. The bottom panel is the filtered stellar density map in the vicinity of \stream.  The region $\phi_1 < -12.5^\circ$ is excluded from Method 1 due to \sgr contamination. (b) Method 2 of signal extraction. The top panel is the \stream component of the model produced by Method 2. The bottom panel is the background and \sgr subtracted stellar density map in the vicinity of \stream. 
   The red shaded regions are the object masks.
   }
   \label{fig:Stream components of both streams}
\end{figure*}

We present the results of Method 1, including the median splines and corresponding $16\%-84\%$ quantile ranges for the stream parameters, in Figure \ref{sfig:gaia splines}. 
We present the resulting model's on-sky distribution and compare it to the filtered stellar density map in Figure \ref{sfig:Gaia stream component}.
Method 1 identifies four peaks in the stellar density at $\sim -6^\circ$, $\sim -1^\circ$, $\sim 2.5^\circ$, and $\sim 7^\circ$ (peaks A, B, C, and D respectively), although the uncertainties are substantial. 
Method 1 also finds \stream to be narrow ($w \leq 0.6^\circ$) with relatively consistent width except for a peak in width at $\phi_1\backsimeq -3^\circ$.
We discuss this peak in greater detail in Section \ref{ssec:discussion}.
Generally, the uncertainties on all of the fit parameters are high and make it challenging to obtain a clear picture of \streamPoss structure.
For example, the $16\% - 84\%$ quantile range of $\exp(\mathcal{I}(-2.5^\circ))$ is $\sim 0.15 \text{ stars/deg}^2$, which is $\sim 71\%$ of the median central stellar density, and the quantile range of $\exp(\mathcal{I}(-1^\circ))$ at the approximate location of peak B is $0.28\text{ stars/deg}^2$ or roughly $89\%$ of the median central stellar density.
We further note that widening the proper motion filters by increasing $N$ allows more \sgr contamination into the filters and can flatten the features while narrowing the filters can accentuate them. This sensitivity is due to the small number counts inherent in Method 1 and the strong contamination in the region. However, changing the exact selection criteria, such as the specific proper motion cut threshold or removing the HB stars, results in similar results within the substantial Poisson uncertainties.

To obtain higher number counts and lower uncertainties, we develop a second method of extracting \streamPoss signal that relies purely on DECaLS photometry and therefore has significantly higher number counts.
We describe this method in the next section (Section \ref{ssec:method 2 double stream model}).
We then compare and discuss our models of \stream extensively in Section \ref{ssec:discussion}.

\section{Fitting Models of Stream Morphology Method 2:\\Filtering Using DECaLS DR9 Data}\label{ssec:method 2 double stream model}

The low number counts and consequentially high uncertainties inherent in Method 1 limit our ability to draw conclusions from that model. 
This motivates the development of a model that does not require kinematic information to filter out \sgrPoss influence, allowing the use of the deeper DECaLS data alone. 
This second method is motivated by two observations. 
First, \cite{Bernard_2016} noticed that \sgrPoss distance gradient has the opposite sign compared to \streamPoss in this region. 
Second, as visible on our matched filter maps (see Figure \ref{sfig:Refined Matched Filter}), the component of \sgr that leaks through the filter is much wider than \streamPoss signal. 

\subsection{Motivation and Overview}

In the following discussion, we define ``\stream matched filter'' as the filter described in Section \ref{sssec:improved matched filter}. Specifically, it is the matched filter with $s = 2/3$ which identifies stars' absolute magnitudes using the \stream distance gradient given in Equation \ref{eq:dist_grad}. This contrasts with the ``\sgr matched filter'' which is a matched filter based on \streamPoss isochrone but that follows the \sgr distance gradient. We discuss this filter in more detail in Section \ref{ssec:empirical characterization of sgr}.

Because of the opposite nature of their distance gradients, much of \sgr that is present in the region and visible in the unfiltered stellar density map (see Figure \ref{sfig:no matched filter}) is removed after filtering (see Figure \ref{sfig:initial matched filter}) despite the two structures' similarity in distance. 
As we later show more explicitly, the only region where \sgr contaminates the map is the vicinity of the two distance gradients' intersection at $\phi_1 \sim -5.1^\circ$ where the matched filter is correctly positioned to also match \sgrPoss overdensity. 

The same effect applies in the other direction. 
If we filter using \sgrPoss distance gradient rather than \streamPoss (i.e. we use the \sgr matched filter rather than the \stream matched filter), we will capture the full extent of \sgr while limiting \streamPoss extent to the small region around the distance gradients' intersection. This result can be seen in Figure \ref{fig:Sgr Stellar Density Map} where \sgr now appears across the field while \stream is limited in extent to within the red box.
Because \sgr is much wider than \stream, it extends to much larger $\phi_2$ values than \stream. 
In these regions, its signal is uncontaminated by \streamPoss presence because \stream is relatively thin per Method 1. 
This is true even in \streamPoss $\phi_1$ range.
Using this wider region, we can fit a model to \sgrPoss morphology by applying a simple mask around the leaking \stream component that is relatively small compared to the \sgr signal of interest.
We can not do the same for \stream, as the necessary \sgr mask would be very large relative to \stream, and \sgr is long and wide enough that its contamination could still be present behind most or all of \stream. 
In short, while \sgr is non-negligible when modeling \stream, it is large enough that if we attempt to model it first, \stream becomes negligible. 
This allows us to model \sgr within the relevant region.

With a \sgr model in hand, we may then account for its influence and model \stream.
However, one difficulty remains.
The \sgr model is fit under a matched filter that uses \sgrPoss distance gradient. 
In most regions, this filter passes a different set of stars than \streamPoss matched filter. 
The resulting models, then, are not directly comparable.
To compare them, we must derive a transformation function that relates the \sgr model as it is seen under \sgrPoss matched filter to how it would be seen under \streamPoss matched filter. 

With this transformation function, we can use the \sgr model as an extremely strong prior on the \sgr component of a joint model of both streams.
This constraint prevents degeneracy between \stream and \sgr and allows us to extract \streamPoss morphology. 

We describe our empirical characterization of \sgr, including our derivation of an empirical distance gradient and stream model, in Section \ref{ssec:empirical characterization of sgr}.
Next, we derive the transformation function in Section \ref{ssec:transformation func}. Finally, we combine the \sgr model and the transformation function to produce and fit a joint model of \sgr and \stream in Section \ref{ssec:2 modeling 300S}.

\subsection{Empirical Characterization of \sgr}\label{ssec:empirical characterization of sgr}

We begin our analysis by empirically characterizing \sgr. 
This way, we may later use our \sgr model to account for its contamination into the \stream stellar density map.
We perform our characterization of \sgr in much the same way as we preliminarily characterized \stream in Section \ref{sec:empirical characterization}.

\subsubsection{Matched Filter Search and Coordinate System}\label{sssec:naive matched filter search and coordinate system}

We use the same matched filter for \sgr as we did for \stream, except we set $s = 1/6$. We do not change the isochrone parameters for our \sgr filter because we want to mimic \sgrPoss influence on \streamPoss signal as closely as possible. We set $s$ lower for \sgr than for \stream to increase the purity of the signal. 
Specifically, we wish to maximally reduce the contamination from \stream into the \sgr density map. 
This contamination occurs in a region around where the two objects' distance gradients intersect.
By reducing the magnitude range of the matched filter, we are limiting where the filter will overlap \streamPoss distance, decreasing the size of the region it contaminates.
Further, as \sgr is much larger than \stream, this thinner filter still passes enough signal to characterize the stream.  

We define a local \sgr coordinate system in a manner similar to our definition of \streamPoss coordinate system in Section \ref{ssec:naive matched filter}. 
We place two endpoints with $(\text{R.A.},\text{Decl.})$ of $(145.0^\circ,19.5^\circ)$ and $(168.0^\circ,16.0^\circ)$ on each end of the visual extent of \sgr in the unfiltered density map (Figure \ref{sfig:no matched filter}). Performing the same transformation as described in Section \ref{ssec:naive matched filter}, we find the transformation matrix

\begin{equation}
    R_{\text{sgr}} = \begin{bmatrix}
        -0.87248158 & 0.37730332 & 0.31051265 \\
 -0.43722902 & -0.88653053 & -0.15130899 \\
 0.21818955 & -0.26777945 & 0.93844951 \\
    \end{bmatrix}
\end{equation}

We describe this coordinate system as $\lambda,b$ coordinates to distinguish it from the common $\Lambda, \beta$ coordinate system used for \sgr \citep[e.g.,][]{Vasiliev_2021}. 
We use our own $\lambda,b$ coordinates rather than $\Lambda,\beta$  because our coordinates are empirically derived specifically on the \sgr response to the \stream matched filter and therefore provide a better picture for how \sgr will contaminate \stream in this particular region. 
Both the \sgr and \stream endpoints and $\phi_2=0^\circ,\,b = 0^\circ$ axes are compared in Figure \ref{sfig:no matched filter}. 

For reference, our $\phi_1,\,\phi_2$ coordinate system for \stream has its origin at $\Lambda,\beta$ of $(129.5^\circ,10.3^\circ)$ on the leading arm of \sgr in the \sgr coordinates of \cite{Vasiliev_2021}. Note that these $\Lambda,\beta$ coordinates differ from those of \citet{Majewski_2003} by the sign of $\Lambda$. Our $\lambda,\,b$ coordinate system for \sgr has its origin at $\Lambda,\beta$ of $(131.5^\circ, 7.9^\circ)$ also on the leading arm and at $(\phi_1,\phi_2) = (-1.4^\circ,  2.7^\circ)$.

\subsubsection{\sgr Distance Gradient Through Its Matched Filter}\label{sssec:sgr distance gradient through matched filter}

\begin{figure}
    \centering
    \includegraphics[width=1\linewidth]{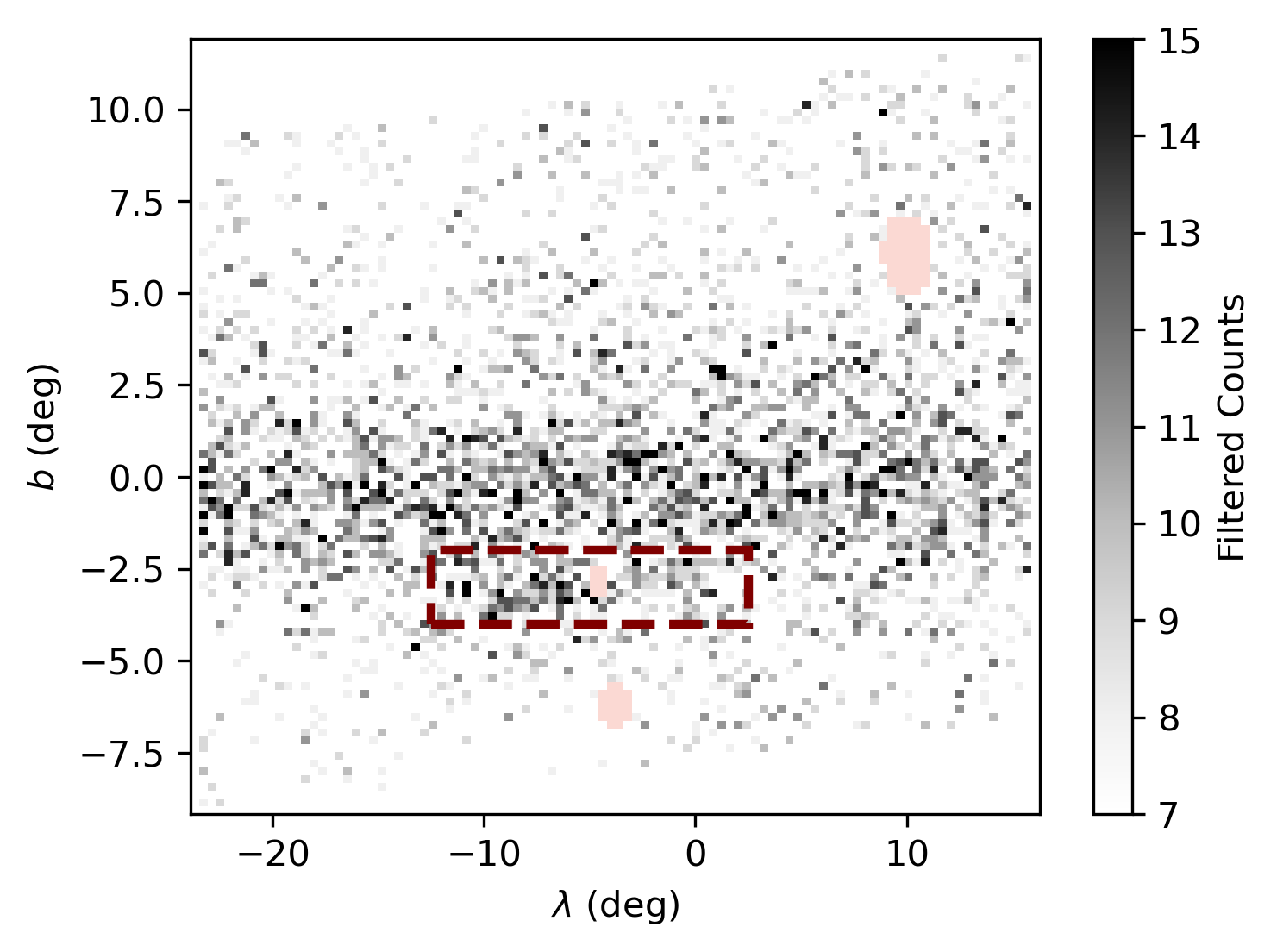}
    \caption{Filtered stellar density map of \sgr using the \sgr distance gradient derived herein and the \sgr matched filter ($s=1/6$).
    The red box represents the \stream mask which prevents cross-contamination.
    \stream can be seen as the localized overdensity within the mask.
    The overlap between the mask and the \sgr overdensity is minimal, suggesting that a strong fit may be achieved despite the mask.
    This allows us to model and account for the \sgr contamination.}
    \label{fig:Sgr Stellar Density Map}
\end{figure}

\begin{figure}
    \centering
    \includegraphics[width=1.0\linewidth]{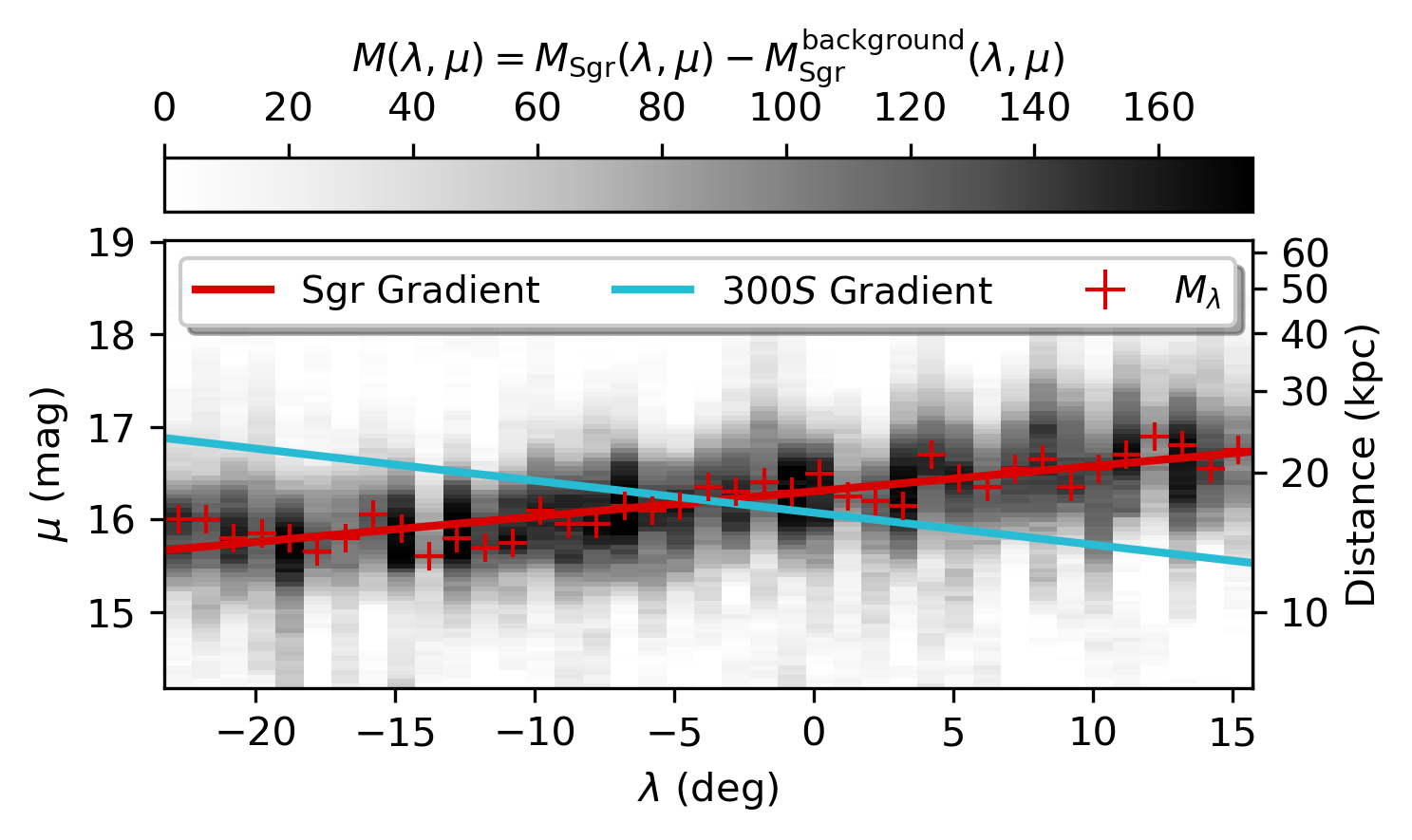}
    \caption{Derivation of empirical \sgr distance gradient. $M(\lambda,\mu)$ represents the background subtracted response to \streamPoss matched filter translated to a given $\mu$ in a given $\lambda$ bin. $M_\lambda = \max_\mu M(\lambda,\mu)$ is the peak of \sgrPoss response to the filter at $\lambda$. We then fit \sgrPoss gradient to these points.} 
    \label{fig:sgr gradient derivation}
\end{figure}

\cite{Ramos_2020} derived a distance gradient for \sgr using RRL stars found in \textit{Gaia} DR3.
However,  we are specifically interested in \sgrPoss response to our \stream matched filter. 
Therefore, a distance gradient that is specifically tuned to the peak response of \sgr to that filter rather than a generalized distance gradient is better for our application. 
As such, we derive a gradient from the \sgr response to the \stream matched filter ($s=2/3$) in this section. 

We begin by selecting an on-stream region with $|b| < 0.5^\circ$. 
Next, we bin on-stream stars based on their $\lambda$ values into bins of width $\Delta \lambda = 1^\circ$.
Generally, by measuring the response of stars in these bins to our \stream matched filter at different distance moduli, we can identify the distance of the peak response versus the $\lambda$ of the bin.
We can then fit a linear function to these peak responses. 
Specifically, we compute the match $M_\text{\sgr}(\lambda,\mu)$ as the number of stars in the $\lambda \rightarrow \lambda + \Delta \lambda$ bin that fall inside the filter's color-magnitude mask when shifted to $\mu$.
We consider $\mu$ values from $14.2$ to $19$ ($6.9 - 63.1\kpc$) with $\Delta \mu = 0.05$ 
for each $\lambda$ bin.
We subtract background by also considering bins with $|b-b_\text{background}|<0.5^\circ$ for $b_\text{background} = -5^\circ,\,+4.5^\circ$. 
These values were chosen to avoid intersection with the masked regions or \stream. 
We then compute $M^\text{(background)}_\text{\sgr}(\lambda,\mu)$ by linearly interpolating the matches in these two background regions to $b=0^\circ$ and obtain a final match $M(\lambda,\mu) = M_\text{\sgr}(\lambda,\mu) - M^\text{(background)}_\text{\sgr}(\lambda,\mu)$.
We then find the peak response by taking $M_\lambda = \max_\mu M(\lambda,\mu)$ for each $\lambda$ and fit a linear function to the points $(\lambda,M_\lambda)$.
This procedure results in a linear function

\begin{equation}
    \mu_\text{sgr}(\lambda) = 0.027\lambda + 16.30
\end{equation}

\noindent which we plot on top of $M(\lambda,\mu)$ and the \stream distance gradient in Figure \ref{fig:sgr gradient derivation}. 
We also show the values of $M_\lambda$, which closely follow the derived gradient.
The \stream gradient in the figure is derived by transforming the set of points $(\lambda,0)$ into $(\phi_1,\phi_2)$ and calculating $\mu_{\stream}(\phi_1)$ using Equation \ref{eq:dist_grad}.  

As seen in Figure \ref{fig:sgr gradient derivation}, the distance moduli overlap at $\lambda \simeq -3.7^\circ$ or $\phi_1 \simeq -5.1^\circ $.
This position is close to the center of the \sgr overdensity visible in the matched filter map of \stream in Figure \ref{sfig:Refined Matched Filter}. 

\begin{figure}[ht!]
    \centering
    \begin{minipage}{1\linewidth}
    \centering \includegraphics[width=1\linewidth]{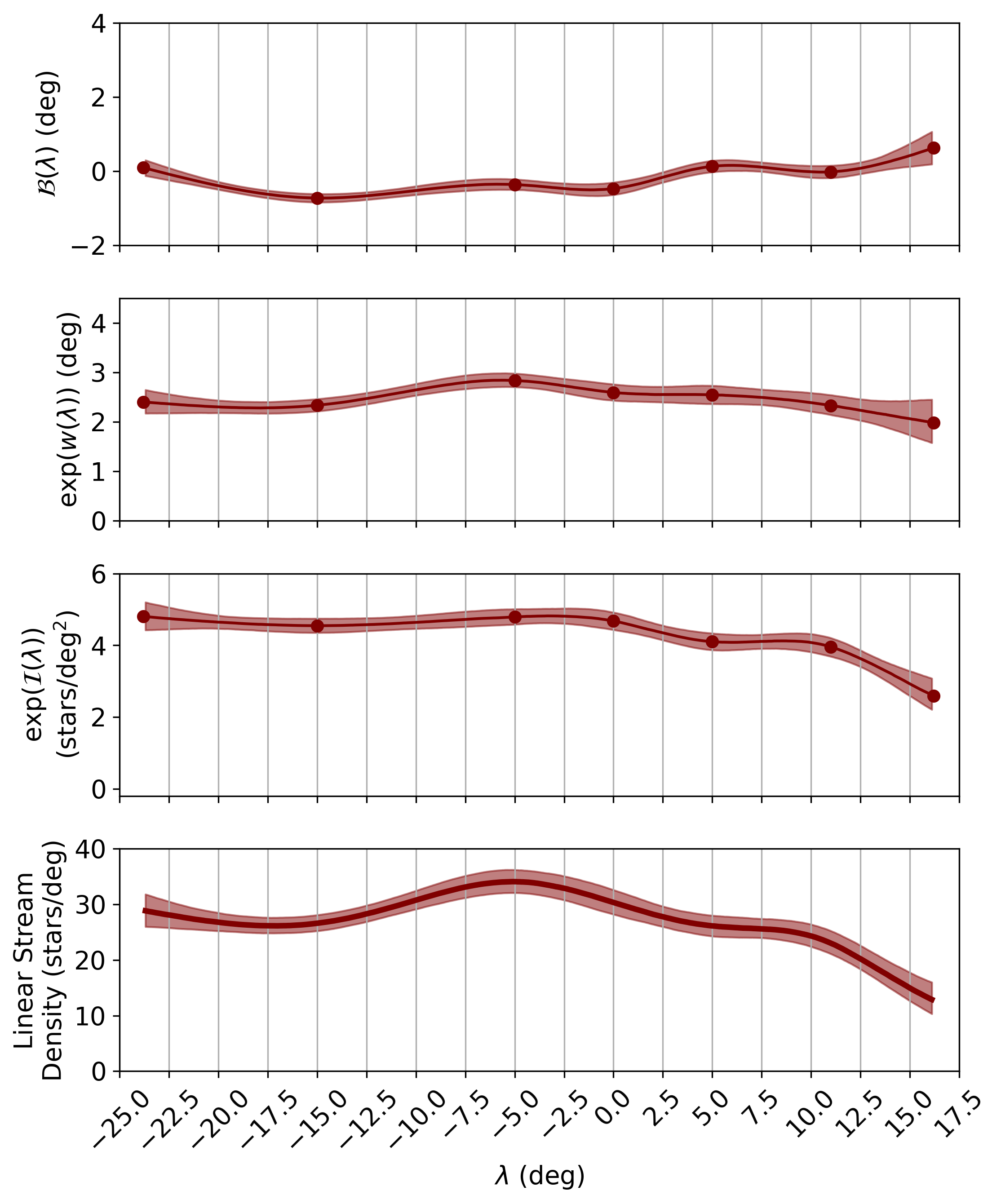}
    \end{minipage}
    \begin{minipage}{1\linewidth}
        \centering
        \includegraphics[width=1.0\linewidth]{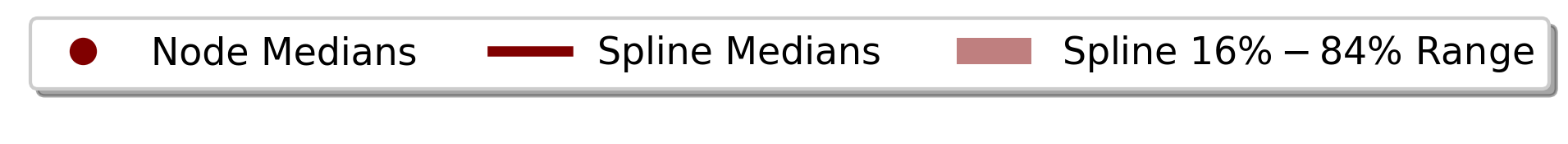}
    \end{minipage}
    \caption{Resulting spline model for \sgr. 
    The first plot corresponds to $\mathcal{B}(\lambda)$, the $b$ position of \sgrPoss track. 
    The second plot demonstrates $\exp (w(\lambda))$ as the Gaussian stream width. 
    The third plot shows $\exp(\mathcal{I}(\lambda))$, the central stellar density of \sgr. 
    As in Figure \ref{fig:Spline Values for the Two Streams}, both $\mathcal{I}$ and $w$ are fit in $\log$ space and their splines and quantile ranges are then transformed to linear space.
    The fourth plot shows the linear stream density.
    We set the y scale of the first plot to be similar to that of Figure \ref{fig:Spline Values for the Two Streams} and set the lower bounds of the remaining plots at zero to emphasize the relative homogeneity of the \sgr model across the relevant region.
    }
    \label{fig:Sgr Splines}
\end{figure}

\subsubsection{Modeling \sgrPoss Morphology}\label{ssec:modeling Sgr morphology}

We model \sgrPoss morphology using a method similar to that which we used to model \streamPoss morphology in Section \ref{ssec:method 1 proper motion filtering}. 
We use the \sgr matched filter ($s=1/6$) and the distance gradient derived in the previous section to create a filtered stellar density map in $\lambda,b$ coordinates with bins of size $\Delta \lambda = 0.4^\circ$, $\Delta b = 0.2^\circ$. To simplify the region, we only consider $\phi_2 > -5^\circ$.
We show this map in Figure \ref{fig:Sgr Stellar Density Map}. 

\stream is visible in this map as the small overdensity at $\lambda \simeq -5^\circ,b \simeq -3^\circ$. 
In order to remove its influence on the \sgr empirical model, we mask the region defined by a rectangle at $\lambda = -12.5^\circ$, $b = -4^\circ$ with a width of $15^\circ$ and height of $2^\circ$. 
This mask is also shown in Figure \ref{fig:Sgr Stellar Density Map}. We additionally mask the objects described in Section \ref{sec:data sources and preparation}. 
As seen in Figure \ref{fig:Sgr Stellar Density Map}, the overlap between the \stream mask and the \sgr overdensity is minimal. This implies that we can still fit a good model of \sgr despite the mask.

We then fit the same cubic spline based model onto this density map as we used in Section \ref{ssec:method 1 proper motion filtering}, except in this case all of the parameters are in terms of $\lambda$ rather than $\phi_1$ and $\Phi_2(\phi_1)$ is replaced by $\mathcal{B}(\lambda)$ as the $b$ position of \sgrPoss track.
In this case, we use 7 nodes for $\mathcal{I}(\lambda)$, $w(\lambda)$, and $\mathcal{B}(\lambda)$.
We use the same priors as \citet{Koposov_2019_orphan} for the \sgr model: $\mathcal{N}(0,2.5)$ for $\mathcal{B}(\lambda)$ and $\mathcal{N}(\log0.9,0.5)$ for $w(\lambda)$.
We use 3 nodes for $\Vec{B}(\lambda)$ to avoid degeneracies between the background and the wide \sgr stream. 
We again run the model for $700$ warmup iterations and $800$ sampling iterations and again find a satisfactory $\hat{R}<1.1$ for all parameters, indicating convergence \citep{Gelman_and_Rubin_92}.\footnote{We again made use of \textsc{stan-splines}, \url{https://zenodo.org/records/14163685}, an implementation of natural cubic splines in \textsc{stan} \citep{Koposov_2019_orphan}.}

\begin{figure}
    \centering
    \includegraphics[width=1\linewidth]{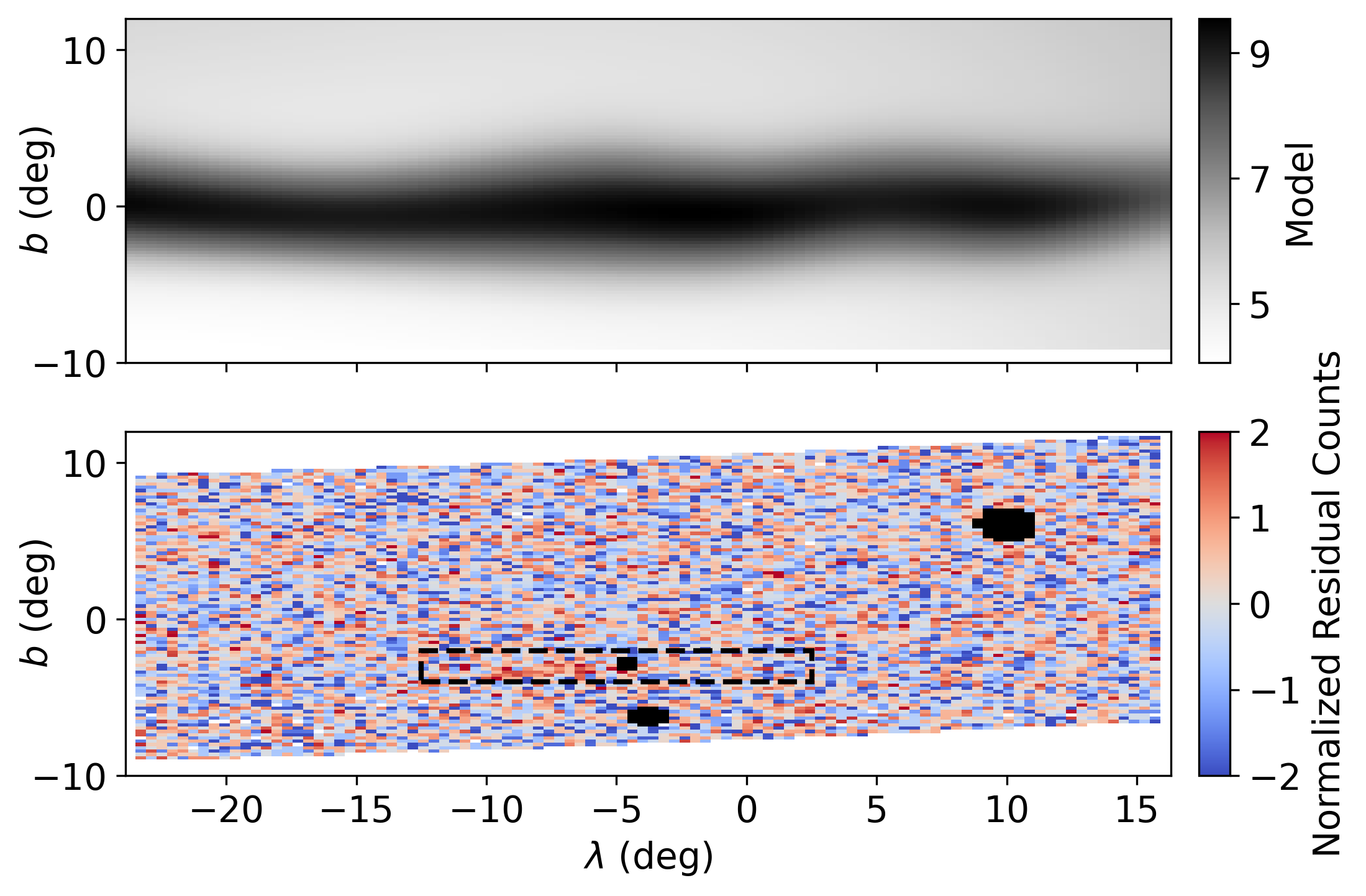}
    \caption{\sgr stream model and residual map. 
    We normalize the residual map by dividing out by each bin's Poisson uncertainty. The dashed box corresponds to the \stream mask used while fitting \sgr, see Section \ref{ssec:modeling Sgr morphology}.
    \stream is clearly visible as the overdensity in the residual map at $\lambda \simeq -5^\circ,\,b \simeq -3^\circ$. The lack of any large structures within the residual map except for \stream indicates that little \sgr structure will be unaccounted for by the model. }
    \label{fig:Sgr Model and Residual Map}
\end{figure}

We show the resulting splines for \sgrPoss model in Figure \ref{fig:Sgr Splines}.
\sgrPoss width is very consistent in the region of interest. 
Between $-15^\circ\leq \lambda \leq 10^\circ$ where \sgr contaminates the \stream map, \sgrPoss width changes by $\sim 0.5^\circ$ and the track remains within $|b|\lesssim 0.7^\circ$. 
For reference, the smallest $|b|$ value of the $\phi_2=0^\circ$ line in this region is $|b|=1.9^\circ$. 
This consistency suggests that interpolating the \sgr model into the masked region is a reasonable assumption. 
This is further supported by the lack of any clear, large overdensities in the residual map except for \stream, as seen in Figure \ref{fig:Sgr Model and Residual Map}.

\subsection{Deriving the Transformation Function Between the Map Under \sgrPoss Distance Gradient and \streamPoss}\label{ssec:transformation func}

To use the \sgr model we derived in Section \ref{ssec:empirical characterization of sgr} to account for \sgr contamination in the \stream density map, we must account for the difference in the filters used to generate the \sgr model and the \stream stellar density map.
In other words, if the empirical model of \sgr is the ``true'' underlying stream, we must know what aspects of it would pass through the \stream matched filter into the \stream density map. 
In this section, we empirically derive a function $T$ that will perform this transformation. 
Specifically, if $\Lambda^\text{\sgr filter}_\text{\sgr}(\lambda,b)$ is the underlying bin-wise Poisson rate component for \sgr under the \sgr matched filter (as described in Section \ref{ssec:modeling Sgr morphology}) and $\Lambda^\text{\stream filter}_\text{\sgr}(\lambda,b)$ is the visible bin-wise Poisson rate for \sgr under the \stream matched filter, we write

\begin{equation}\label{eq:transformation function definition}
    \Lambda^\text{\stream Filter}_\text{\sgr} = T(\lambda,b) \times \Lambda^\text{\sgr Filter}_\text{\sgr}
\end{equation}

\noindent where $T(\lambda,b)$ is some function. We can further constrain $T(\lambda,b)$. We assume that \sgrPoss normalized distribution in color-magnitude space is constant -- or that its overall distribution is constant up to scalar -- within a given $\lambda$ cross section. 
One reason this assumption could fail is a metallicity gradient in \sgr \citep[e.g.,][]{Hayes_2020, Limberg_2023}, but we do not expect this to be significant.
We anticipate internal stellar population variations in \sgr to be negligible for our purposes.
Using a large sample of RGB stars, \cite{Cunningham_2024} fit a metallicity gradient to \sgr in both $\Lambda$ and $\beta$. 
On the leading arm in the north where the origins of both of our coordinate systems are, they found $\nabla \feh = [-0.00125, -0.0178] \dex/\deg$. Over the $\sim 6^\circ$ in $\beta$ within a given $\lambda$ cross-section, \sgrPoss metallicity changes on the order of $0.1\dex$. The corresponding changes in the color-magnitude distribution are negligible compared to the width of our matched filters, especially along the main sequence.

To use this assumption to constrain $T(\lambda)$, consider a set of bins at some $\lambda$. By assumption, the density of \sgr is fixed up to scalar multiplier in color-magnitude space at this $\lambda$. To be precise, let $D_\lambda(g-r,g)$ be the normalized distribution. Then we may write the distribution within a specific $(\lambda,b)$ bin as $S_\lambda(b)D_\lambda(g-r,g)$ where $S_\lambda$ is the $b$ dependent scalar. Note that the operations of multiplying by the scaler $S_\lambda(b)$ and applying a matched filter $F_\mu$ commute, i.e., the density of stars $\Lambda^F_\text{Sgr}$ matched to the filter is $\Lambda^F_\text{Sgr} = F_\mu\bigl (S_\lambda(b)D_\lambda(g-r,g)\bigr ) = S_\lambda(b) F_\mu\bigl(D_\lambda(g-r,g))$. This implies that, given a $(\lambda,b)$ bin, Equation \ref{eq:transformation function definition} may be rewritten as:

\begin{align}
    \nonumber T(\lambda,b) = \frac{\Lambda^\text{\stream Filter}_\text{\sgr}}{\Lambda^\text{\sgr Filter}_\text{\sgr}} = \frac{F_{\mu(\phi_1)}^\text{\stream}\bigl(S_\lambda(b)D_\lambda(g-r,g))}{F_{\mu'(\lambda)}^\text{\sgr}\bigl(S_\lambda(b)D_\lambda(g-r,g))}\\
    \nonumber= \frac{S_\lambda(b)}{S_\lambda(b)}\cdot \frac{F_{\mu(\phi_1)}^\text{\stream}\bigl(D_\lambda(g-r,g))}{F_{\mu'(\lambda)}^\text{\sgr}\bigl(D_\lambda(g-r,g))} 
    = \frac{F_{\mu(\phi_1)}^\text{\stream}\bigl(D_\lambda(g-r,g))}{F_{\mu'(\lambda)}^\text{\sgr}\bigl(D_\lambda(g-r,g))}\\
\end{align}

Further, the only spatial dependence the filters have is their vertical translations $\mu(\phi_1)$ and $\mu'(\lambda)$ due to their respective distance gradients. This implies that $F_\mu\bigl(D_\lambda(g-r,g))$ only depends on $D_\lambda$'s relation to $\lambda$; and the filters' dependence on the distance gradients. So the variables involved are $\lambda$, $\mu'(\lambda) = \mu_\text{\sgr}(\lambda)$, and $\mu(\phi_1) = \mu_{\stream}(\phi_1(\lambda,b))$. 

Because of the similarity in the stream tracks, we may further assume that $\phi_1$ is a function purely of $\lambda$ near $b=0^\circ$. Along the $b=0^\circ$ line, this assumption is exact.
In the region $\lambda = -25^\circ$ to $\lambda = 15^\circ$, the quantity $\left |\phi_1(\lambda,b_1)-\phi_1(\lambda,b_2)\right |$ is $<0.31^\circ$ $\forall\, b_1,b_2\in [-2^\circ,2^\circ]$. This corresponds to a difference in $\mu_{\stream}$ of $0.01$. 
When $b_1,b_2 \in $ $[-6^\circ,2^\circ]$ which includes \stream, this quantity is bounded by $0.63^\circ$ corresponding to $\Delta \mu_{\stream} = 0.02$. In both cases, $\Delta \mu_{\stream}$ is well under the width of \streamPoss matched filter.

Therefore, it is reasonable to assume $F_{\mu(\phi_1)}^\text{\stream}\bigl(D_\lambda(g-r,g))$ and $F_{\mu'(\lambda)}^\text{\sgr}\bigl(D_\lambda(g-r,g))$ are functions only of $\lambda$. By the above discussion, it is then reasonable to assume that $T$ only depends on $\lambda$ in this region. Therefore, we may write $T(\lambda,b) = T(\lambda)$.\footnote{
With these assumptions, there is a closely related description. By the assumption that the color-magnitude distribution is constant up to scaler in $b$, we can write the filtered signal 
$$\Lambda_\text{\sgr}^\text{Filter} = I_\text{\sgr}(\lambda,b) \times f^\text{Filter}_\text{\sgr}(\lambda)$$ 
\noindent where $I_\text{\sgr}(\lambda,b)$ represents the integrated stellar density in a bin and $f^\text{Filter}_\text{\sgr}(\lambda)$ represents the fraction of the \sgr signal that passes through the filter at $\lambda$. In this depiction, $f$ contains information about both the filter and the shape of the color-magnitude distribution at $\lambda$. The intensity $I_\text{\sgr}$ contains information about the scaling of the distribution. Then again we find
$$T = \frac{\Lambda_\text{\sgr}^\text{\stream Filter}}{\Lambda_\text{\sgr}^\text{\sgr Filter}} = \frac{I_\text{\sgr}(\lambda,b) \times f^\text{\stream Filter}_\text{\sgr}(\lambda)}{I_\text{\sgr}(\lambda,b) \times f^\text{\sgr Filter}_\text{\sgr}(\lambda)}$$
\noindent is purely a function of $\lambda$.
} 

\begin{table}[]
    \centering
    \caption{Coefficients for the Fit of $T(\lambda)$.}
    \begin{tabular}{c|c|c|c|c}
        $a_0$ & $a_1$ & $a_2$ & $a_3$ \\
        \hline
        $3.10620$ & $-0.04286$  & $-0.00908$ & $-0.00008$ \\
    \end{tabular}
    \label{tab:T fit}
\end{table}

\begin{figure}
    \centering
    \includegraphics[width=1\linewidth]{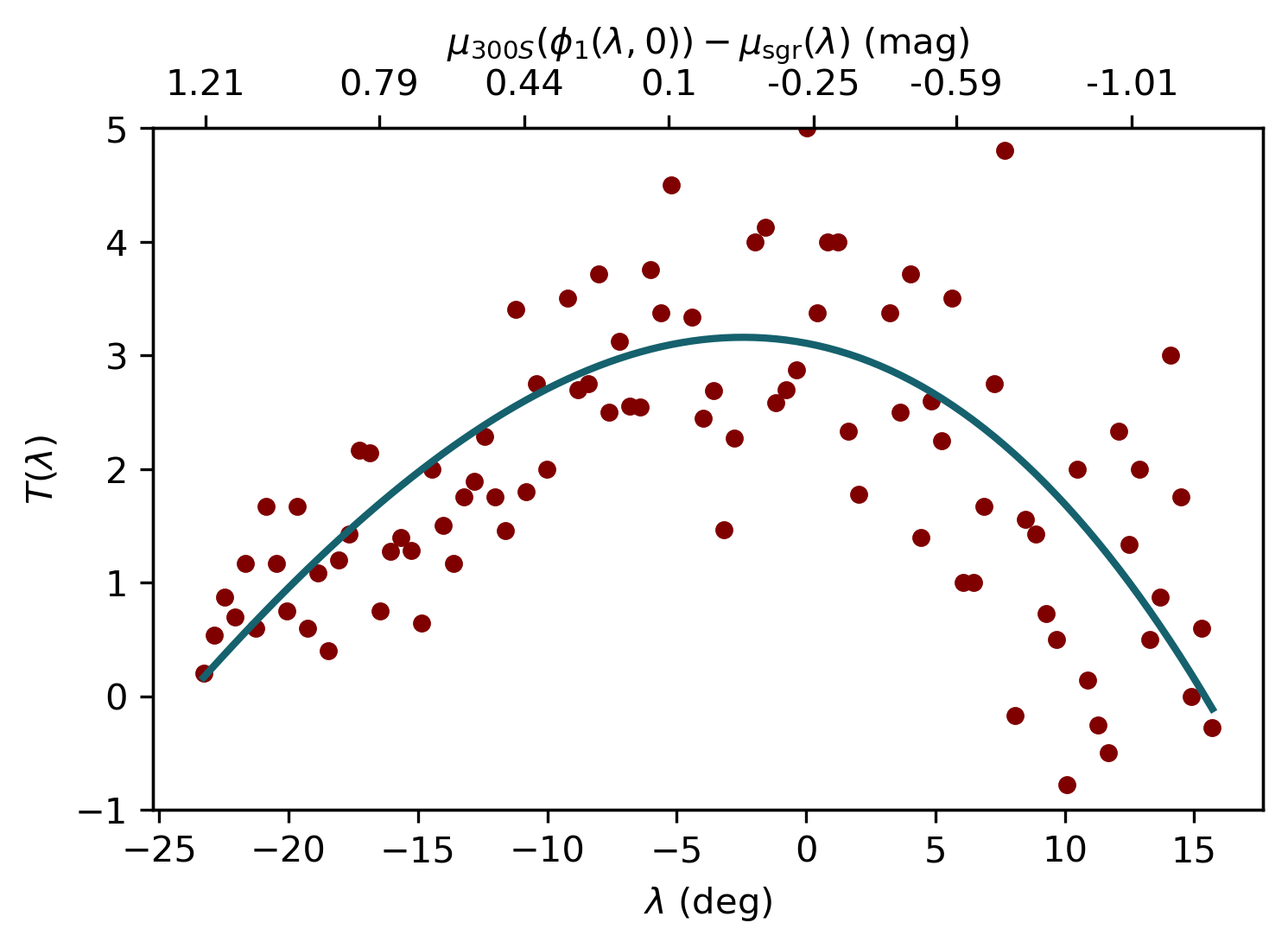}
    \caption{Fit of transformation function $T(\lambda)$. The red circles represent the empirical ratio $\Lambda_\text{\sgr}^\text{\stream Filter}/\Lambda_\text{\sgr}^\text{\sgr Filter}$ calculated using Equation \ref{eq:transformation function use}. 
    By multiplying the \sgr model by our fit of $T(\lambda)$, we are able to mimic how \sgr leaks through \streamPoss matched filter.
    For visualization purposes, we exclude five outlier points with $T(\lambda) > 5$ from this plot.
    These outliers and the points with negative $T(\lambda)$ likely originate from abnormal values of either the foreground or background. 
    }
    \label{fig:transformation fit}
\end{figure}
We empirically derive $T(\lambda)$ by comparing the stellar density map of \sgr as produced by \streamPoss matched filter to that produced by \sgrPoss filter.
To begin, we compute a comparable density map for \streamPoss filter by binning matching stars into the same $(\lambda,b)$ bins that we used for \sgrPoss density map in Section \ref{ssec:empirical characterization of sgr}. 
Next, we perform the following reduction for both maps. Given a $\lambda$ value, we take the median count over the corresponding bins with $\left |b_\text{bin}\right | < 1.5^\circ$ to reduce the impact of stochastic density fluctuations in the ratio $T$.
We call these foreground medians $C_\text{\stream}(\lambda)$ and $C_\text{\sgr}(\lambda)$ for \stream and \sgrPoss density maps respectively.
We also compute off-\sgr backgrounds as the average of the backgrounds at two locations. 
Specifically, we calculate the medians of same-$\lambda$ bins for  $\left | b_\text{bin} + b_\text{background}\right | < 1.5^\circ$ with the two constraints $b_\text{background} = 6^\circ,\,-6^\circ$ for each map.
We then calculate the average of the binned counts between these two values of $b_\text{background}$ to obtain a $\lambda$-dependent background for both \stream and \sgrPoss maps. We call these background values $B_\text{\stream}(\lambda)$ and $B_\text{\sgr}(\lambda)$ respectively. Then we can calculate the ratio 
\begin{equation}
    \label{eq:transformation function use}
    T(\lambda) = \frac{\Lambda_\text{\sgr}^\text{\stream Filter}}{\Lambda_\text{\sgr}^\text{\sgr Filter}} = \frac{C_\text{\stream}-B_\text{\stream}}{C_\text{\sgr}-B_\text{\sgr}}
\end{equation} 
at each $\lambda$-bin. 

We fit a cubic polynomial to this ratio for use in our final \stream model.
The fit is shown in Figure \ref{fig:transformation fit} and the coefficients are provided in Table \ref{tab:T fit}. 

\begin{figure*}[ht!]
    \centering
    \subfigure[\label{fig:sgr transformed model}]{\includegraphics[width = 0.45\linewidth]{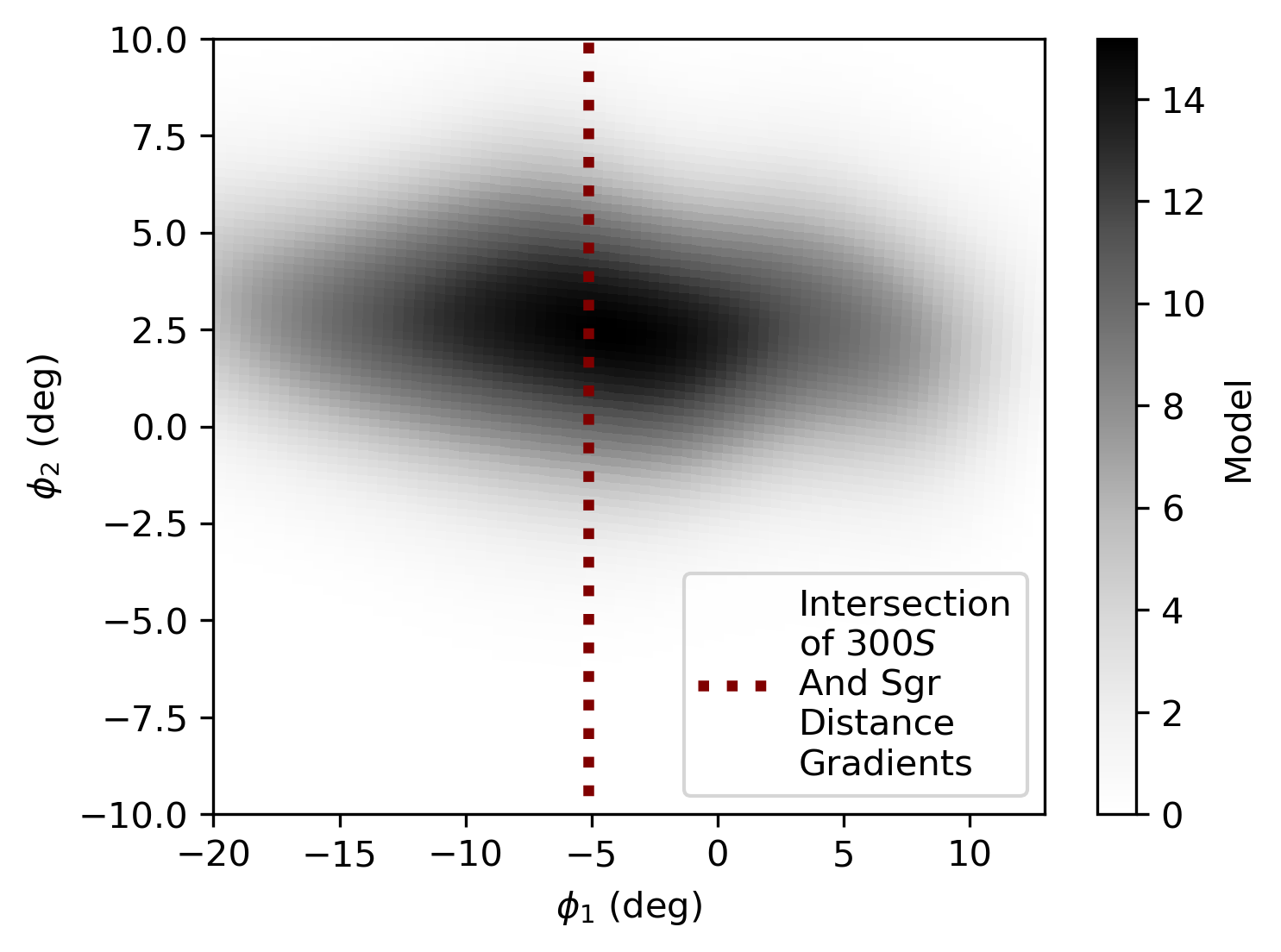}}
    \subfigure[\label{fig:sgr subtracted intensity}]{\includegraphics[width = 0.45\linewidth]{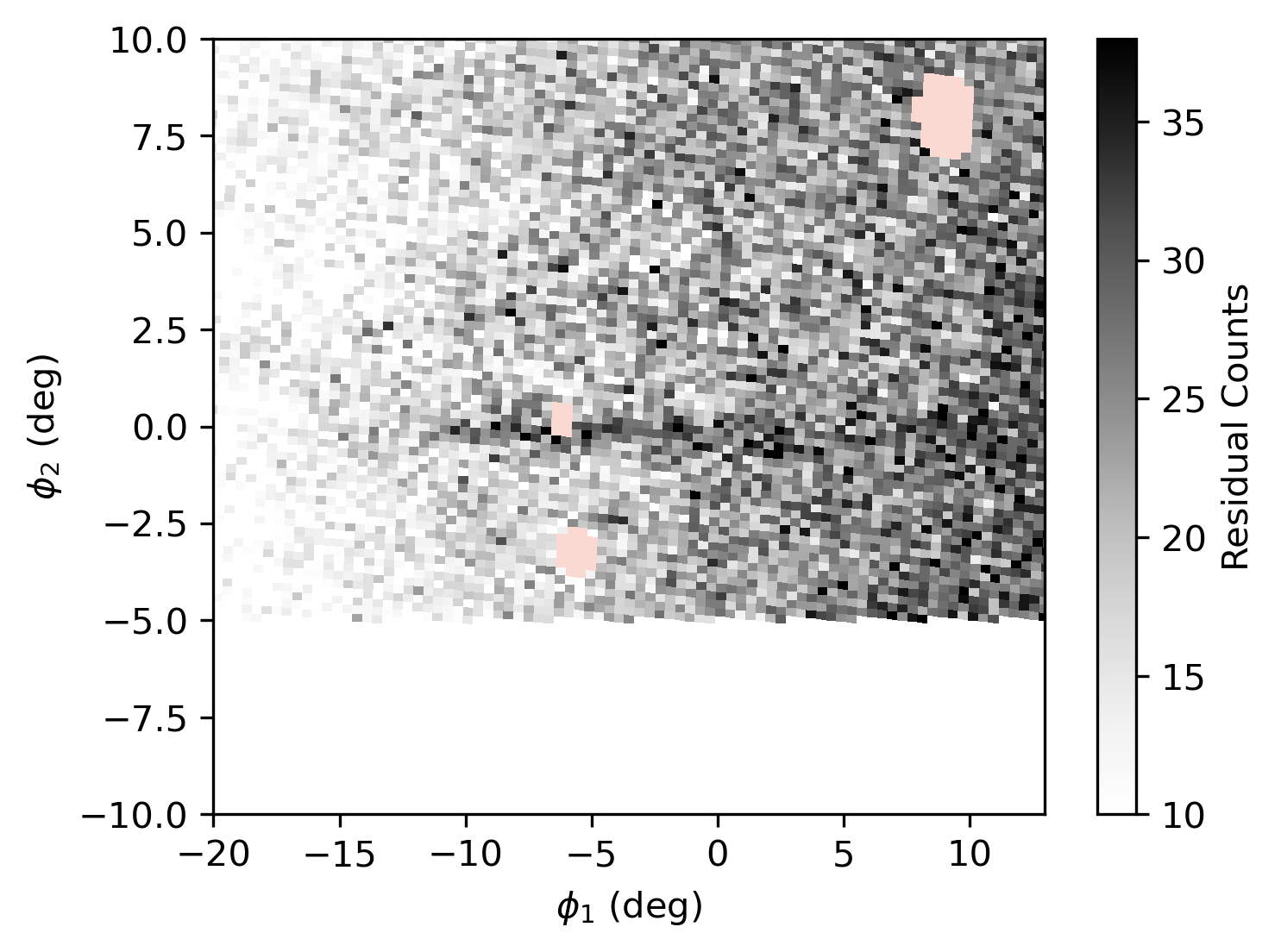}}
    \subfigure[\label{fig:sgr and background subtracted intensity}]{\includegraphics[width = 0.45\linewidth]{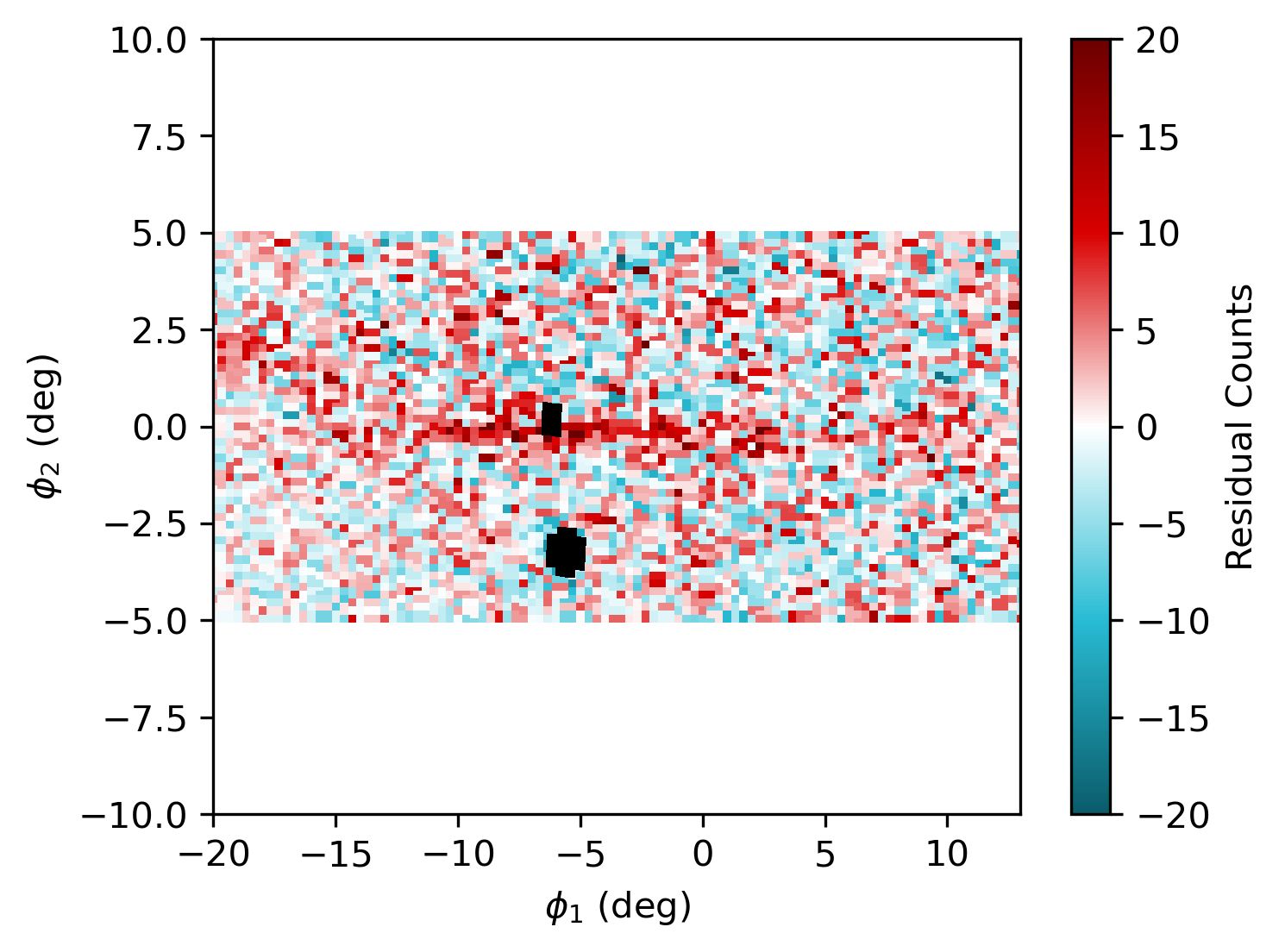}}
    \subfigure[\label{fig:close view sgr and background subtracted}]{%
    \begin{minipage}{1\linewidth}
    \begin{center}
        \includegraphics[width = 1\linewidth]{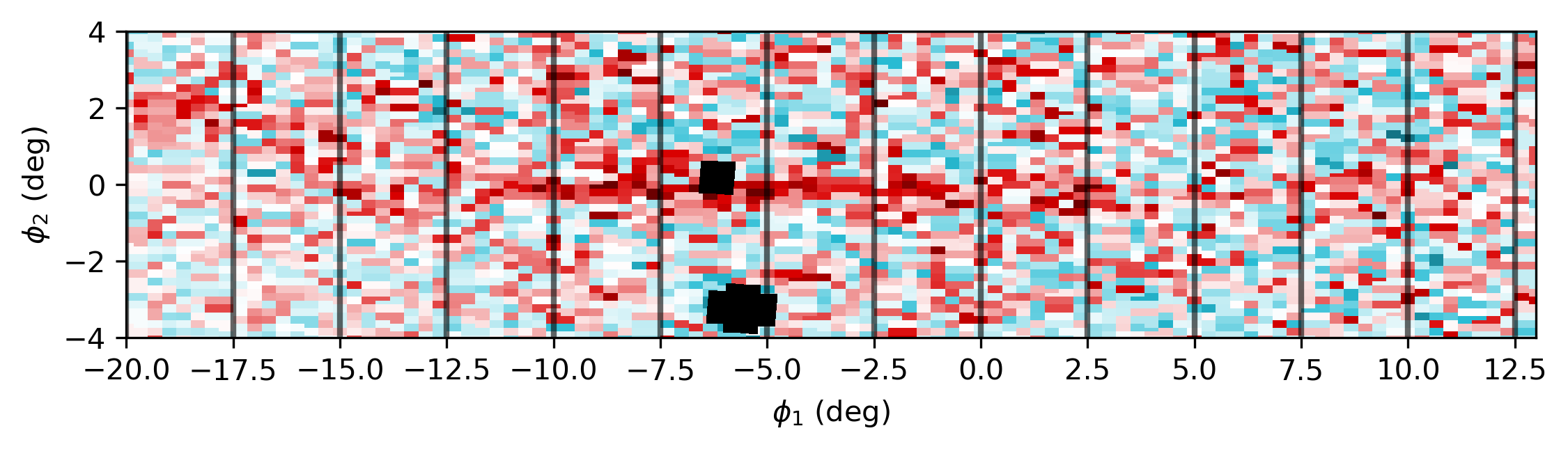}
    \end{center}
    \end{minipage}}
    \caption{Impact of $T(\lambda)$ on \sgr and subsequent subtraction. (a) The \sgr model (as shown in Figure \ref{fig:Sgr Model and Residual Map}) after being transformed using $T(\lambda)$. Note how the intensity peaks near the intersection of \stream and \sgrPoss distance gradients and falls off on either side as the gradients diverge. 
    (b) The stellar density map of \stream with the transformed \sgr model subtracted out.
    \stream remains clearly present in this density map as an overdensity along $\phi_2=0^\circ$ while the \sgr overdensity present in Figure \ref{sfig:Refined Matched Filter} has been removed. 
    The region below $\phi_2 = -5^\circ$ has been removed because it is excluded from the \sgr model fit. 
    The red shaded regions are the object masks.
    We keep the proportions of Figure \ref{sfig:Refined Matched Filter} in this figure for reference. 
    (c) The interpolated stellar density map of \stream after both $\Lambda_\text{background}$ and the transformed $\Lambda_\text{\sgr}$ as fit in Section \ref{ssec:2 modeling 300S} have been subtracted out. 
    In (c) and (d), we apply nearest neighbor interpolation to transform the $\lambda,b$ bins into $\phi_1,\phi_2$ bins for easier comparison.
    The black overlays are the bin masks due to the object masks applied to the datasets in Section \ref{sec:data sources and preparation}.
    As with the \sgr model, extraneous regions are removed in the fit to emphasize the stream signal. 
    Only $\phi_2 \in [-5^\circ,5^\circ]$ is shown to match the region \streamPoss model is fit upon.
    Again, we maintain the proportions of the map for easy comparison.
    (d) A closer look at the resulting stream signal after both $\Lambda_\text{background}$ and the transformed $\Lambda_\text{\sgr}$ have been subtracted out.
    The bin coloration is the same as in (c).
    \stream is now clear against the residual map.
    No other large structures within the residual map are visible, indicating that the contamination in the region is being correctly described. 
    Further, many of the major stream features are clearly visible in this map.
    The four peaks in the stream density are visible at $\phi_1 \simeq -6^\circ,\,-1^\circ,\,2.5^\circ,\,8^\circ$ (peaks A, B, C, and D respectively). 
    The possible gap can be seen as the blue region between $\phi_1 \simeq 2.5^\circ$ and $\phi_1 \simeq 5^\circ-7.5^\circ$
    The kink is visible as the angled overdensity visible between $\phi_1 = -20^\circ$ to $-12.5^\circ$.
    }
    \label{fig:transformed and subtracted sgr}
\end{figure*}

We can now virtually pass the \sgr model through the \stream matched filter by multiplying it by $T(\lambda)$. 
The result of this transformation on the \sgr model can be seen in Figure \ref{fig:sgr transformed model}. 
Here, the \sgr model appears similarly to the \sgr overdensity in Figure \ref{sfig:Refined Matched Filter}.
It peaks around $\phi_1 \simeq -5^\circ$ and decays in both directions as $\left |\Delta \mu\right |$ increases. 
This $\phi_1$ is close to the intersection of the distance gradients, as can be seen in Figure \ref{fig:sgr transformed model}.
Subtracting this transformed model of \sgr from \streamPoss stellar density map, we obtain the map in Figure \ref{fig:sgr subtracted intensity}. In this map, we can clearly see \stream as an elongated overdensity around $\phi_2 = 0^\circ$ and \sgrPoss contamination is no longer clearly visible. \stream is especially visible when we also subtract the background, as seen in Figures \ref{fig:sgr and background subtracted intensity} and \ref{fig:close view sgr and background subtracted}.

\subsection{Modeling \streamPoss morphology}\label{ssec:2 modeling 300S}

Using the transformation of the \sgr model from the previous section, we are able to model \stream itself.
Like in our model using \textit{Gaia} in Section \ref{ssec:method 1 proper motion filtering}, we define $\mathcal{I}(\phi_1)$, $\Phi_2(\phi_1)$, and $w(\phi_1)$ as the log central stellar density, $\phi_2$ position, and log Gaussian width respectively for \stream.
Using these components, we write the stream density $\Lambda_\text{\stream}(\phi_1)$ using Equation \ref{eq:300S Density}.
We similarly continue to use a linear background model, $\Lambda_\text{background}(\phi_1)$ with $\Vec{B}\in \mathbb{R}^2$. 

To account for \sgr, we add to this model a \sgr component, $\Lambda_\text{\sgr}(\lambda)$, like the one we fit in Section \ref{ssec:empirical characterization of sgr}.
This model is parameterized by $\mathcal{I}_\text{sgr}(\lambda)$, $\mathcal{B}_\text{sgr}(\lambda)$, and $w_\text{sgr}(\lambda)$ as \sgrPoss log central stellar density, $b$ position, and log Gaussian width respectively.
To avoid the substantial degeneracy between \sgrPoss components and \streamPoss, we use the pure \sgr model and transformation function we derived in the previous sections to constrain the \sgr component.
Using the \sgr model, we place smooth priors of $\mathcal{N}(\mu_\text{\sgr},\sigma_\text{\sgr})$ on each node value where $\mu_\text{\sgr}$ and $\sigma_\text{\sgr}$ are the median and standard deviation of the corresponding node's value in the pure \sgr model as fit in Section \ref{ssec:empirical characterization of sgr}. 
We additionally require that this value is within $50\sigma_\text{\sgr}$ of the corresponding value in the \sgr model. 
Then, we include the transformation function within the model to write the full density as

\begin{align}
    \nonumber\Lambda(\phi_1,\phi_2) = \Lambda_\text{\stream}(\phi_1,\phi_2) \\+ \nonumber\Lambda_\text{background}(\phi_1,\phi_2) \\+ T(\lambda(\phi_1,\phi_2))\times\Lambda_\text{\sgr}(\lambda(\phi_1,\phi_2),b(\phi_1,\phi_2))
\end{align}

We use the same priors for the \stream nodes as in Section \ref{ssec:method 1 proper motion filtering}: $\mathcal{N}(0,1)$ for $\Phi_2(\phi_1)$ and $\mathcal{N}(\log 0.4,0.5)$ for $w(\phi_1)$. 
Like with our other stream models, we assume that the stellar density is binwise Poisson distributed with a spatially varying rate.
We then define the likelihood using Equation \ref{eq:loglik} and sample the posterior distribution using \textsc{stan}. 
We perform this sampling using the same bins that we have used in the rest of this section (described in Section \ref{ssec:empirical characterization of sgr}) to maintain consistency between the \sgr model, $T(\lambda)$, and the \stream model.
We use $16$ nodes for the stream parameters $\mathcal{I}(\phi_1)$, $w(\phi_1)$, and $\Phi_2(\phi_1)$, the same nodes as in Section \ref{ssec:empirical characterization of sgr} for $\mathcal{I}_\text{\sgr}(\lambda)$, $w_\text{\sgr}(\lambda)$, and $\mathcal{B}_\text{\sgr}(\lambda)$, and $6$ nodes for the background parameter $\Vec{B}$. 
The specific choice of node positions does not significantly alter the resulting fit.
To simplify the background, we constrain the region to $\phi_2 \in [-5^\circ,5^\circ]$. 
We run the model for 700 warm-up iterations and 800 sampling iterations using 4 chains.\footnote{We again made use of \textsc{stan-splines}: \url{https://zenodo.org/records/14163685} \citep{Koposov_2019_orphan}.}
As with our other models, we achieve a satisfactory $\hat{R}<1.1$ on each parameter, indicating convergence \citep{Gelman_and_Rubin_92}. 

We present the resulting median splines and corresponding $16\% - 84\%$ quantile ranges for the stream parameters extracted using our second method in Figure \ref{sfig:sub splines}.
We compare the stream component of the model with the background and \sgr subtracted density map in Figure \ref{sfig:sub stream component}
We discuss this model and compare it to the \textit{Gaia}-based model in Section \ref{ssec:discussion} and Figure \ref{fig:stream results figure}.

\section{Discussion of Models}\label{ssec:discussion}

\begin{figure*}
    \centering
    \begin{minipage}{1\linewidth}
    \subfigure[\label{sfig:empirical models onsky distribution}]{\includegraphics[width = 1\linewidth]{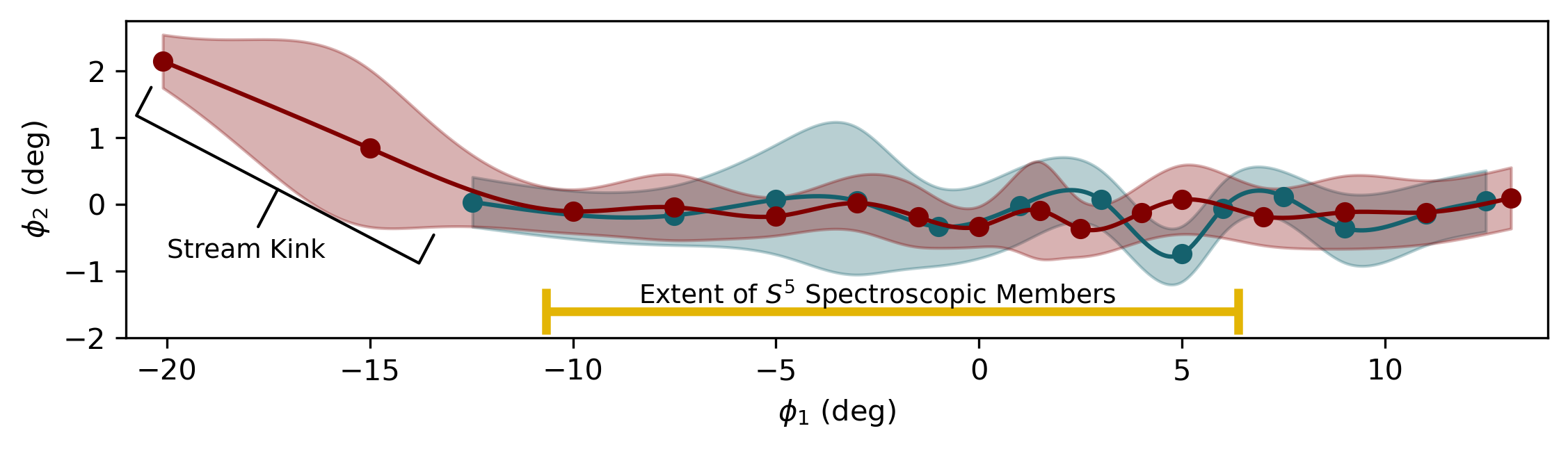}}
    \subfigure[\label{sfig:empirical models intensity}]{\includegraphics[width = 1\linewidth]{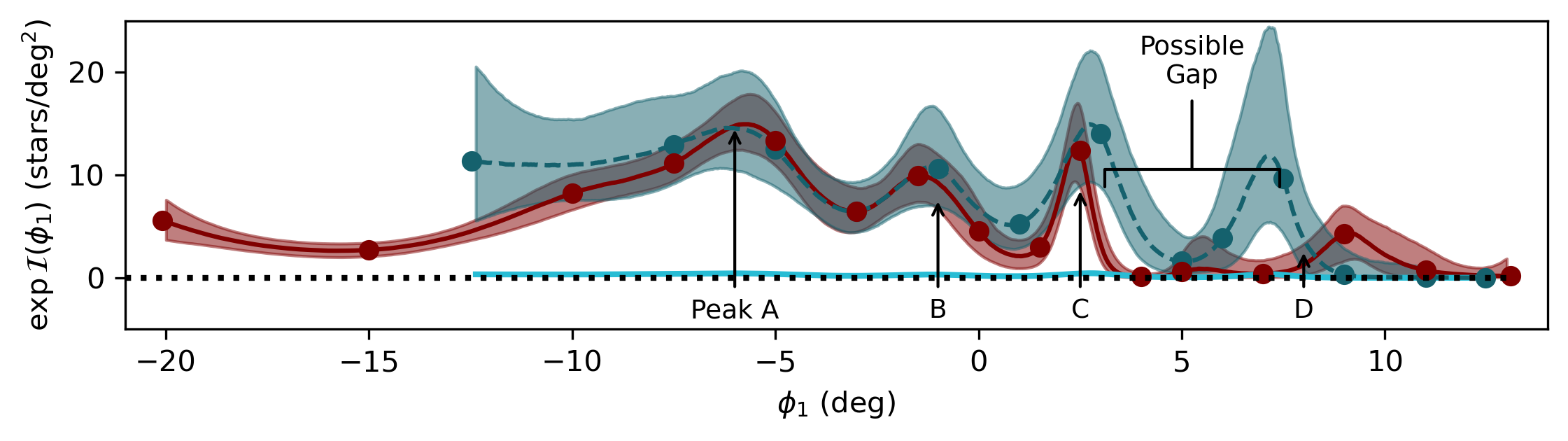}}
    \end{minipage}
    \begin{minipage}{1\linewidth}
    \centering
        \includegraphics[width = 0.77\linewidth]{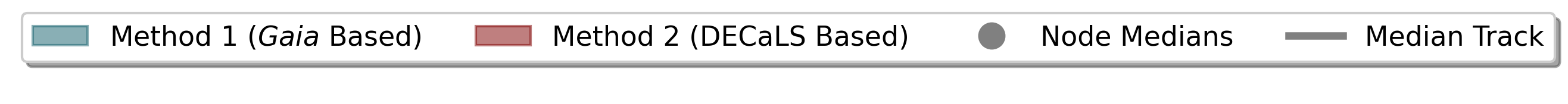}
    \end{minipage}
    \caption{Comparison of modeled stream morphologies.
    (a) The on-sky image of the stream models using the median spline values shown in Figure \ref{fig:Spline Values for the Two Streams}. 
    The shaded regions represent the stream's full width at half maximum as identified by the respective model.
    We further include the $\phi_1$ range of the \SSSSS members we use to calculate the distance gradient and the proper motion filters as the yellow line.
    \streamPoss kink is labeled.
    (b) The $\exp\mathcal{I}(\phi_1)$ splines overplotted for both models. 
    The positions of peaks A, B, C, and D as well as the potential gap are labeled.
    The dotted horizontal line is at $0$ for reference.
    As seen in Figure \ref{fig:Spline Values for the Two Streams}, the two central stellar density splines have inherently different characteristic magnitudes. To compare their structures here, we scale the Method 1 spline by $\max_{\phi_1}
(\exp\mathcal{I}_2)/\max_{\phi_1}(\exp\mathcal{I}_1)$ to force the corresponding peaks of the splines to line up. 
The Method 1 spline with its original magnitude can be seen as the solid light blue line. 
The similarity in structure of these two splines despite their different derivations indicates that the resulting stream density profile is robust and that our second method of signal extraction (Section \ref{ssec:method 2 double stream model}) was successful. 
    }
    \label{fig:stream results figure}
\end{figure*}

Generally, the two methods of signal extraction that we apply to model \stream lead to agreeing central stellar densities along the majority of the track, as can be seen in the comparison in Figure \ref{fig:stream results figure} and the splines in Figure \ref{fig:Spline Values for the Two Streams}.\footnote{We include tables of the \stream spline nodes' $\phi_1$ positions and their corresponding values in Appendices \ref{ap:Method 1 Spline Values} and \ref{ap:Method 2 Spline Values} the end of this paper. 
We also include this information for the \sgr spline nodes in Appendix \ref{ap:Sgr Spline Values}.
We share the complete model results and machine-readable tables at \url{https://zenodo.org/records/15391938}.
}
This agreement is especially apparent in the region of highest stellar density from $\phi_1 \simeq -7^\circ$ to $\simeq 2.5^\circ$. 
They also lead to agreeing tracks and half-widths for the majority of the stream.
Because the two methods used different datasets and filtering methodologies, this agreement indicates that our fits are robust against contamination.

\subsection{Stream Features Present Under Both Methods} \label{ssec:features present in both models}

Generally, the models agree that \stream is an angularly long (though relatively physically short) thin stream, whose track closely traces a great circle. 
Stream length is related to both the age and energy distribution of the stream \citep{Johnston_2001}.
Method 1 detects \stream between $-12.5^\circ$ and $\sim 8^\circ$ and Method 2 extends that measurement to identify the stream within $\sim-20^\circ$ and $\sim12^\circ$. At $32^\circ$ and a distance of $17\kpc$, \stream has a physical length of $9.7\kpc$.
This is much shorter than the Jet and GD-1 streams, the other known globular cluster streams on retrograde orbits (which have lengths of $\sim 16\kpc$ and $\sim 15.4\kpc$ respectively, \citealp{Ferguson_2021} and \citealp{Malhan_2018,Price_Whelan_2018,Webb_2019,de_Boer_2020}). 

The models also agree that \stream is thin and its track is close to a great circle. In the region $\phi_1 \in [-12.5^\circ,13.0^\circ]$, Method 2's median Gaussian width is $\lesssim 0.6^\circ$ and its $|\phi_2|\lesssim0.4^\circ$. At a distance of $17\kpc$, this width corresponds to $178\pc$ which is reasonable for a globular cluster stream  ($w\sim 100\pc$) and extreme for a dwarf galaxy stream \citep[$w \gtrsim 500\pc$,][]{Patrick_2022}. 
This further verifies the conclusions of both \cite{Li_2022}'s chemical and kinematic analysis and \cite{Usman_2024}'s further chemical analysis that \streamPoss progenitor is a globular cluster.

We also identify oscillations in $\mathcal{I}(\phi_1)$ with four peaks at $\phi_1 \simeq -6^\circ$, $-1^\circ$, $2.5^\circ,$ and $8^\circ$ (peaks A, B, C, and D respectively). The positions of these peaks are shown in Figure \ref{fig:stream results figure} with arrows. 
The specific coordinates are identified by eye. 
These peaks are similar to the peaks found in ATLAS \citep{Li_2021}, another globular cluster stream. 
We do note that the Orphan-Chenab (OC) stream's track \citep{Koposov2023} intersects the $\phi_2 = 0^\circ$ line at $\phi_1 = -2^\circ$ and the intersection of the OC stream's full width at half maximum is $-3^\circ <\phi_1<-1^\circ$.
In the OC stream coordinate system \citep{Koposov_2019_orphan}, this intersection occurs at $\phi_1^\text{(OC)} = 83^\circ$.
However, it is unlikely that the OC stream is making a significant impact on our results.
At that location, \cite{Koposov2023}'s distance spline indicates that the OC stream's distance is $25.2\kpc$ while our distance gradient places \stream at $16.5\kpc$.
Moreover, the $\phi_1$ of the objects' intersection ranges over both part of peak B and part of the trough between peaks A and B. 
If the OC stream were leaking through the filter, we would not expect a trough in that $\phi_1$ range. 
Finally, the OC stream is not visible on the residual maps in Figure \ref{fig:transformed and subtracted sgr}, indicating that it has a minimal influence. 

One possible explanation of these oscillations is the influence of epicycles, which can lead to peaks in stellar density \citep{Kupper_2012} and are predicted to exist in globular cluster streams \citep[e.g.,][]{Weatherford_2024}. 
\cite{Ibata_2020_gd1} use epicycles to explain similar peaks in density found in the GD-1 stellar stream.
As epicycles are a potential explanation for \streamPoss density peaks, are useful for constraining stream mass \citep{Ibata_2020_gd1}, and are useful for indicating the progenitor's position because they appear equally spaced from it \citep{Kupper_2008}, we briefly investigate whether these density variations could be epicyclic in origin.
\cite{Kupper_2008} derived the following relation for the spatial period of epicyclic over-densities for a progenitor on a circular orbit

\begin{equation}
y_c= \frac{4\pi \Omega}{\kappa}\left ( 1-\frac{4\Omega^2}{\kappa^2}\right )x_L
\end{equation}

\noindent where $\Omega$ is the angular velocity of the progenitor, $\kappa$ is the epicyclic frequency, and $x_L = \left ( GM/(4\Omega^2-\kappa^2)\right )^{1/3}$ is the radius of the Lagrange point. For a Milky Way potential, \cite{Kupper_2012} reduces this relation to $y_C \approx 3\pi x_L$ using $\kappa \approx 1.4\Omega$ in the Milky Way potential \citep{Just_2009}. Then, $y_c \approx 3\pi \left ( GM/(2\Omega^2) \right )^{1/3}$.
Using these equations, we may check whether epicycles are a reasonable explanation of \streamPoss periodic nature. 

At present, \streamPoss angular velocity $\Omega$ is  $5.7\rads$. This $\Omega$ is around two thirds of the stable circular angular velocity at \streamPoss present day Galactic radius because of the high eccentricity of its orbit. Assuming a stellar mass of $\log_{10} (M/M_\odot) = 4.5$, the lower bound of \cite{Usman_2024}'s mass bounds, the epicyclic spatial period $y_c$ is $1.2\kpc$. 
At $17\kpc$, this corresponds to $4.1^\circ$, which is of the correct order to explain \streamPoss density peaks whose average separation is $\sim4.7^\circ$.\footnote{Using the slightly higher lower mass bound of $\log_{10}(M/\msun)  =4.7$ found in Section \ref{ssec:lower limit mass}, we obtain an epicyclic angular period of $4.7\deg$.}
This indicates that epicycles are a plausible explanation of \streamPoss interesting density variations.

Unfortunately, because there is no analytic treatment of epicycles for eccentric orbits \citep{Kupper_2012} and \streamPoss orbit is highly eccentric (\citealp{Fu_2018} and \citealp{Li_2022}, also see Section \ref{sssec:orbital characteristics}), these calculations are quite approximate. 
\cite{Kupper_2012}'s simulations found that $y_c$ shrinks at orbital apocenter and then grows as the cluster reaches pericenter with orbit eccentricities of $0.25$ and $0.5$. 
This effect would likely be more considerable for \stream due to its higher eccentricity. 
Further, as discussed in Section \ref{ssec:lower limit mass}, \streamPoss progenitor's mass was likely higher than $\log_{10} (M/M_\odot) = 4.5$, which would act to increase $y_c$.
\streamPoss density oscillations could also be residue from the complex disruption of globular clusters \citep{Malhan_2020} or other environmental interactions.
Further dynamical simulations of \streamPoss density peaks  are necessary to formulate a conclusive explanation for the stream's structure by shedding light on the specifics of \streamPoss disruption and decisively clarifying the plausibility of an epicyclic description. 

Finally, depending on the exact position of peak D and shape of peak C (see the next section), the models agree on the presence of a $\sim 3^\circ-\sim6^\circ$ gap in the stream. 
This gap is on a similar angular scale as the $4^\circ$ gap found in the Jet stream by \cite{Ferguson_2021} but is around two-thirds the physical length.
At $17\kpc$, \streamPoss gap is $\sim1.4\kpc$ long assuming an angular length equal to the average peak separation of $4.7^\circ$. 
Such a gap could be formed through a variety of processes including interactions with small-scale dark matter subhalos \citep[e.g.,][]{Bonaca_2019} or other environmental features, interactions with the progenitor's environment \citep[e.g.,][]{Malhan_2020}, or the process of total progenitor disruption \citep[e.g.,][]{Webb_2019}. 
\begin{figure*}[ht!]
    \begin{minipage}{1\linewidth}
        \centering
        \includegraphics[width = 1\linewidth]{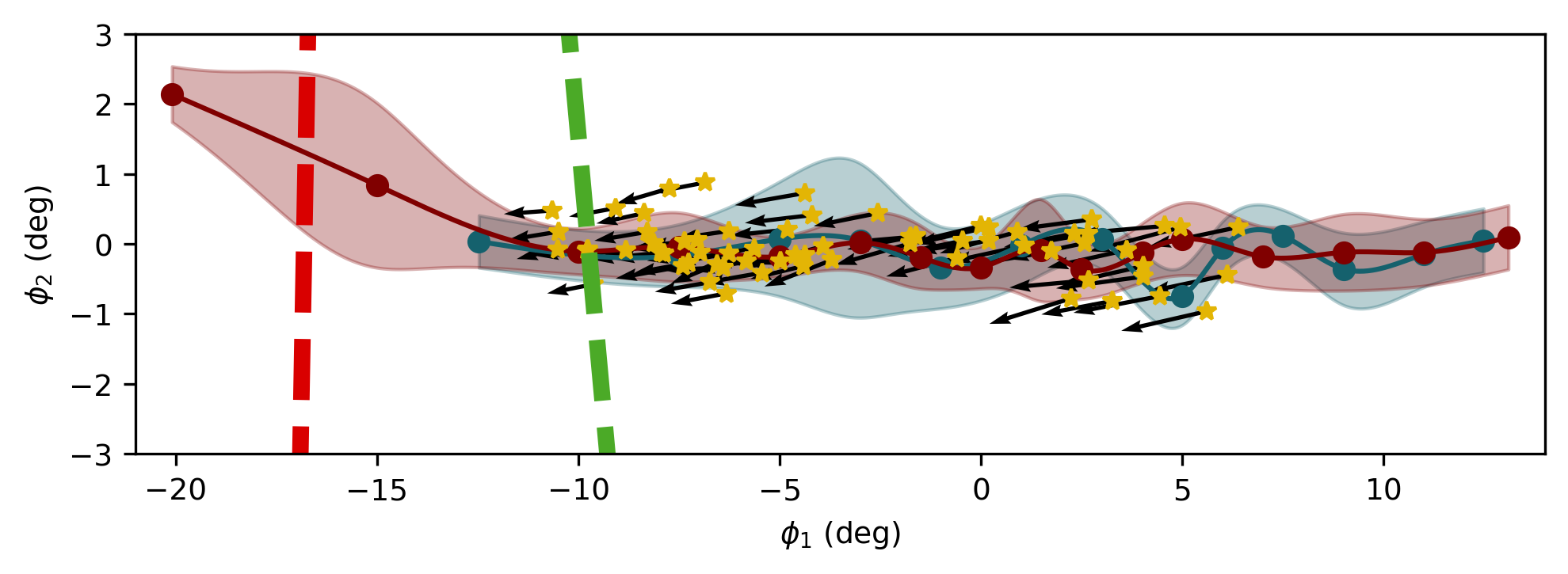}
    \end{minipage}
    \begin{minipage}{1\linewidth}
        \centering
        \includegraphics[width = 0.72\linewidth]{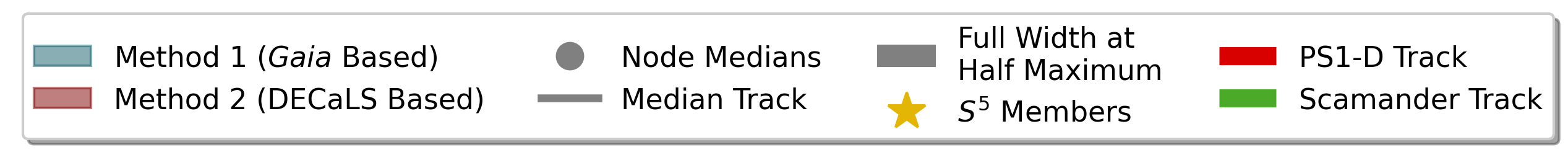}
    \end{minipage}
    \caption{The positions and reflex-corrected proper motions of the \SSSSS member stars overplotted on the stream models. The shaded regions represent the full widths at half maximum as identified by the respective model. 
    Reflex corrections were performed assuming the same solar motion as described in Section \ref{ssec:preparing the dynamical model}.
    Note that the members are clustered near the 3 central peaks in stellar density at peaks A, B, and C. The proper motions of the \SSSSS members are slightly misaligned with both models' stream tracks. 
    This indicates that \stream experienced a significant interaction over its lifetime. As we discuss in Section \ref{sec:dynamical modeling}, we find that a close interaction with the LMC is relatively consistent with the current kinematics of \stream, as well as the kink we measure at $\phi_1 \simeq -12.5^\circ$. The \textsc{galstreams} tracks for Scamander and PS1-D are also included. Neither track has the correct orientation to explain the kink.}
    \label{fig:S5 Member Proper Motions}
\end{figure*}

\subsection{Tensions Between the Methods}

Although the models generally agree on the morphology of the stream and its features, there are a few tensions.
First, although both models identify peaks A, B, and C in approximately the same location and peak D between $\phi_1\sim 7^\circ$ and $\phi_1\sim 10^\circ$, Method 2 finds the latter structure to be at a larger $\phi_1$.
It is possible that Method 1 does not fully characterize that feature due to a combination of the low count statistics inherent in the method and the relatively low peak stellar density of the feature.
In fact, the bump in Method 1's spline is due to only a few bins in the vicinity of peak D which contain only $\sim 1-2$ stars each. 
As the $S^5$ catalog does not cover this structure, further spectroscopic analysis in the region will be necessary to confirm it's morphology.
However, given that both methods pick up a substantive structure in this region, we believe that it is a real component of the stream. 

Second, despite the methods' agreement on the presence and relative height of peak C, they disagree slightly on how rapidly the density decays as $\phi_1$ increases past $2.5^\circ$. 
As seen in Figure \ref{fig:S5 Member Proper Motions}, \SSSSS finds stream members in this region, indicating that the stream extends there. 
We find it unlikely that Method 2 fails to capture this behavior due to poor \sgr subtraction, as $T(\lambda)$ strongly suppresses \sgr in this region (see Figures \ref{fig:transformation fit} and \ref{fig:sgr transformed model}) and there is no pattern in the residuals near \stream in Figure \ref{fig:sgr subtracted intensity}.
Moreover, we find that alternative node placements can lead Method 1 to have lower central stellar density in the region in closer agreement with Method 2, although they also increase the amplitude of Method 1's peak C within the substantial Poisson uncertainties.
We also note that without the inclusion of the horizontal branch candidates (see Section \ref{ssec:gaia filtering}) and with a corresponding slight increase in the proper motion filter width to make up for the lost candidates, the two models also achieve closer agreement on the width and depth of the gap.
Therefore, we are inconclusive on the specifics of the stream density's decay to the right of peak C, although there is likely a gap in this region. This gap is visually apparent in Figures \ref{fig:sgr and background subtracted intensity} and \ref{fig:close view sgr and background subtracted}. 

Third, despite the models' agreement on the stream track and width over the majority of the stream, Method 1 identifies a widening in the stream at $-5^\circ\lesssim\phi_1\lesssim -2.5^\circ$. This widening is due to the small collection of stars within $1^\circ\lesssim\phi_2\lesssim3^\circ$ in the relevant region, as seen in Figure \ref{fig:p12 Gaia}. 
These stars are on the side of \stream closer to \sgrPoss stream track, meaning that they are potentially contamination from the complex \sgr background population.
This is reinforced by the fact that Method 2 does not identify widening in this region. 
The smoothly varying \sgr background used in Method 2 should not be eliminating this signal if were strongly present.
Moreover, a stronger prior on the width in Method 1 of $\mathcal{N}(\log0.4,0.3)$ substantially reduces the widening in this region.
However, these stars do pass through both the proper motion and isochrone filters described in Section \ref{ssec:method 1 proper motion filtering}. 
As these stars are outside of the $\SSSSS$ footprint, we can not test their radial velocities against \streamPoss distinct $\sim300\kms$. 
Therefore, further spectroscopic followup is necessary to fully characterize this potential feature.

\subsection{Kink Visible Under Method 2}\label{ssec:kink empirical}

Method 2 detects a ``kink,'' or sudden bend away from the $\phi_2 = 0^\circ$ line, in the stream track at $\phi_1 \lesssim -12.5^\circ$. 
As this kink is at the end of the stream where the signal is weakest, and we are unable to verify it with Method 1 due to the \sgr contamination, it could be an artifact of background mismodeling.
On the other hand, as \SSSSS did not cover that region and therefore there are no members there, we must extrapolate our distance gradient into it. 
This could imply that our matched filter may be poorly extracting the kink's signal, resulting in our under-modeling it. 

The kink does not align with any other streams in the region.
Of the streams listed in the \textsc{galstreams} catalog \citep{Mateu_2023}, only Scamander \citep{Grillmair_2017} and PS1-D \citep{Bernard_2016} are near the kink spatially and at a similar distance. Specifically, Scamander intersects \stream at $\phi_1\sim-9.7^\circ$ and is at a distance of $21\kpc$ \citep{Mateu_2023}. 
Our distance gradient places \stream at $18.7\pm 1.6\kpc$ there.
PS1-D intersects \stream at $\phi_1\sim-16.8^\circ$ and is at a distance of $23\kpc$ \citep{Mateu_2023}.
\stream is at $21.0\pm 2.0\kpc$ there.
As seen in Figure \ref{fig:S5 Member Proper Motions}, neither of these candidates' \textsc{galstreams} tracks \citep{Mateu_2023} are close to being in the correct orientation to explain the overdensity (also see Figure 1(d) in \citealp{Grillmair_2017}).

In addition, we find that \streamPoss track is misaligned with the proper motions of the \SSSSS members. 
This is visible in Figure \ref{fig:S5 Member Proper Motions}.
Such a proper motion-stream track misalignment was also found in the OC stream \citep{Koposov_2019_orphan} and there was due to a strong interaction with the LMC.  
We investigate the possibility of an \stream-LMC interaction in the next section (Section \ref{sec:dynamical modeling}) and find that such an interaction can both produce a kink that aligns with our empirical model and reproduce the kinematics of the \SSSSS members. 
The dynamical model provides a strong motivation for additional spectroscopic followup in the region of the kink to better clarify whether it is a real component of the stream.

\subsection{The Total Mass of \stream}\label{ssec:lower limit mass}

\begin{figure}
    \centering
    \includegraphics[width=1.0\linewidth]{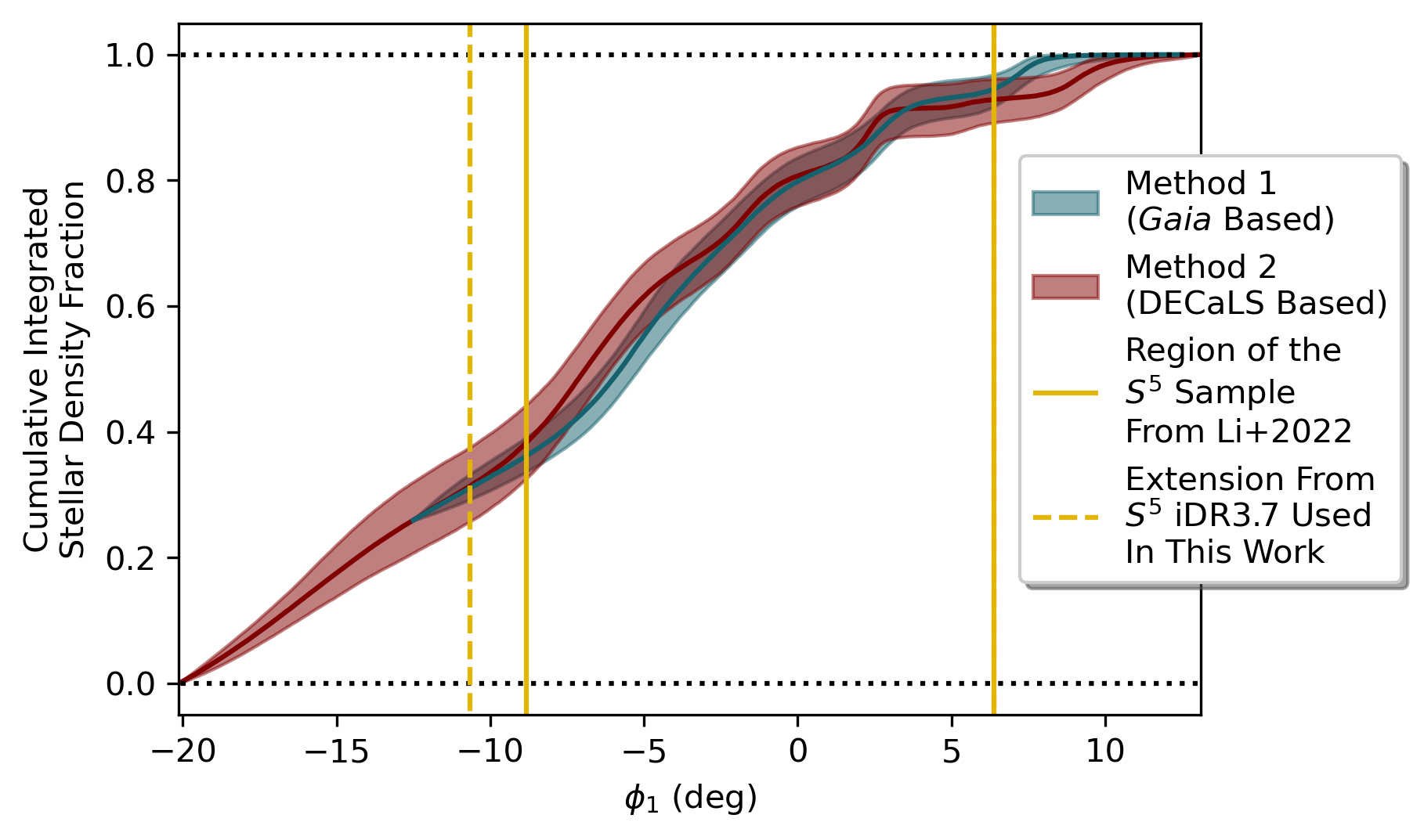}
    \caption{Integrated stellar density under both methods. The yellow lines represent the $\phi_1$ range of the $S^5$ members. The solid lines show the range of members identified by \cite{Li_2022} and used by \cite{Usman_2024}. The dashed line shows the extension of that region available in the \SSSSS iDR3.7 sample used in this work. 
    The shaded regions represent the $16\%-84\%$ quantile range of the integrated stellar density. To account for the lower coverage of Method 1 and make the two methods comparable, we set Method 1's $\phi_1 = -12.5^\circ$ starting point to the integrated density fraction found by Method 2 at that $\phi_1$. Under Method 2, the $S^5$ members encapsulate $61\%$ of the stream going from $\sim31\%$ to $\sim93\%$.}
    \label{fig:integrated light}
\end{figure}

Using our maps of \stream, we may also consider integrated stellar density over the stream.
We show the cumulative distributions in Figure \ref{fig:integrated light}. As Method 1 does not cover the entire footprint of the stream due to the $\phi_1 > -12.5^\circ$ requirement, we assume for the sake of comparison that if Method 1 could analyze the $\phi_1 \leq -12.5^\circ$ region, it would find the same proportion of stars as were found using Method 2.
As can be seen in Figure \ref{fig:integrated light}, once this adjustment is made, the two methods produce mostly agreeing distributions. This is as expected from the similarities in their $\mathcal{I}(\phi_1)$ splines. 
The primary difference in these distributions is in the positive $\phi_1$ region where the models disagree on the exact shape of peak D. There is also worse agreement around $ \phi_1\sim -5^\circ$ in the vicinity of the stream's widening under Method 1.

Using Method 2, we find that the $S^5$ members used in this work have a footprint which encapsulates around $61\%$ of the stream stars. Around $31^{+6}_{-6}\%$ -- where the reported uncertainties are the $16\%$ and $84\%$ quantiles respectively -- are cutoff on the left where $\phi_1 \leq -10.7^\circ$. $7^{+4}_{-3}\%$ are cutoff on the right with $\phi_1 \geq 6.4^\circ$. The members used by \cite{Usman_2024} from \cite{Li_2022} have a slightly smaller footprint which includes around $55\%$ of the stream stars, cutting off around $38^{+6}_{-6}\%$ on the left with $\phi_1 \leq -8.8^\circ$ and $7^{+4}_{-3}\%$ on the right with $\phi_1 \geq 6.4^\circ$.

\cite{Usman_2024} found a lower limit mass of the progenitor of $\log_{10} (M/M_\odot) = 4.5$ assuming the ratio of observable to total stars within their footprint is equal to the same ratio for a Salpeter initial mass function \citep{Salpeter_1955}. As they used the members of \cite{Li_2022}, they further assume for the sake of a lower limit that the sample is spatially complete and do not consider stars outside of \cite{Li_2022}'s footprint. 
Using the coverage we identify with our models, we can improve this lower bound. 
For the sake of a lower bound, we consider the $16\%$ quantile on the left and the $84\%$ quantile on the right. 
Using these values, we find the footprint to encapsulate around $63\%$ of stream stars. Then we may increase this lower limit to $\log_{10} (M/M_\odot) = 4.7$. 
Together with \cite{Usman_2024}'s upper limit, we find \streamPoss progenitor mass to be in the range $10^{4.7-4.9}M_\odot$.

\begin{figure*}[ht!]
    \centering
    \begin{minipage}{1\linewidth}
    \centering
    \subfigure[Without LMC \label{sfig:onsky distribution no LMC}]{\includegraphics[width = 1\linewidth]{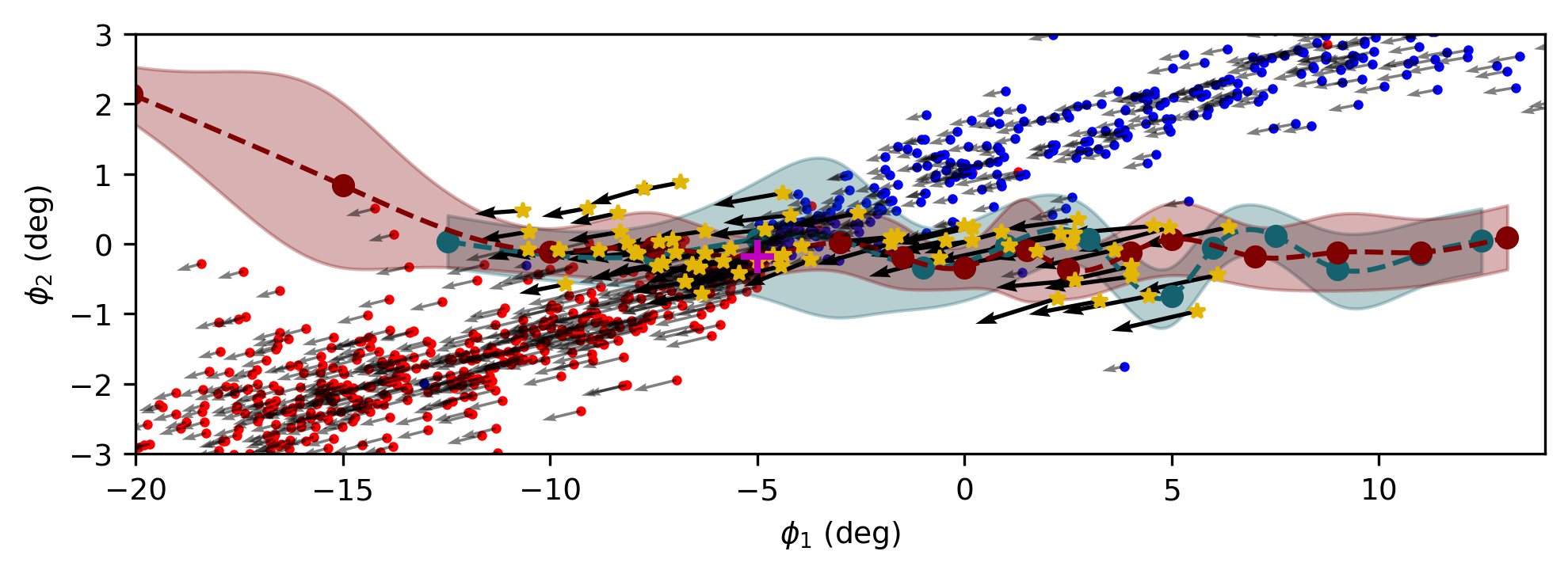}}
    \subfigure[With LMC \label{sfig:onsky distribution LMC}]{\includegraphics[width = 1\linewidth]{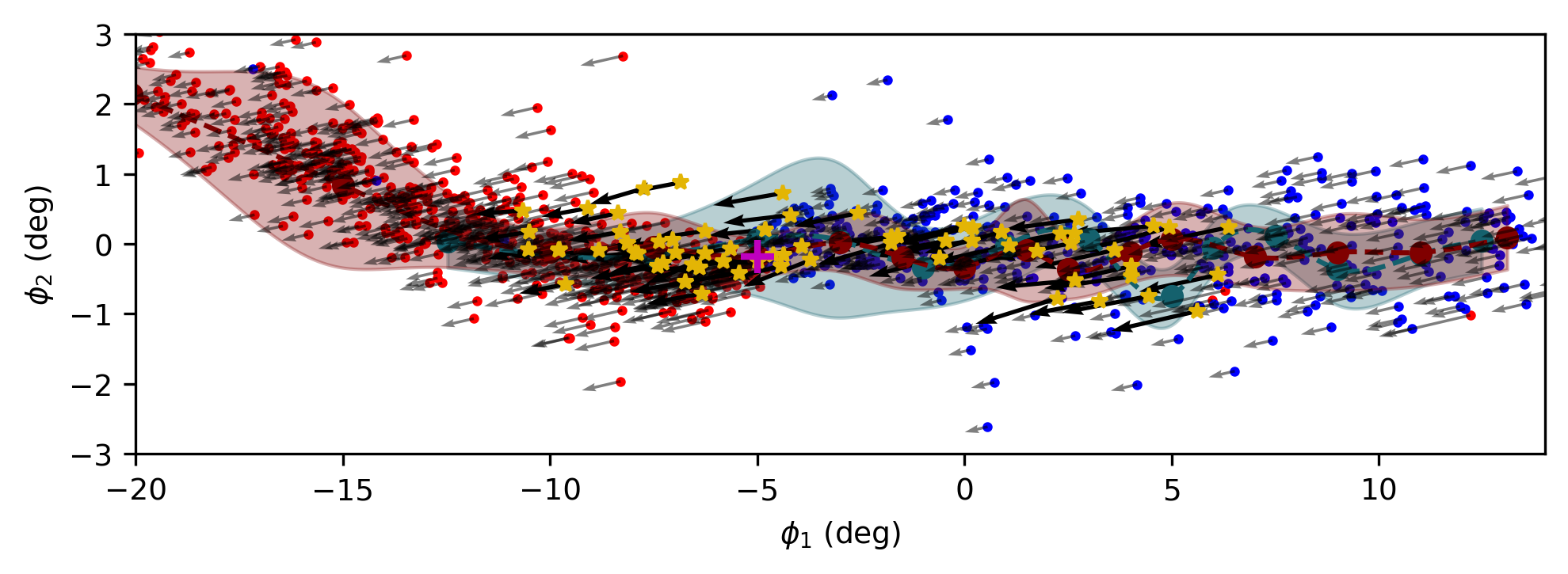}}
    \end{minipage}
    \begin{minipage}{1\linewidth}
    \centering
    \includegraphics[width = 0.9\linewidth]{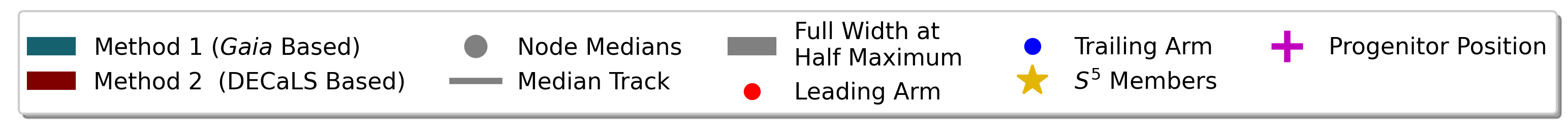}
    \end{minipage}
    \caption{On-sky distribution of particles in the dynamical simulation of \stream overplotted onto the empirical stream models. (a) The results of the simulation with the same progenitor initial conditions while excluding the influence of the LMC. 
    It is very challenging to find progenitor initial conditions which lead to simulated particles matching both the \SSSSS kinematics and the on-sky distribution of the tail of \stream. 
    (b) The results of the simulation including the influence of the LMC. From these simulations, it is clear that the LMC had a substantial influence on the present day morphology of \stream, resulting in a rotation of the stream track by $\sim 11^\circ$. The simulation reproduces the position and slope of the kink, indicating it to be a real structure.}
    \label{fig:dynamical simulation onsky distribution}
\end{figure*}

\section{Dynamical Modeling}\label{sec:dynamical modeling}

To better understand the origins of the kink that Method 2 identifies in \stream (see Figure \ref{fig:stream results figure}), we consider whether \stream was influenced by the LMC. 
In this section, we prepare a simple dynamical model of \stream as an initial investigation of the \stream-LMC interaction.
We find that the LMC likely had a strong influence on the formation of \stream. 
Further, our model indicates that the LMC interaction with \stream can form a kink in the stream whose angle and position relative to the remainder of the stream track matches our empirical result. 
This is evidence that the kink Method 2 identifies is a real feature of the stream.

\subsection{Preparing the Dynamical Model}\label{ssec:preparing the dynamical model}

We utilize \cite{Chen_2024}'s implementation of their particle spray technique in \textsc{galpy} \citep{Bovy_2015} for our dynamical model.
This technique was designed specifically with globular cluster streams in mind. It produces an accurate description of test objects in phase space by sampling tracer particles from distributions found through N-body simulations that are robust against changes in progenitor and orbit parameters \citep{Chen_2024}.
We set the Sun's current position and kinematics using the \textsc{astropy} v4.0 \citep{astropy} values with $d_\odot = 8.122\kpc$ \citep{Gravity_2018} and $v_\odot = (12.9, 245.6, 7.78)\kms$ \citep{Reid_2004,Drimmel_2018,Gravity_2018}.
We use \textsc{galpy}'s \textsc{MWPotential2014} setting $R_0$ according to the \textsc{astropy} v4.0 parameters and $V_c(R_0) = 236\kms$. 
This setting of $V_c$ leads to a halo virial mass of $1.05\times 10^{12}\msun$ which is the mean virial mass found in recent studies of the MW \citep{Bobylev_2023} and leads to a reasonable value of $V_\odot$ given the uncertainties on the quantity \citep{BlandHawthorn_2016,Ding_2019}.
We also find this increase in mass better reproduces the tracks of other stellar streams.
In general, however, our results do not depend on the specific value of $V_c$, as expected given the flexibility of \textsc{MWPotential2014} under reasonable values of $R_0$ and $V_c$ \citep{Bovy_2016}.

We treat the LMC potential as a Hernquist potential \citep{Hernquist_1990} as done in \cite{Erkal_2019}'s investigation of the OC stream's LMC interaction. We fix the LMC mass as $18.4\times 10^{10}\msun$ as derived in \cite{Shipp_2021} and use a scale radius of $19.91\kpc$ such that the circular velocity at a radius of $8.7\kpc$ matches the measurement of \cite{van_der_Marel_2014} of $91.7\kms$ as was assumed in \cite{Shipp_2021}. 
We note that in a more recent study, \cite{Cullinane_2020} found a circular velocity of $88\kms$ at a radius of $10.5\kpc$. 
Using this value instead does not dramatically change our results.
However, we present results using the circular velocity of \cite{van_der_Marel_2014} to maintain consistency with \cite{Shipp_2021}.
The current day 6D phase space coordinate of the LMC used in this simulation is given in Table \ref{tab:dynamical simulation parameters}.

\begin{figure*}[ht!]
    \centering
    \begin{minipage}{1\linewidth}
        \subfigure[Distance Moduli\label{sfig:dynamical simulation distances}]{\includegraphics[width = 0.5\textwidth]{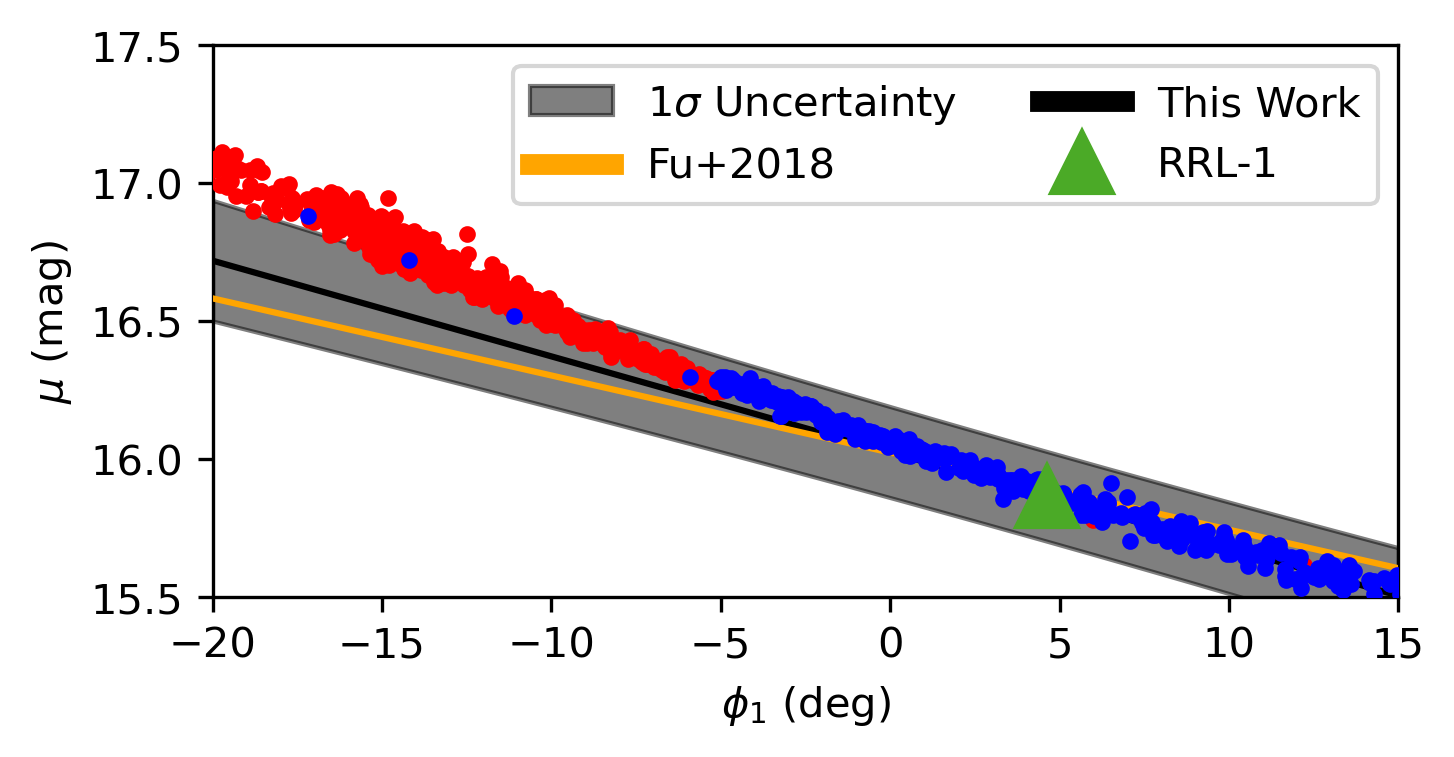}}
        \subfigure[Radial Velocities\label{sfig:dynamical simulation rvs}]{\includegraphics[width = 0.5\textwidth]{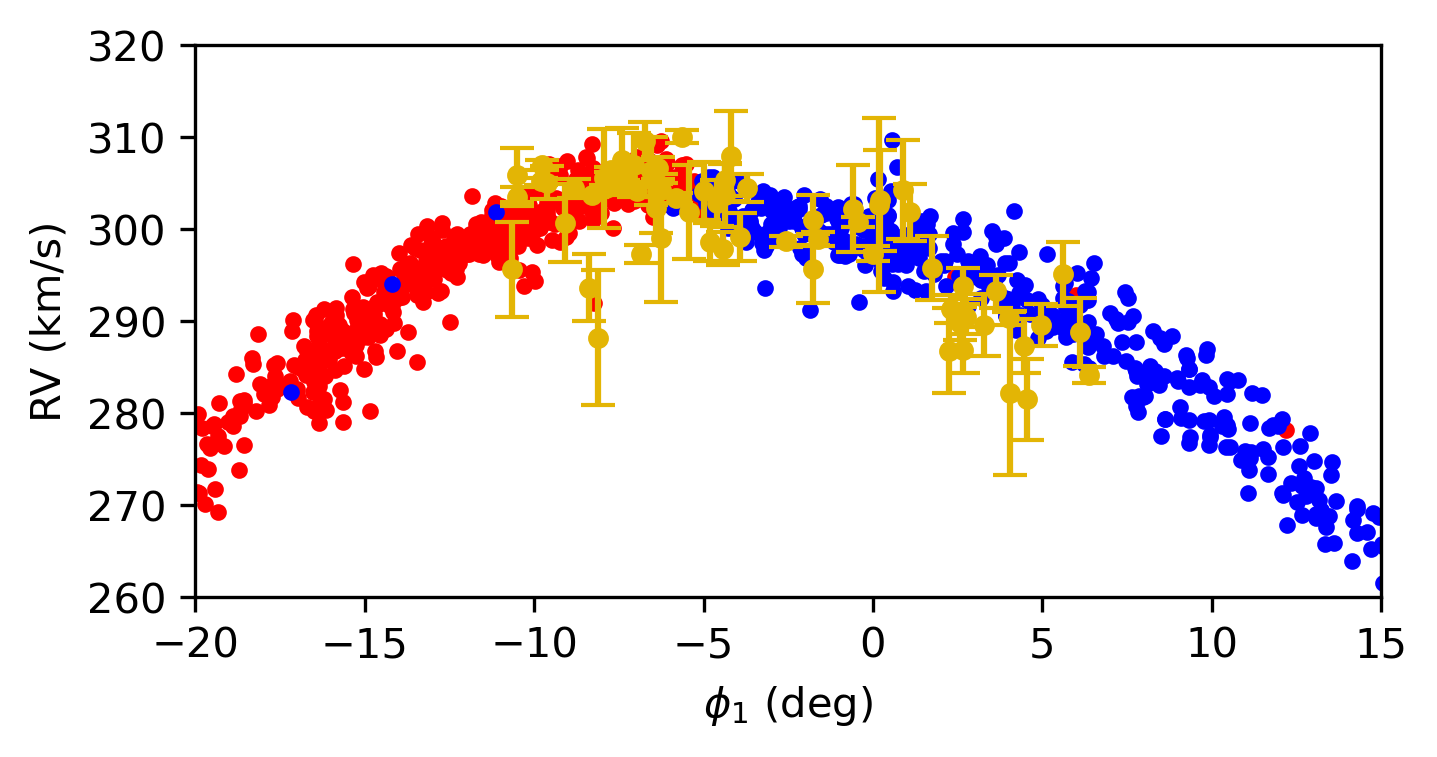}}
    \end{minipage}
    \\
    \begin{minipage}{1\linewidth}
        \subfigure[Proper Motions in $\phi_1$ \label{sfig:dynamical simulation pm1}]{\includegraphics[width = 0.5\textwidth]{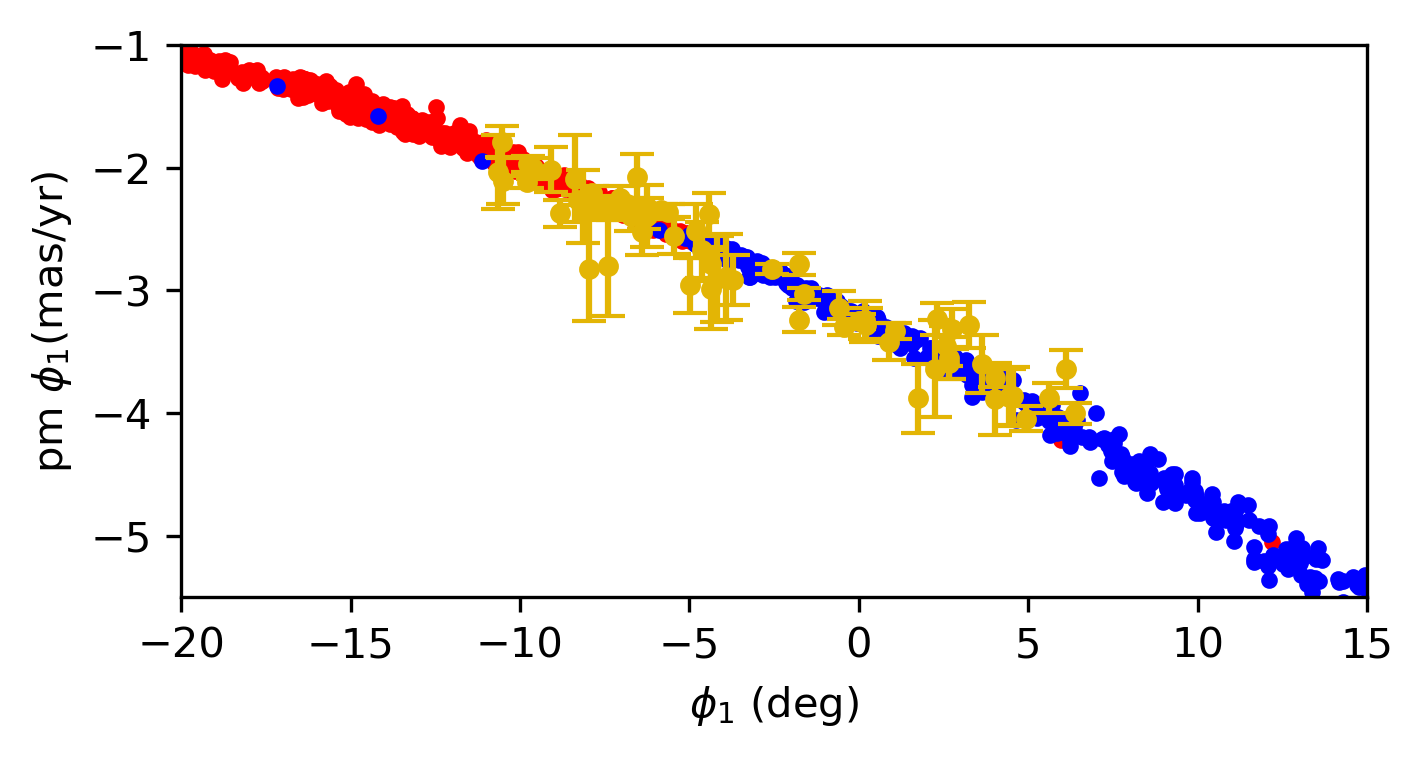}}
        \subfigure[Proper Motions in $\phi_2$\label{sfig:dynamical simulation pm2}]{\includegraphics[width = 0.5\textwidth]{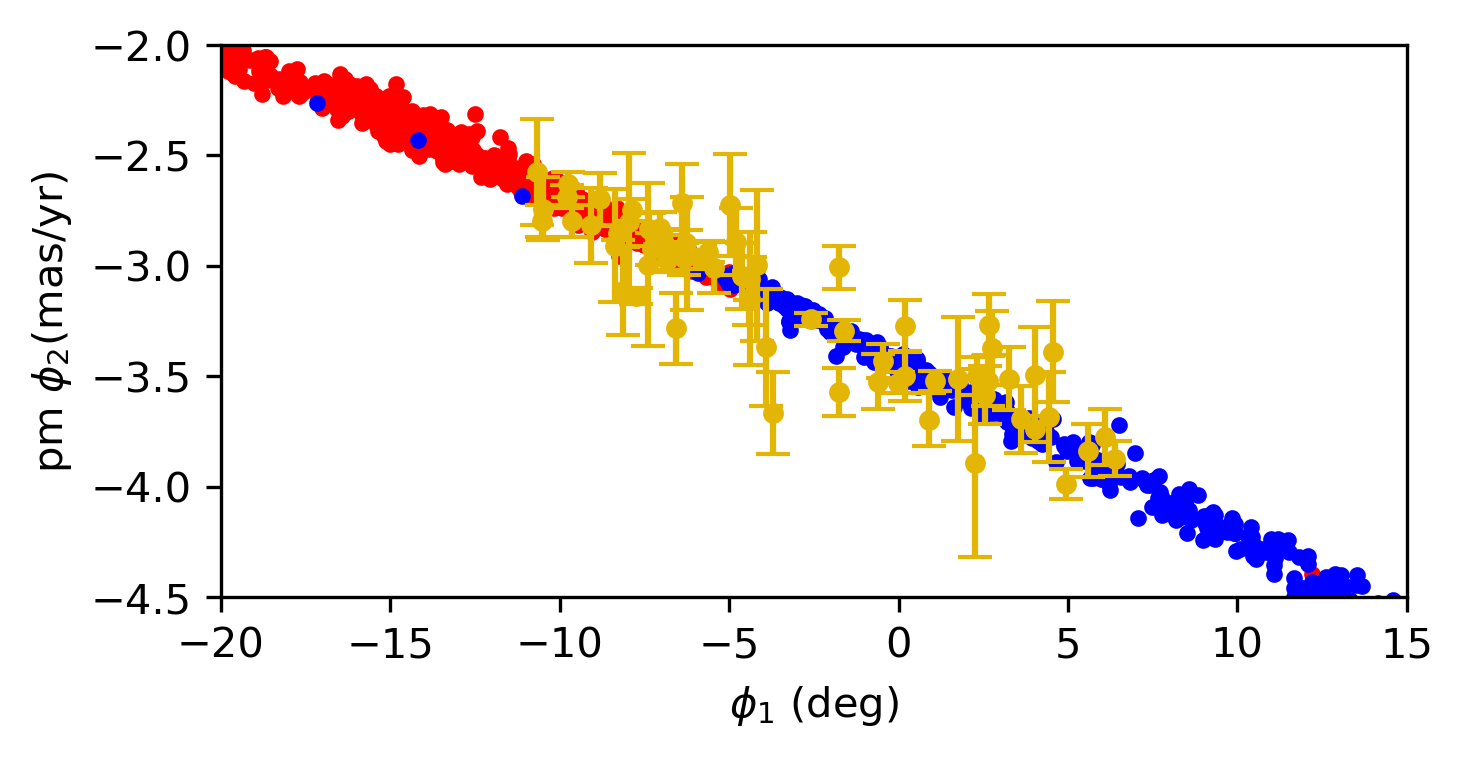}}
    \end{minipage}
    \caption{Comparison between the result of the dynamical simulation which includes the LMC and additional empirical properties of \stream. The three free variables -- progenitor position, distance offset, and scale radius -- were tuned by hand to match the data. The other variables were set using these parameters and tracks derived from the \SSSSS members and our empirical models. See Section \ref{ssec:preparing the dynamical model} (a) Comparison between \streamPoss distance gradient derived in Section \ref{ssec:distance gradient}, the transformed distance gradient of \cite{Fu_2018} (see Section \ref{ssec:distance gradient} and Equation \ref{eq:Fu Transformed}), and the dynamical simulation distances. The dynamical simulation predicts a nonlinear distance gradient similar to the one observed in the data in Section \ref{ssec:nonlinear distance gradient}. The dynamical simulation's gradient at \rrlPoss $\phi_1$ is close to the distance of \rrl. (b) A comparison between the dynamical model's radial velocities and those of the \SSSSS member stars. (c) A comparison between the dynamical model's proper motions in $\phi_1$ and those of the \SSSSS member stars. (d) A comparison between the dynamical model's proper motions in $\phi_2$ and those of the \SSSSS member stars. 
    The kinematic comparisons are not reflex corrected per the discussion in Section \ref{sssec:agreement between the dynamical model and the empirical results}.
    }
    \label{fig:dynamical simulation offsky features}
\end{figure*}

\begin{figure*}[p!]
    \centering
    \begin{minipage}{1\linewidth}
       \subfigure[]{\includegraphics[width = 0.33\textwidth]{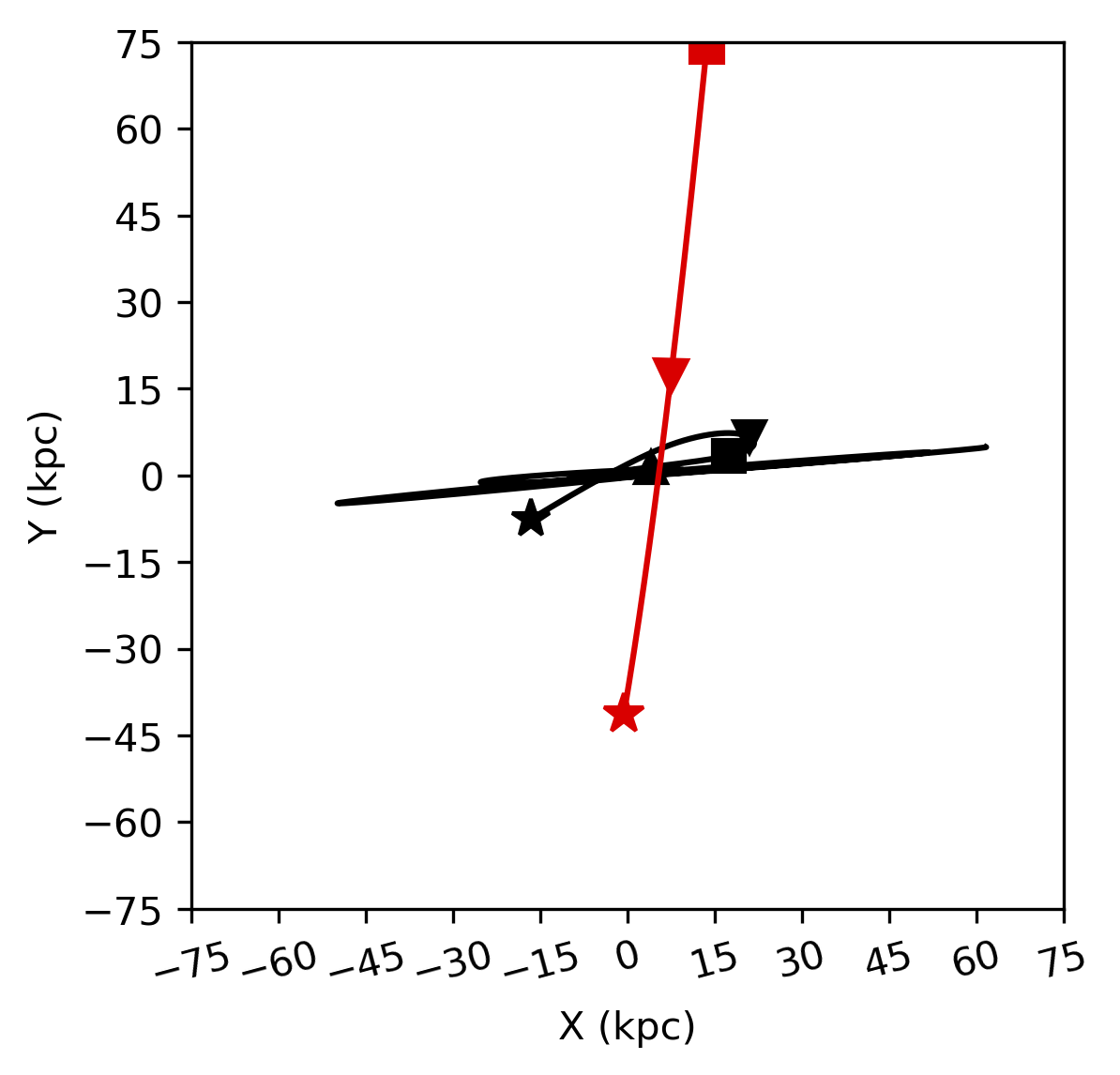}}
        \subfigure[]{\includegraphics[width = 0.33\textwidth]{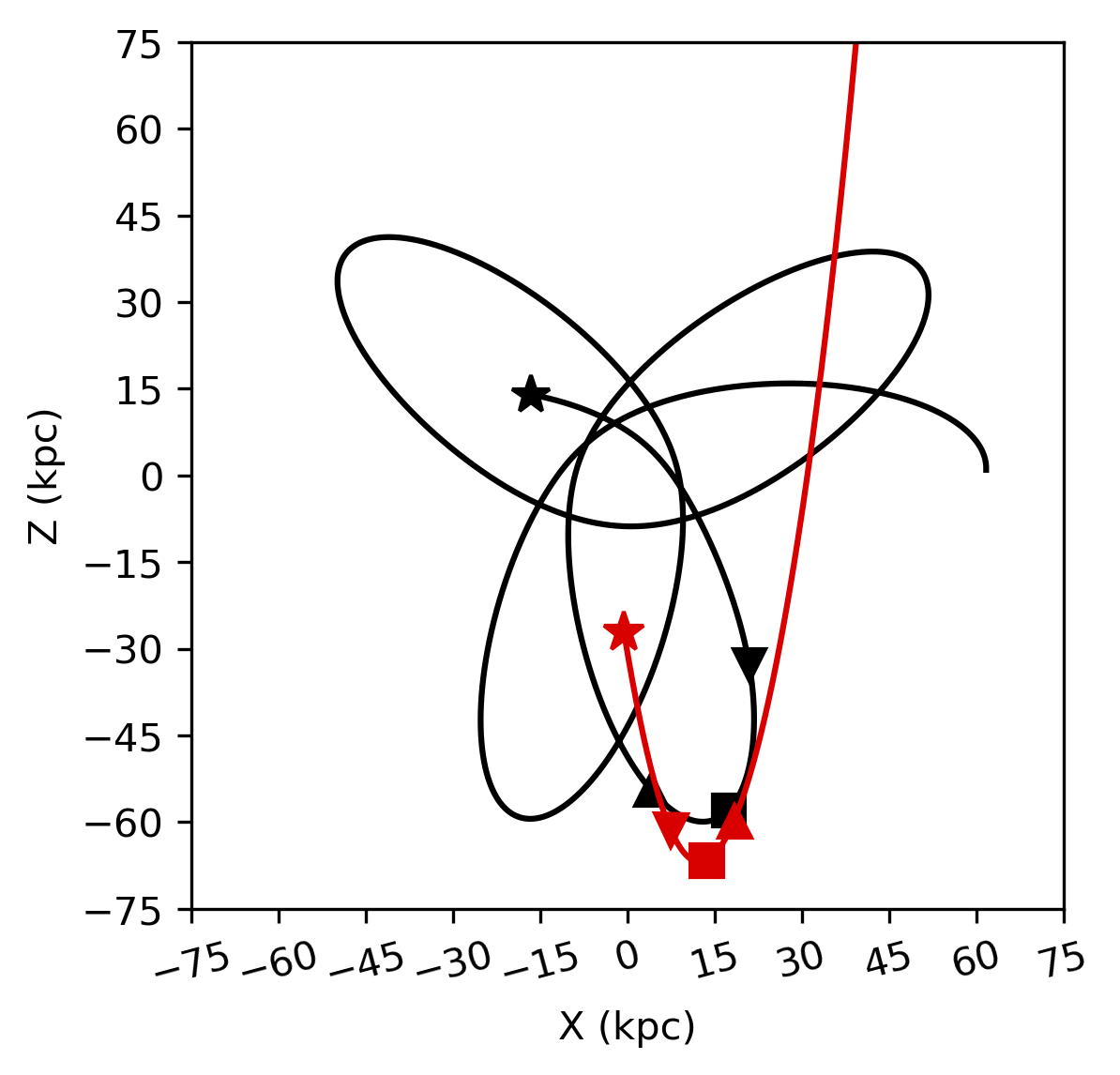}}
        \subfigure[]{\includegraphics[width = 0.33\textwidth]{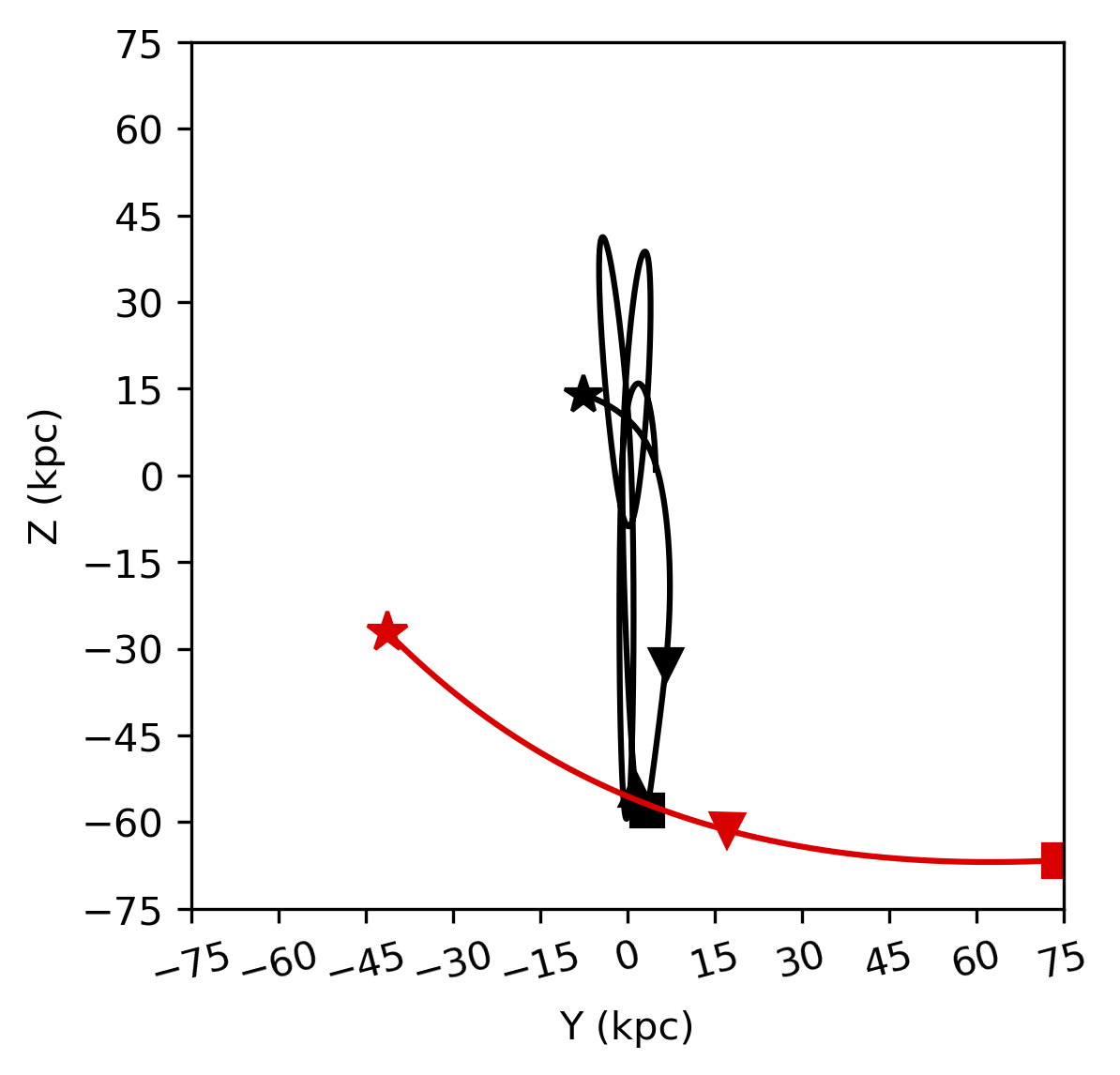}} 
    \end{minipage}
    \\
    \begin{minipage}{1\linewidth}
        \subfigure[]{\includegraphics[width = 0.33\textwidth]{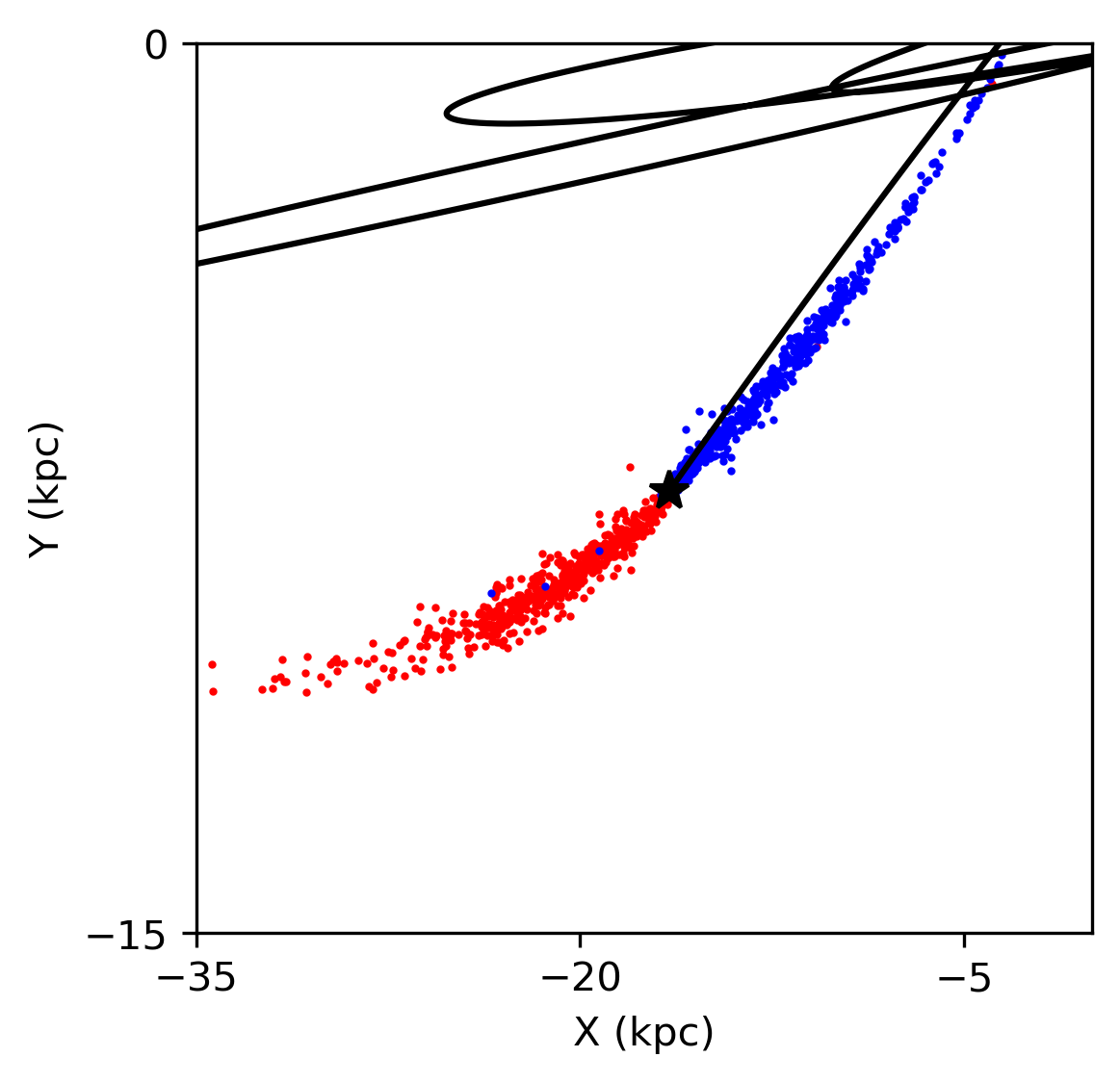}}
        \subfigure[]{\includegraphics[width = 0.33\textwidth]{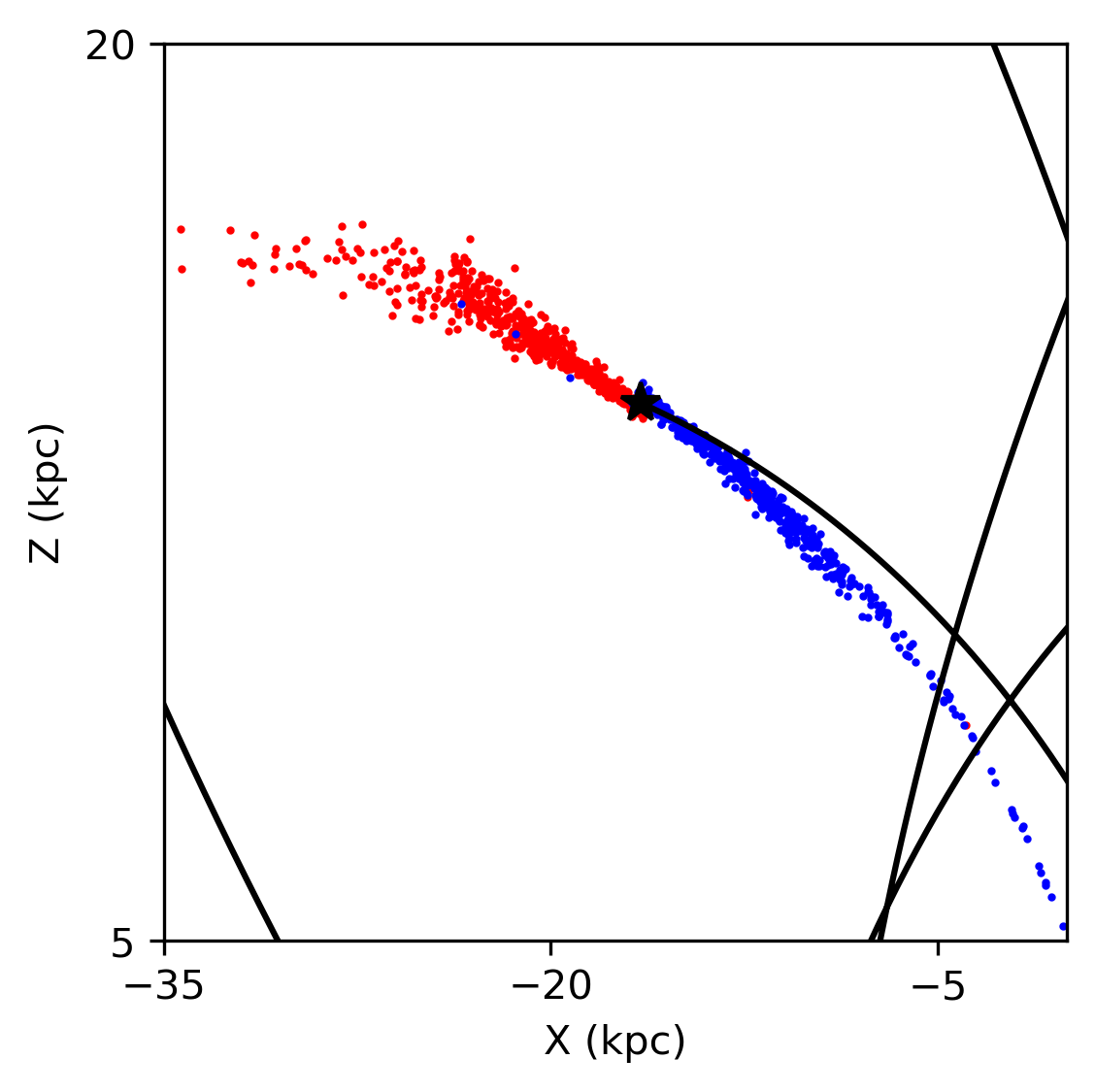}}
        \subfigure[]{\includegraphics[width = 0.33\textwidth]{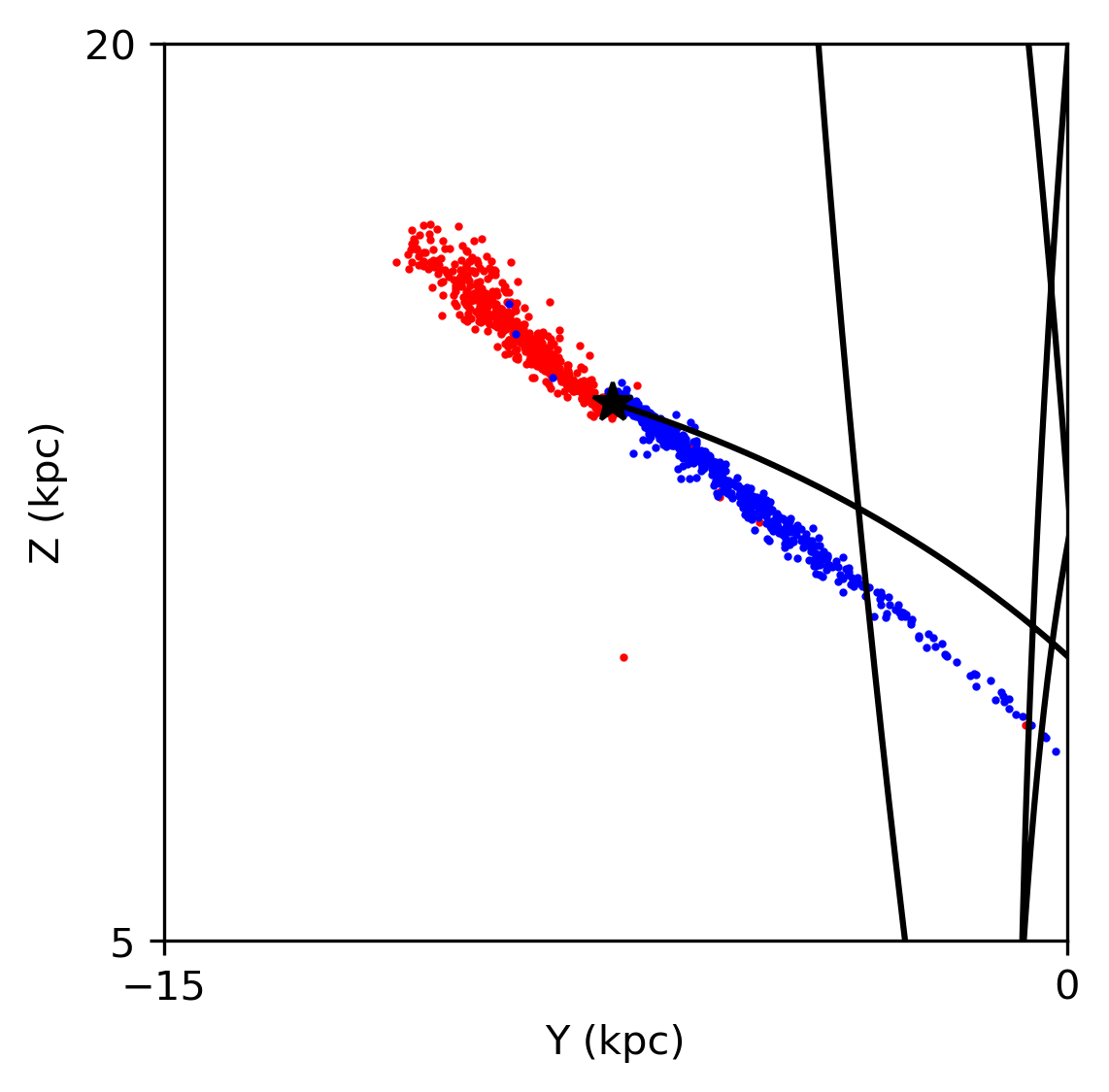}}
    \end{minipage}
    \begin{minipage}{1\linewidth}
        \centering
        \includegraphics[width = 0.7\linewidth]{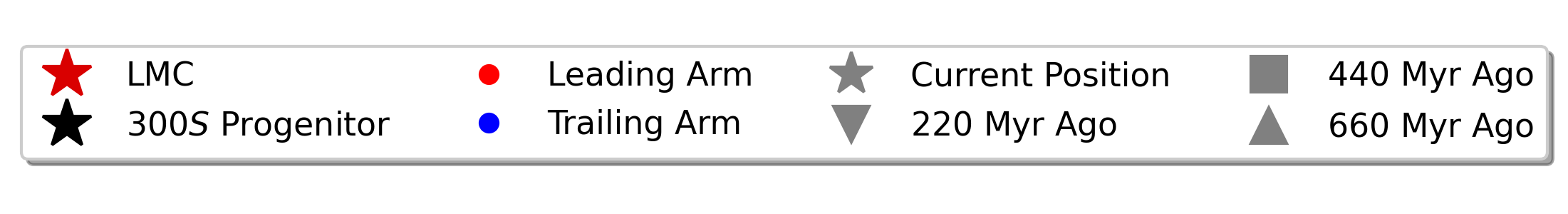}
    \end{minipage}
    \caption{\streamPoss orbit in the dynamical simulations with the LMC. \stream was integrated back $4\gyr$ in these simulations.
    (a-c) Full views of the orbit in galactocentric coordinates using the same axis proportions as Figure 12 in \cite{Fu_2018}. \streamPoss progenitor's present day location is marked by a black star.
    The LMC's trajectory is represented by the red line with the object's present day position marked as a red star. 
    The objects' positions at previous times are also shown. 
    \streamPoss time of closest approach to the LMC at $\sim220\myr$ ago is labeled with the upside down triangle.  
    (d-f) Zoomed in views of \streamPoss orbit showing the offset between the red and blue simulated stream particles and the black orbital track. 
    The red points represent the leading arm of the stream and the blue represent the trailing arm.
    The clear misalignment between \stream and its orbital track is primarily due to the influence of the LMC.}
    \label{fig:dynamical simulation orbit}
\end{figure*}

We treat both the LMC and MW as particulate sources for their respective potentials. For ease of implementation, we pre-compute the trajectory of the LMC and the corresponding non-inertial reference frame of the MW. We incorporate dynamical friction using the \textsc{galpy} implementation of \textsc{ChandrasekharDynamicalFrictionForce} \citep{Bovy_2015}.

Given the free parameter $\phi_1$, we assume a present day progenitor position of $(\phi_1,\Phi_2(\phi_1))$ where $\Phi_2(\phi_1)$ is Method 2's track position. We set the progenitor distance according to \streamPoss distance gradient with a free offset to account for the nonlinearity of the distance gradient described in Section \ref{ssec:nonlinear distance gradient}. We set the progenitor's kinematics using quadratic fits to the $S^5$ member proper motions and radial velocities (see Table \ref{tab:Proper Motion Quadratic Coeffs} for the coefficients). We use a progenitor mass of $\log_{10} (M/M_\odot) = 4.7$ -- the new lower limit mass discussed in Section \ref{ssec:lower limit mass} -- and set \streamPoss mass profile as a Plummer profile \citep{Plummer_1911} with a free scale radius that we set to $12.5\pc$ to match the width of the stream. We assume that tidal stripping began $4\gyr$ ago. The current day 6D phase space coordinate of \streamPoss progenitor is also given in Table \ref{tab:dynamical simulation parameters}.

\begin{table*}[]
    \centering
    \caption{Current Day Phase Space Information for the LMC and \stream Used in the Dynamical Model.}
    \begin{tabular}{lc|lcc}
        \hline
        Parameter & Stream & Parameter &LMC &Note \\
        \hline
        $\phi_{1,\text{prog}}$ $(\deg)$ & $-5$ & $\alpha_{\text{LMC}}$ $(\deg)$ & $80.8942$ & \cite{Kaisina_2012}\\
        $\phi_{2,\text{prog}}$ $(\deg)$ & $-0.18$ & $\delta_{\text{LMC}}$ $(\deg)$  & $-69.7561$ & \cite{Kaisina_2012}\\
        $v_{r,\text{prog}}$ $(\kms)$ & $303.5$& $v_{r,\text{LMC}}$ $(\kms)$  & $262.2$ &\cite{vanderMarel2002} \\
        $d_{r,\text{prog}}$ $(\kpc)$ &$17.4 + d_\text{offset} = 18.0$ & $d_{r,\text{LMC}}$ $(\kpc)$  &$49.97$& \cite{Pietrzyński2013}\\
        $\mu_{\phi_1 *,\text{prog}}$ $(\masyr)$ & $-2.6$&  $\mu_{\alpha *,\text{LMC}}$ $(\masyr)$ & $1.91$ & \cite{Kallivayalil_2013}\\
        $\mu_{\phi_2,\text{prog}}$ $(\masyr)$ &$-3.1$ & $\mu_{\delta,\text{LMC}}$ $(\masyr)$ & $0.229$&\cite{Kallivayalil_2013}\\
        $M_\text{prog}$ $(\msun)$ & $10^{4.7}$& $M_\text{LMC}$ $(10^{10}\msun)$ & $18.4$&\cite{Shipp_2021}\\
        $r_\text{prog}$ $(\pc)$ & $12.5$ &\multicolumn{1}{c}{---} &\multicolumn{1}{c}{---}&\multicolumn{1}{c}{---} \\
        \hline
    \end{tabular}
    \label{tab:dynamical simulation parameters}
\end{table*}

In order to understand whether a LMC interaction can reproduce the kink Method 2 identifies, we set the free parameters of the $\phi_1$ progenitor position and the distance offset from \streamPoss distance gradient by hand to maximize agreement between the modeled stream's on-sky track and the empirical on-sky tracks. We find that $\phi_1 = -5^\circ$ and a distance offset of $0.6\kpc$ produce good agreement with the empirical stream track. 
As visible in Figures \ref{sfig:dynamical simulation distances} and \ref{fig:300S Distance Response}, a distance offset of $0.6\kpc$ also fits the empirical distance gradient at the location of \rrl and roughly matches the stream overdensity in $\phi_1-\mu$ space. 
Further work should attempt to fit these parameters more carefully, perhaps in a joint model with the LMC mass as in, e.g., \cite{Erkal_2019}, \cite{Shipp_2021}, and \cite{Koposov2023}.

\newpage

\subsection{Discussion of the Dynamical Model}

We present the results of the dynamical model in Figures \ref{fig:dynamical simulation onsky distribution}, \ref{fig:dynamical simulation offsky features}, and \ref{fig:dynamical simulation orbit}.\footnote{We also include an animation showcasing \streamPoss evolution with time and the effect of the LMC perturbation. 
There are two versions of the animation. 
The first can be found at \url{https://youtu.be/7MozYSgV5wQ} and just showcases \streamPoss evolution in galactocentric coordinates. 
The second can be found at \url{https://youtu.be/s-lxWk-ZruE} and also displays \streamPoss evolution in projections onto the $x-y$, $x-z$, and $y-z$ planes as well as an on-sky projection of \stream with coordinates transformed such that the $x$-axis is tangent to \streamPoss orbit as projected onto the sky.
We also include the particles' present day phase space information at \url{https://zenodo.org/records/15391938}.
}
In Figure \ref{fig:dynamical simulation onsky distribution}, we show the on-sky particle distribution from the simulation with and without the LMC interaction overplotted onto the empirical stream tracks. In Figure \ref{fig:dynamical simulation offsky features}, we compare the simulated distance, proper motion, and radial velocity tracks from the model with the LMC against those of the \SSSSS member stars. Finally, in Figure \ref{fig:dynamical simulation orbit} we show the orbit of \stream and its spatial distribution in galactocentric coordinates from the simulation with the LMC.

\subsubsection{Agreement Between the Dynamical Model and the Empirical Results}\label{sssec:agreement between the dynamical model and the empirical results}

The dynamical model's on-sky tracks reveal the importance of the LMC interaction in explaining \streamPoss modern day morphology. Figure \ref{sfig:onsky distribution no LMC} shows the result of the simulation without the influence of the LMC using the same progenitor current day phase space information. Except for their intersection at the progenitor's location, the simulated track and empirical tracks differ substantially in this case. In fact, the trailing arm -- considering particles with $\phi_1 < 15^\circ$ where both the majority of LMC model's trailing arm particles $(\sim73\%)$, and the primary on-sky signal, are located -- makes a $10.3^\circ$ angle with the $\phi_2 = 0^\circ$ line. 
Indeed, due to the mismatch between the proper motion slopes and the stream track, it is very challenging to reproduce both the on-sky angle of the trailing arm and the \SSSSS member kinematics without the LMC.
This result indicates that a substantial perturbative influence is necessary to explain \streamPoss modern day track and kinematics. 

Once the LMC is included in the dynamical model, the two on-sky tracks agree. 
Figure \ref{sfig:onsky distribution LMC} shows the simulation's track when the LMC is included. 
Now the track roughly follows the empirical on-sky track, with the angle between the tail and the $\phi_2 = 0^\circ$ line being only $-0.2^\circ$. This implies that the LMC interaction results in a $\sim 11^\circ$ rotation of the stream track. 
Further, the dynamical model reproduces the kink that Method 2 identifies with the correct slope.
This supports the kink being a real structure rather than an artifact of \sgr mismodeling. 
As both the stream's overall on-sky angle and the presence of the kink are dependent on the LMC interaction, \streamPoss track has the potential to be a strong constraint on the LMC's present mass.
The LMC induced misalignment between the stream track and the orbital track is clearly visible in galactocentric coordinates in Figure \ref{fig:dynamical simulation orbit} as well.

\begin{figure*}[ht]
    \centering
    \begin{minipage}{1\linewidth}
    \begin{center}
        \includegraphics[width = 1\linewidth]{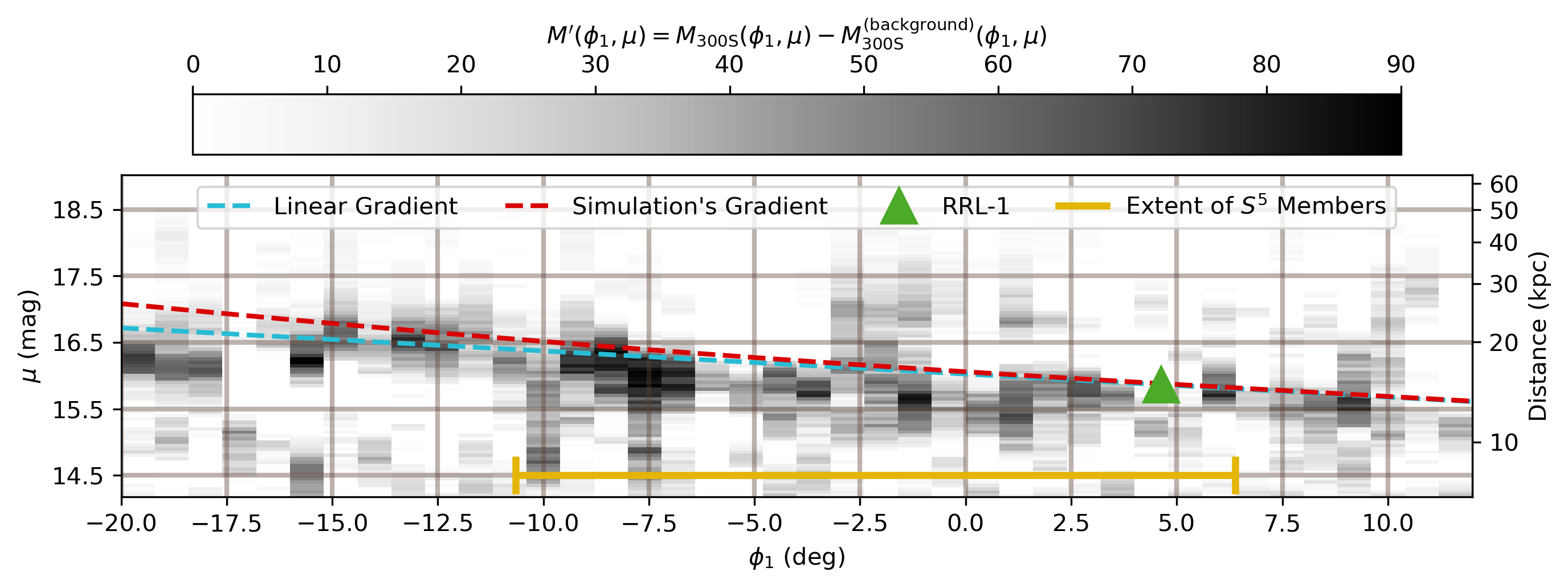}
    \end{center}
    \end{minipage}
    \caption{
    The background subtracted $\phi_1 - \mu$ distribution for \stream. The yellow line represents the extent of the \SSSSS members that were used in Section \ref{ssec:distance gradient} to compute \streamPoss distance gradient. \rrl is shown as a green triangle. \streamPoss gradient itself is shown as a blue dashed line. Although the gradient approximately follows the stream overdensity, the stream distance appears to curve upward (i.e. towards larger distances) faster than the gradient in negative $\phi_1$ and appears flatter than the gradient in positive $\phi_1$. The red dashed line is the distance function we find through our dynamical simulations of \stream. Its curvature matches \streamPoss and captures the upturn at $\phi_1 \lesssim -5^\circ$, although it does not account for the overdensity at $\phi_1 \sim -20^\circ$.
    }
    \label{fig:300S Distance Response}
\end{figure*}

The dynamical simulation produces a nonlinear distance gradient which we compare with the empirical distance gradient derived in Section \ref{ssec:distance gradient} and with the gradient of \cite{Fu_2018} in Figure \ref{sfig:dynamical simulation distances}. 
This figure makes clear why a progenitor distance offset was necessary in the instantiation of the dynamical model.
Although the simulation gradient agrees with the empirical one in positive $\phi_1$, it grows faster as $\phi_1$ decreases. 
The distance offset makes up for this deviation at \streamPoss progenitor's location in the simulation at $\phi_1 = -5^\circ$.
The dynamical model's gradient is still within $\sim1.7\sigma$ of the linear gradient over $\phi_1 \in [-20^\circ,15^\circ]$. 
It identifies a stream distance of $15.0\kpc$ at the position of \rrl which agrees with \streamPoss distance gradient at that position. 
Finally, the dynamical model's gradient agrees with \streamPoss overdensity in $\phi_1 - \mu$ space as described in Section \ref{ssec:nonlinear distance gradient} and seen in Figure \ref{fig:300S Distance Response}. 

The dynamical model proper motion tracks agree with the track formed by the \SSSSS member stars as seen in Figures \ref{sfig:dynamical simulation pm1} and \ref{sfig:dynamical simulation pm2}.
The radial velocity track also matches the \SSSSS member track as seen in Figure \ref{sfig:dynamical simulation rvs}.
These kinematics are not reflex corrected in order to remain independent of the particulars of the distance gradient. 

Overall, the dynamical simulation and the empirical models are in good agreement in their spatial distributions and kinematics. 
This agreement includes the kink identified empirically by Method 2.
It implies that a strong interaction with the LMC can produce the feature while being consistent with most of the stream's other kinematic and spatial attributes.
From this analysis, we conclude that the kink is possibly real and, if so, its formation and morphology could provide a strong constraint on the LMC's mass.
Further investigations and spectroscopic followup in the region will be necessary to fully confirm or reject the feature.

\begin{figure}
    \centering
    \makebox[\columnwidth]{
    \includegraphics[width=1\linewidth]{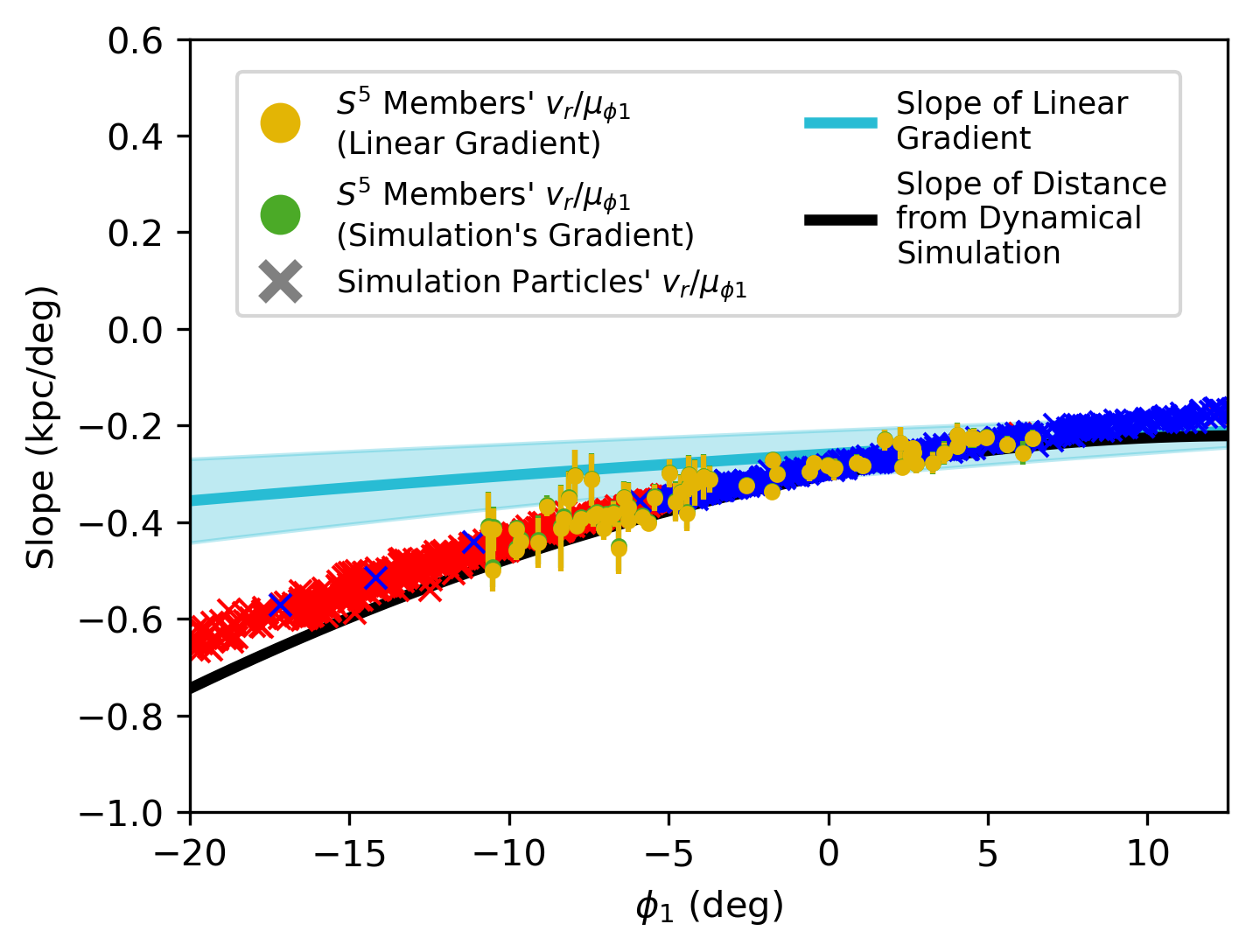}   
    }
    \caption{Relation between radial velocity and distance gradient for \stream. The yellow dots and associated uncertainties represent the \SSSSS member velocity slope after reflex correction using the \stream distance gradient derived in Section \ref{ssec:distance gradient}. The green points represent the velocity slope after reflex correction using the distance gradient derived from the dynamical simulation. $1\sigma$ uncertainty in the linear gradient slope is shown in the shaded region.}
    \label{fig:radial velocity proper motion slope}
\end{figure}

\subsubsection{Orbital Characteristics}\label{sssec:orbital characteristics}

The dynamical simulation identifies \streamPoss orbit to have an apocenter of $62.0\kpc$ and a pericenter of $7.1\kpc$ with an eccentricity of $0.80$. This is around the same eccentricity as found by \cite{Li_2022}, but with a larger pericenter and apocenter. \cite{Li_2022} found an apocenter of $45.8\kpc$ and a pericenter of $5.8\kpc$. Most of this difference is not due to the influence of the LMC. Without the LMC, we find \streamPoss apocenter is $58.4\kpc$ and its pericenter is $6.8\kpc$. Rather, the difference is likely due to modeling choices. For instance, \cite{Li_2022} use the MW potential derived by \cite{McMillan_2017} which has a virial mass around $30\%$ larger than \textsc{MWPotential2014} with our settings for $R_0$ and $V_c(R_0)$. 
Our results are similar to those of \cite{Fu_2018} who found a pericenter/apocenter of $4.1/\sim60 \kpc$.
Currently, \streamPoss radius to the Galactic center is $23\kpc$ and it last passed pericenter $69\myr$ ago. This is $\sim 8\%$ of \streamPoss orbital period of $861\myr$ in our model.

\subsubsection{Nonlinear Distance Gradient}\label{ssec:nonlinear distance gradient}

Figure \ref{sfig:dynamical simulation distances} shows that the dynamical simulation of \stream produces a nonlinear distance gradient that agrees with our empirically derived gradient in positive $\phi_1$ but diverges slightly at negative $\phi_1$. In this section, we analyze the possibility of \stream having a nonlinear distance gradient.

To begin, we use our \sgr model and our matched filter definitions to investigate \streamPoss distribution in $\phi_1 - \mu$ space. We derive its $M_{\stream}(\phi_1,\mu)$ distribution as we did for \sgr in Section \ref{sssec:sgr distance gradient through matched filter} and Figure $\ref{fig:sgr gradient derivation}$. We again begin by selecting an on-stream region. To capture the kink, we consider quadrilateral bins with corners $\phi_2 = \Phi_2(\phi_1) \pm 0.75^\circ$ where $\Phi_2(\phi_1)$ is the Method 2 stream track. We set $\Delta \phi_1$ to a constant $0.8^\circ$. We compute $M_{\stream}(\phi_1,\mu)$ by again considering $\mu$ values from $14.2$ to $19$ $(6.9 - 63.1\kpc)$ with $\Delta \mu = 0.05$. We use a new matched filter defined as in Section \ref{ssec:naive matched filter} (Equations \ref{eq:lower bound} and \ref{eq:upper bound}) with $s = 1/3$. 
The resulting $M_{\stream}(\phi_1,\mu)$ is heavily contaminated by \sgr. 
To obtain any useful result from this distribution, it is once again necessary to compute \sgrPoss contribution and subtract it out. 

In order to subtract \sgrPoss contribution, we consider the stars within the reflection of the bins across \sgrPoss stream track at their respective $\lambda$ coordinates. We then compute a new $M^\text{(background)}_{\stream}(\phi_1,\mu)$ using stars which fall in the corresponding reflected bins. Finally, we compute the background subtracted distribution $M'(\phi_1,\mu) = M_{\stream}(\phi_1,\mu) - M^\text{(background)}_{\stream}(\phi_1,\mu)$. 
We show $M'(\lambda,\mu)$ in Figure \ref{fig:300S Distance Response}. 

We first note that our subtraction of \sgrPoss influence appears successful. In Figure \ref{fig:300S Distance Response}, the four peaks in \streamPoss stellar density are visible at the correct $\phi_1$ values.
We also find that the extracted overdensity matches the distance to \rrl at its $\phi_1$ position. 

Interestingly, the gradient of the overdensity in Figure \ref{fig:300S Distance Response} appears nonlinear. 
We overplot the distance gradient derived from the \SSSSS member stars in Section \ref{ssec:distance gradient} as the light blue dashed line. 
Although the overdensity generally agrees with this gradient, there are some deviations. 
In $\phi_1\gtrsim0^\circ$ the gradient decreases faster than the overdensity and the former's slope has a larger magnitude. In $\phi_1\lesssim0^\circ$, the linear gradient's slope appears to have a smaller magnitude and varies slower. 

Although \streamPoss overdensity appears to have a more extreme slope than the distance gradient between $\phi_1 \sim -15^\circ$ and $\sim-5^\circ$, it bends down again between $\phi_1 \sim -20^\circ$ and $\sim-16^\circ$. 
It is unclear whether this bend is real. 
\streamPoss kink causes its track to approach \sgrPoss track in the relevant region, making it more difficult to fully remove \sgrPoss contamination. 
The overdensity at $-20^\circ$ is near the distance of \sgr in the region, see Figure \ref{fig:sgr gradient derivation}, implying that it could be contamination.
Nevertheless, the complex behavior of \streamPoss distance indicates a possible under-modeling of \stream in those regions.

As an additional check, we compare the velocity of the stream with its spatial distribution. Generally, streams that are unperturbed travel in the same direction as they are extended \citep[e.g.,][]{Erkal_2019,de_Boer_2020,Li_2021,Shipp_2021}. As can be seen in Figure \ref{fig:radial velocity proper motion slope}, the \SSSSS members and dynamical simulation particles' $v_r/\mu_{\phi1}$ are in strong agreement and are both misaligned with the linear stream slope, especially in the vicinity of the kink. This pattern is independent of the specific distance gradient used to compute the solar reflex correction. The ratio more closely mimics the slope of the nonlinear gradient identified by the dynamical simulation. 
However, perturbations can also form offsets between the distance gradient slope and $v_r/\mu_{\phi1}$ \citep[e.g.,][]{Shipp_2021}.

Overall, we find tentative evidence that \stream has a nonlinear distance gradient. However, we considered a nonlinear fit to both the \SSSSS members and the horizontal branch candidates identified using Equation \ref{eq:Horizontal Branch Filter} and found that neither dramatically altered the resulting empirical models. So we leave additional study to future work. Additional simulation work would help pin down a precise functional form for \streamPoss gradient that describes any nonlinearity, and additional spectroscopic members will help constrain the gradient robustly against \sgr.

\begin{figure*}
    \subfigure[\label{sfig:dynamical kinematics time}]{\includegraphics[width = 0.475\linewidth]{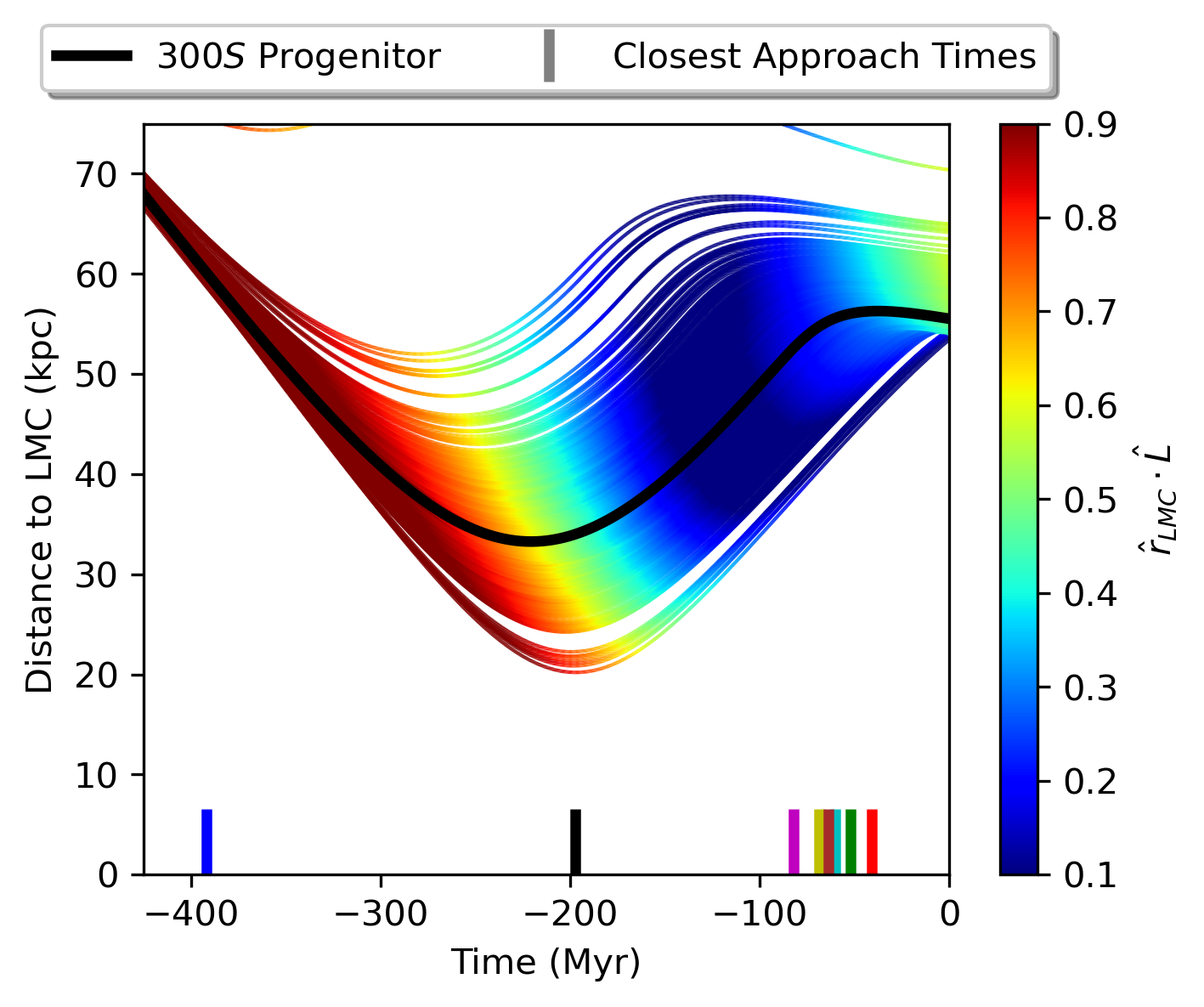}}
    \subfigure[\label{sfig:temporal kinematics}]{\includegraphics[width = 0.52\linewidth]{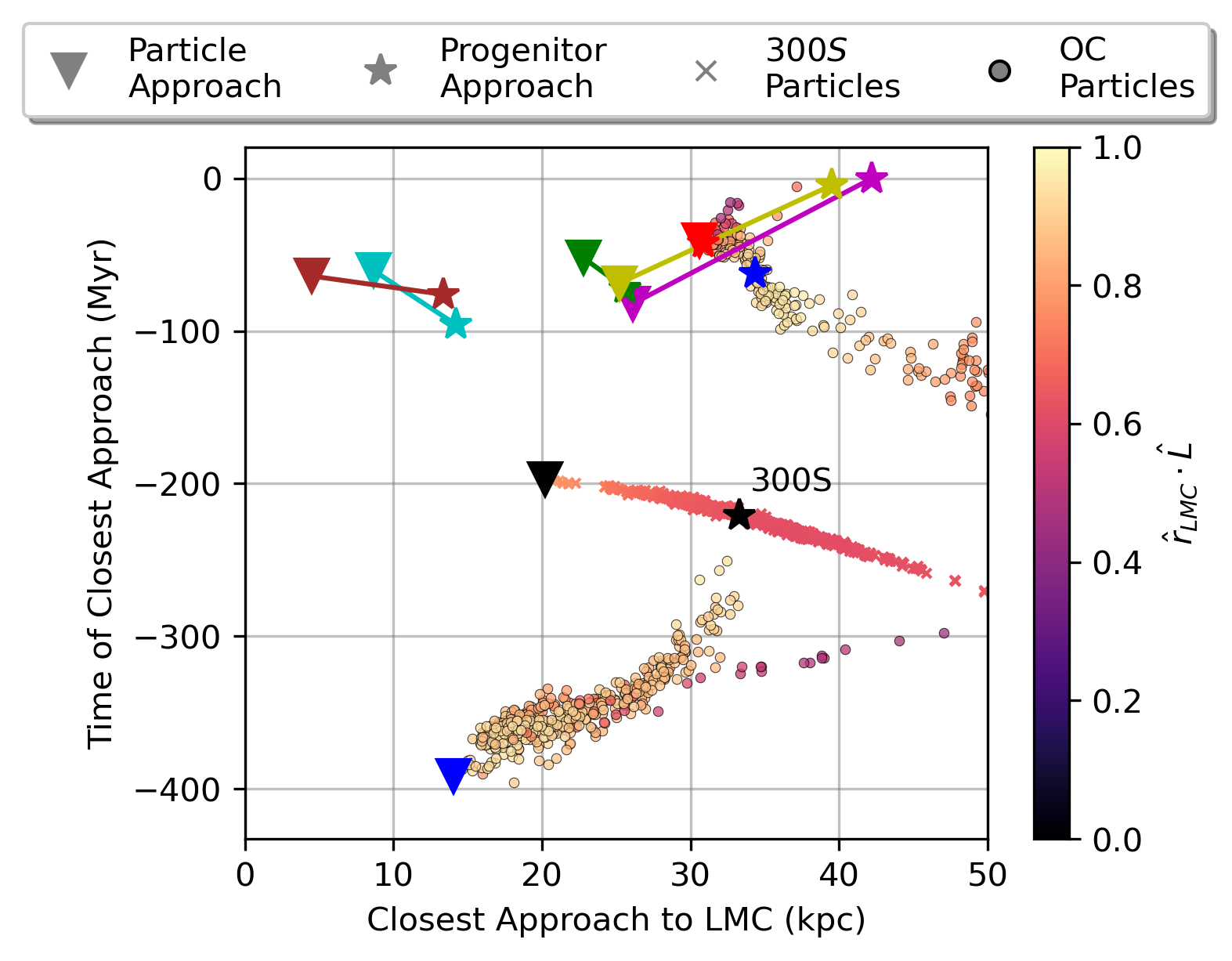}} 
    
    \begin{minipage}{0.7\linewidth}
        \rightline{
        \subfigure[\label{sfig:dynamical kinematics shipp}]{\includegraphics[width = 0.75\linewidth]{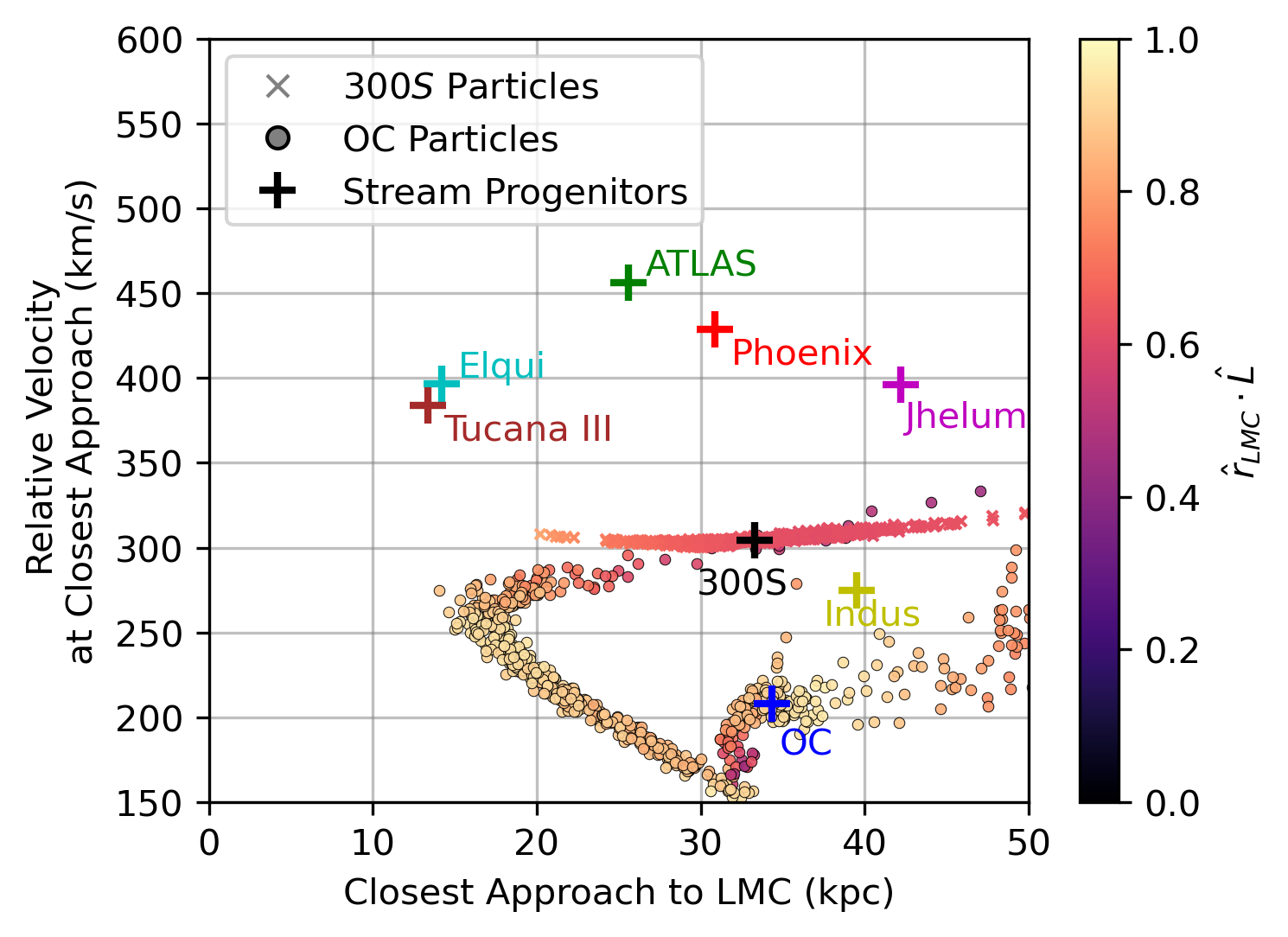}}}
    \end{minipage}\begin{minipage}{0.2\linewidth}
        \centering
        \includegraphics[width = 0.9\linewidth]{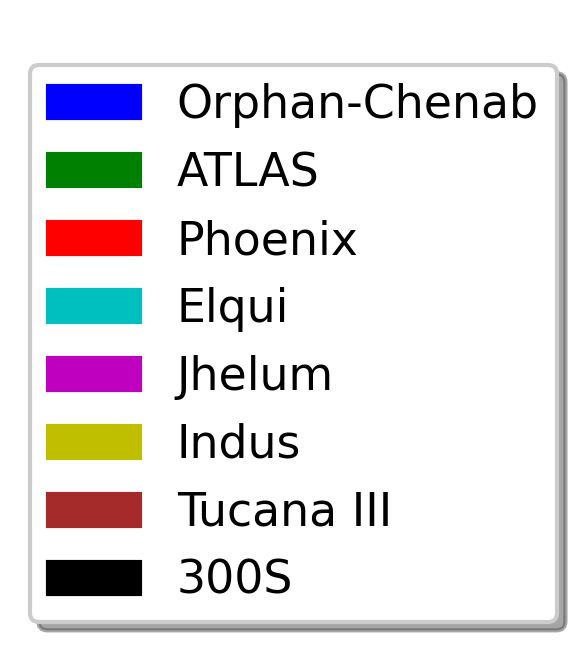}
    \end{minipage}
    
    \caption{Summary of kinematic relation between \stream and the LMC. (a) The time dependence of the \stream-LMC interaction. The multicolored lines are the trajectories of \streamPoss particles. The color represents $\hat{r}_\text{LMC}\cdot \hat{L}$ as a metric for the on-sky visibility of the perturbation. For more details, see \cite{Shipp_2021}. The black line shows the trajectory of \streamPoss progenitor. Note especially the gradient in $\hat{r}_\text{LMC}\cdot \hat{L}$ over time and its high value on \streamPoss approach towards the LMC. The colored vertical ticks represent the times of closest particle approach for the streams simulated by \cite{Shipp_2021}. 
    (b) A comparison between \streamPoss closest approach time and distance with other streams' closest approach times and distances. The triangles represent the time and distance of closest particle approach while the stars represent the time and distance of closest progenitor approach. Streams without plotted particles have their markers connected for easier comparison. Note how \stream captures a unique period around $220\myr$ ago which is not probed by any stream other than OC, whose closest particle approach and closest progenitor approach straddle \stream.
    (c) \streamPoss position in closest approach distance-relative velocity space compared to other streams. Stream particles are included for the OC stream and \stream in (b) and (c).
    }
    \label{fig:kinematics of dynamical model}
\end{figure*}

\subsubsection{The Influence of the LMC}\label{ssec:influence of lmc}

Figure \ref{fig:dynamical simulation onsky distribution} visually demonstrates how the LMC strongly influenced the formation and morphology of \stream. In this section, we quantify the strength of this relationship. 

\cite{Shipp_2021} utilized dynamical models of stellar streams to infer the mass of the LMC. 
To compare the relative strength of different streams in this inference, they considered the streams' relative velocities and distances compared to the LMC at their closest approach. 
Generally, the closer a stream's approach and the lower its relative velocity to the LMC, the stronger its perturbation. 
They noted, however, that not all perturbations are created equal. 
Perturbations in the radial direction lead to measurable differences in distance gradient and radial velocity, while perturbations in the on-sky plane lead to measurable differences in angular position and proper motion. 
As the distance gradient is difficult to measure with accuracy, the former type of perturbation leads to weaker constraints. 
To quantify this ``measurability'' of an LMC-induced perturbation, \cite{Shipp_2021} introduced the metric $\hat{r}_\text{LMC}\cdot \hat{L}$ as the dot product of the unit vector between a segment of the stream and the LMC with the stream segment's normalized angular momentum. 
Approximately, the closer this metric is to $1$, the more the perturbation is realized as measurable on-sky effects. 

We provide \streamPoss position within closest approach distance-relative velocity space in Figure \ref{sfig:dynamical kinematics shipp}. To avoid biasing the results with particles that are emitted after closest approach, we only consider particles released over $220\myr$ ago. \stream has a median $\hat{r}_\text{LMC}\cdot \hat{L}$ of $0.6$. Interestingly, this value changed substantially over \streamPoss interaction with the LMC, as seen in Figure \ref{sfig:dynamical kinematics time}. During \streamPoss approach towards the LMC, $\hat{r}_\text{LMC}\cdot \hat{L}$ was $\geq 0.9$ and on the same order as most of the OC stream, the stellar stream leading to the best constraint on the LMC mass of those analyzed by \cite{Shipp_2021}. However, during the \stream-LMC interaction, $\hat{r}_\text{LMC}\cdot \hat{L}$ decreased to below $0.1$. This indicates that much of the on-sky perturbation of \stream occurred during its approach towards the LMC. 

We note that, due to the LMC interaction, \stream  also has a close interaction with the progenitor of \sgr. Specifically, we integrate the orbit of the \sgr progenitor in our model assuming \sgrPoss current day position and velocity in galactocentric coordinates are $\vec x_\text{\sgr} = \left\{17.9,2.6,-6.6\right\}\kpc$ and $\vec v_\text{\sgr} = \left\{239.5,-29.6,213.5\right\}\kms$ as used by \cite{Vasiliev_2021}. In our simulation, the \stream progenitor has a closest approach of $9\kpc$ to \sgrPoss progenitor around $2\gyr$ ago. The leading arm of \stream gets closer, with a closest particle approach of $1.2\kpc$. Without the LMC, \streamPoss progenitor only reaches a closest approach of $\sim 23\kpc$, now around $3\gyr$ ago. We leave further study of the \stream and \sgr interaction to future work.

To make \streamPoss results comparable to those of other streams, we simulate the streams considered by \cite{Shipp_2021} within our simulation environment using the particle spray model of \cite{Chen_2024} which we use for \stream. We utilize the stream progenitor current day parameters described in \cite{Shipp_2021}'s Table A.1 for each of the streams they considered. Differences between our results and \cite{Shipp_2021}'s are due to the sensitivity of stellar streams to model hyperparameters \citep[e.g.][]{Koposov2023}. However, we recover their general results. We compare \stream and the other streams in Figure \ref{fig:kinematics of dynamical model}.

Compared to the streams analyzed by \cite{Shipp_2021}, \stream has a reasonably small velocity relative to the LMC upon closest approach. The relative velocity of \streamPoss progenitor at closest approach was $304\kms$, which is smaller than $5$ of the $7$ streams measured by \cite{Shipp_2021}. \stream has a distance to the LMC upon its closest approach in the middle of the other streams. Its progenitor's closest approach was $33\kpc$, which is closer than $3$ of the $7$ streams measured in \cite{Shipp_2021}. The trailing arm of \stream approaches closer, at around $25\kpc$.

Additionally, \stream has a unique time of closest approach. 
As seen in Figure \ref{sfig:dynamical kinematics time}, \streamPoss progenitor passed through its closest approach to the LMC $221\myr$ ago. 
\streamPoss closest particle to the LMC passed its closest approach around $197\myr$ ago. This is nearly $200\myr$ later than the approach of the closest OC stream particle at $392\myr$ ago. That time is shown in Figure \ref{sfig:dynamical kinematics time} as a vertical, blue tick. 
Other than the OC stream and \stream, each stream has a time of closest progenitor approach and closest particle approach $\lesssim 100\myr$ ago.
\streamPoss closest approach time of $221\myr$ fits in the middle between the OC stream's closest particle approach and the other streams' approaches. Due to the difference in time of closest approach between \stream and these other streams, \stream could act to probe the LMC at a different and less accessible moment in its temporal evolution. This new moment could act to better constrain the LMC's mass evolution and its orbit, as done in \cite{Koposov2023}. However, the measurability of more recent parts of \streamPoss LMC interaction may be limited, as indicated by the decreasing $\hat{r}_\text{LMC}\cdot \hat{L}$ over the interaction. 

Overall, the \stream-LMC interaction is both relatively strong and observable. The interaction acts to both rotate \streamPoss on-sky distribution and produce a kink in \stream. \stream has a very high $\hat{r}_\text{LMC}\cdot \hat{L}$ on its approach although its $\hat{r}_\text{LMC}\cdot \hat{L}$ decreases over the interaction. In relation to other streams, \stream has a small relative velocity upon closest approach. These features motivate additional study of \streamPoss utility as a constraint on the LMC's mass.

\section{Conclusions}\label{sec:conclusions}

In this paper, we present our study of the \stream stellar stream's morphology and formation. 
Understanding these attributes facilitates both additional follow-up observations of \stream and further studies of the stream's complex development. 
They also lay the groundwork for using \stream to constrain the structure of our Galaxy and, potentially, the LMC. 

\sgr is the primary difficulty in understanding \streamPoss morphology because it acts to contaminate the majority of the relevant field. 
To extract \streamPoss morphology against this contamination, we identified a coordinate system for \stream and re-derived its distance gradient with a method robust against \sgrPoss influence (Section \ref{ssec:naive matched filter} and \ref{ssec:distance gradient} respectively). 
We then applied two methods to extract \streamPoss signal and fit empirical models to its morphology. 
Method 1 used kinematic information to separate the signals (Section \ref{ssec:method 1 proper motion filtering}) while Method 2 used photometric information to simultaneously model \stream and \sgr (Section \ref{ssec:method 2 double stream model}). 
Motivated by these empirical results, we performed dynamical modeling to study the formation of \stream and its morphological features (Section \ref{sec:dynamical modeling}). 

The primary results of our empirical analysis can be found in Figures \ref{fig:distance gradients}, \ref{fig:Spline Values for the Two Streams}, \ref{fig:close view sgr and background subtracted}, and \ref{fig:stream results figure}. From our empirical analysis, we make the following conclusions.

\begin{itemize}
    
    \item We rederive a distance gradient robustly against \sgr contamination (see Section \ref{ssec:distance gradient} and Figure \ref{fig:distance gradients}).

    \item The empirical models generated using our two methods agree on their overlap and together describe \stream across $\sim 32^\circ$, extending the known track by as much as $7^\circ$ (see Figures \ref{fig:Spline Values for the Two Streams} and \ref{fig:stream results figure}). 
    
    \item Overall, \stream is a thin stellar stream that closely follows a great circle except for its kink (see Section \ref{ssec:features present in both models}).
    
    \item Although \stream has a long angular extent, its physical length is relatively short at $9.7\kpc$ (see Section \ref{ssec:features present in both models}).
    
    \item Within the newly identified stream regions, we find that \stream has a possible gap and a kink (see Sections \ref{ssec:features present in both models} and \ref{ssec:kink empirical} and Figure \ref{fig:stream results figure}). 
    
    \item Within the primary stream region, we find three peaks labeled A, B, and C in stellar density. 
    A fourth peak D has a less certain center and structure.
    
    \item The separation and periodicity of \streamPoss peaks are consistent with epicycles; but could also be related to either the complex disruption of globular clusters, or to some other perturbative influence on \stream (see Section \ref{ssec:features present in both models} and Figure \ref{fig:stream results figure}). 
    
    \item Using the integrated stellar density over our empirical models, we improve \cite{Usman_2024}'s lower limit on stream mass from $\log_{10} (M/M_\odot) = 4.5$ to $\log_{10} (M/M_\odot) = 4.7$  (see Section \ref{ssec:lower limit mass} and Figure \ref{fig:integrated light}). 
\end{itemize}

The primary results from our dynamical simulation can be found in Figures \ref{fig:dynamical simulation onsky distribution}, \ref{fig:dynamical simulation offsky features}, \ref{fig:dynamical simulation orbit}, and \ref{fig:kinematics of dynamical model}. From our dynamical simulation we make the following conclusions.

\begin{itemize}
    \item Our dynamical simulation indicates that the LMC strongly influenced \streamPoss formation, rotating the stream by approximately $11^\circ$ (see Sections \ref{sssec:agreement between the dynamical model and the empirical results} and \ref{ssec:influence of lmc} and Figures \ref{fig:dynamical simulation onsky distribution} and \ref{fig:kinematics of dynamical model}). 
    
    \item The dynamical simulation also reproduces \streamPoss kink, and indicates that it may be explained by a strong interaction with the LMC (see Section \ref{sssec:agreement between the dynamical model and the empirical results} and Figure \ref{fig:dynamical simulation onsky distribution}). 

    \item We find possible evidence that \stream has a nonlinear distance gradient with a decreasing slope over increasing $\phi_1$ (see Section \ref{ssec:nonlinear distance gradient} and Figure \ref{fig:300S Distance Response}). 

    \item Compared to other stellar streams used for constraining the LMC's mass \citep{Shipp_2021}, \streamPoss closest approach with the LMC occurred at $221\myr$ ago, a time distinct from other streams' closest approaches (see Section \ref{ssec:influence of lmc} and Figure \ref{sfig:dynamical kinematics time}).
    
\end{itemize}

There are multiple paths for further investigation of \stream. 
Improved dynamical simulations to better understand the \stream-LMC interaction would allow for a clearer picture of both \streamPoss morphology itself and the LMC's mass. 
Dynamical simulations could better constrain the formation of \streamPoss peaks and kink, as well as its interaction with \sgrPoss progenitor. 
Further, because \streamPoss interaction with the LMC leads to dramatic changes in its on-sky orientation, \stream has the potential to be another strong constraint on the LMC's mass. 
Due to its unique time of closest approach, further analysis of \streamPoss interaction with the LMC could provide a look into the time evolution of the LMC's mass profile. 
Because \stream is extremely sensitive to small scale dark matter subhalos \citep{Lu_2025} and on a retrograde orbit, further simulation studies could also examine whether \streamPoss structure hints at interactions with these elusive objects.

Empirical follow-up on the nature and morphology of \streamPoss kink -- to confirm its existence and angular relation to the rest of the stream -- would also be useful for this analysis.
Additional spectroscopically confirmed members and improvements to the precision of their distances could also help pin down the form of the distance gradient and of peak D.
Further investigations into \streamPoss density peaks could provide a better understanding of the necessary disruption physics or to possible perturbers. 
Similarly, if peak D is confirmed by followup investigations, \streamPoss gap provides another source for studies of the structure of \streamPoss progenitor, globular cluster disruption, and of the MW potential. 
Overall, \stream is a rich source for further investigations.

\section*{Acknowledgments}

B.C. acknowledges support from the University of Chicago Provost Scholarship and Provost Scholars Program.
A.P.J. acknowledges support from the National Science Foundation under grants AST-2206264 and AST-2307599.
S.K. acknowledges support from the Science \& Technology Facilities Council (STFC) grant ST/Y001001/1.
For the purpose of open access, the author has applied a Creative Commons Attribution (CC BY) license to any Author Accepted Manuscript version arising from this submission.
G.L. acknowledges funding from KICP/UChicago through a KICP Postdoctoral Fellowship.
A.P.L. and T.S.L. acknowledge financial support from the Natural Sciences and Engineering Research Council of Canada (NSERC) through grant RGPIN-2022-04794.
Y.Y. and G.F.L. acknowledge support from the Australian Research Council through the Discovery Program grant DP220102254.    
This research benefited from the Dwarf Galaxies, Star Clusters, and Streams Workshop hosted by the Kavli Institute for Cosmological Physics.
This work was supported by the Australian Research Council Centre of Excellence for All Sky Astrophysics in 3 Dimensions (ASTRO 3D), through project number CE170100013. S.L.M. acknowledges the support of the Australian Research Council through Discovery Project grant DP180101791, and the support of the UNSW Scientia Fellowship Program.
We thank the anonymous referee for their valuable comments.

The paper includes data obtained with the Anglo-Australian Telescope in Australia. We acknowledge the traditional owners of the land on which the AAT stands, the Gamilaraay people, and pay our respects to elders past, present and emerging.

The Legacy Surveys consist of three individual and complementary projects: the Dark Energy Camera Legacy Survey (DECaLS; Proposal ID \#2014B-0404; PIs: David Schlegel and Arjun Dey), the Beijing-Arizona Sky Survey (BASS; NOAO Prop. ID \#2015A-0801; PIs: Zhou Xu and Xiaohui Fan), and the Mayall z-band Legacy Survey (MzLS; Prop. ID \#2016A-0453; PI: Arjun Dey). DECaLS, BASS and MzLS together include data obtained, respectively, at the Blanco telescope, Cerro Tololo Inter-American Observatory, NSF’s NOIRLab; the Bok telescope, Steward Observatory, University of Arizona; and the Mayall telescope, Kitt Peak National Observatory, NOIRLab. Pipeline processing and analyses of the data were supported by NOIRLab and the Lawrence Berkeley National Laboratory (LBNL). The Legacy Surveys project is honored to be permitted to conduct astronomical research on Iolkam Du’ag (Kitt Peak), a mountain with particular significance to the Tohono O’odham Nation.

NOIRLab is operated by the Association of Universities for Research in Astronomy (AURA) under a cooperative agreement with the National Science Foundation. LBNL is managed by the Regents of the University of California under contract to the U.S. Department of Energy.

This project used data obtained with the Dark Energy Camera (DECam), which was constructed by the Dark Energy Survey (DES) collaboration. Funding for the DES Projects has been provided by the U.S. Department of Energy, the U.S. National Science Foundation, the Ministry of Science and Education of Spain, the Science and Technology Facilities Council of the United Kingdom, the Higher Education Funding Council for England, the National Center for Supercomputing Applications at the University of Illinois at Urbana-Champaign, the Kavli Institute of Cosmological Physics at the University of Chicago, Center for Cosmology and Astro-Particle Physics at the Ohio State University, the Mitchell Institute for Fundamental Physics and Astronomy at Texas A\&M University, Financiadora de Estudos e Projetos, Fundacao Carlos Chagas Filho de Amparo, Financiadora de Estudos e Projetos, Fundacao Carlos Chagas Filho de Amparo a Pesquisa do Estado do Rio de Janeiro, Conselho Nacional de Desenvolvimento Cientifico e Tecnologico and the Ministerio da Ciencia, Tecnologia e Inovacao, the Deutsche Forschungsgemeinschaft and the Collaborating Institutions in the Dark Energy Survey. The Collaborating Institutions are Argonne National Laboratory, the University of California at Santa Cruz, the University of Cambridge, Centro de Investigaciones Energeticas, Medioambientales y Tecnologicas-Madrid, the University of Chicago, University College London, the DES-Brazil Consortium, the University of Edinburgh, the Eidgenossische Technische Hochschule (ETH) Zurich, Fermi National Accelerator Laboratory, the University of Illinois at Urbana-Champaign, the Institut de Ciencies de l’Espai (IEEC/CSIC), the Institut de Fisica d’Altes Energies, Lawrence Berkeley National Laboratory, the Ludwig Maximilians Universitat Munchen and the associated Excellence Cluster Universe, the University of Michigan, NSF’s NOIRLab, the University of Nottingham, the Ohio State University, the University of Pennsylvania, the University of Portsmouth, SLAC National Accelerator Laboratory, Stanford University, the University of Sussex, and Texas A\&M University.

BASS is a key project of the Telescope Access Program (TAP), which has been funded by the National Astronomical Observatories of China, the Chinese Academy of Sciences (the Strategic Priority Research Program “The Emergence of Cosmological Structures” Grant \# XDB09000000), and the Special Fund for Astronomy from the Ministry of Finance. The BASS is also supported by the External Cooperation Program of Chinese Academy of Sciences (Grant \# 114A11KYSB20160057), and Chinese National Natural Science Foundation (Grant \# 12120101003, \# 11433005).

The Legacy Survey team makes use of data products from the Near-Earth Object Wide-field Infrared Survey Explorer (NEOWISE), which is a project of the Jet Propulsion Laboratory/California Institute of Technology. NEOWISE is funded by the National Aeronautics and Space Administration.

The Legacy Surveys imaging of the DESI footprint is supported by the Director, Office of Science, Office of High Energy Physics of the U.S. Department of Energy under Contract No. DE-AC02-05CH1123, by the National Energy Research Scientific Computing Center, a DOE Office of Science User Facility under the same contract; and by the U.S. National Science Foundation, Division of Astronomical Sciences under Contract No. AST-0950945 to NOAO.

This work has made use of data from the European Space Agency (ESA) mission
{\it Gaia} (\url{https://www.cosmos.esa.int/gaia}), processed by the {\it Gaia}
Data Processing and Analysis Consortium (DPAC,
\url{https://www.cosmos.esa.int/web/gaia/dpac/consortium}). Funding for the DPAC
has been provided by national institutions, in particular the institutions
participating in the {\it Gaia} Multilateral Agreement.

This work made use of \texttt{numpy} \citep{numpy}, \texttt{matplotlib} \citep{matplotlib}, \texttt{scipy} \citep{scipy}, \texttt{pandas} \citep{pandas,pandas_zenodo}, \texttt{ipython} \citep{ipython}, \texttt{statsmodels} \citep{statsmodels}, \texttt{shapely} \citep{Gillies_Shapely_2022}, \texttt{gala} \citep{gala,gala_zenodo}, and \texttt{galpy}\footnote{\url{http://github.com/jobovy/galpy}} \citep{Bovy_2015}.
%
%
\begin{appendix}
    
    \section{Spline Results for Method 1}\label{ap:Method 1 Spline Values}
    In this appendix, we include spline node positions and values for \stream as found using Method 1 (Section \ref{ssec:method 1 proper motion filtering}). 
    Here and in the other appendices, the provided node parameters may be used as inputs to a natural cubic spline interpolation scheme, such as \verb|CubicSpline| from the Python library \verb|scipy.interpolate| with \verb|bc_type='natural'|.
    We additionally provide machine readable versions of these tables, as well the complete set of samples for the models and the present day states of the dynamical simulation particles, at \url{https://zenodo.org/records/15391938}.
    
    \begin{table}[H]
        \centering
        \caption{Spline Nodes for $\Phi_2(\phi_1)$ for Method 1}
        \label{tab:Method 1 Phi2}

    \begin{tabular}{c|c|c|c}
    $\phi_1$ (deg) &Median& $16\%$ & $84\%$ \\
    \hline
    $-12.4722$ & $0.03451$ & $-0.15458$ & $0.23195$ \\
    $-7.5$ & $-0.16588$ & $-0.2858$ & $-0.05416$ \\
    $-5$ & $0.07033$ & $-0.13633$ & $0.25153$ \\
    $-3$ & $0.04836$ & $-0.28184$ & $0.3611$ \\
    $-1$ & $-0.3357$ & $-0.5696$ & $-0.11528$ \\
    $1$ & $-0.01952$ & $-0.23315$ & $0.23527$ \\
    $3$ & $0.06788$ & $-0.07103$ & $0.20049$ \\
    $5$ & $-0.7419$ & $-1.06453$ & $-0.15301$ \\
    $6$ & $-0.05864$ & $-0.44178$ & $0.1797$ \\
    $7.5$ & $0.11682$ & $-0.13757$ & $0.33846$ \\
    $9$ & $-0.35772$ & $-0.93933$ & $0.49332$ \\
    $11$ & $-0.14704$ & $-1.1424$ & $0.862$ \\
    $12.47615$ & $0.04826$ & $-0.9441$ & $1.00607$ \\
    \end{tabular}
        
    \end{table}   

    \begin{table}[H]
        \centering
        \caption{Spline Nodes for $ w(\phi_1)$ for Method 1}
        \label{tab:Method 1 Widths}

        \begin{tabular}{c|c|c|c}
        $\phi_1$ (deg) &Median& $16\%$ & $84\%$ \\
        \hline
        $-12.4722$ & $-1.1482$ & $-1.56019$ & $-0.749$ \\
        $-7.5$ & $-0.96866$ & $-1.22905$ & $-0.74144$ \\
        $-5$ & $-0.359$ & $-0.61178$ & $-0.12468$ \\
        $-3$ & $-0.06608$ & $-0.42604$ & $0.17956$ \\
        $-1$ & $-0.69863$ & $-1.09264$ & $-0.38308$ \\
        $1$ & $-0.73258$ & $-1.2275$ & $-0.2789$ \\
        $3$ & $-0.9855$ & $-1.21629$ & $-0.74314$ \\
        $5$ & $-1.04113$ & $-1.55225$ & $-0.51345$ \\
        $6$ & $-1.1174$ & $-1.58302$ & $-0.66265$ \\
        $7.5$ & $-1.1564$ & $-1.58847$ & $-0.67537$ \\
        $9$ & $-0.82212$ & $-1.30515$ & $-0.36033$ \\
        $11$ & $-0.91997$ & $-1.42016$ & $-0.38506$ \\
        $12.47615$ & $-0.94538$ & $-1.45144$ & $-0.43219$ \\
        \end{tabular}
        
    \end{table}    

    \begin{table}[H]
        \centering
        \caption{Spline Nodes for $\mathcal{I}(\phi_1)$ for Method 1}
        \label{tab:Method 1 Central Stellar Density}

        \begin{tabular}{c|c|c|c}
        $\phi_1$ (deg) &Median& $16\%$ & $84\%$ \\
        \hline
        $-12.4722$ & $-1.09481$ & $-1.83362$ & $-0.47423$ \\
        $-7.5$ & $-0.96614$ & $-1.30942$ & $-0.65287$ \\
        $-5$ & $-0.99868$ & $-1.34493$ & $-0.67723$ \\
        $-3$ & $-1.65766$ & $-2.03268$ & $-1.28925$ \\
        $-1$ & $-1.1614$ & $-1.58938$ & $-0.72899$ \\
        $1$ & $-1.86545$ & $-2.46176$ & $-1.31733$ \\
        $3$ & $-0.88536$ & $-1.34635$ & $-0.47797$ \\
        $5$ & $-3.06207$ & $-4.28368$ & $-2.10239$ \\
        $6$ & $-2.17293$ & $-3.62987$ & $-1.25492$ \\
        $7.5$ & $-1.2604$ & $-2.07768$ & $-0.55286$ \\
        $9$ & $-4.67722$ & $-10.00582$ & $-2.48122$ \\
        $11$ & $-6.93234$ & $-12.24508$ & $-3.64041$ \\
        $12.47615$ & $-8.99249$ & $-13.13426$ & $-4.54915$ \\
        \end{tabular}
        
    \end{table}

    \section{Spline Results for Sgr Model}\label{ap:Sgr Spline Values}
    
    In this appendix, we include the spline node positions and values for \sgr as found in Section \ref{ssec:empirical characterization of sgr}.

    \begin{table}[H]
        \centering
        \caption{Spline Nodes for $\mathcal{B}(\lambda)$ for \sgr}
        \label{tab:Method 1 Phi2}

        \begin{tabular}{c|c|c|c}
        $\lambda$ (deg) &Median& $16\%$ & $84\%$ \\
        \hline
        $-23.78772$ & $0.09363$ & $-0.11451$ & $0.31663$ \\
        $-15$ & $-0.72274$ & $-0.8373$ & $-0.61206$ \\
        $-5$ & $-0.35984$ & $-0.49998$ & $-0.21786$ \\
        $0$ & $-0.46852$ & $-0.63451$ & $-0.30133$ \\
        $5$ & $0.13257$ & $-0.02154$ & $0.28308$ \\
        $11$ & $-0.018$ & $-0.18006$ & $0.1528$ \\
        $16.20932$ & $0.632$ & $0.19356$ & $1.09284$ \\
        \end{tabular}
        
    \end{table}   

    \begin{table}[H]
        \centering
        \caption{Spline Nodes for $ w(\lambda)$ for \sgr}
        \label{tab:Method 1 Widths}

        \begin{tabular}{c|c|c|c}
        $\lambda$ (deg) &Median& $16\%$ & $84\%$ \\
        \hline
        $-23.78772$ & $0.87547$ & $0.77479$ & $0.97608$ \\
        $-15$ & $0.84795$ & $0.79479$ & $0.90084$ \\
        $-5$ & $1.04225$ & $0.99478$ & $1.09105$ \\
        $0$ & $0.95295$ & $0.88879$ & $1.0153$ \\
        $5$ & $0.93433$ & $0.85965$ & $1.00481$ \\
        $11$ & $0.84543$ & $0.7609$ & $0.933$ \\
        $16.20932$ & $0.68329$ & $0.4453$ & $0.89756$ \\
        \end{tabular}
        
    \end{table}    

    \begin{table}[H]
        \centering
        \caption{Spline Nodes for $ \mathcal{I}(\lambda)$ for \sgr}
        \label{tab:Method 1 Central Stellar Density}

        \begin{tabular}{c|c|c|c}
        $\lambda$ (deg) &Median& $16\%$ & $84\%$ \\
        \hline
        $-23.78772$ & $1.56917$ & $1.48627$ & $1.65099$ \\
        $-15$ & $1.5144$ & $1.47073$ & $1.55763$ \\
        $-5$ & $1.56675$ & $1.52205$ & $1.61157$ \\
        $0$ & $1.54271$ & $1.48907$ & $1.59289$ \\
        $5$ & $1.4115$ & $1.35292$ & $1.46659$ \\
        $11$ & $1.375$ & $1.30466$ & $1.43639$ \\
        $16.20932$ & $0.95117$ & $0.77894$ & $1.11815$ \\
        \end{tabular}
        
    \end{table}

    \section{Spline Results for Method 2}\label{ap:Method 2 Spline Values}
    
    In this appendix, we include spline node positions and values for \stream as found using Method 2 (Section \ref{ssec:method 2 double stream model}).

    \begin{table}[H]
        \centering
        \caption{Spline Nodes for $\Phi_2(\phi_1)$ for Method 2}
        \label{tab:Method 2 Phi2}

        \begin{tabular}{c|c|c|c}
        $\phi_1$ (deg) &Median& $16\%$ & $84\%$ \\
        \hline
        $-20.09482$ & $2.13973$ & $1.94689$ & $2.35251$ \\
        $-15$ & $0.83754$ & $0.53236$ & $1.10646$ \\
        $-10$ & $-0.10692$ & $-0.16822$ & $-0.03979$ \\
        $-7.5$ & $-0.04552$ & $-0.11765$ & $0.02836$ \\
        $-5$ & $-0.18016$ & $-0.23413$ & $-0.12755$ \\
        $-3$ & $0.01807$ & $-0.08089$ & $0.12465$ \\
        $-1.5$ & $-0.19026$ & $-0.31072$ & $-0.0808$ \\
        $0$ & $-0.33377$ & $-0.49751$ & $-0.17492$ \\
        $1.5$ & $-0.0927$ & $-0.40269$ & $0.23555$ \\
        $2.5$ & $-0.36475$ & $-0.47459$ & $-0.25023$ \\
        $4$ & $-0.1276$ & $-0.76882$ & $0.63328$ \\
        $5$ & $0.06985$ & $-0.60708$ & $0.72525$ \\
        $7$ & $-0.18322$ & $-0.84495$ & $0.46682$ \\
        $9$ & $-0.12043$ & $-0.54863$ & $0.21882$ \\
        $11$ & $-0.12115$ & $-0.60963$ & $0.54154$ \\
        $13.09569$ & $0.09346$ & $-1.02696$ & $1.04144$ \\
        \end{tabular}
        
    \end{table}   

    \begin{table}[H]
        \centering
        \caption{Spline Nodes for $ w(\phi_1)$ for Method 2}
        \label{tab:Method 2 Widths}

        \begin{tabular}{c|c|c|c}
        $\phi_1$ (deg) &Median& $16\%$ & $84\%$ \\
        \hline
        $-20.09482$ & $-1.08362$ & $-1.41119$ & $-0.74395$ \\
        $-15$ & $-0.00022$ & $-0.37753$ & $0.31449$ \\
        $-10$ & $-1.27464$ & $-1.48191$ & $-1.0512$ \\
        $-7.5$ & $-0.86829$ & $-1.08127$ & $-0.70401$ \\
        $-5$ & $-1.40426$ & $-1.57312$ & $-1.22967$ \\
        $-3$ & $-1.04102$ & $-1.3476$ & $-0.72157$ \\
        $-1.5$ & $-0.95999$ & $-1.21266$ & $-0.72737$ \\
        $0$ & $-1.35799$ & $-1.69143$ & $-0.991$ \\
        $1.5$ & $-0.47983$ & $-0.91903$ & $-0.05595$ \\
        $2.5$ & $-1.02566$ & $-1.24726$ & $-0.77864$ \\
        $4$ & $-1.0223$ & $-1.52709$ & $-0.51001$ \\
        $5$ & $-0.82289$ & $-1.32445$ & $-0.36727$ \\
        $7$ & $-1.00148$ & $-1.47787$ & $-0.51148$ \\
        $9$ & $-0.7672$ & $-1.22214$ & $-0.39706$ \\
        $11$ & $-0.90397$ & $-1.33514$ & $-0.44385$ \\
        $13.09569$ & $-0.94416$ & $-1.43329$ & $-0.43366$ \\
        \end{tabular}
            
    \end{table}    

    \begin{table}[H]
        \centering
        \caption{Spline Nodes for $ \mathcal{I}(\phi_1)$ for Method 2}
        \label{tab:Method 2 Central Stellar Density}

        \begin{tabular}{c|c|c|c}
        $\phi_1$ (deg) &Median& $16\%$ & $84\%$ \\
        \hline
        $-20.09482$ & $1.72369$ & $1.32105$ & $2.06118$ \\
        $-15$ & $0.99816$ & $0.73675$ & $1.22487$ \\
        $-10$ & $2.10921$ & $1.89082$ & $2.30911$ \\
        $-7.5$ & $2.41012$ & $2.24694$ & $2.56922$ \\
        $-5$ & $2.59217$ & $2.39074$ & $2.78161$ \\
        $-3$ & $1.86333$ & $1.50099$ & $2.17118$ \\
        $-1.5$ & $2.29467$ & $2.00034$ & $2.56773$ \\
        $0$ & $1.51388$ & $0.85506$ & $1.98447$ \\
        $1.5$ & $1.10012$ & $0.36338$ & $1.60074$ \\
        $2.5$ & $2.51922$ & $2.16935$ & $2.81322$ \\
        $4$ & $-2.38898$ & $-3.98931$ & $-0.80396$ \\
        $5$ & $-0.5341$ & $-3.26221$ & $0.87144$ \\
        $7$ & $-0.93168$ & $-3.15648$ & $0.5347$ \\
        $9$ & $1.45176$ & $0.4942$ & $1.94699$ \\
        $11$ & $-0.33036$ & $-3.19487$ & $1.06812$ \\
        $13.09569$ & $-1.7112$ & $-3.9308$ & $0.75085$ \\
        \end{tabular}
        
    \end{table}

\end{appendix}

\newpage

\bibliography{main}{}

@article{Geha_2009,
   title={THE LEAST-LUMINOUS GALAXY: SPECTROSCOPY OF THE MILKY WAY SATELLITE SEGUE 1},
   volume={692},
   ISSN={1538-4357},
   url={http://dx.doi.org/10.1088/0004-637X/692/2/1464},
   DOI={10.1088/0004-637x/692/2/1464},
   number={2},
   journal={The Astrophysical Journal},
   publisher={American Astronomical Society},
   author={Geha, Marla and Willman, Beth and Simon, Joshua D. and Strigari, Louis E. and Kirby, Evan N. and Law, David R. and Strader, Jay},
   year={2009},
   month=feb, pages={1464–1475} }

@article{Niederste_Ostholt_2009,
   title={The origin of Segue 1},
   volume={398},
   ISSN={1365-2966},
   url={http://dx.doi.org/10.1111/j.1365-2966.2009.15287.x},
   DOI={10.1111/j.1365-2966.2009.15287.x},
   number={4},
   journal={Monthly Notices of the Royal Astronomical Society},
   publisher={Oxford University Press (OUP)},
   author={Niederste-Ostholt, M. and Belokurov, V. and Evans, N. W. and Gilmore, G. and Wyse, R. F. G. and Norris, J. E.},
   year={2009},
   month=oct, pages={1771–1781} }

@article{Norris_2010,
   title={CHEMICAL ENRICHMENT IN THE FAINTEST GALAXIES: THE CARBON AND IRON ABUNDANCE SPREADS IN THE BOÖTES I DWARF SPHEROIDAL GALAXY AND THE SEGUE 1 SYSTEM},
   volume={723},
   ISSN={1538-4357},
   url={http://dx.doi.org/10.1088/0004-637X/723/2/1632},
   DOI={10.1088/0004-637x/723/2/1632},
   number={2},
   journal={The Astrophysical Journal},
   publisher={American Astronomical Society},
   author={Norris, John E. and Wyse, Rosemary F. G. and Gilmore, Gerard and Yong, David and Frebel, Anna and Wilkinson, Mark I. and Belokurov, V. and Zucker, Daniel B.},
   year={2010},
   month=oct, pages={1632–1650} }

@article{Belokurov_2007,
   title={Cats and Dogs, Hair and a Hero: A Quintet of New Milky Way Companions},
   volume={654},
   ISSN={1538-4357},
   url={http://dx.doi.org/10.1086/509718},
   DOI={10.1086/509718},
   number={2},
   journal={The Astrophysical Journal},
   publisher={American Astronomical Society},
   author={Belokurov, V. and Zucker, D. B. and Evans, N. W. and Kleyna, J. T. and Koposov, S. and Hodgkin, S. T. and Irwin, M. J. and Gilmore, G. and Wilkinson, M. I. and Fellhauer, M. and Bramich, D. M. and Hewett, P. C. and Vidrih, S. and De Jong, J. T. A. and Smith, J. A. and Rix, H.‐W. and Bell, E. F. and Wyse, R. F. G. and Newberg, H. J. and Mayeur, P. A. and Yanny, B. and Rockosi, C. M. and Gnedin, O. Y. and Schneider, D. P. and Beers, T. C. and Barentine, J. C. and Brewington, H. and Brinkmann, J. and Harvanek, M. and Kleinman, S. J. and Krzesinski, J. and Long, D. and Nitta, A. and Snedden, S. A.},
   year={2007},
   month=jan, pages={897–906} }

@article{Simon_2011,
   title={A COMPLETE SPECTROSCOPIC SURVEY OF THE MILKY WAY SATELLITE SEGUE 1: THE DARKEST GALAXY},
   volume={733},
   ISSN={1538-4357},
   url={http://dx.doi.org/10.1088/0004-637X/733/1/46},
   DOI={10.1088/0004-637x/733/1/46},
   number={1},
   journal={The Astrophysical Journal},
   publisher={American Astronomical Society},
   author={Simon, Joshua D. and Geha, Marla and Minor, Quinn E. and Martinez, Gregory D. and Kirby, Evan N. and Bullock, James S. and Kaplinghat, Manoj and Strigari, Louis E. and Willman, Beth and Choi, Philip I. and Tollerud, Erik J. and Wolf, Joe},
   year={2011},
   month=may, 
    pages={46} }

@article{Frebel_2013,
   title={THE 300 km s–1STELLAR STREAM NEAR SEGUE 1: INSIGHTS FROM HIGH-RESOLUTION SPECTROSCOPY OF ITS BRIGHTEST STAR},
   volume={771},
   ISSN={1538-4357},
   url={http://dx.doi.org/10.1088/0004-637X/771/1/39},
   DOI={10.1088/0004-637x/771/1/39},
   number={1},
   journal={The Astrophysical Journal},
   publisher={American Astronomical Society},
   author={Frebel, Anna and Lunnan, Ragnhild and Casey, Andrew R. and Norris, John E. and Wyse, Rosemary F. G. and Gilmore, Gerard},
   year={2013},
   month=jun, pages={39} }

@article{Grillmair_2013, 
    title={A Matched-Filter Map of the 300 km/s Stream}, 
    volume={9}, 
    DOI={10.1017/S1743921313006728}, 
    number={S298}, 
    journal={Proceedings of the International Astronomical Union}, 
    author={Grillmair, C. J.}, 
    year={2013}, 
    pages={405–405}
}

@article{Bernard_2016,
   title={A synoptic map of halo substructures from the Pan-STARRS1 3π survey},
   volume={463},
   ISSN={1365-2966},
   url={http://dx.doi.org/10.1093/mnras/stw2134},
   DOI={10.1093/mnras/stw2134},
   number={2},
   journal={Monthly Notices of the Royal Astronomical Society},
   publisher={Oxford University Press (OUP)},
   author={Bernard, Edouard J. and Ferguson, Annette M. N. and Schlafly, Edward F. and Martin, Nicolas F. and Rix, Hans-Walter and Bell, Eric F. and Finkbeiner, Douglas P. and Goldman, Bertrand and Martínez-Delgado, David and Sesar, Branimir and Wyse, Rosemary F. G. and Burgett, William S. and Chambers, Kenneth C. and Draper, Peter W. and Hodapp, Klaus W. and Kaiser, Nicholas and Kudritzki, Rolf-Peter and Magnier, Eugene A. and Metcalfe, Nigel and Wainscoat, Richard J. and Waters, Christopher},
   year={2016},
   month=aug, pages={1759–1768} }

@article{Carlin_2012,
   title={THE ORIGIN OF THE VIRGO STELLAR SUBSTRUCTURE},
   volume={753},
   ISSN={1538-4357},
   url={http://dx.doi.org/10.1088/0004-637X/753/2/145},
   DOI={10.1088/0004-637x/753/2/145},
   number={2},
   journal={The Astrophysical Journal},
   publisher={American Astronomical Society},
   author={Carlin, Jeffrey L. and Yam, William and Casetti-Dinescu, Dana I. and Willett, Benjamin A. and Newberg, Heidi J. and Majewski, Steven R. and Girard, Terrence M.},
   year={2012},
   month=jun, pages={145} }

@article{Perottoni_2022,
   title={The Unmixed Debris of Gaia-Sausage/Enceladus in the Form of a Pair of Halo Stellar Overdensities},
   volume={936},
   ISSN={2041-8213},
   url={http://dx.doi.org/10.3847/2041-8213/ac88d6},
   DOI={10.3847/2041-8213/ac88d6},
   number={1},
   journal={The Astrophysical Journal Letters},
   publisher={American Astronomical Society},
   author={Perottoni, Hélio D. and Limberg, Guilherme and Amarante, João A. S. and Rossi, Silvia and Queiroz, Anna B. A. and Santucci, Rafael M. and Pérez-Villegas, Angeles and Chiappini, Cristina},
   year={2022},
   month=aug, pages={L2} }

@article{Li_2022,
   title={S5: The Orbital and Chemical Properties of One Dozen Stellar Streams},
   volume={928},
   ISSN={1538-4357},
   url={http://dx.doi.org/10.3847/1538-4357/ac46d3},
   DOI={10.3847/1538-4357/ac46d3},
   number={1},
   journal={The Astrophysical Journal},
   publisher={American Astronomical Society},
   author={Li, Ting S. and Ji, Alexander P. and Pace, Andrew B. and Erkal, Denis and Koposov, Sergey E. and Shipp, Nora and Da Costa, Gary S. and Cullinane, Lara R. and Kuehn, Kyler and Lewis, Geraint F. and Mackey, Dougal and Simpson, Jeffrey D. and Zucker, Daniel B. and Ferguson, Peter S. and Martell, Sarah L. and Bland-Hawthorn, Joss and Balbinot, Eduardo and Tavangar, Kiyan and Drlica-Wagner, Alex and De Silva, Gayandhi M. and Simon, Joshua D.},
   year={2022},
   month=mar, pages={30} }

@article{Fu_2018,
   title={The Origin of the 300 km s−1 Stream near Segue 1},
   volume={866},
   ISSN={1538-4357},
   url={http://dx.doi.org/10.3847/1538-4357/aad9f9},
   DOI={10.3847/1538-4357/aad9f9},
   number={1},
   journal={The Astrophysical Journal},
   publisher={American Astronomical Society},
   author={Fu, Sal Wanying and Simon, Joshua D. and Shetrone, Matthew and Bovy, Jo and Beers, Timothy C. and Fernández-Trincado, J. G. and Placco, Vinicius M. and Zamora, Olga and Allende Prieto, Carlos and García-Hernández, D. A. and Harding, Paul and Ivans, Inese and Lane, Richard and Nitschelm, Christian and Roman-Lopes, Alexandre and Sobeck, Jennifer},
   year={2018},
   month=oct, pages={42} }

@misc{Usman_2024,
      title={Multiple Populations and a CH Star Found in the 300S Globular Cluster Stellar Stream}, 
      author={Sam A. Usman and Alexander P. Ji and Ting S. Li and Andrew B. Pace and Lara R. Cullinane and Gary S. Da Costa and Sergey E. Koposov and Geraint F. Lewis and Daniel B. Zucker and Vasily Belokurov and Joss Bland-Hawthorn and Peter S. Ferguson and Terese T. Hansen and Guilherme Limberg and Sarah L. Martell and Madeleine McKenzie and Joshua D. Simon},
      year={2024},
      eprint={2401.02476},
      archivePrefix={arXiv},
      primaryClass={astro-ph.GA},
      url={https://arxiv.org/abs/2401.02476}, 
}

@misc{Ibata_2023,
      title={Charting the Galactic acceleration field II. A global mass model of the Milky Way from the STREAMFINDER Atlas of Stellar Streams detected in Gaia DR3}, 
      author={Rodrigo Ibata and Khyati Malhan and Wassim Tenachi and Anke Ardern-Arentsen and Michele Bellazzini and Paolo Bianchini and Piercarlo Bonifacio and Elisabetta Caffau and Foivos Diakogiannis and Raphael Errani and Benoit Famaey and Salvatore Ferrone and Nicolas Martin and Paola di Matteo and Giacomo Monari and Florent Renaud and Else Starkenburg and Guillaume Thomas and Akshara Viswanathan and Zhen Yuan},
      year={2023},
      eprint={2311.17202},
      archivePrefix={arXiv},
      primaryClass={astro-ph.GA},
      url={https://arxiv.org/abs/2311.17202} 
}

@article{Ibata_2021,
   title={Charting the Galactic Acceleration Field. I. A Search for Stellar Streams with Gaia DR2 and EDR3 with Follow-up from ESPaDOnS and UVES},
   volume={914},
   ISSN={1538-4357},
   url={http://dx.doi.org/10.3847/1538-4357/abfcc2},
   DOI={10.3847/1538-4357/abfcc2},
   number={2},
   journal={The Astrophysical Journal},
   publisher={American Astronomical Society},
   author={Ibata, Rodrigo and Malhan, Khyati and Martin, Nicolas and Aubert, Dominique and Famaey, Benoit and Bianchini, Paolo and Monari, Giacomo and Siebert, Arnaud and Thomas, Guillaume F. and Bellazzini, Michele and Bonifacio, Piercarlo and Caffau, Elisabetta and Renaud, Florent},
   year={2021},
   month=jun, pages={123} }

@article{Martin_2022,
   title={The Pristine survey – XVI. The metallicity of 26 stellar streams around the Milky Way detected with the <tt>STREAMFINDER</tt> in Gaia EDR3},
   volume={516},
   ISSN={1365-2966},
   url={http://dx.doi.org/10.1093/mnras/stac2426},
   DOI={10.1093/mnras/stac2426},
   number={4},
   journal={Monthly Notices of the Royal Astronomical Society},
   publisher={Oxford University Press (OUP)},
   author={Martin, Nicolas F and Ibata, Rodrigo A and Starkenburg, Else and Yuan, Zhen and Malhan, Khyati and Bellazzini, Michele and Viswanathan, Akshara and Aguado, David and Arentsen, Anke and Bonifacio, Piercarlo and Carlberg, Ray and González Hernández, Jonay I and Hill, Vanessa and Jablonka, Pascale and Kordopatis, Georges and Lardo, Carmela and McConnachie, Alan W and Navarro, Julio and Sánchez-Janssen, Rubén and Sestito, Federico and Thomas, Guillaume F and Venn, Kim A and Vitali, Sara and Voggel, Karina T},
   year={2022},
   month=aug, pages={5331–5354} }

@article{Malhan_2022,
   title={The Global Dynamical Atlas of the Milky Way Mergers: Constraints from Gaia EDR3–based Orbits of Globular Clusters, Stellar Streams, and Satellite Galaxies},
   volume={926},
   ISSN={1538-4357},
   url={http://dx.doi.org/10.3847/1538-4357/ac4d2a},
   DOI={10.3847/1538-4357/ac4d2a},
   number={2},
   journal={The Astrophysical Journal},
   publisher={American Astronomical Society},
   author={Malhan, Khyati and Ibata, Rodrigo A. and Sharma, Sanjib and Famaey, Benoit and Bellazzini, Michele and Carlberg, Raymond G. and D’Souza, Richard and Yuan, Zhen and Martin, Nicolas F. and Thomas, Guillaume F.},
   year={2022},
   month=feb, pages={107} }

@article{Forbes_2020,
   title={Reverse engineering the Milky Way},
   volume={493},
   ISSN={1365-2966},
   url={http://dx.doi.org/10.1093/mnras/staa245},
   DOI={10.1093/mnras/staa245},
   number={1},
   journal={Monthly Notices of the Royal Astronomical Society},
   publisher={Oxford University Press (OUP)},
   author={Forbes, Duncan A},
   year={2020},
   month=jan, pages={847–854} }

@article{Dotter_2016,
   title={MESA ISOCHRONES AND STELLAR TRACKS (MIST) 0: METHODS FOR THE CONSTRUCTION OF STELLAR ISOCHRONES},
   volume={222},
   ISSN={1538-4365},
   url={http://dx.doi.org/10.3847/0067-0049/222/1/8},
   DOI={10.3847/0067-0049/222/1/8},
   number={1},
   journal={The Astrophysical Journal Supplement Series},
   publisher={American Astronomical Society},
   author={Dotter, Aaron},
   year={2016},
   month=jan, pages={8} }

@article{Limberg_2023,
doi = {10.3847/1538-4357/acb694},
url = {https://dx.doi.org/10.3847/1538-4357/acb694},
year = {2023},
month = {mar},
publisher = {The American Astronomical Society},
volume = {946},
number = {2},
pages = {66},
author = {Guilherme Limberg and Anna B. A. Queiroz and Hélio D. Perottoni and Silvia Rossi and João A. S. Amarante and Rafael M. Santucci and Cristina Chiappini and Angeles Pérez-Villegas and Young Sun Lee},
title = {Phase-space Properties and Chemistry of the Sagittarius Stellar Stream Down to the Extremely Metal-poor ([Fe/H] ≲ −3) Regime},
journal = {The Astrophysical Journal}
}

@article{Cunningham_2024,
doi = {10.3847/1538-4357/ad187b},
url = {https://dx.doi.org/10.3847/1538-4357/ad187b},
year = {2024},
month = {mar},
publisher = {The American Astronomical Society},
volume = {963},
number = {2},
pages = {95},
author = {Cunningham, Emily C. and Hunt, Jason A. S. and Price-Whelan, Adrian M. and Johnston, Kathryn V. and Ness, Melissa K. and Lu, Yuxi (Lucy) and Escala, Ivanna and Stelea, Ioana A.},
title = {Chemical Cartography of the Sagittarius Stream with Gaia},
journal = {The Astrophysical Journal}
}

@article{Hayes_2020,
doi = {10.3847/1538-4357/ab62ad},
url = {https://dx.doi.org/10.3847/1538-4357/ab62ad},
year = {2020},
month = {jan},
publisher = {The American Astronomical Society},
volume = {889},
number = {1},
pages = {63},
author = {Hayes, Christian R. and Majewski, Steven R. and Hasselquist, Sten and Anguiano, Borja and Shetrone, Matthew and Law, David R. and Schiavon, Ricardo P. and Cunha, Katia and Smith, Verne V. and Beaton, Rachael L. and Price-Whelan, Adrian M. and Allende Prieto, Carlos and Battaglia, Giuseppina and Bizyaev, Dmitry and Brownstein, Joel R. and Cohen, Roger E. and Frinchaboy, Peter M. and García-Hernández, D. A. and Lacerna, Ivan and Lane, Richard R. and Mészáros, Szabolcs and Bidin, Christian Moni and Mũnoz, Ricardo R. and Nidever, David L. and Oravetz, Audrey and Oravetz, Daniel and Pan, Kaike and Roman-Lopes, Alexandre and Sobeck, Jennifer and Stringfellow, Guy},
title = {Metallicity and α-Element Abundance Gradients along the Sagittarius Stream as Seen by APOGEE},
journal = {The Astrophysical Journal}
}

@ARTICLE{Vasiliev_2021,
       author = {{Vasiliev}, Eugene and {Belokurov}, Vasily and {Erkal}, Denis},
        title = "{Tango for three: Sagittarius, LMC, and the Milky Way}",
      journal = {\mnras},
         year = 2021,
        month = feb,
       volume = {501},
       number = {2},
        pages = {2279-2304},
          doi = {10.1093/mnras/staa3673},
archivePrefix = {arXiv},
       eprint = {2009.10726},
 primaryClass = {astro-ph.GA},
       adsurl = {https://ui.adsabs.harvard.edu/abs/2021MNRAS.501.2279V}
}

@article{Majewski_2003,
   title={A Two Micron All Sky Survey View of the Sagittarius Dwarf Galaxy. I. Morphology of the Sagittarius Core and Tidal Arms},
   volume={599},
   ISSN={1538-4357},
   url={http://dx.doi.org/10.1086/379504},
   DOI={10.1086/379504},
   number={2},
   journal={The Astrophysical Journal},
   publisher={American Astronomical Society},
   author={Majewski, Steven R. and Skrutskie, M. F. and Weinberg, Martin D. and Ostheimer, James C.},
   year={2003},
   month=dec, pages={1082–1115} }

@article{Ramos_2020,
   title={Full 5D characterisation of the Sagittarius stream with Gaia DR2 RR Lyrae},
   volume={638},
   ISSN={1432-0746},
   url={http://dx.doi.org/10.1051/0004-6361/202037819},
   DOI={10.1051/0004-6361/202037819},
   journal={Astronomy \&amp; Astrophysics},
   publisher={EDP Sciences},
   author={Ramos, P. and Mateu, C. and Antoja, T. and Helmi, A. and Castro-Ginard, A. and Balbinot, E. and Carrasco, J. M.},
   year={2020},
   month=jun, pages={A104} }

@article{Grillmair_2017,
   title={TAILS FROM THE ORPHANAGE},
   volume={834},
   ISSN={1538-4357},
   url={http://dx.doi.org/10.3847/1538-4357/834/2/98},
   DOI={10.3847/1538-4357/834/2/98},
   number={2},
   journal={The Astrophysical Journal},
   publisher={American Astronomical Society},
   author={Grillmair, Carl J.},
   year={2017},
   month=jan, pages={98} }

@article{Koposov_2019_orphan,
    author = {Koposov, S E and Belokurov, V and Li, T S and Mateu, C and Erkal, D and Grillmair, C J and Hendel, D and Price-Whelan, A M and Laporte, C F P and Hawkins, K and Sohn, S T and del Pino, A and Evans, N W and Slater, C T and Kallivayalil, N and Navarro, J F},
    title = "{Piercing the Milky Way: an all-sky view of the Orphan Stream}",
    journal = {Monthly Notices of the Royal Astronomical Society},
    volume = {485},
    number = {4},
    pages = {4726-4742},
    year = {2019},
    month = {02},
    issn = {0035-8711},
    doi = {10.1093/mnras/stz457},
    url = {https://doi.org/10.1093/mnras/stz457},
    eprint = {https://academic.oup.com/mnras/article-pdf/485/4/4726/28235223/stz457.pdf},
}

@article{Erkal_2017,
   title={A sharper view of Pal 5’s tails: discovery of stream perturbations with a novel non-parametric technique},
   volume={470},
   ISSN={1365-2966},
   url={http://dx.doi.org/10.1093/mnras/stx1208},
   DOI={10.1093/mnras/stx1208},
   number={1},
   journal={Monthly Notices of the Royal Astronomical Society},
   publisher={Oxford University Press (OUP)},
   author={Erkal, Denis and Koposov, Sergey E. and Belokurov, Vasily},
   year={2017},
   month=may, pages={60–84} }

@article{Ibata_2020_gd1,
   title={Detection of Strong Epicyclic Density Spikes in the GD-1 Stellar Stream: An Absence of Evidence for the Influence of Dark Matter Subhalos?},
   volume={891},
   ISSN={1538-4357},
   url={http://dx.doi.org/10.3847/1538-4357/ab7303},
   DOI={10.3847/1538-4357/ab7303},
   number={2},
   journal={The Astrophysical Journal},
   publisher={American Astronomical Society},
   author={Ibata, Rodrigo and Thomas, Guillaume and Famaey, Benoit and Malhan, Khyati and Martin, Nicolas and Monari, Giacomo},
   year={2020},
   month=mar, pages={161} }

@article{Webb_2019,
    author = {Webb, Jeremy J and Bovy, Jo},
    title = "{Searching for the GD-1 stream progenitor in Gaia DR2 with direct N-body simulations}",
    journal = {Monthly Notices of the Royal Astronomical Society},
    volume = {485},
    number = {4},
    pages = {5929-5938},
    year = {2019},
    month = {03},
    issn = {0035-8711},
    doi = {10.1093/mnras/stz867},
    url = {https://doi.org/10.1093/mnras/stz867},
    eprint = {https://academic.oup.com/mnras/article-pdf/485/4/5929/28271840/stz867.pdf},
}

@article{Price_Whelan_2018,
   title={Off the Beaten Path: Gaia Reveals GD-1 Stars outside of the Main Stream},
   volume={863},
   ISSN={2041-8213},
   url={http://dx.doi.org/10.3847/2041-8213/aad7b5},
   DOI={10.3847/2041-8213/aad7b5},
   number={2},
   journal={The Astrophysical Journal Letters},
   publisher={American Astronomical Society},
   author={Price-Whelan, Adrian M. and Bonaca, Ana},
   year={2018},
   month=aug, pages={L20} }

@article{Bonaca_2019,
   title={The Spur and the Gap in GD-1: Dynamical Evidence for a Dark Substructure in the Milky Way Halo},
   volume={880},
   ISSN={1538-4357},
   url={http://dx.doi.org/10.3847/1538-4357/ab2873},
   DOI={10.3847/1538-4357/ab2873},
   number={1},
   journal={The Astrophysical Journal},
   publisher={American Astronomical Society},
   author={Bonaca, Ana and Hogg, David W. and Price-Whelan, Adrian M. and Conroy, Charlie},
   year={2019},
   month=jul, pages={38} }

@article{Li_2021,
   title={Broken into Pieces: ATLAS and Aliqa Uma as One Single Stream},
   volume={911},
   ISSN={1538-4357},
   url={http://dx.doi.org/10.3847/1538-4357/abeb18},
   DOI={10.3847/1538-4357/abeb18},
   number={2},
   journal={The Astrophysical Journal},
   publisher={American Astronomical Society},
   author={Li, Ting S. and Koposov, Sergey E. and Erkal, Denis and Ji, Alexander P. and Shipp, Nora and Pace, Andrew B. and Hilmi, Tariq and Kuehn, Kyler and Lewis, Geraint F. and Mackey, Dougal and Simpson, Jeffrey D. and Wan, Zhen and Zucker, Daniel B. and Bland-Hawthorn, Joss and Cullinane, Lara R. and Da Costa, Gary S. and Drlica-Wagner, Alex and Hattori, Kohei and Martell, Sarah L. and Sharma, Sanjib},
   year={2021},
   month=apr, pages={149} }

@article{Ferguson_2021,
   title={DELVE-ing into the Jet: A Thin Stellar Stream on a Retrograde Orbit at 30 kpc},
   volume={163},
   ISSN={1538-3881},
   url={http://dx.doi.org/10.3847/1538-3881/ac3492},
   DOI={10.3847/1538-3881/ac3492},
   number={1},
   journal={The Astronomical Journal},
   publisher={American Astronomical Society},
   author={Ferguson, P. S. and Shipp, N. and Drlica-Wagner, A. and Li, T. S. and Cerny, W. and Tavangar, K. and Pace, A. B. and Marshall, J. L. and Riley, A. H. and Adamów, M. and Carlin, J. L. and Choi, Y. and Erkal, D. and James, D. J. and Koposov, Sergey E. and Kuropatkin, N. and Martínez-Vázquez, C. E. and Mau, S. and Mutlu-Pakdil, B. and Olsen, K. A. G. and Sakowska, J. D. and Stringfellow, G. S. and Yanny, B.},
   year={2021},
   month=dec, pages={18} }

@article{Patrick_2022,
    author = {Patrick, Jeffrey M and Koposov, Sergey E and Walker, Matthew G},
    title = "{Uniform modelling of the stellar density of thirteen tidal streams within the Galactic halo}",
    journal = {Monthly Notices of the Royal Astronomical Society},
    volume = {514},
    number = {2},
    pages = {1757-1781},
    year = {2022},
    month = {05},
    issn = {0035-8711},
    doi = {10.1093/mnras/stac1478},
    url = {https://doi.org/10.1093/mnras/stac1478},
    eprint = {https://academic.oup.com/mnras/article-pdf/514/2/1757/44055422/stac1478.pdf},
}

@article{Mateu_2023,
   title={galstreams: A library of Milky Way stellar stream footprints and tracks},
   volume={520},
   ISSN={1365-2966},
   url={http://dx.doi.org/10.1093/mnras/stad321},
   DOI={10.1093/mnras/stad321},
   number={4},
   journal={Monthly Notices of the Royal Astronomical Society},
   publisher={Oxford University Press (OUP)},
   author={Mateu, Cecilia},
   year={2023},
   month=jan, pages={5225–5258} }

@article{Shipp_2018,
   title={Stellar Streams Discovered in the Dark Energy Survey},
   volume={862},
   ISSN={1538-4357},
   url={http://dx.doi.org/10.3847/1538-4357/aacdab},
   DOI={10.3847/1538-4357/aacdab},
   number={2},
   journal={The Astrophysical Journal},
   publisher={American Astronomical Society},
   author={Shipp, N. and Drlica-Wagner, A. and Balbinot, E. and Ferguson, P. and Erkal, D. and Li, T. S. and Bechtol, K. and Belokurov, V. and Buncher, B. and Carollo, D. and Kind, M. Carrasco and Kuehn, K. and Marshall, J. L. and Pace, A. B. and Rykoff, E. S. and Sevilla-Noarbe, I. and Sheldon, E. and Strigari, L. and Vivas, A. K. and Yanny, B. and Zenteno, A. and Abbott, T. M. C. and Abdalla, F. B. and Allam, S. and Avila, S. and Bertin, E. and Brooks, D. and Burke, D. L. and Carretero, J. and Castander, F. J. and Cawthon, R. and Crocce, M. and Cunha, C. E. and D’Andrea, C. B. and da Costa, L. N. and Davis, C. and Vicente, J. De and Desai, S. and Diehl, H. T. and Doel, P. and Evrard, A. E. and Flaugher, B. and Fosalba, P. and Frieman, J. and García-Bellido, J. and Gaztanaga, E. and Gerdes, D. W. and Gruen, D. and Gruendl, R. A. and Gschwend, J. and Gutierrez, G. and Hartley, W. and Honscheid, K. and Hoyle, B. and James, D. J. and Johnson, M. D. and Krause, E. and Kuropatkin, N. and Lahav, O. and Lin, H. and Maia, M. A. G. and March, M. and Martini, P. and Menanteau, F. and Miller, C. J. and Miquel, R. and Nichol, R. C. and Plazas, A. A. and Romer, A. K. and Sako, M. and Sanchez, E. and Santiago, B. and Scarpine, V. and Schindler, R. and Schubnell, M. and Smith, M. and Smith, R. C. and Sobreira, F. and Suchyta, E. and Swanson, M. E. C. and Tarle, G. and Thomas, D. and Tucker, D. L. and Walker, A. R. and Wechsler, R. H.},
   year={2018},
   month=jul, pages={114} }

@article{Soker_2001,
   title={The “second parameter”: a memory from the globular cluster formation epoch},
   volume={324},
   ISSN={1365-2966},
   url={http://dx.doi.org/10.1046/j.1365-8711.2001.04310.x},
   DOI={10.1046/j.1365-8711.2001.04310.x},
   number={1},
   journal={Monthly Notices of the Royal Astronomical Society},
   publisher={Oxford University Press (OUP)},
   author={Soker, N. and Hadar, R.},
   year={2001},
   month=jun, pages={213–217} }

@article{Deason_2011,
   title={The Milky Way stellar halo out to 40 kpc: squashed, broken but smooth: The Milky Way stellar halo},
   volume={416},
   ISSN={0035-8711},
   url={http://dx.doi.org/10.1111/j.1365-2966.2011.19237.x},
   DOI={10.1111/j.1365-2966.2011.19237.x},
   number={4},
   journal={Monthly Notices of the Royal Astronomical Society},
   publisher={Oxford University Press (OUP)},
   author={Deason, A. J. and Belokurov, V. and Evans, N. W.},
   year={2011},
   month=jul, pages={2903–2915} }

@article{Belokurov_2015,
   title={Stellar streams around the Magellanic Clouds},
   volume={456},
   ISSN={1365-2966},
   url={http://dx.doi.org/10.1093/mnras/stv2688},
   DOI={10.1093/mnras/stv2688},
   number={1},
   journal={Monthly Notices of the Royal Astronomical Society},
   publisher={Oxford University Press (OUP)},
   author={Belokurov, Vasily and Koposov, Sergey E.},
   year={2015},
   month=dec, pages={602–616} }

@article{Clementini_2023,
   title={GaiaData Release 3: Specific processing and validation of all-sky RR Lyrae and Cepheid stars: The RR Lyrae sample},
   volume={674},
   ISSN={1432-0746},
   url={http://dx.doi.org/10.1051/0004-6361/202243964},
   DOI={10.1051/0004-6361/202243964},
   journal={Astronomy \&amp; Astrophysics},
   publisher={EDP Sciences},
   author={Clementini, G. and Ripepi, V. and Garofalo, A. and Molinaro, R. and Muraveva, T. and Leccia, S. and Rimoldini, L. and Holl, B. and Jevardat de Fombelle, G. and Sartoretti, P. and Marchal, O. and Audard, M. and Nienartowicz, K. and Andrae, R. and Marconi, M. and Szabados, L. and Evans, D. W. and Lecoeur-Taibi, I. and Mowlavi, N. and Musella, I. and Eyer, L.},
   year={2023},
   month=jun, pages={A18} }

@misc{Bonaca_2024,
      title={Stellar Streams in the Gaia Era}, 
      author={Ana Bonaca and Adrian M. Price-Whelan},
      year={2024},
      eprint={2405.19410},
      archivePrefix={arXiv},
      primaryClass={astro-ph.GA},
      url={https://arxiv.org/abs/2405.19410}, 
}

@ARTICLE{Salpeter_1955,
       author = {{Salpeter}, Edwin E.},
        title = "{The Luminosity Function and Stellar Evolution.}",
      journal = {\apj},
         year = 1955,
        month = jan,
       volume = {121},
        pages = {161},
          doi = {10.1086/145971},
       adsurl = {https://ui.adsabs.harvard.edu/abs/1955ApJ...121..161S}
}

@article{Johnston_2001,
   title={Interpreting Debris from Satellite Disruption in External Galaxies},
   volume={557},
   ISSN={1538-4357},
   url={http://dx.doi.org/10.1086/321644},
   DOI={10.1086/321644},
   number={1},
   journal={The Astrophysical Journal},
   publisher={American Astronomical Society},
   author={Johnston, Kathryn V. and Sackett, Penny D. and Bullock, James S.},
   year={2001},
   month=aug, pages={137–149} }

@article{Just_2009,
   title={Quantitative analysis of clumps in the tidal tails of star clusters},
   volume={392},
   ISSN={1365-2966},
   url={http://dx.doi.org/10.1111/j.1365-2966.2008.14099.x},
   DOI={10.1111/j.1365-2966.2008.14099.x},
   number={3},
   journal={Monthly Notices of the Royal Astronomical Society},
   publisher={Oxford University Press (OUP)},
   author={Just, A. and Berczik, P. and Petrov, M. I. and Ernst, A.},
   year={2009},
   month=jan, pages={969–981} }

@misc{Weatherford_2024,
      title={Stellar Escape from Globular Clusters. II. Clusters May Eat Their Own Tails}, 
      author={Newlin C. Weatherford and Frederic A. Rasio and Sourav Chatterjee and Giacomo Fragione and Fulya Kıroğlu and Kyle Kremer},
      year={2024},
      eprint={2310.01485},
      archivePrefix={arXiv},
      primaryClass={astro-ph.GA},
      url={https://arxiv.org/abs/2310.01485}, 
}

@article{Kupper_2012,
   title={More on the structure of tidal tails: More on the structure of tidal tails},
   volume={420},
   ISSN={0035-8711},
   url={http://dx.doi.org/10.1111/j.1365-2966.2011.20242.x},
   DOI={10.1111/j.1365-2966.2011.20242.x},
   number={3},
   journal={Monthly Notices of the Royal Astronomical Society},
   publisher={Oxford University Press (OUP)},
   author={Küpper, Andreas H. W. and Lane, Richard R. and Heggie, Douglas C.},
   year={2012},
   month=jan, pages={2700–2714} }

@article{Kupper_2008,
   title={On the structure of tidal tails},
   volume={387},
   ISSN={1365-2966},
   url={http://dx.doi.org/10.1111/j.1365-2966.2008.13323.x},
   DOI={10.1111/j.1365-2966.2008.13323.x},
   number={3},
   journal={Monthly Notices of the Royal Astronomical Society},
   publisher={Oxford University Press (OUP)},
   author={Küpper, Andreas H. W. and Macleod, Andrew and Heggie, Douglas C.},
   year={2008},
   month=jul, pages={1248–1252} }

@article{Malhan_2020,
    author = {Malhan, Khyati and Valluri, Monica and Freese, Katherine},
    title = "{Probing the nature of dark matter with accreted globular cluster streams}",
    journal = {Monthly Notices of the Royal Astronomical Society},
    volume = {501},
    number = {1},
    pages = {179-200},
    year = {2020},
    month = {11},
    issn = {0035-8711},
    doi = {10.1093/mnras/staa3597},
    url = {https://doi.org/10.1093/mnras/staa3597},
    eprint = {https://academic.oup.com/mnras/article-pdf/501/1/179/34934485/staa3597.pdf},
}

@misc{Pearson_2017,
      title={Gaps and length asymmetry in the stellar stream Palomar 5 as effects of Galactic bar rotation}, 
      author={Sarah Pearson and Adrian M. Price-Whelan and Kathryn V. Johnston},
      year={2017},
      eprint={1703.04627},
      archivePrefix={arXiv},
      primaryClass={astro-ph.GA},
      url={https://arxiv.org/abs/1703.04627}, 
}

@article{Banik_2021,
   title={Novel constraints on the particle nature of dark matter from stellar streams},
   volume={2021},
   ISSN={1475-7516},
   url={http://dx.doi.org/10.1088/1475-7516/2021/10/043},
   DOI={10.1088/1475-7516/2021/10/043},
   number={10},
   journal={Journal of Cosmology and Astroparticle Physics},
   publisher={IOP Publishing},
   author={Banik, Nilanjan and Bovy, Jo and Bertone, Gianfranco and Erkal, Denis and de Boer, T.J.L.},
   year={2021},
   month=oct, pages={043} }

@book{CarlinNewberg_2016,
    author = {Jeffery L. Carlin and Heidi J. Newberg},
    title = {Tidal Streams in the Local Group and Beyond},
    publisher = {Springer Cham},
    year = {2016},
    isbn = {978-3-319-19335-9}
}

@article{Erkal_2019,
   title={The total mass of the Large Magellanic Cloud from its perturbation on the Orphan stream},
   volume={487},
   ISSN={1365-2966},
   url={http://dx.doi.org/10.1093/mnras/stz1371},
   DOI={10.1093/mnras/stz1371},
   number={2},
   journal={Monthly Notices of the Royal Astronomical Society},
   publisher={Oxford University Press (OUP)},
   author={Erkal, D and Belokurov, V and Laporte, C F P and Koposov, S E and Li, T S and Grillmair, C J and Kallivayalil, N and Price-Whelan, A M and Evans, N W and Hawkins, K and Hendel, D and Mateu, C and Navarro, J F and del Pino, A and Slater, C T and Sohn, S T},
   year={2019},
   month=may, pages={2685–2700} }

@article{Shipp_2021,
   title={Measuring the Mass of the Large Magellanic Cloud with Stellar Streams Observed by S
               5},
   volume={923},
   ISSN={1538-4357},
   url={http://dx.doi.org/10.3847/1538-4357/ac2e93},
   DOI={10.3847/1538-4357/ac2e93},
   number={2},
   journal={The Astrophysical Journal},
   publisher={American Astronomical Society},
   author={Shipp, Nora and Erkal, Denis and Drlica-Wagner, Alex and Li, Ting S. and Pace, Andrew B. and Koposov, Sergey E. and Cullinane, Lara R. and Da Costa, Gary S. and Ji, Alexander P. and Kuehn, Kyler and Lewis, Geraint F. and Mackey, Dougal and Simpson, Jeffrey D. and Wan, Zhen and Zucker, Daniel B. and Bland-Hawthorn, Joss and Ferguson, Peter S. and Lilleengen, Sophia},
   year={2021},
   month=dec, pages={149} }

@article{Koposov2023,
   title={S5: Probing the Milky Way and Magellanic Clouds potentials with the 6D map of the Orphan–Chenab stream},
   volume={521},
   ISSN={1365-2966},
   url={http://dx.doi.org/10.1093/mnras/stad551},
   DOI={10.1093/mnras/stad551},
   number={4},
   journal={Monthly Notices of the Royal Astronomical Society},
   publisher={Oxford University Press (OUP)},
   author={Koposov, Sergey E and Erkal, Denis and Li, Ting S and Da Costa, Gary S and Cullinane, Lara R and Ji, Alexander P and Kuehn, Kyler and Lewis, Geraint F and Pace, Andrew B and Shipp, Nora and Zucker, Daniel B and Bland-Hawthorn, Joss and Lilleengen, Sophia and Martell, Sarah L},
   year={2023},
   month=feb, pages={4936–4962} }

@article{Bonaca_2014,
   title={MILKY WAY MASS AND POTENTIAL RECOVERY USING TIDAL STREAMS IN A REALISTIC HALO},
   volume={795},
   ISSN={1538-4357},
   url={http://dx.doi.org/10.1088/0004-637X/795/1/94},
   DOI={10.1088/0004-637x/795/1/94},
   number={1},
   journal={The Astrophysical Journal},
   publisher={American Astronomical Society},
   author={Bonaca, Ana and Geha, Marla and Küpper, Andreas H. W. and Diemand, Jürg and Johnston, Kathryn V. and Hogg, David W.},
   year={2014},
   month=oct, pages={94} }

@article{Ji_2021 ,
   title={Kinematics of Antlia 2 and Crater 2 from the Southern Stellar Stream Spectroscopic Survey (S
               5)},
   volume={921},
   ISSN={1538-4357},
   url={http://dx.doi.org/10.3847/1538-4357/ac1869},
   DOI={10.3847/1538-4357/ac1869},
   number={1},
   journal={The Astrophysical Journal},
   publisher={American Astronomical Society},
   author={Ji, Alexander P. and Koposov, Sergey E. and Li, Ting S. and Erkal, Denis and Pace, Andrew B. and Simon, Joshua D. and Belokurov, Vasily and Cullinane, Lara R. and Da Costa, Gary S. and Kuehn, Kyler and Lewis, Geraint F. and Mackey, Dougal and Shipp, Nora and Simpson, Jeffrey D. and Zucker, Daniel B. and Hansen, Terese T. and Bland-Hawthorn, Joss},
   year={2021},
   month=oct, pages={32} }

@article{Li_2019,
    author = {Li, T S and Koposov, S E and Zucker, D B and Lewis, G F and Kuehn, K and Simpson, J D and Ji, A P and Shipp, N and Mao, Y-Y and Geha, M and Pace, A B and Mackey, A D and Allam, S and Tucker, D L and Da Costa, G S and Erkal, D and Simon, J D and Mould, J R and Martell, S L and Wan, Z and De Silva, G M and Bechtol, K and Balbinot, E and Belokurov, V and Bland-Hawthorn, J and Casey, A R and Cullinane, L and Drlica-Wagner, A and Sharma, S and Vivas, A K and Wechsler, R H and Yanny, B and (S5 Collaboration)},
    title = "{The southern stellar stream spectroscopic survey (S5): Overview, target selection, data reduction, validation, and early science}",
    journal = {Monthly Notices of the Royal Astronomical Society},
    volume = {490},
    number = {3},
    pages = {3508-3531},
    year = {2019},
    month = {10},
    issn = {0035-8711},
    doi = {10.1093/mnras/stz2731},
    url = {https://doi.org/10.1093/mnras/stz2731},
    eprint = {https://academic.oup.com/mnras/article-pdf/490/3/3508/30338172/stz2731.pdf},
}

@software{Koposov_2019_rvspecfit,
  author = {Sergey E. Koposov},
  title = {RVSpecFit: Radial velocity and stellar atmospheric parameter fitting},
  url = {https://ascl.net/1907.013},
  year = {2019}
}

@article{Riello_2021,
   title={GaiaEarly Data Release 3: Photometric content and validation},
   volume={649},
   ISSN={1432-0746},
   url={http://dx.doi.org/10.1051/0004-6361/202039587},
   DOI={10.1051/0004-6361/202039587},
   journal={Astronomy \&amp; Astrophysics},
   publisher={EDP Sciences},
   author={Riello, M. and De Angeli, F. and Evans, D. W. and Montegriffo, P. and Carrasco, J. M. and Busso, G. and Palaversa, L. and Burgess, P. W. and Diener, C. and Davidson, M. and Rowell, N. and Fabricius, C. and Jordi, C. and Bellazzini, M. and Pancino, E. and Harrison, D. L. and Cacciari, C. and van Leeuwen, F. and Hambly, N. C. and Hodgkin, S. T. and Osborne, P. J. and Altavilla, G. and Barstow, M. A. and Brown, A. G. A. and Castellani, M. and Cowell, S. and De Luise, F. and Gilmore, G. and Giuffrida, G. and Hidalgo, S. and Holland, G. and Marinoni, S. and Pagani, C. and Piersimoni, A. M. and Pulone, L. and Ragaini, S. and Rainer, M. and Richards, P. J. and Sanna, N. and Walton, N. A. and Weiler, M. and Yoldas, A.},
   year={2021},
   month=apr, pages={A3} }

@MISC{Gaia_docs_ch20,
       author = {{Hambly}, N. and {Andrae}, R. and {De Angeli}, F. and {Antonio}, M. and {Arenou}, F. and {Audard}, M. and {Babusiaux}, C. and {Bailer-Jones}, C. and {Bakker}, J. and {Bastian}, U. and {Bauchet}, N. and {Bellas-Velidis}, I. and {Blomme}, R. and {Bombrun}, A. and {Brouillet}, N. and {Brugaletta}, E. and {de Bruijne}, J. and {Busonero}, D. and {Busso}, G. and {Carballo}, R. and {Carnerero}, M.~I. and {Clementini}, G. and {Creevey}, O.~L. and {Damerdji}, Y. and {Delchambre}, L. and {Distefano}, E. and {Drimmel}, R. and {Ducourant}, C. and {Duran}, J. and {Fabricius}, C. and {Eyer}, L. and {Faigler}, S. and {Findeisen}, K. and {Jevardat de Fombelle}, G. and {Fouesneau}, M. and {Fr{\'e}mat}, Y. and {Galluccio}, L. and {Garabato}, D. and {Gavras}, P. and {Giuffrida}, G. and {Gomel}, R. and {Gonz{\'a}lez}, {\'A}. and {Gonz{\'a}lez-N{\'u}{\~n}ez}, J. and {Gosset}, {\'E}. and {Gracia-Abril}, G. and {Halbwachs}, J. -L. and {Harrison}, D.~L. and {Heiter}, U. and {Hernandez}, J. and {Hestroffer}, D. and {Hobbs}, D. and {Hodgkin}, S. and {Holl}, B. and {Hutton}, A. and {Katz}, D. and {Klioner}, S. and {Leccia}, S. and {Lebreton}, Y. and {Lecoeur-Ta{\"\i}bi}, I. and {van Leeuwen}, M. and {Lindegren}, L. and {Lobel}, A. and {Luri}, X. and {Mantelet}, G. and {Marrese}, P.~M. and {Marinoni}, S. and {Marshall}, D.~J. and {Masana}, E. and {Mazeh}, T. and {Michalik}, D. and {Molinaro}, R. and {Mora}, A. and {Mowlavi}, N. and {Nienartowicz}, K. and {Ordenovic}, C. and {Panahi}, A. and {Pancino}, E. and {Pauwels}, T. and {Pichon}, B. and {Portell}, J. and {Pourbaix}, D. and {Raiteri}, C.~M. and {Recio-Blanco}, A. and {De Ridder}, J. and {Riello}, M. and {Rimoldini}, L. and {Ripepi}, V. and {Rixon}, G. and {Robin}, A.~C. and {Rybizki}, J. and {Sartoretti}, P. and {Sarro Baro}, L.~M. and {Seabroke}, G. and {Segovia Serrato}, J.~C. and {Siopis}, C. and {Smart}, R. and {Soubiran}, C. and {Sozzetti}, A. and {Spoto}, F. and {Tanga}, P. and {Teyssier}, D. and {Utrilla}, E. and {Masip Vela}, A. and {Wyrzykowski}, {\L}. and {Zucker}, S.},
        title = "{Gaia DR3 documentation Chapter 20: Datamodel description}",
 howpublished = {Gaia DR3 documentation, European Space Agency; Gaia Data Processing and Analysis Consortium. Online at \url{https://gea.esac.esa.int/archive/documentation/GDR3/index.html''>https://gea.esac.esa.int/archive/documentation/GDR3/index.html}},
         year = 2022,
        month = jun,
          eid = {20},
        pages = {20},
       adsurl = {https://ui.adsabs.harvard.edu/abs/2022gdr3.reptE..20H}
}

@MISC{Gaia_docs_ch7,
       author = {{Pourbaix}, D. and {Arenou}, F. and {Gavras}, P. and {Gosset}, {\'E}. and {Halbwachs}, J. -L. and {Siopis}, C. and {Sozzetti}, A. and {Bauchet}, N. and {Damerdji}, Y. and {Delchambre}, L. and {Delisle}, J. -B. and {Giacobbe}, P. and {Holl}, B. and {Jorissen}, A. and {Lattanzi}, M.~G. and {Leclerc}, N. and {Morel}, T. and {Sadowski}, G. and {Sahlmann}, J. and {Segransan}, D.},
        title = "{Gaia DR3 documentation Chapter 7: Non-single stars}",
 howpublished = {Gaia DR3 documentation, European Space Agency; Gaia Data Processing and Analysis Consortium.},
         year = 2022,
        month = jun,
          eid = {7},
        pages = {7},
       adsurl = {https://ui.adsabs.harvard.edu/abs/2022gdr3.reptE...7P}
}

@article{Martin_2008,
   title={A Comprehensive Maximum Likelihood Analysis of the Structural Properties of Faint Milky Way Satellites},
   volume={684},
   ISSN={1538-4357},
   url={http://dx.doi.org/10.1086/590336},
   DOI={10.1086/590336},
   number={2},
   journal={The Astrophysical Journal},
   publisher={American Astronomical Society},
   author={Martin, Nicolas F. and de Jong, Jelte T. A. and Rix, Hans‐Walter},
   year={2008},
   month=sep, pages={1075–1092} }

@ARTICLE{Carpenter_2017,
  title    = "Stan: A Probabilistic Programming Language",
  author   = "Carpenter, Bob and Gelman, Andrew and Hoffman, Matthew D and Lee,
              Daniel and Goodrich, Ben and Betancourt, Michael and Brubaker,
              Marcus A and Guo, Jiqiang and Li, Peter and Riddell, Allen",
  journal  = "J Stat Softw",
  volume   =  76,
  month    =  jan,
  year     =  2017,
  address  = "United States",
  language = "en"
}

@misc{STAN_2018,
title = {{The Stan Core Library}},
author = {{Stan Development Team}},
note = {Version 2.18.0},
year = {2018},
url = {http://mc-stan.org/ 19},
}

@incollection{Neal_2011,
   booktitle={Handbook of Markov Chain Monte Carlo},
   title = {MCMC Using Hamiltonian Dynamics},
   ISBN={9780429138508},
   url={http://dx.doi.org/10.1201/b10905},
   DOI={10.1201/b10905},
   publisher={Chapman and Hall/CRC},
   author={Radford M. Neal},
   year={2011},
   month=may}

@misc{Hoffman_2011,
      title={The No-U-Turn Sampler: Adaptively Setting Path Lengths in Hamiltonian Monte Carlo}, 
      author={Matthew D. Hoffman and Andrew Gelman},
      year={2011},
      eprint={1111.4246},
      archivePrefix={arXiv},
      primaryClass={stat.CO},
      url={https://arxiv.org/abs/1111.4246} 
}

@article{Gelman_and_Rubin_92,
author = {Andrew Gelman and Donald B. Rubin},
title = {{Inference from Iterative Simulation Using Multiple Sequences}},
volume = {7},
journal = {Statistical Science},
number = {4},
publisher = {Institute of Mathematical Statistics},
pages = {457 -- 472},
year = {1992},
doi = {10.1214/ss/1177011136},
URL = {https://doi.org/10.1214/ss/1177011136}
}

@article{Bovy_2015,
   title={galpy: A python LIBRARY FOR GALACTIC DYNAMICS},
   volume={216},
   ISSN={1538-4365},
   url={http://dx.doi.org/10.1088/0067-0049/216/2/29},
   DOI={10.1088/0067-0049/216/2/29},
   number={2},
   journal={The Astrophysical Journal Supplement Series},
   publisher={American Astronomical Society},
   author={Bovy, Jo},
   year={2015},
   month=feb, pages={29} }

@ARTICLE{McMillan_2017,
       author = {{McMillan}, Paul J.},
        title = "{The mass distribution and gravitational potential of the Milky Way}",
      journal = {\mnras},
         year = 2017,
        month = feb,
       volume = {465},
       number = {1},
        pages = {76-94},
          doi = {10.1093/mnras/stw2759},
archivePrefix = {arXiv},
       eprint = {1608.00971},
 primaryClass = {astro-ph.GA},
       adsurl = {https://ui.adsabs.harvard.edu/abs/2017MNRAS.465...76M}
}

@ARTICLE{Hernquist_1990,
       author = {{Hernquist}, Lars},
        title = "{An Analytical Model for Spherical Galaxies and Bulges}",
      journal = {\apj},
         year = 1990,
        month = jun,
       volume = {356},
        pages = {359},
          doi = {10.1086/168845},
       adsurl = {https://ui.adsabs.harvard.edu/abs/1990ApJ...356..359H}
}

@article{gala,
  doi = {10.21105/joss.00388},
  url = {https://doi.org/10.21105%2Fjoss.00388},
  year = 2017,
  month = {oct},
  publisher = {The Open Journal},
  volume = {2},
  number = {18},
  author = {Adrian M. Price-Whelan},
  title = {Gala: A Python package for galactic dynamics},
  journal = {The Journal of Open Source Software}}

@misc{Chen_2024,
      title={Improved particle spray algorithm for modeling globular cluster streams}, 
      author={Yingtian Chen and Monica Valluri and Oleg Y. Gnedin and Neil Ash},
      year={2024},
      eprint={2408.01496},
      archivePrefix={arXiv},
      primaryClass={astro-ph.GA},
      url={https://arxiv.org/abs/2408.01496}, 
}

@article{Plummer_1911,
       author = {{Plummer}, H.~C.},
        title = "{On the problem of distribution in globular star clusters}",
      journal = {\mnras},
         year = 1911,
        month = mar,
       volume = {71},
        pages = {460-470},
          doi = {10.1093/mnras/71.5.460},
       adsurl = {https://ui.adsabs.harvard.edu/abs/1911MNRAS..71..460P}
}

@ARTICLE{Gaia_DR3,
       author = {{Gaia Collaboration} and {Vallenari}, A. and {Brown}, A.~G.~A. and {Prusti}, T. and {de Bruijne}, J.~H.~J. and {Arenou}, F. and {Babusiaux}, C. and {Biermann}, M. and {Creevey}, O.~L. and {Ducourant}, C. and {Evans}, D.~W. and {Eyer}, L. and {Guerra}, R. and {Hutton}, A. and {Jordi}, C. and {Klioner}, S.~A. and {Lammers}, U.~L. and {Lindegren}, L. and {Luri}, X. and {Mignard}, F. and {Panem}, C. and {Pourbaix}, D. and {Randich}, S. and {Sartoretti}, P. and {Soubiran}, C. and {Tanga}, P. and {Walton}, N.~A. and {Bailer-Jones}, C.~A.~L. and {Bastian}, U. and {Drimmel}, R. and {Jansen}, F. and {Katz}, D. and {Lattanzi}, M.~G. and {van Leeuwen}, F. and {Bakker}, J. and {Cacciari}, C. and {Casta{\~n}eda}, J. and {De Angeli}, F. and {Fabricius}, C. and {Fouesneau}, M. and {Fr{\'e}mat}, Y. and {Galluccio}, L. and {Guerrier}, A. and {Heiter}, U. and {Masana}, E. and {Messineo}, R. and {Mowlavi}, N. and {Nicolas}, C. and {Nienartowicz}, K. and {Pailler}, F. and {Panuzzo}, P. and {Riclet}, F. and {Roux}, W. and {Seabroke}, G.~M. and {Sordo}, R. and {Th{\'e}venin}, F. and {Gracia-Abril}, G. and {Portell}, J. and {Teyssier}, D. and {Altmann}, M. and {Andrae}, R. and {Audard}, M. and {Bellas-Velidis}, I. and {Benson}, K. and {Berthier}, J. and {Blomme}, R. and {Burgess}, P.~W. and {Busonero}, D. and {Busso}, G. and {C{\'a}novas}, H. and {Carry}, B. and {Cellino}, A. and {Cheek}, N. and {Clementini}, G. and {Damerdji}, Y. and {Davidson}, M. and {de Teodoro}, P. and {Nu{\~n}ez Campos}, M. and {Delchambre}, L. and {Dell'Oro}, A. and {Esquej}, P. and {Fern{\'a}ndez-Hern{\'a}ndez}, J. and {Fraile}, E. and {Garabato}, D. and {Garc{\'\i}a-Lario}, P. and {Gosset}, E. and {Haigron}, R. and {Halbwachs}, J. -L. and {Hambly}, N.~C. and {Harrison}, D.~L. and {Hern{\'a}ndez}, J. and {Hestroffer}, D. and {Hodgkin}, S.~T. and {Holl}, B. and {Jan{\ss}en}, K. and {Jevardat de Fombelle}, G. and {Jordan}, S. and {Krone-Martins}, A. and {Lanzafame}, A.~C. and {L{\"o}ffler}, W. and {Marchal}, O. and {Marrese}, P.~M. and {Moitinho}, A. and {Muinonen}, K. and {Osborne}, P. and {Pancino}, E. and {Pauwels}, T. and {Recio-Blanco}, A. and {Reyl{\'e}}, C. and {Riello}, M. and {Rimoldini}, L. and {Roegiers}, T. and {Rybizki}, J. and {Sarro}, L.~M. and {Siopis}, C. and {Smith}, M. and {Sozzetti}, A. and {Utrilla}, E. and {van Leeuwen}, M. and {Abbas}, U. and {{\'A}brah{\'a}m}, P. and {Abreu Aramburu}, A. and {Aerts}, C. and {Aguado}, J.~J. and {Ajaj}, M. and {Aldea-Montero}, F. and {Altavilla}, G. and {{\'A}lvarez}, M.~A. and {Alves}, J. and {Anders}, F. and {Anderson}, R.~I. and {Anglada Varela}, E. and {Antoja}, T. and {Baines}, D. and {Baker}, S.~G. and {Balaguer-N{\'u}{\~n}ez}, L. and {Balbinot}, E. and {Balog}, Z. and {Barache}, C. and {Barbato}, D. and {Barros}, M. and {Barstow}, M.~A. and {Bartolom{\'e}}, S. and {Bassilana}, J. -L. and {Bauchet}, N. and {Becciani}, U. and {Bellazzini}, M. and {Berihuete}, A. and {Bernet}, M. and {Bertone}, S. and {Bianchi}, L. and {Binnenfeld}, A. and {Blanco-Cuaresma}, S. and {Blazere}, A. and {Boch}, T. and {Bombrun}, A. and {Bossini}, D. and {Bouquillon}, S. and {Bragaglia}, A. and {Bramante}, L. and {Breedt}, E. and {Bressan}, A. and {Brouillet}, N. and {Brugaletta}, E. and {Bucciarelli}, B. and {Burlacu}, A. and {Butkevich}, A.~G. and {Buzzi}, R. and {Caffau}, E. and {Cancelliere}, R. and {Cantat-Gaudin}, T. and {Carballo}, R. and {Carlucci}, T. and {Carnerero}, M.~I. and {Carrasco}, J.~M. and {Casamiquela}, L. and {Castellani}, M. and {Castro-Ginard}, A. and {Chaoul}, L. and {Charlot}, P. and {Chemin}, L. and {Chiaramida}, V. and {Chiavassa}, A. and {Chornay}, N. and {Comoretto}, G. and {Contursi}, G. and {Cooper}, W.~J. and {Cornez}, T. and {Cowell}, S. and {Crifo}, F. and {Cropper}, M. and {Crosta}, M. and {Crowley}, C. and {Dafonte}, C. and {Dapergolas}, A. and {David}, M. and {David}, P. and {de Laverny}, P. and {De Luise}, F. and {De March}, R.},
        title = "{Gaia Data Release 3. Summary of the content and survey properties}",
      journal = {\aap},
         year = 2023,
        month = jun,
       volume = {674},
          eid = {A1},
        pages = {A1},
          doi = {10.1051/0004-6361/202243940},
archivePrefix = {arXiv},
       eprint = {2208.00211},
 primaryClass = {astro-ph.GA},
       adsurl = {https://ui.adsabs.harvard.edu/abs/2023A\&A...674A...1G}
}

@ARTICLE{Gaia_Mission,
       author = {{Gaia Collaboration} and {Prusti}, T. and {de Bruijne}, J.~H.~J. and {Brown}, A.~G.~A. and {Vallenari}, A. and {Babusiaux}, C. and {Bailer-Jones}, C.~A.~L. and {Bastian}, U. and {Biermann}, M. and {Evans}, D.~W. and {Eyer}, L. and {Jansen}, F. and {Jordi}, C. and {Klioner}, S.~A. and {Lammers}, U. and {Lindegren}, L. and {Luri}, X. and {Mignard}, F. and {Milligan}, D.~J. and {Panem}, C. and {Poinsignon}, V. and {Pourbaix}, D. and {Randich}, S. and {Sarri}, G. and {Sartoretti}, P. and {Siddiqui}, H.~I. and {Soubiran}, C. and {Valette}, V. and {van Leeuwen}, F. and {Walton}, N.~A. and {Aerts}, C. and {Arenou}, F. and {Cropper}, M. and {Drimmel}, R. and {H{\o}g}, E. and {Katz}, D. and {Lattanzi}, M.~G. and {O'Mullane}, W. and {Grebel}, E.~K. and {Holland}, A.~D. and {Huc}, C. and {Passot}, X. and {Bramante}, L. and {Cacciari}, C. and {Casta{\~n}eda}, J. and {Chaoul}, L. and {Cheek}, N. and {De Angeli}, F. and {Fabricius}, C. and {Guerra}, R. and {Hern{\'a}ndez}, J. and {Jean-Antoine-Piccolo}, A. and {Masana}, E. and {Messineo}, R. and {Mowlavi}, N. and {Nienartowicz}, K. and {Ord{\'o}{\~n}ez-Blanco}, D. and {Panuzzo}, P. and {Portell}, J. and {Richards}, P.~J. and {Riello}, M. and {Seabroke}, G.~M. and {Tanga}, P. and {Th{\'e}venin}, F. and {Torra}, J. and {Els}, S.~G. and {Gracia-Abril}, G. and {Comoretto}, G. and {Garcia-Reinaldos}, M. and {Lock}, T. and {Mercier}, E. and {Altmann}, M. and {Andrae}, R. and {Astraatmadja}, T.~L. and {Bellas-Velidis}, I. and {Benson}, K. and {Berthier}, J. and {Blomme}, R. and {Busso}, G. and {Carry}, B. and {Cellino}, A. and {Clementini}, G. and {Cowell}, S. and {Creevey}, O. and {Cuypers}, J. and {Davidson}, M. and {De Ridder}, J. and {de Torres}, A. and {Delchambre}, L. and {Dell'Oro}, A. and {Ducourant}, C. and {Fr{\'e}mat}, Y. and {Garc{\'\i}a-Torres}, M. and {Gosset}, E. and {Halbwachs}, J. -L. and {Hambly}, N.~C. and {Harrison}, D.~L. and {Hauser}, M. and {Hestroffer}, D. and {Hodgkin}, S.~T. and {Huckle}, H.~E. and {Hutton}, A. and {Jasniewicz}, G. and {Jordan}, S. and {Kontizas}, M. and {Korn}, A.~J. and {Lanzafame}, A.~C. and {Manteiga}, M. and {Moitinho}, A. and {Muinonen}, K. and {Osinde}, J. and {Pancino}, E. and {Pauwels}, T. and {Petit}, J. -M. and {Recio-Blanco}, A. and {Robin}, A.~C. and {Sarro}, L.~M. and {Siopis}, C. and {Smith}, M. and {Smith}, K.~W. and {Sozzetti}, A. and {Thuillot}, W. and {van Reeven}, W. and {Viala}, Y. and {Abbas}, U. and {Abreu Aramburu}, A. and {Accart}, S. and {Aguado}, J.~J. and {Allan}, P.~M. and {Allasia}, W. and {Altavilla}, G. and {{\'A}lvarez}, M.~A. and {Alves}, J. and {Anderson}, R.~I. and {Andrei}, A.~H. and {Anglada Varela}, E. and {Antiche}, E. and {Antoja}, T. and {Ant{\'o}n}, S. and {Arcay}, B. and {Atzei}, A. and {Ayache}, L. and {Bach}, N. and {Baker}, S.~G. and {Balaguer-N{\'u}{\~n}ez}, L. and {Barache}, C. and {Barata}, C. and {Barbier}, A. and {Barblan}, F. and {Baroni}, M. and {Barrado y Navascu{\'e}s}, D. and {Barros}, M. and {Barstow}, M.~A. and {Becciani}, U. and {Bellazzini}, M. and {Bellei}, G. and {Bello Garc{\'\i}a}, A. and {Belokurov}, V. and {Bendjoya}, P. and {Berihuete}, A. and {Bianchi}, L. and {Bienaym{\'e}}, O. and {Billebaud}, F. and {Blagorodnova}, N. and {Blanco-Cuaresma}, S. and {Boch}, T. and {Bombrun}, A. and {Borrachero}, R. and {Bouquillon}, S. and {Bourda}, G. and {Bouy}, H. and {Bragaglia}, A. and {Breddels}, M.~A. and {Brouillet}, N. and {Br{\"u}semeister}, T. and {Bucciarelli}, B. and {Budnik}, F. and {Burgess}, P. and {Burgon}, R. and {Burlacu}, A. and {Busonero}, D. and {Buzzi}, R. and {Caffau}, E. and {Cambras}, J. and {Campbell}, H. and {Cancelliere}, R. and {Cantat-Gaudin}, T. and {Carlucci}, T. and {Carrasco}, J.~M. and {Castellani}, M. and {Charlot}, P. and {Charnas}, J. and {Charvet}, P. and {Chassat}, F. and {Chiavassa}, A. and {Clotet}, M. and {Cocozza}, G. and {Collins}, R.~S. and {Collins}, P. and {Costigan}, G.},
        title = "{The Gaia mission}",
      journal = {\aap},
         year = 2016,
        month = nov,
       volume = {595},
          eid = {A1},
        pages = {A1},
          doi = {10.1051/0004-6361/201629272},
archivePrefix = {arXiv},
       eprint = {1609.04153},
 primaryClass = {astro-ph.IM},
       adsurl = {https://ui.adsabs.harvard.edu/abs/2016A\&A...595A...1G}
}

@article{matplotlib,
  title={Matplotlib: A 2D graphics environment},
  author={Hunter, John D},
  journal={Computing in science \& engineering},
  volume={9},
  number={3},
  pages={90--95},
  year={2007},
  publisher={IEEE}
}

@ARTICLE{numpy,
  author  = {Harris, Charles R. and Millman, K. Jarrod and van der Walt, Stéfan J and Gommers, Ralf and Virtanen, Pauli and Cournapeau, David and Wieser, Eric and Taylor, Julian and Berg, Sebastian and Smith, Nathaniel J. and Kern, Robert and Picus, Matti and Hoyer, Stephan and van Kerkwijk, Marten H. and Brett, Matthew and Haldane, Allan and Fernández del Río, Jaime and Wiebe, Mark and Peterson, Pearu and Gérard-Marchant, Pierre and Sheppard, Kevin and Reddy, Tyler and Weckesser, Warren and Abbasi, Hameer and Gohlke, Christoph and Oliphant, Travis E.},
  title   = {Array programming with {NumPy}},
  journal = {Nature},
  year    = {2020},
  volume  = {585},
  pages   = {357–362},
  doi     = {10.1038/s41586-020-2649-2}
}

@inproceedings{statsmodels,
  title={statsmodels: Econometric and statistical modeling with python},
  author={Seabold, Skipper and Perktold, Josef},
  booktitle={9th Python in Science Conference},
  year={2010},
}

@article{astropy,
  title={The Astropy Project: Building an open-science project and status of the v2. 0 core package},
  author={Price-Whelan, Adrian M and Sip{\H{o}}cz, BM and G{\"u}nther, HM and Lim, PL and Crawford, SM and Conseil, S and Shupe, DL and Craig, MW and Dencheva, N and Ginsburg, A and others},
  journal={The Astronomical Journal},
  volume={156},
  number={3},
  pages={123},
  year={2018},
  publisher={IOP Publishing}
}

@ARTICLE{scipy,
  author  = {Virtanen, Pauli and Gommers, Ralf and Oliphant, Travis E. and Haberland, Matt and Reddy, Tyler and Cournapeau, David and Burovski, Evgeni and Peterson, Pearu and Weckesser, Warren and Bright, Jonathan and {van der Walt}, St{\'e}fan J. and Brett, Matthew and Wilson, Joshua and Millman, K. Jarrod and Mayorov, Nikolay and Nelson, Andrew R. J. and Jones, Eric and Kern, Robert and Larson, Eric and Carey, C J and Polat, {\.I}lhan and Feng, Yu and Moore, Eric W. and {VanderPlas}, Jake and Laxalde, Denis and Perktold, Josef and Cimrman, Robert and Henriksen, Ian and Quintero, E. A. and Harris, Charles R. and Archibald, Anne M. and Ribeiro, Ant{\^o}nio H. and Pedregosa, Fabian and {van Mulbregt}, Paul and {SciPy 1.0 Contributors}},
  title   = {{{SciPy} 1.0: Fundamental Algorithms for Scientific Computing in Python}},
  journal = {Nature Methods},
  year    = {2020},
  volume  = {17},
  pages   = {261--272},
  adsurl  = {https://rdcu.be/b08Wh},
  doi     = {10.1038/s41592-019-0686-2},
}

@inproceedings{pandas,
  title={Data structures for statistical computing in python},
  author={McKinney, Wes and others},
  booktitle={Proceedings of the 9th Python in Science Conference},
  volume={445},
  pages={51--56},
  year={2010},
  organization={Austin, TX}
}

@article{ipython,
  title={IPython: a system for interactive scientific computing},
  author={P{\'e}rez, Fernando and Granger, Brian E},
  journal={Computing in Science \& Engineering},
  volume={9},
  number={3},
  year={2007},
  publisher={IEEE}
}

@software{gala_zenodo,
  author       = {Adrian Price-Whelan and
                  Harrison Souchereau and
                  Tom Wagg and
                  Brigitta Sipőcz and
                  Nathaniel Starkman and
                  Yingtian Chen and
                  Sophia Lilleengen and
                  NGC and
                  Daniel Lenz and
                  Johnny Greco and
                  Akeem Hart and
                  AlexKurek and
                  Clément Robert and
                  Dan Foreman-Mackey and
                  HNLala and
                  P. L. Lim and
                  Semyeong Oh and
                  Sergey Koposov and
                  Zhaozhou Li},
  title        = {adrn/gala: v1.9.1},
  month        = aug,
  year         = 2024,
  publisher    = {Zenodo},
  version      = {v1.9.1},
  doi          = {10.5281/zenodo.13377376},
  url          = {https://doi.org/10.5281/zenodo.13377376},
}

@software{pandas_zenodo,
  author       = {The pandas development team},
  title        = {pandas-dev/pandas: Pandas},
  month        = jun,
  year         = 2023,
  publisher    = {Zenodo},
  version      = {v2.0.3},
  doi          = {10.5281/zenodo.8092754},
  url          = {https://doi.org/10.5281/zenodo.8092754},
}

@article{van_der_Marel_2014,
   title={THIRD-EPOCH MAGELLANIC CLOUD PROPER MOTIONS. II. THE LARGE MAGELLANIC CLOUD ROTATION FIELD IN THREE DIMENSIONS},
   volume={781},
   ISSN={1538-4357},
   url={http://dx.doi.org/10.1088/0004-637X/781/2/121},
   DOI={10.1088/0004-637x/781/2/121},
   number={2},
   journal={The Astrophysical Journal},
   publisher={American Astronomical Society},
   author={van der Marel, Roeland P. and Kallivayalil, Nitya},
   year={2014},
   month=jan, pages={121} }

@Article{Pietrzyński2013,
author={Pietrzy{\'{n}}ski, G.
and Graczyk, D.
and Gieren, W.
and Thompson, I. B.
and Pilecki, B.
and Udalski, A.
and Soszy{\'{n}}ski, I.
and Koz{\l}owski, S.
and Konorski, P.
and Suchomska, K.
and Bono, G.
and Moroni, P. G. Prada
and Villanova, S.
and Nardetto, N.
and Bresolin, F.
and Kudritzki, R. P.
and Storm, J.
and Gallenne, A.
and Smolec, R.
and Minniti, D.
and Kubiak, M.
and Szyma{\'{n}}ski, M. K.
and Poleski, R.
and Wyrzykowski, {\L}
and Ulaczyk, K.
and Pietrukowicz, P.
and G{\'o}rski, M.
and Karczmarek, P.},
title={An eclipsing-binary distance to the Large Magellanic Cloud accurate to two per cent},
journal={Nature},
year={2013},
month={Mar},
day={01},
volume={495},
number={7439},
pages={76-79},
issn={1476-4687},
doi={10.1038/nature11878},
url={https://doi.org/10.1038/nature11878}
}

@ARTICLE{vanderMarel2002,
       author = {{van der Marel}, Roeland P. and {Alves}, David R. and {Hardy}, Eduardo and {Suntzeff}, Nicholas B.},
        title = "{New Understanding of Large Magellanic Cloud Structure, Dynamics, and Orbit from Carbon Star Kinematics}",
      journal = {\aj},
         year = 2002,
        month = nov,
       volume = {124},
       number = {5},
        pages = {2639-2663},
          doi = {10.1086/343775},
archivePrefix = {arXiv},
       eprint = {astro-ph/0205161},
 primaryClass = {astro-ph},
       adsurl = {https://ui.adsabs.harvard.edu/abs/2002AJ....124.2639V}
}

@misc{Lu_2025,
      title={Detectability of dark matter subhalo impacts in Milky Way stellar streams}, 
      author={Junyang Lu and Tongyan Lin and Mukul Sholapurkar and Ana Bonaca},
      year={2025},
      eprint={2502.07781},
      archivePrefix={arXiv},
      primaryClass={astro-ph.GA},
      url={https://arxiv.org/abs/2502.07781}, 
}

@article{Pace_2022,
doi = {10.3847/1538-4357/ac997b},
url = {https://dx.doi.org/10.3847/1538-4357/ac997b},
year = {2022},
month = {nov},
publisher = {The American Astronomical Society},
volume = {940},
number = {2},
pages = {136},
author = {Pace, Andrew B. and Erkal, Denis and Li, Ting S.},
title = {Proper Motions, Orbits, and Tidal Influences of Milky Way Dwarf Spheroidal Galaxies},
journal = {The Astrophysical Journal}
}

@article{Tavangar_2022,
doi = {10.3847/1538-4357/ac399b},
url = {https://dx.doi.org/10.3847/1538-4357/ac399b},
year = {2022},
month = {jan},
publisher = {The American Astronomical Society},
volume = {925},
number = {2},
pages = {118},
author = {Tavangar, K. and Ferguson, P. and Shipp, N. and Drlica-Wagner, A. and Koposov, S. and Erkal, D. and Balbinot, E. and García-Bellido, J. and Kuehn, K. and Lewis, G. F. and Li, T. S. and Mau, S. and Pace, A. B. and Riley, A. H. and Abbott, T. M. C. and Aguena, M. and Allam, S. and Andrade-Oliveira, F. and Annis, J. and Bertin, E. and Brooks, D. and Burke, D. L. and Carnero Rosell, A. and Carrasco Kind, M. and Carretero, J. and Costanzi, M. and da Costa, L. N. and Pereira, M. E. S. and De Vicente, J. and Diehl, H. T. and Everett, S. and Ferrero, I. and Flaugher, B. and Frieman, J. and Gaztanaga, E. and Gerdes, D. W. and Gruen, D. and Gruendl, R. A. and Gschwend, J. and Gutierrez, G. and Hinton, S. R. and Hollowood, D. L. and Honscheid, K. and James, D. J. and Kuropatkin, N. and Maia, M. A. G. and Marshall, J. L. and Menanteau, F. and Miquel, R. and Morgan, R. and Ogando, R. L. C. and Palmese, A. and Paz-Chinchón, F. and Pieres, A. and Plazas Malagón, A. A. and Rodriguez-Monroy, M. and Sanchez, E. and Scarpine, V. and Serrano, S. and Sevilla-Noarbe, I. and Smith, M. and Suchyta, E. and Swanson, M. E. C. and Tarle, G. and To, C. and Varga, T. N. and Walker, A. R. and (DES Collaboration)},
title = {From the Fire: A Deeper Look at the Phoenix Stream},
journal = {The Astrophysical Journal}
}

@ARTICLE{Dey_2019,
       author = {{Dey}, Arjun and {Schlegel}, David J. and {Lang}, Dustin and {Blum}, Robert and {Burleigh}, Kaylan and {Fan}, Xiaohui and {Findlay}, Joseph R. and {Finkbeiner}, Doug and {Herrera}, David and {Juneau}, St{\'e}phanie and {Landriau}, Martin and {Levi}, Michael and {McGreer}, Ian and {Meisner}, Aaron and {Myers}, Adam D. and {Moustakas}, John and {Nugent}, Peter and {Patej}, Anna and {Schlafly}, Edward F. and {Walker}, Alistair R. and {Valdes}, Francisco and {Weaver}, Benjamin A. and {Y{\`e}che}, Christophe and {Zou}, Hu and {Zhou}, Xu and {Abareshi}, Behzad and {Abbott}, T.~M.~C. and {Abolfathi}, Bela and {Aguilera}, C. and {Alam}, Shadab and {Allen}, Lori and {Alvarez}, A. and {Annis}, James and {Ansarinejad}, Behzad and {Aubert}, Marie and {Beechert}, Jacqueline and {Bell}, Eric F. and {BenZvi}, Segev Y. and {Beutler}, Florian and {Bielby}, Richard M. and {Bolton}, Adam S. and {Brice{\~n}o}, C{\'e}sar and {Buckley-Geer}, Elizabeth J. and {Butler}, Karen and {Calamida}, Annalisa and {Carlberg}, Raymond G. and {Carter}, Paul and {Casas}, Ricard and {Castander}, Francisco J. and {Choi}, Yumi and {Comparat}, Johan and {Cukanovaite}, Elena and {Delubac}, Timoth{\'e}e and {DeVries}, Kaitlin and {Dey}, Sharmila and {Dhungana}, Govinda and {Dickinson}, Mark and {Ding}, Zhejie and {Donaldson}, John B. and {Duan}, Yutong and {Duckworth}, Christopher J. and {Eftekharzadeh}, Sarah and {Eisenstein}, Daniel J. and {Etourneau}, Thomas and {Fagrelius}, Parker A. and {Farihi}, Jay and {Fitzpatrick}, Mike and {Font-Ribera}, Andreu and {Fulmer}, Leah and {G{\"a}nsicke}, Boris T. and {Gaztanaga}, Enrique and {George}, Koshy and {Gerdes}, David W. and {Gontcho}, Satya Gontcho A. and {Gorgoni}, Claudio and {Green}, Gregory and {Guy}, Julien and {Harmer}, Diane and {Hernandez}, M. and {Honscheid}, Klaus and {Huang}, Lijuan Wendy and {James}, David J. and {Jannuzi}, Buell T. and {Jiang}, Linhua and {Joyce}, Richard and {Karcher}, Armin and {Karkar}, Sonia and {Kehoe}, Robert and {Kneib}, Jean-Paul and {Kueter-Young}, Andrea and {Lan}, Ting-Wen and {Lauer}, Tod R. and {Le Guillou}, Laurent and {Le Van Suu}, Auguste and {Lee}, Jae Hyeon and {Lesser}, Michael and {Perreault Levasseur}, Laurence and {Li}, Ting S. and {Mann}, Justin L. and {Marshall}, Robert and {Mart{\'\i}nez-V{\'a}zquez}, C.~E. and {Martini}, Paul and {du Mas des Bourboux}, H{\'e}lion and {McManus}, Sean and {Meier}, Tobias Gabriel and {M{\'e}nard}, Brice and {Metcalfe}, Nigel and {Mu{\~n}oz-Guti{\'e}rrez}, Andrea and {Najita}, Joan and {Napier}, Kevin and {Narayan}, Gautham and {Newman}, Jeffrey A. and {Nie}, Jundan and {Nord}, Brian and {Norman}, Dara J. and {Olsen}, Knut A.~G. and {Paat}, Anthony and {Palanque-Delabrouille}, Nathalie and {Peng}, Xiyan and {Poppett}, Claire L. and {Poremba}, Megan R. and {Prakash}, Abhishek and {Rabinowitz}, David and {Raichoor}, Anand and {Rezaie}, Mehdi and {Robertson}, A.~N. and {Roe}, Natalie A. and {Ross}, Ashley J. and {Ross}, Nicholas P. and {Rudnick}, Gregory and {Safonova}, Sasha and {Saha}, Abhijit and {S{\'a}nchez}, F. Javier and {Savary}, Elodie and {Schweiker}, Heidi and {Scott}, Adam and {Seo}, Hee-Jong and {Shan}, Huanyuan and {Silva}, David R. and {Slepian}, Zachary and {Soto}, Christian and {Sprayberry}, David and {Staten}, Ryan and {Stillman}, Coley M. and {Stupak}, Robert J. and {Summers}, David L. and {Sien Tie}, Suk and {Tirado}, H. and {Vargas-Maga{\~n}a}, Mariana and {Vivas}, A. Katherina and {Wechsler}, Risa H. and {Williams}, Doug and {Yang}, Jinyi and {Yang}, Qian and {Yapici}, Tolga and {Zaritsky}, Dennis and {Zenteno}, A. and {Zhang}, Kai and {Zhang}, Tianmeng and {Zhou}, Rongpu and {Zhou}, Zhimin},
        title = "{Overview of the DESI Legacy Imaging Surveys}",
      journal = {\aj},
         year = 2019,
        month = may,
       volume = {157},
       number = {5},
          eid = {168},
        pages = {168},
          doi = {10.3847/1538-3881/ab089d},
archivePrefix = {arXiv},
       eprint = {1804.08657},
 primaryClass = {astro-ph.IM},
       adsurl = {https://ui.adsabs.harvard.edu/abs/2019AJ....157..168D}
}

@article{Flaugher_2015,
doi = {10.1088/0004-6256/150/5/150},
url = {https://dx.doi.org/10.1088/0004-6256/150/5/150},
year = {2015},
month = {oct},
publisher = {The American Astronomical Society},
volume = {150},
number = {5},
pages = {150},
author = {Flaugher, B. and Diehl, H. T. and Honscheid, K. and Abbott, T. M. C. and Alvarez, O. and Angstadt, R. and Annis, J. T. and Antonik, M. and Ballester, O. and Beaufore, L. and Bernstein, G. M. and Bernstein, R. A. and Bigelow, B. and Bonati, M. and Boprie, D. and Brooks, D. and Buckley-Geer, E. J. and Campa, J. and Cardiel-Sas, L. and Castander, F. J. and Castilla, J. and Cease, H. and Cela-Ruiz, J. M. and Chappa, S. and Chi, E. and Cooper, C. and da Costa, L. N. and Dede, E. and Derylo, G. and DePoy, D. L. and de Vicente, J. and Doel, P. and Drlica-Wagner, A. and Eiting, J. and Elliott, A. E. and Emes, J. and Estrada, J. and Fausti Neto, A. and Finley, D. A. and Flores, R. and Frieman, J. and Gerdes, D. and Gladders, M. D. and Gregory, B. and Gutierrez, G. R. and Hao, J. and Holland, S. E. and Holm, S. and Huffman, D. and Jackson, C. and James, D. J. and Jonas, M. and Karcher, A. and Karliner, I. and Kent, S. and Kessler, R. and Kozlovsky, M. and Kron, R. G. and Kubik, D. and Kuehn, K. and Kuhlmann, S. and Kuk, K. and Lahav, O. and Lathrop, A. and Lee, J. and Levi, M. E. and Lewis, P. and Li, T. S. and Mandrichenko, I. and Marshall, J. L. and Martinez, G. and Merritt, K. W. and Miquel, R. and Muñoz, F. and Neilsen, E. H. and Nichol, R. C. and Nord, B. and Ogando, R. and Olsen, J. and Palaio, N. and Patton, K. and Peoples, J. and Plazas, A. A. and Rauch, J. and Reil, K. and Rheault, J.-P. and Roe, N. A. and Rogers, H. and Roodman, A. and Sanchez, E. and Scarpine, V. and Schindler, R. H. and Schmidt, R. and Schmitt, R. and Schubnell, M. and Schultz, K. and Schurter, P. and Scott, L. and Serrano, S. and Shaw, T. M. and Smith, R. C. and Soares-Santos, M. and Stefanik, A. and Stuermer, W. and Suchyta, E. and Sypniewski, A. and Tarle, G. and Thaler, J. and Tighe, R. and Tran, C. and Tucker, D. and Walker, A. R. and Wang, G. and Watson, M. and Weaverdyck, C. and Wester, W. and Woods, R. and Yanny, B. and (The DES Collaboration)},
title = {THE DARK ENERGY CAMERA},
journal = {The Astronomical Journal}
}

@article{Garfalo_2022,
    author = {Garofalo, A and Delgado, H E and Sarro, L M and Clementini, G and Muraveva, T and Marconi, M and Ripepi, V},
    title = {New LZ and PW(Z) relations of RR Lyrae stars calibrated with Gaia EDR3 parallaxes},
    journal = {Monthly Notices of the Royal Astronomical Society},
    volume = {513},
    number = {1},
    pages = {788-806},
    year = {2022},
    month = {03},
    issn = {0035-8711},
    doi = {10.1093/mnras/stac735},
    url = {https://doi.org/10.1093/mnras/stac735},
    eprint = {https://academic.oup.com/mnras/article-pdf/513/1/788/43462573/stac735.pdf},
}

@INPROCEEDINGS{Rockosi_2003,
       author = {{Rockosi}, C.~M. and {Beers}, T.~C. and {Allende Prieto}, C. and {Wilhelm}, R. and {Sloan Digital Sky Survey Collaboration}},
        title = "{A Large, ``Fair'', Sample of Halo and Thick Disk Stars from the SDSS}",
    booktitle = {American Astronomical Society Meeting Abstracts},
         year = 2003,
       series = {American Astronomical Society Meeting Abstracts},
       volume = {203},
        month = dec,
          eid = {112.10},
        pages = {112.10},
       adsurl = {https://ui.adsabs.harvard.edu/abs/2003AAS...20311210R}
}

@ARTICLE{Odenkirchen_2001,
       author = {{Odenkirchen}, Michael and {Grebel}, Eva K. and {Rockosi}, Constance M. and {Dehnen}, Walter and {Ibata}, Rodrigo and {Rix}, Hans-Walter and {Stolte}, Andrea and {Wolf}, Christian and {Anderson}, Jr., John E. and {Bahcall}, Neta A. and {Brinkmann}, Jon and {Csabai}, Istv{\'a}n and {Hennessy}, G. and {Hindsley}, Robert B. and {Ivezi{\'c}}, {\v{Z}}eljko and {Lupton}, Robert H. and {Munn}, Jeffrey A. and {Pier}, Jeffrey R. and {Stoughton}, Chris and {York}, Donald G.},
        title = "{Detection of Massive Tidal Tails around the Globular Cluster Palomar 5 with Sloan Digital Sky Survey Commissioning Data}",
      journal = {\apjl},
         year = 2001,
        month = feb,
       volume = {548},
       number = {2},
        pages = {L165-L169},
          doi = {10.1086/319095},
archivePrefix = {arXiv},
       eprint = {astro-ph/0012311},
 primaryClass = {astro-ph},
       adsurl = {https://ui.adsabs.harvard.edu/abs/2001ApJ...548L.165O}
}

@article{Cullinane_2020,
    author = {Cullinane, L R and Mackey, A D and Da Costa, G S and Koposov, S E and Belokurov, V and Erkal, D and Koch, A and Kunder, A and Nataf, D M},
    title = {The Magellanic Edges Survey I: Description and first results},
    journal = {Monthly Notices of the Royal Astronomical Society},
    volume = {497},
    number = {3},
    pages = {3055-3075},
    year = {2020},
    month = {07},
    issn = {0035-8711},
    doi = {10.1093/mnras/staa2048},
    url = {https://doi.org/10.1093/mnras/staa2048},
    eprint = {https://academic.oup.com/mnras/article-pdf/497/3/3055/33647241/staa2048.pdf},
}

@article{Liu_2020,
doi = {10.3847/1538-4365/ab72f8},
url = {https://dx.doi.org/10.3847/1538-4365/ab72f8},
year = {2020},
month = {apr},
publisher = {The American Astronomical Society},
volume = {247},
number = {2},
pages = {68},
author = {Liu, G.-C. and Huang, Y. and Zhang, H.-W. and Xiang, M.-S. and Ren, J.-J. and Chen, B.-Q. and Yuan, H.-B. and Wang, C. and Yang, Y. and Tian, Z.-J. and Wang, F. and Liu, X.-W.},
title = {Probing the Galactic Halo with RR Lyrae Stars. I. The Catalog},
journal = {The Astrophysical Journal Supplement Series}
}

@article{Malhan_2018,
    author = {Malhan, Khyati and Ibata, Rodrigo A},
    title = {STREAMFINDER – I. A new algorithm for detecting stellar streams},
    journal = {Monthly Notices of the Royal Astronomical Society},
    volume = {477},
    number = {3},
    pages = {4063-4076},
    year = {2018},
    month = {04},
    issn = {0035-8711},
    doi = {10.1093/mnras/sty912},
    url = {https://doi.org/10.1093/mnras/sty912},
    eprint = {https://academic.oup.com/mnras/article-pdf/477/3/4063/24816623/sty912.pdf},
}

@ARTICLE{Bonaca_2021,
       author = {{Bonaca}, Ana and {Naidu}, Rohan P. and {Conroy}, Charlie and {Caldwell}, Nelson and {Cargile}, Phillip A. and {Han}, Jiwon Jesse and {Johnson}, Benjamin D. and {Kruijssen}, J.~M. Diederik and {Myeong}, G.~C. and {Speagle}, Joshua S. and {Ting}, Yuan-Sen and {Zaritsky}, Dennis},
        title = "{Orbital Clustering Identifies the Origins of Galactic Stellar Streams}",
      journal = {\apjl},
         year = 2021,
        month = mar,
       volume = {909},
       number = {2},
          eid = {L26},
        pages = {L26},
          doi = {10.3847/2041-8213/abeaa9},
archivePrefix = {arXiv},
       eprint = {2012.09171},
 primaryClass = {astro-ph.GA},
       adsurl = {https://ui.adsabs.harvard.edu/abs/2021ApJ...909L..26B}
}

@ARTICLE{Belokurov_2018,
       author = {{Belokurov}, V. and {Erkal}, D. and {Evans}, N.~W. and {Koposov}, S.~E. and {Deason}, A.~J.},
        title = "{Co-formation of the disc and the stellar halo}",
      journal = {\mnras},
         year = 2018,
        month = jul,
       volume = {478},
       number = {1},
        pages = {611-619},
          doi = {10.1093/mnras/sty982},
archivePrefix = {arXiv},
       eprint = {1802.03414},
 primaryClass = {astro-ph.GA},
       adsurl = {https://ui.adsabs.harvard.edu/abs/2018MNRAS.478..611B}
}

@ARTICLE{Helmi_2018,
       author = {{Helmi}, Amina and {Babusiaux}, Carine and {Koppelman}, Helmer H. and {Massari}, Davide and {Veljanoski}, Jovan and {Brown}, Anthony G.~A.},
        title = "{The merger that led to the formation of the Milky Way's inner stellar halo and thick disk}",
      journal = {\nat},
         year = 2018,
        month = oct,
       volume = {563},
       number = {7729},
        pages = {85-88},
          doi = {10.1038/s41586-018-0625-x},
archivePrefix = {arXiv},
       eprint = {1806.06038},
 primaryClass = {astro-ph.GA},
       adsurl = {https://ui.adsabs.harvard.edu/abs/2018Natur.563...85H}
}

@ARTICLE{Haywood_2018,
       author = {{Haywood}, M. and {Di Matteo}, P. and {Lehnert}, M.~D. and {Snaith}, O. and {Khoperskov}, S. and {G{\'o}mez}, A.},
        title = "{In Disguise or Out of Reach: First Clues about In Situ and Accreted Stars in the Stellar Halo of the Milky Way from Gaia DR2}",
      journal = {\apj},
         year = 2018,
        month = aug,
       volume = {863},
       number = {2},
          eid = {113},
        pages = {113},
          doi = {10.3847/1538-4357/aad235},
archivePrefix = {arXiv},
       eprint = {1805.02617},
 primaryClass = {astro-ph.GA},
       adsurl = {https://ui.adsabs.harvard.edu/abs/2018ApJ...863..113H}
}

@ARTICLE{Simion_2019,
       author = {{Simion}, Iulia T. and {Belokurov}, Vasily and {Koposov}, Sergey E.},
        title = "{Common origin for Hercules-Aquila and Virgo Clouds in Gaia DR2}",
      journal = {\mnras},
         year = 2019,
        month = jan,
       volume = {482},
       number = {1},
        pages = {921-928},
          doi = {10.1093/mnras/sty2744},
archivePrefix = {arXiv},
       eprint = {1807.01335},
 primaryClass = {astro-ph.GA},
       adsurl = {https://ui.adsabs.harvard.edu/abs/2019MNRAS.482..921S}
}

@ARTICLE{Limberg_2022,
       author = {{Limberg}, Guilherme and {Souza}, Stefano O. and {P{\'e}rez-Villegas}, Angeles and {Rossi}, Silvia and {Perottoni}, H{\'e}lio D. and {Santucci}, Rafael M.},
        title = "{Reconstructing the Disrupted Dwarf Galaxy Gaia-Sausage/Enceladus Using Its Stars and Globular Clusters}",
      journal = {\apj},
         year = 2022,
        month = aug,
       volume = {935},
       number = {2},
          eid = {109},
        pages = {109},
          doi = {10.3847/1538-4357/ac8159},
archivePrefix = {arXiv},
       eprint = {2206.10505},
 primaryClass = {astro-ph.GA},
       adsurl = {https://ui.adsabs.harvard.edu/abs/2022ApJ...935..109L}
}

@ARTICLE{Ruiz-Macias_2021,
       author = {{Ruiz-Macias}, Omar and {Zarrouk}, Pauline and {Cole}, Shaun and {Baugh}, Carlton M. and {Norberg}, Peder and {Lucey}, John and {Dey}, Arjun and {Eisenstein}, Daniel J. and {Doel}, Peter and {Gazta{\~n}aga}, Enrique and {Hahn}, ChangHoon and {Kehoe}, Robert and {Kitanidis}, Ellie and {Landriau}, Martin and {Lang}, Dustin and {Moustakas}, John and {Myers}, Adam D. and {Prada}, Francisco and {Schubnell}, Michael and {Weinberg}, David H. and {Wilson}, M.~J.},
        title = "{Characterizing the target selection pipeline for the Dark Energy Spectroscopic Instrument Bright Galaxy Survey}",
      journal = {\mnras},
         year = 2021,
        month = apr,
       volume = {502},
       number = {3},
        pages = {4328-4349},
          doi = {10.1093/mnras/stab292},
archivePrefix = {arXiv},
       eprint = {2007.14950},
 primaryClass = {astro-ph.GA},
       adsurl = {https://ui.adsabs.harvard.edu/abs/2021MNRAS.502.4328R}
}

@ARTICLE{Bonaca_2012,
       author = {{Bonaca}, Ana and {Geha}, Marla and {Kallivayalil}, Nitya},
        title = "{A Cold Milky Way Stellar Stream in the Direction of Triangulum}",
      journal = {\apjl},
         year = 2012,
        month = nov,
       volume = {760},
       number = {1},
          eid = {L6},
        pages = {L6},
          doi = {10.1088/2041-8205/760/1/L6},
archivePrefix = {arXiv},
       eprint = {1209.5391},
 primaryClass = {astro-ph.GA},
       adsurl = {https://ui.adsabs.harvard.edu/abs/2012ApJ...760L...6B}
}

@ARTICLE{Musella_2012,
       author = {{Musella}, Ilaria and {Ripepi}, Vincenzo and {Marconi}, Marcella and {Clementini}, Gisella and {Dall'Ora}, Massimo and {Scowcroft}, Victoria and {Moretti}, Maria Ida and {Di Fabrizio}, Luca and {Greco}, Claudia and {Coppola}, Giuseppina and {Bersier}, David and {Catelan}, M{\'a}rcio and {Grado}, Aniello and {Limatola}, Luca and {Smith}, Horace A. and {Kinemuchi}, Karen},
        title = "{Stellar Archeology in the Galactic Halo with Ultra-faint Dwarfs. VII. Hercules}",
      journal = {\apj},
         year = 2012,
        month = sep,
       volume = {756},
       number = {2},
          eid = {121},
        pages = {121},
          doi = {10.1088/0004-637X/756/2/121},
archivePrefix = {arXiv},
       eprint = {1206.4031},
 primaryClass = {astro-ph.GA},
       adsurl = {https://ui.adsabs.harvard.edu/abs/2012ApJ...756..121M}
}

@ARTICLE{Garofalo_2013,
       author = {{Garofalo}, Alessia and {Cusano}, Felice and {Clementini}, Gisella and {Ripepi}, Vincenzo and {Dall'Ora}, Massimo and {Moretti}, Maria Ida and {Coppola}, Giuseppina and {Musella}, Ilaria and {Marconi}, Marcella},
        title = "{Variable Stars in the Ultra-faint Dwarf Spheroidal Galaxy Ursa Major I}",
      journal = {\apj},
         year = 2013,
        month = apr,
       volume = {767},
       number = {1},
          eid = {62},
        pages = {62},
          doi = {10.1088/0004-637X/767/1/62},
archivePrefix = {arXiv},
       eprint = {1302.3230},
 primaryClass = {astro-ph.GA},
       adsurl = {https://ui.adsabs.harvard.edu/abs/2013ApJ...767...62G}
}

@ARTICLE{Vivas_2020,
       author = {{Vivas}, A. Katherina and {Mart{\'\i}nez-V{\'a}zquez}, Clara and {Walker}, Alistair R.},
        title = "{Gaia RR Lyrae Stars in Nearby Ultra-faint Dwarf Satellite Galaxies}",
      journal = {\apjs},
         year = 2020,
        month = mar,
       volume = {247},
       number = {1},
          eid = {35},
        pages = {35},
          doi = {10.3847/1538-4365/ab67c0},
archivePrefix = {arXiv},
       eprint = {2001.01107},
 primaryClass = {astro-ph.GA},
       adsurl = {https://ui.adsabs.harvard.edu/abs/2020ApJS..247...35V}
}

@ARTICLE{Bastian_2018,
       author = {{Bastian}, Nate and {Lardo}, Carmela},
        title = "{Multiple Stellar Populations in Globular Clusters}",
      journal = {\araa},
         year = 2018,
        month = sep,
       volume = {56},
        pages = {83-136},
          doi = {10.1146/annurev-astro-081817-051839},
archivePrefix = {arXiv},
       eprint = {1712.01286},
 primaryClass = {astro-ph.SR},
       adsurl = {https://ui.adsabs.harvard.edu/abs/2018ARA\&A..56...83B}
}

@ARTICLE{Kupper_2010,
       author = {{K{\"u}pper}, Andreas H.~W. and {Kroupa}, Pavel and {Baumgardt}, Holger and {Heggie}, Douglas C.},
        title = "{Tidal tails of star clusters}",
      journal = {\mnras},
         year = 2010,
        month = jan,
       volume = {401},
       number = {1},
        pages = {105-120},
          doi = {10.1111/j.1365-2966.2009.15690.x},
archivePrefix = {arXiv},
       eprint = {0909.2619},
 primaryClass = {astro-ph.SR},
       adsurl = {https://ui.adsabs.harvard.edu/abs/2010MNRAS.401..105K}
}

@ARTICLE{Johnston_98,
       author = {{Johnston}, Kathryn V.},
        title = "{A Prescription for Building the Milky Way's Halo from Disrupted Satellites}",
      journal = {\apj},
         year = 1998,
        month = mar,
       volume = {495},
       number = {1},
        pages = {297-308},
          doi = {10.1086/305273},
archivePrefix = {arXiv},
       eprint = {astro-ph/9710007},
 primaryClass = {astro-ph},
       adsurl = {https://ui.adsabs.harvard.edu/abs/1998ApJ...495..297J}
}

@ARTICLE{Bovy_2016,
       author = {{Bovy}, Jo and {Bahmanyar}, Anita and {Fritz}, Tobias K. and {Kallivayalil}, Nitya},
        title = "{The Shape of the Inner Milky Way Halo from Observations of the Pal 5 and GD--1 Stellar Streams}",
      journal = {\apj},
         year = 2016,
        month = dec,
       volume = {833},
       number = {1},
          eid = {31},
        pages = {31},
          doi = {10.3847/1538-4357/833/1/31},
archivePrefix = {arXiv},
       eprint = {1609.01298},
 primaryClass = {astro-ph.GA},
       adsurl = {https://ui.adsabs.harvard.edu/abs/2016ApJ...833...31B}
}

@ARTICLE{Kupper_2015,
       author = {{K{\"u}pper}, Andreas H.~W. and {Balbinot}, Eduardo and {Bonaca}, Ana and {Johnston}, Kathryn V. and {Hogg}, David W. and {Kroupa}, Pavel and {Santiago}, Basilio X.},
        title = "{Globular Cluster Streams as Galactic High-Precision Scales{\textemdash}the Poster Child Palomar 5}",
      journal = {\apj},
         year = 2015,
        month = apr,
       volume = {803},
       number = {2},
          eid = {80},
        pages = {80},
          doi = {10.1088/0004-637X/803/2/80},
archivePrefix = {arXiv},
       eprint = {1502.02658},
 primaryClass = {astro-ph.GA},
       adsurl = {https://ui.adsabs.harvard.edu/abs/2015ApJ...803...80K}
}

@ARTICLE{Hattori_2016,
       author = {{Hattori}, Kohei and {Erkal}, Denis and {Sanders}, Jason L.},
        title = "{Shepherding tidal debris with the Galactic bar: the Ophiuchus stream}",
      journal = {\mnras},
         year = 2016,
        month = jul,
       volume = {460},
       number = {1},
        pages = {497-512},
          doi = {10.1093/mnras/stw1006},
archivePrefix = {arXiv},
       eprint = {1512.04536},
 primaryClass = {astro-ph.GA},
       adsurl = {https://ui.adsabs.harvard.edu/abs/2016MNRAS.460..497H}
}

@ARTICLE{Bobylev_2023,
       author = {{Bobylev}, V.~V. and {Baykova}, A.~T.},
        title = "{Modern Estimates of the Mass of the Milky Way}",
      journal = {Astronomy Reports},
         year = 2023,
        month = aug,
       volume = {67},
       number = {8},
        pages = {812-823},
          doi = {10.1134/S1063772923080024},
       adsurl = {https://ui.adsabs.harvard.edu/abs/2023ARep...67..812B}
}

@ARTICLE{Reid_2004,
       author = {{Reid}, M.~J. and {Brunthaler}, A.},
        title = "{The Proper Motion of Sagittarius A*. II. The Mass of Sagittarius A*}",
      journal = {\apj},
         year = 2004,
        month = dec,
       volume = {616},
       number = {2},
        pages = {872-884},
          doi = {10.1086/424960},
archivePrefix = {arXiv},
       eprint = {astro-ph/0408107},
 primaryClass = {astro-ph},
       adsurl = {https://ui.adsabs.harvard.edu/abs/2004ApJ...616..872R}
}

@ARTICLE{Drimmel_2018,
       author = {{Drimmel}, Ronald and {Poggio}, Eloisa},
        title = "{On the Solar Velocity}",
      journal = {Research Notes of the American Astronomical Society},
         year = 2018,
        month = nov,
       volume = {2},
       number = {4},
          eid = {210},
        pages = {210},
          doi = {10.3847/2515-5172/aaef8b},
       adsurl = {https://ui.adsabs.harvard.edu/abs/2018RNAAS...2..210D}
}

@ARTICLE{Gravity_2018,
       author = {{GRAVITY Collaboration} and {Abuter}, R. and {Amorim}, A. and {Anugu}, N. and {Baub{\"o}ck}, M. and {Benisty}, M. and {Berger}, J.~P. and {Blind}, N. and {Bonnet}, H. and {Brandner}, W. and {Buron}, A. and {Collin}, C. and {Chapron}, F. and {Cl{\'e}net}, Y. and {Coud{\'e} Du Foresto}, V. and {de Zeeuw}, P.~T. and {Deen}, C. and {Delplancke-Str{\"o}bele}, F. and {Dembet}, R. and {Dexter}, J. and {Duvert}, G. and {Eckart}, A. and {Eisenhauer}, F. and {Finger}, G. and {F{\"o}rster Schreiber}, N.~M. and {F{\'e}dou}, P. and {Garcia}, P. and {Garcia Lopez}, R. and {Gao}, F. and {Gendron}, E. and {Genzel}, R. and {Gillessen}, S. and {Gordo}, P. and {Habibi}, M. and {Haubois}, X. and {Haug}, M. and {Hau{\ss}mann}, F. and {Henning}, Th. and {Hippler}, S. and {Horrobin}, M. and {Hubert}, Z. and {Hubin}, N. and {Jimenez Rosales}, A. and {Jochum}, L. and {Jocou}, K. and {Kaufer}, A. and {Kellner}, S. and {Kendrew}, S. and {Kervella}, P. and {Kok}, Y. and {Kulas}, M. and {Lacour}, S. and {Lapeyr{\`e}re}, V. and {Lazareff}, B. and {Le Bouquin}, J. -B. and {L{\'e}na}, P. and {Lippa}, M. and {Lenzen}, R. and {M{\'e}rand}, A. and {M{\"u}ler}, E. and {Neumann}, U. and {Ott}, T. and {Palanca}, L. and {Paumard}, T. and {Pasquini}, L. and {Perraut}, K. and {Perrin}, G. and {Pfuhl}, O. and {Plewa}, P.~M. and {Rabien}, S. and {Ram{\'\i}rez}, A. and {Ramos}, J. and {Rau}, C. and {Rodr{\'\i}guez-Coira}, G. and {Rohloff}, R. -R. and {Rousset}, G. and {Sanchez-Bermudez}, J. and {Scheithauer}, S. and {Sch{\"o}ller}, M. and {Schuler}, N. and {Spyromilio}, J. and {Straub}, O. and {Straubmeier}, C. and {Sturm}, E. and {Tacconi}, L.~J. and {Tristram}, K.~R.~W. and {Vincent}, F. and {von Fellenberg}, S. and {Wank}, I. and {Waisberg}, I. and {Widmann}, F. and {Wieprecht}, E. and {Wiest}, M. and {Wiezorrek}, E. and {Woillez}, J. and {Yazici}, S. and {Ziegler}, D. and {Zins}, G.},
        title = "{Detection of the gravitational redshift in the orbit of the star S2 near the Galactic centre massive black hole}",
      journal = {\aap},
         year = 2018,
        month = jul,
       volume = {615},
          eid = {L15},
        pages = {L15},
          doi = {10.1051/0004-6361/201833718},
archivePrefix = {arXiv},
       eprint = {1807.09409},
 primaryClass = {astro-ph.GA},
       adsurl = {https://ui.adsabs.harvard.edu/abs/2018A\&A...615L..15G}
}

@inproceedings{Smith_2004,
author = {Greg A. Smith and Will Saunders and Terry Bridges and Vladimir Churilov and Allan Lankshear and John Dawson and David Correll and Lew Waller and Roger Haynes and Gabriella Frost},
title = {{AAOmega: a multipurpose fiber-fed spectrograph for the AAT}},
volume = {5492},
booktitle = {Ground-based Instrumentation for Astronomy},
editor = {Alan F. M. Moorwood and Masanori Iye},
organization = {International Society for Optics and Photonics},
publisher = {SPIE},
pages = {410 -- 420},
year = {2004},
doi = {10.1117/12.551013},
URL = {https://doi.org/10.1117/12.551013}
}

@ARTICLE{Lewis_2002,
       author = {{Lewis}, I.~J. and {Cannon}, R.~D. and {Taylor}, K. and {Glazebrook}, K. and {Bailey}, J.~A. and {Baldry}, I.~K. and {Barton}, J.~R. and {Bridges}, T.~J. and {Dalton}, G.~B. and {Farrell}, T.~J. and {Gray}, P.~M. and {Lankshear}, A. and {McCowage}, C. and {Parry}, I.~R. and {Sharples}, R.~M. and {Shortridge}, K. and {Smith}, G.~A. and {Stevenson}, J. and {Straede}, J.~O. and {Waller}, L.~G. and {Whittard}, J.~D. and {Wilcox}, J.~K. and {Willis}, K.~C.},
        title = "{The Anglo-Australian Observatory 2dF facility}",
      journal = {\mnras},
         year = 2002,
        month = jun,
       volume = {333},
       number = {2},
        pages = {279-299},
          doi = {10.1046/j.1365-8711.2002.05333.x},
archivePrefix = {arXiv},
       eprint = {astro-ph/0202175},
 primaryClass = {astro-ph},
       adsurl = {https://ui.adsabs.harvard.edu/abs/2002MNRAS.333..279L}
}

@software{AAO_2015,
       author = {{AAO software team}},
        title = "{2dfdr: Data reduction software}",
 howpublished = {Astrophysics Source Code Library, record ascl:1505.015},
         year = 2015,
        month = may,
          eid = {ascl:1505.015},
       adsurl = {https://ui.adsabs.harvard.edu/abs/2015ascl.soft05015A}
}

@ARTICLE{Erkal_2016,
       author = {{Erkal}, Denis and {Belokurov}, Vasily and {Bovy}, Jo and {Sanders}, Jason L.},
        title = "{The number and size of subhalo-induced gaps in stellar streams}",
      journal = {\mnras},
         year = 2016,
        month = nov,
       volume = {463},
       number = {1},
        pages = {102-119},
          doi = {10.1093/mnras/stw1957},
archivePrefix = {arXiv},
       eprint = {1606.04946},
 primaryClass = {astro-ph.GA},
       adsurl = {https://ui.adsabs.harvard.edu/abs/2016MNRAS.463..102E}
}

@ARTICLE{Kaisina_2012,
       author = {{Kaisina}, E.~I. and {Makarov}, D.~I. and {Karachentsev}, I.~D. and {Kaisin}, S.~S.},
        title = "{Observational database for studies of nearby universe}",
      journal = {Astrophysical Bulletin},
         year = 2012,
        month = jan,
       volume = {67},
       number = {1},
        pages = {115-122},
          doi = {10.1134/S1990341312010105},
       adsurl = {https://ui.adsabs.harvard.edu/abs/2012AstBu..67..115K}
}

@ARTICLE{Kallivayalil_2013,
       author = {{Kallivayalil}, Nitya and {van der Marel}, Roeland P. and {Besla}, Gurtina and {Anderson}, Jay and {Alcock}, Charles},
        title = "{Third-epoch Magellanic Cloud Proper Motions. I. Hubble Space Telescope/WFC3 Data and Orbit Implications}",
      journal = {\apj},
         year = 2013,
        month = feb,
       volume = {764},
       number = {2},
          eid = {161},
        pages = {161},
          doi = {10.1088/0004-637X/764/2/161},
archivePrefix = {arXiv},
       eprint = {1301.0832},
 primaryClass = {astro-ph.CO},
       adsurl = {https://ui.adsabs.harvard.edu/abs/2013ApJ...764..161K}
}

@ARTICLE{de_Boer_2020,
       author = {{de Boer}, T.~J.~L. and {Erkal}, D. and {Gieles}, M.},
        title = "{A closer look at the spur, blob, wiggle, and gaps in GD-1}",
      journal = {\mnras},
         year = 2020,
        month = jun,
       volume = {494},
       number = {4},
        pages = {5315-5332},
          doi = {10.1093/mnras/staa917},
archivePrefix = {arXiv},
       eprint = {1911.05745},
 primaryClass = {astro-ph.GA},
       adsurl = {https://ui.adsabs.harvard.edu/abs/2020MNRAS.494.5315D}
}

@ARTICLE{Chambers_2016,
       author = {{Chambers}, K.~C. and {Magnier}, E.~A. and {Metcalfe}, N. and {Flewelling}, H.~A. and {Huber}, M.~E. and {Waters}, C.~Z. and {Denneau}, L. and {Draper}, P.~W. and {Farrow}, D. and {Finkbeiner}, D.~P. and {Holmberg}, C. and {Koppenhoefer}, J. and {Price}, P.~A. and {Rest}, A. and {Saglia}, R.~P. and {Schlafly}, E.~F. and {Smartt}, S.~J. and {Sweeney}, W. and {Wainscoat}, R.~J. and {Burgett}, W.~S. and {Chastel}, S. and {Grav}, T. and {Heasley}, J.~N. and {Hodapp}, K.~W. and {Jedicke}, R. and {Kaiser}, N. and {Kudritzki}, R. -P. and {Luppino}, G.~A. and {Lupton}, R.~H. and {Monet}, D.~G. and {Morgan}, J.~S. and {Onaka}, P.~M. and {Shiao}, B. and {Stubbs}, C.~W. and {Tonry}, J.~L. and {White}, R. and {Ba{\~n}ados}, E. and {Bell}, E.~F. and {Bender}, R. and {Bernard}, E.~J. and {Boegner}, M. and {Boffi}, F. and {Botticella}, M.~T. and {Calamida}, A. and {Casertano}, S. and {Chen}, W. -P. and {Chen}, X. and {Cole}, S. and {Deacon}, N. and {Frenk}, C. and {Fitzsimmons}, A. and {Gezari}, S. and {Gibbs}, V. and {Goessl}, C. and {Goggia}, T. and {Gourgue}, R. and {Goldman}, B. and {Grant}, P. and {Grebel}, E.~K. and {Hambly}, N.~C. and {Hasinger}, G. and {Heavens}, A.~F. and {Heckman}, T.~M. and {Henderson}, R. and {Henning}, T. and {Holman}, M. and {Hopp}, U. and {Ip}, W. -H. and {Isani}, S. and {Jackson}, M. and {Keyes}, C.~D. and {Koekemoer}, A.~M. and {Kotak}, R. and {Le}, D. and {Liska}, D. and {Long}, K.~S. and {Lucey}, J.~R. and {Liu}, M. and {Martin}, N.~F. and {Masci}, G. and {McLean}, B. and {Mindel}, E. and {Misra}, P. and {Morganson}, E. and {Murphy}, D.~N.~A. and {Obaika}, A. and {Narayan}, G. and {Nieto-Santisteban}, M.~A. and {Norberg}, P. and {Peacock}, J.~A. and {Pier}, E.~A. and {Postman}, M. and {Primak}, N. and {Rae}, C. and {Rai}, A. and {Riess}, A. and {Riffeser}, A. and {Rix}, H.~W. and {R{\"o}ser}, S. and {Russel}, R. and {Rutz}, L. and {Schilbach}, E. and {Schultz}, A.~S.~B. and {Scolnic}, D. and {Strolger}, L. and {Szalay}, A. and {Seitz}, S. and {Small}, E. and {Smith}, K.~W. and {Soderblom}, D.~R. and {Taylor}, P. and {Thomson}, R. and {Taylor}, A.~N. and {Thakar}, A.~R. and {Thiel}, J. and {Thilker}, D. and {Unger}, D. and {Urata}, Y. and {Valenti}, J. and {Wagner}, J. and {Walder}, T. and {Walter}, F. and {Watters}, S.~P. and {Werner}, S. and {Wood-Vasey}, W.~M. and {Wyse}, R.},
        title = "{The Pan-STARRS1 Surveys}",
      journal = {arXiv e-prints},
         year = 2016,
        month = dec,
          eid = {arXiv:1612.05560},
        pages = {arXiv:1612.05560},
          doi = {10.48550/arXiv.1612.05560},
archivePrefix = {arXiv},
       eprint = {1612.05560},
 primaryClass = {astro-ph.IM},
       adsurl = {https://ui.adsabs.harvard.edu/abs/2016arXiv161205560C}
}

@ARTICLE{Abbott_2018,
       author = {{Abbott}, T.~M.~C. and {Abdalla}, F.~B. and {Allam}, S. and {Amara}, A. and {Annis}, J. and {Asorey}, J. and {Avila}, S. and {Ballester}, O. and {Banerji}, M. and {Barkhouse}, W. and {Baruah}, L. and {Baumer}, M. and {Bechtol}, K. and {Becker}, M.~R. and {Benoit-L{\'e}vy}, A. and {Bernstein}, G.~M. and {Bertin}, E. and {Blazek}, J. and {Bocquet}, S. and {Brooks}, D. and {Brout}, D. and {Buckley-Geer}, E. and {Burke}, D.~L. and {Busti}, V. and {Campisano}, R. and {Cardiel-Sas}, L. and {Carnero Rosell}, A. and {Carrasco Kind}, M. and {Carretero}, J. and {Castander}, F.~J. and {Cawthon}, R. and {Chang}, C. and {Chen}, X. and {Conselice}, C. and {Costa}, G. and {Crocce}, M. and {Cunha}, C.~E. and {D'Andrea}, C.~B. and {da Costa}, L.~N. and {Das}, R. and {Daues}, G. and {Davis}, T.~M. and {Davis}, C. and {De Vicente}, J. and {DePoy}, D.~L. and {DeRose}, J. and {Desai}, S. and {Diehl}, H.~T. and {Dietrich}, J.~P. and {Dodelson}, S. and {Doel}, P. and {Drlica-Wagner}, A. and {Eifler}, T.~F. and {Elliott}, A.~E. and {Evrard}, A.~E. and {Farahi}, A. and {Fausti Neto}, A. and {Fernandez}, E. and {Finley}, D.~A. and {Flaugher}, B. and {Foley}, R.~J. and {Fosalba}, P. and {Friedel}, D.~N. and {Frieman}, J. and {Garc{\'\i}a-Bellido}, J. and {Gaztanaga}, E. and {Gerdes}, D.~W. and {Giannantonio}, T. and {Gill}, M.~S.~S. and {Glazebrook}, K. and {Goldstein}, D.~A. and {Gower}, M. and {Gruen}, D. and {Gruendl}, R.~A. and {Gschwend}, J. and {Gupta}, R.~R. and {Gutierrez}, G. and {Hamilton}, S. and {Hartley}, W.~G. and {Hinton}, S.~R. and {Hislop}, J.~M. and {Hollowood}, D. and {Honscheid}, K. and {Hoyle}, B. and {Huterer}, D. and {Jain}, B. and {James}, D.~J. and {Jeltema}, T. and {Johnson}, M.~W.~G. and {Johnson}, M.~D. and {Kacprzak}, T. and {Kent}, S. and {Khullar}, G. and {Klein}, M. and {Kovacs}, A. and {Koziol}, A.~M.~G. and {Krause}, E. and {Kremin}, A. and {Kron}, R. and {Kuehn}, K. and {Kuhlmann}, S. and {Kuropatkin}, N. and {Lahav}, O. and {Lasker}, J. and {Li}, T.~S. and {Li}, R.~T. and {Liddle}, A.~R. and {Lima}, M. and {Lin}, H. and {L{\'o}pez-Reyes}, P. and {MacCrann}, N. and {Maia}, M.~A.~G. and {Maloney}, J.~D. and {Manera}, M. and {March}, M. and {Marriner}, J. and {Marshall}, J.~L. and {Martini}, P. and {McClintock}, T. and {McKay}, T. and {McMahon}, R.~G. and {Melchior}, P. and {Menanteau}, F. and {Miller}, C.~J. and {Miquel}, R. and {Mohr}, J.~J. and {Morganson}, E. and {Mould}, J. and {Neilsen}, E. and {Nichol}, R.~C. and {Nogueira}, F. and {Nord}, B. and {Nugent}, P. and {Nunes}, L. and {Ogando}, R.~L.~C. and {Old}, L. and {Pace}, A.~B. and {Palmese}, A. and {Paz-Chinch{\'o}n}, F. and {Peiris}, H.~V. and {Percival}, W.~J. and {Petravick}, D. and {Plazas}, A.~A. and {Poh}, J. and {Pond}, C. and {Porredon}, A. and {Pujol}, A. and {Refregier}, A. and {Reil}, K. and {Ricker}, P.~M. and {Rollins}, R.~P. and {Romer}, A.~K. and {Roodman}, A. and {Rooney}, P. and {Ross}, A.~J. and {Rykoff}, E.~S. and {Sako}, M. and {Sanchez}, M.~L. and {Sanchez}, E. and {Santiago}, B. and {Saro}, A. and {Scarpine}, V. and {Scolnic}, D. and {Serrano}, S. and {Sevilla-Noarbe}, I. and {Sheldon}, E. and {Shipp}, N. and {Silveira}, M.~L. and {Smith}, M. and {Smith}, R.~C. and {Smith}, J.~A. and {Soares-Santos}, M. and {Sobreira}, F. and {Song}, J. and {Stebbins}, A. and {Suchyta}, E. and {Sullivan}, M. and {Swanson}, M.~E.~C. and {Tarle}, G. and {Thaler}, J. and {Thomas}, D. and {Thomas}, R.~C. and {Troxel}, M.~A. and {Tucker}, D.~L. and {Vikram}, V. and {Vivas}, A.~K. and {Walker}, A.~R. and {Wechsler}, R.~H. and {Weller}, J. and {Wester}, W. and {Wolf}, R.~C. and {Wu}, H. and {Yanny}, B. and {Zenteno}, A. and {Zhang}, Y. and {Zuntz}, J. and {DES Collaboration} and {Juneau}, S. and {Fitzpatrick}, M. and {Nikutta}, R.},
        title = "{The Dark Energy Survey: Data Release 1}",
      journal = {\apjs},
         year = 2018,
        month = dec,
       volume = {239},
       number = {2},
          eid = {18},
        pages = {18},
          doi = {10.3847/1538-4365/aae9f0},
archivePrefix = {arXiv},
       eprint = {1801.03181},
 primaryClass = {astro-ph.IM},
       adsurl = {https://ui.adsabs.harvard.edu/abs/2018ApJS..239...18A}
}

@ARTICLE{Majewski_2017,
       author = {{Majewski}, Steven R. and {Schiavon}, Ricardo P. and {Frinchaboy}, Peter M. and {Allende Prieto}, Carlos and {Barkhouser}, Robert and {Bizyaev}, Dmitry and {Blank}, Basil and {Brunner}, Sophia and {Burton}, Adam and {Carrera}, Ricardo and {Chojnowski}, S. Drew and {Cunha}, K{\'a}tia and {Epstein}, Courtney and {Fitzgerald}, Greg and {Garc{\'\i}a P{\'e}rez}, Ana E. and {Hearty}, Fred R. and {Henderson}, Chuck and {Holtzman}, Jon A. and {Johnson}, Jennifer A. and {Lam}, Charles R. and {Lawler}, James E. and {Maseman}, Paul and {M{\'e}sz{\'a}ros}, Szabolcs and {Nelson}, Matthew and {Nguyen}, Duy Coung and {Nidever}, David L. and {Pinsonneault}, Marc and {Shetrone}, Matthew and {Smee}, Stephen and {Smith}, Verne V. and {Stolberg}, Todd and {Skrutskie}, Michael F. and {Walker}, Eric and {Wilson}, John C. and {Zasowski}, Gail and {Anders}, Friedrich and {Basu}, Sarbani and {Beland}, Stephane and {Blanton}, Michael R. and {Bovy}, Jo and {Brownstein}, Joel R. and {Carlberg}, Joleen and {Chaplin}, William and {Chiappini}, Cristina and {Eisenstein}, Daniel J. and {Elsworth}, Yvonne and {Feuillet}, Diane and {Fleming}, Scott W. and {Galbraith-Frew}, Jessica and {Garc{\'\i}a}, Rafael A. and {Garc{\'\i}a-Hern{\'a}ndez}, D. An{\'\i}bal and {Gillespie}, Bruce A. and {Girardi}, L{\'e}o and {Gunn}, James E. and {Hasselquist}, Sten and {Hayden}, Michael R. and {Hekker}, Saskia and {Ivans}, Inese and {Kinemuchi}, Karen and {Klaene}, Mark and {Mahadevan}, Suvrath and {Mathur}, Savita and {Mosser}, Beno{\^\i}t and {Muna}, Demitri and {Munn}, Jeffrey A. and {Nichol}, Robert C. and {O'Connell}, Robert W. and {Parejko}, John K. and {Robin}, A.~C. and {Rocha-Pinto}, Helio and {Schultheis}, Matthias and {Serenelli}, Aldo M. and {Shane}, Neville and {Silva Aguirre}, Victor and {Sobeck}, Jennifer S. and {Thompson}, Benjamin and {Troup}, Nicholas W. and {Weinberg}, David H. and {Zamora}, Olga},
        title = "{The Apache Point Observatory Galactic Evolution Experiment (APOGEE)}",
      journal = {\aj},
         year = 2017,
        month = sep,
       volume = {154},
       number = {3},
          eid = {94},
        pages = {94},
          doi = {10.3847/1538-3881/aa784d},
archivePrefix = {arXiv},
       eprint = {1509.05420},
 primaryClass = {astro-ph.IM},
       adsurl = {https://ui.adsabs.harvard.edu/abs/2017AJ....154...94M}
}

@ARTICLE{Yanny_2009,
       author = {{Yanny}, Brian and {Rockosi}, Constance and {Newberg}, Heidi Jo and {Knapp}, Gillian R. and {Adelman-McCarthy}, Jennifer K. and {Alcorn}, Bonnie and {Allam}, Sahar and {Allende Prieto}, Carlos and {An}, Deokkeun and {Anderson}, Kurt S.~J. and {Anderson}, Scott and {Bailer-Jones}, Coryn A.~L. and {Bastian}, Steve and {Beers}, Timothy C. and {Bell}, Eric and {Belokurov}, Vasily and {Bizyaev}, Dmitry and {Blythe}, Norm and {Bochanski}, John J. and {Boroski}, William N. and {Brinchmann}, Jarle and {Brinkmann}, J. and {Brewington}, Howard and {Carey}, Larry and {Cudworth}, Kyle M. and {Evans}, Michael and {Evans}, N.~W. and {Gates}, Evalyn and {G{\"a}nsicke}, B.~T. and {Gillespie}, Bruce and {Gilmore}, Gerald and {Nebot Gomez-Moran}, Ada and {Grebel}, Eva K. and {Greenwell}, Jim and {Gunn}, James E. and {Jordan}, Cathy and {Jordan}, Wendell and {Harding}, Paul and {Harris}, Hugh and {Hendry}, John S. and {Holder}, Diana and {Ivans}, Inese I. and {Ivezi{\v{c}}}, {\v{Z}}eljko and {Jester}, Sebastian and {Johnson}, Jennifer A. and {Kent}, Stephen M. and {Kleinman}, Scot and {Kniazev}, Alexei and {Krzesinski}, Jurek and {Kron}, Richard and {Kuropatkin}, Nikolay and {Lebedeva}, Svetlana and {Lee}, Young Sun and {French Leger}, R. and {L{\'e}pine}, S{\'e}bastien and {Levine}, Steve and {Lin}, Huan and {Long}, Daniel C. and {Loomis}, Craig and {Lupton}, Robert and {Malanushenko}, Olena and {Malanushenko}, Viktor and {Margon}, Bruce and {Martinez-Delgado}, David and {McGehee}, Peregrine and {Monet}, Dave and {Morrison}, Heather L. and {Munn}, Jeffrey A. and {Neilsen}, Jr., Eric H. and {Nitta}, Atsuko and {Norris}, John E. and {Oravetz}, Dan and {Owen}, Russell and {Padmanabhan}, Nikhil and {Pan}, Kaike and {Peterson}, R.~S. and {Pier}, Jeffrey R. and {Platson}, Jared and {Re Fiorentin}, Paola and {Richards}, Gordon T. and {Rix}, Hans-Walter and {Schlegel}, David J. and {Schneider}, Donald P. and {Schreiber}, Matthias R. and {Schwope}, Axel and {Sibley}, Valena and {Simmons}, Audrey and {Snedden}, Stephanie A. and {Allyn Smith}, J. and {Stark}, Larry and {Stauffer}, Fritz and {Steinmetz}, M. and {Stoughton}, C. and {SubbaRao}, Mark and {Szalay}, Alex and {Szkody}, Paula and {Thakar}, Aniruddha R. and {Sivarani}, Thirupathi and {Tucker}, Douglas and {Uomoto}, Alan and {Vanden Berk}, Dan and {Vidrih}, Simon and {Wadadekar}, Yogesh and {Watters}, Shannon and {Wilhelm}, Ron and {Wyse}, Rosemary F.~G. and {Yarger}, Jean and {Zucker}, Dan},
        title = "{SEGUE: A Spectroscopic Survey of 240,000 Stars with g = 14-20}",
      journal = {\aj},
         year = 2009,
        month = may,
       volume = {137},
       number = {5},
        pages = {4377-4399},
          doi = {10.1088/0004-6256/137/5/4377},
archivePrefix = {arXiv},
       eprint = {0902.1781},
 primaryClass = {astro-ph.GA},
       adsurl = {https://ui.adsabs.harvard.edu/abs/2009AJ....137.4377Y}
}

@ARTICLE{Yang_2025,
       author = {{Yang}, Yong and {Lewis}, Geraint F. and {Erkal}, Denis and {Li}, Ting S. and {Li}, Andrew P. and {Martell}, Sarah L. and {Cullinane}, Lara R. and {Limberg}, Guilherme and {Zucker}, Daniel B. and {Bland-Hawthorn}, Joss and {Pace}, Andrew B. and {Da Costa}, Gary S. and {Ji}, Alexander P. and {Koposov}, Sergey E. and {Kuehn}, Kyler and {Shipp}, Nora and {Pearson}, Miles and {Usman}, Sam A. and {S}},
        title = "{Flipping of the Tidal Tails of the Ophiuchus Stream due to the Decelerating Galactic Bar}",
      journal = {\apj},
         year = 2025,
        month = may,
       volume = {984},
       number = {2},
          eid = {189},
        pages = {189},
          doi = {10.3847/1538-4357/adc57c},
archivePrefix = {arXiv},
       eprint = {2503.19221},
 primaryClass = {astro-ph.GA},
       adsurl = {https://ui.adsabs.harvard.edu/abs/2025ApJ...984..189Y}
}

@ARTICLE{Massari_2019,
       author = {{Massari}, D. and {Koppelman}, H.~H. and {Helmi}, A.},
        title = "{Origin of the system of globular clusters in the Milky Way}",
      journal = {\aap},
         year = 2019,
        month = oct,
       volume = {630},
          eid = {L4},
        pages = {L4},
          doi = {10.1051/0004-6361/201936135},
archivePrefix = {arXiv},
       eprint = {1906.08271},
 primaryClass = {astro-ph.GA},
       adsurl = {https://ui.adsabs.harvard.edu/abs/2019A\&A...630L...4M}
}

@ARTICLE{Ding_2019,
       author = {{Ding}, Ping-Jie and {Zhu}, Zi and {Liu}, Jia-Cheng},
        title = "{Local standard of rest based on Gaia DR2 catalog}",
      journal = {Research in Astronomy and Astrophysics},
         year = 2019,
        month = may,
       volume = {19},
       number = {5},
          eid = {068},
        pages = {068},
          doi = {10.1088/1674-4527/19/5/68},
       adsurl = {https://ui.adsabs.harvard.edu/abs/2019RAA....19...68D}
}

@ARTICLE{BlandHawthorn_2016,
       author = {{Bland-Hawthorn}, Joss and {Gerhard}, Ortwin},
        title = "{The Galaxy in Context: Structural, Kinematic, and Integrated Properties}",
      journal = {\araa},
         year = 2016,
        month = sep,
       volume = {54},
        pages = {529-596},
          doi = {10.1146/annurev-astro-081915-023441},
archivePrefix = {arXiv},
       eprint = {1602.07702},
 primaryClass = {astro-ph.GA},
       adsurl = {https://ui.adsabs.harvard.edu/abs/2016ARA\&A..54..529B}
}

@software{Gillies_Shapely_2022,
author = {Gillies, Sean and van der Wel, Casper and Van den Bossche, Joris and Taves, Mike W. and Arnott, Joshua and Ward, Brendan C. and {others}},
doi = {10.5281/zenodo.5597138},
license = {BSD-3-Clause},
month = dec,
title = {{Shapely}},
url = {https://github.com/shapely/shapely},
version = {2.0.0},
year = {2022}
}
\bibliographystyle{aasjournal}

\end{document}